\documentclass[12pt]{iopart}
\usepackage{iopams,epsf}

\newfont{\mbm}{msbm10 scaled \magstep 1}
\newfont{\smbm}{msbm10}
\newfont{\ssmbm}{msbm10 at 6pt}
\def\bb#1{\hbox{\mbm #1}}
\def\sbb#1{\hbox{\smbm #1}}
\def\ssbb#1{\hbox{\ssmbm #1}}
\newcommand{\be}{\begin{equation}}
\newcommand{\ee}{\end{equation}}
\newcommand{\ba}{\begin{eqnarray}}
\newcommand{\ea}{\end{eqnarray}}
\def\newsec#1{\expandafter{\section{#1}}\markboth{{#1}}{{#1}}}
\newcommand{\secRef}{\expandafter{\section*{References}}\markboth{References}{References}}

\begin{document}

\rightline{CERN-TH/2002-025}
\rightline{ROM2F-2002/08}
\rightline{LPTENS 02/14}
\rightline{CPHT RR 020.0202}
\rightline{hep-th/0204089}

\title[Open Strings]{Open  Strings}

\author{C Angelantonj\dag\ and A Sagnotti\ddag}

\address{\dag\ Theory Division -- CERN, 1211 Geneva 23, Switzerland}

\address{\ddag\ Dipartimento di Fisica, Universit\`a di Roma ``Tor
Vergata'', INFN - Sezione di Roma ``Tor Vergata'', Via della Ricerca
Scientifica 1, I-00133 Roma}

\begin{abstract}
 This review is devoted to open strings, and in particular to the
often surprising features of their spectra. It follows and summarizes
developments that took place mainly at the University of Rome ``Tor
Vergata'' over the last decade, and centred on world-sheet
aspects of the constructions now commonly referred to as 
``orientifolds''. Our presentation aims to bridge
the gap between the world-sheet analysis, that first exhibited
many of the novel features of these systems, and their
geometric description in terms of extended objects, 
D-branes and O-planes, contributed by many 
other colleagues, and most notably by J. Polchinski. We
therefore proceed through a number of prototype examples, starting
from the bosonic string and moving on to ten-dimensional
fermionic strings and their toroidal and orbifold compactifications,
in an attempt to guide the reader in a self-contained journey to the
more recent developments related to the breaking of supersymmetry.
\end{abstract}
\vskip 3.5cm
\begin{center}
{\it Dedicated to John H. Schwarz \\
on the occasion of his sixtieth birthday}
\end{center}
\vskip 3cm

\newpage
\tableofcontents
\newpage

\newsec{Introduction and summary}

The celebrated Veneziano formula \cite{veneziano} for open-string
tachyons, that marked the birth of String Theory in the form
of ``dual models'' for hadron resonances, was shortly followed
by the Shapiro-Virasoro formula for closed-string
tachyons \cite{sv} and by their multi-particle generalizations \cite{kn}, 
as well as by the Neveu-Schwarz-Ramond
fermionic string \cite{nsr}.  The early work of the 
following decade provided the foundations for the subject \cite{jacob}, 
whose very scope took
a sharp turn toward its current interpretation as a theory of
the fundamental interactions only at the end of the seventies, some time
after Scherk 
and Schwarz and Yoneya \cite{ssy} elucidated the close link between the 
low-energy behaviour of string amplitudes on the one hand, and 
higher-dimensional gauge theories and gravity on the other. 
The work of Green and Schwarz, that 
finally resulted in their celebrated anomaly cancellation mechanism
\cite{gs}, opened the way to the string construction of four-dimensional
chiral spectra free of the usual ultraviolet divergences of point-particle
gravity \cite{divgrav}. This originally rested on 
Calabi-Yau compactifications \cite{chsw,gsw} of the low-energy 
supergravity \cite{sugra} 
of heterotic strings \cite{heterotic}, that for many years have been at 
the heart of string phenomenology. Most of the efforts
were then related to the ${\rm E}_8 \times {\rm E}_8$ heterotic 
model, naturally connected to four-dimensional low-energy physics, and
it was indeed the prominence of exceptional gauge groups \cite{tm}, 
generated by charges spread over closed strings \cite{affine}, together 
with the impossibility of realizing them in open strings \cite{cp,cp2,cp3}, 
that stimulated this intense activity \cite{hetphen}.
The seminal work of Gliozzi, Scherk 
and Olive (GSO) \cite{gso} somehow laid the ground for a
string description of these phenomena,
since it first showed how a na{\"\i}ve string spectrum could be
naturally projected to a supersymmetric one. 
Why these projections {\it should} generally
be present, however, became
clear only in the late eighties, when they were given a
{\it raison d'\^etre} in the geometric constraint of modular
invariance \cite{cardy1} of the underlying conformal field theory
\cite{bpz,fms,cftrev}, a property that the bosonic string had manifested
long before \cite{shapmod}.
By then, one had attained a precise dictionary relating
world-sheet constructions to their space-time counterparts, albeit
limitedly to the case of oriented closed strings, and some of these could
be related, via suitable compactifications, to chiral four-dimensional
matter coupled to ${\cal N}=1$ supergravity \cite{abk,4d,bert}. 
A basic entry in
this respect was provided by the idea of orbifolds \cite{dhvw}, that not only
allowed to extend string constructions beyond the toroidal
case \cite{narain} but, more importantly, 
endowed a wide class of
GSO projections with a geometric interpretation, linking them to
singular limits of Calabi-Yau
reductions. Discrete symmetries play a pivotal r\^ole in this
context, while the orbifold structure permeates the whole of
Conformal Field Theory \cite{cftrev}.

The work of the ``Tor Vergata'' group summarized in this review began
in the second half of the eighties.  
The realization of consistent GSO projections for open strings then emerged
as a major open problem, since
standard ideas based on modular invariance failed to apply directly
to world sheets with boundaries. The main insights were provided 
by the absence of short-distance singularities in 
the SO(32) superstring \cite{gsop}, ultimately responsible 
for its anomaly cancellation, and by 
a similar behaviour of the SO(8192) bosonic string. This
had been exhibited by three rather distinct methods: direct calculation of
one-point functions \cite{douglas1}, factorization of tachyon
amplitudes \cite{weinberg} and singular limits of vacuum amplitudes 
\cite{marcus86,bs88}. The difference between the two
types of phenomena was elucidated in \cite{pc}, where the absence
of space-time
anomalies in the SO(32) superstring was related to the behaviour of 
its R-R sector. In both
cases, however, one knew neither how to break the gauge group, nor how
to attain any non-trivial compactification. Orbifolds
provided again the proper setting, once extended to discrete
symmetries mixing left and right modes \cite{cargese}, and this 
generalization, now commonly termed an ``orientifold'', linked
the closed and open bosonic strings in twenty-six dimensions and
the type I and type IIB superstrings in ten dimensions.

A few other groups \cite{govaerts,horava1} soon elaborated on the 
proposal of \cite{cargese}, while others were considering
similar issues from an apparently different viewpoint. Their work
marked the birth of D-branes, that emerged from the behaviour
under T-duality of
open-string toroidal backgrounds \cite{dlp,horava2,green}. These also
made an early appearance in \cite{ps}, in an analysis of
$\bb{Z}_2$ orbifolds stimulated by the low-energy considerations in
\cite{hm}, but the emphasis fell solely on their spectrum that, 
however, clearly revealed the r\^ole of Neumann-Neumann, 
Neumann-Dirichlet and Dirichlet-Dirichlet strings and their mutual
consistency. 
Once more, the basic ingredients were long known \cite{venturi}, while the 
novelties were the rules enforcing the proper GSO projections. 
Our later efforts \cite{rmg} were stimulated by the fermionic constructions of
four-dimensional superstrings \cite{abk,4d} and by related properties of
lattices \cite{bert}, in an attempt to constrain the GSO projections
from the residual higher-loop modular invariance, but soon the seminal
paper of Cardy on Boundary Conformal Field Theory \cite{cardy2} 
allowed a precise algebraic construction of boundaries respecting
a given symmetry \cite{bs,bs2}. This promptly resulted in
new classes of ten-dimensional orientifolds, the 0A and 0B descendants, 
with rich patterns of gauge symmetry, and in new surprising
six-dimensional models with (1,0) supersymmetry that, in sharp
contrast with heterotic ones, contain variable numbers
of (anti)self-dual two-forms. Their presence was an early success for
the proposal of \cite{cargese}, as we soon realized \cite{as92},
since the two-forms, remnants of the 21 type IIB ones
of the $T^4/\bb{Z}_2$ orbifold, play a crucial 
r\^ole in a generalized version of the Green-Schwarz anomaly 
cancellation mechanism. The six-dimensional supergravity models
associated to these generalized Green-Schwarz terms are also of interest
in their own right, since they display
singularities in the gauge couplings, first
noticed in \cite{as92}, that can be associated to a novel type of phase 
transition whereby a soliton of the model, a string, becomes 
tensionless \cite{tensionless}. 
Six-dimensional string models obtained
from compactifications on group lattices also exhibited peculiar
rank reductions of the Chan-Paton gauge group, that could be linked
to quantized values of the NS-NS two-form $B_{ab}$ \cite{bps,kab,cab}. 
Our subsequent efforts were aimed at a better understanding of the
underlying boundary (and crosscap) Conformal Field Theory, 
first in diagonal minimal
models \cite{bps2}, where the Cardy prescription was extended to
the Klein bottle and M\"obius amplitudes, 
and then in WZW models, where new 
structures emerged and, perhaps more importantly, where we learned how 
to modify Klein-bottle projections \cite{fps,pss,pss2}. This soon
resulted in an interesting
application: a ten-dimensional 0B orientifold completely free of tachyons, 
now commonly termed $0'{\rm B}$ string \cite{susy95,c0b,bfl1,berlin0b}. 

Polchinski's paper on the R-R charge of D-branes and 
O-planes \cite{pol95} gave rise to an
upsurge of interest in these constructions, as well as in the
r\^ole of open strings in non-perturbative aspects of closed-string
physics, since it tied a number of world-sheet results to a pervasive
space-time picture involving solitonic extended objects, with a key r\^ole
in the web of string dualities \cite{mtheory} \footnote{ 
The non-derivative couplings present 
in the $(-{1\over 2},-{3\over 2})$ asymmetric ghost picture,
originally noticed in \cite{bps}, are the world-sheet manifestation of the
R-R charge of D-branes and O-planes.}. Many
started working actively on D-branes and orientifolds, and
new developments followed. 
Our work summarized here has led to the first instance of a 
four-dimensional model 
with three generations of chiral matter \cite{abpss}, the starting 
point for a number of subsequent constructions
\cite{z3}, to a better understanding of the peculiar
current algebra associated to the generalized Green-Schwarz mechanism 
\cite{fabio} and, more recently, to novel realizations of 
supersymmetry breaking by Scherk-Schwarz deformations \cite{ss} in string
vacua allowed by the presence of open strings \cite{bd,ads1,adds,bg} 
and by the simultaneous presence of branes and antibranes 
\cite{2benot2b,au,aadds,abg}.
This work extended the original closed-string constructions of
\cite{kounnas} to the case of open strings, exhibiting the new
phenomenon of ``brane supersymmetry breaking'' 
\cite{bsb,cab,au,bmp,aadds,abg}, met independently
in the USp(32) ten-dimensional type I model in \cite{sugimoto}.
More recently, stimulated by the proposal of \cite{wito32,bachasmag} on
magnetic supersymmetry breaking, 
we have also  studied instanton-like \cite{witsmall} magnetic deformations 
yielding new supersymmetric vacua with gauge groups of reduced rank and
multiple matter sectors \cite{magnetic,magnetic2}. 
These constructions may be regarded as a
realization in type I vacua of proposals related to systems of
branes at angles \cite{douglas2}, a viewpoint widely 
pursued by other groups in attempts to construct brane realizations of 
the Standard Model \cite{modbuild}. 

In writing this review, we have made a selection of the topics that we have
touched upon over the years, in an
attempt to guide the reader, hopefully in a self-contained and pedagogical
fashion, through a number of examples, drawn mostly from toroidal 
and orbifold models, that are meant to illustrate the wide 
variety of phenomena brought about by these generalized GSO 
projections in their simplest occurrences. As a result, our discussion is
centred on the key features of the open-string partition functions and of
the underlying Boundary Conformal Field Theory, at the expense of other
interesting topics, to wit the low-energy effective field
theory and the applications to model building, that are left out.
We thus begin with the bosonic string and its orientifolds,
and proceed to ten-dimensional fermionic strings and their toroidal
and orbifold compactifications, with a slight diversion at the end
to display some general properties of the D-branes allowed in
the ten-dimensional string models. The concluding section highlights
some general aspects of (rational) Boundary Conformal Field Theory, showing 
in particular how orientifolds can also prove useful tools to
extract D-brane spectra and how one can formulate ``completeness'' 
conditions \cite{pss2} for boundaries or, equivalently, for brane types. 
We shall emphasize throughout how
the partition functions of abstract Conformal Field Theories, even beyond
their applications to String Theory, if properly formulated,
encode clearly all relevant phenomena. The review of Dudas \cite{emrev} 
on phenomenological aspects of type I vacua and the more recent
review of Stanev \cite{yasrev} on Boundary Conformal Field Theory have some 
overlap, both in spirit and in contents, with the present one, while a
number of previous short reviews have also touched upon some
of these issues \cite{extratov}.

It is a pleasure to dedicate this review 
article to John H. Schwarz on the occasion
of his sixtieth birthday. His work pervades the whole of String Theory,
and in particular the developments summarized
here, while his example inspired, directly or indirectly, both us and
our friends and collaborators educated at the University of Rome 
``Tor Vergata''. 

\newsec{The bosonic string}

In this section we describe some generic features of open-string
constructions, using the bosonic string as an example. In particular,
we review the basic structure of the Polyakov expansion and some  general
properties of Chan-Paton groups, including their relation to fermionic
modes living at the ends of open strings. We shall confine our
attention to the light-cone quantization method, sufficient to
describe string spectra in most circumstances. Here we shall meet the four
vacuum amplitudes with vanishing Euler character that determine the
spectrum of these models: torus, Klein bottle, annulus and
M\"obius strip. Finally,  in this simple setting we shall also
make our first encounter with a tadpole condition, that determines a
special choice for the open-string gauge group.

\vskip 12pt
\subsection{The Polyakov expansion}

Models of oriented closed strings have the simple and remarkable
feature of receiving
one contribution at each order of perturbation theory \cite{pol}. These
correspond to closed orientable Riemann surfaces with increasing
numbers of handles $h$ \cite{fk}, and their perturbative series is weighted 
by $g_s^{-\chi}$, where the Euler character
$\chi$ is
\be
\chi = 2 - 2 h \, ,
\ee 
and where the string coupling $g_s$ is determined by the vacuum
expectation value of a ubiquitous massless scalar mode of 
closed strings, the {\it dilaton} $\varphi$, according to
\be 
g_s = e^{\langle \varphi \rangle} \,. \label{dilatoncouplingconst}
\ee

\begin{figure}
\begin{center}
\epsfbox{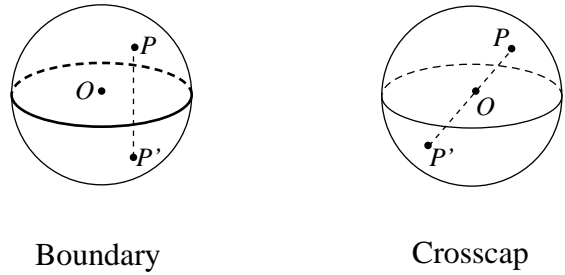}
\end{center}
\caption{Boundary and crosscap.}
\label{fig1}
\end{figure}

The models of interest in this review are actually more complicated.
Their closed strings are unoriented, while their spectra usually
include additional sectors with unoriented open strings. As a result,
their Polyakov expansions involve additional Riemann surfaces, that
contain variable numbers of two new structures: holes surrounded by
{\it boundaries}, $b$, and {\it crosscaps}, $c$ \cite{oalv}. 
The Euler character for a surface with $h$ handles, $b$ holes and 
$c$ crosscaps is
\be
\chi = 2 - 2 h - b - c \, ,
\ee 
and therefore the perturbation series now includes both even and
odd powers of $g_s$. Boundaries are easily pictured, and their
simplest occurrence is found in a surface of Euler character $\chi=1$,
the {\it disk}. This is doubly covered by a sphere, from which it may be
retrieved identifying pair-wise points of opposite
latitude, as in figure \ref{fig1}. The upper hemisphere then corresponds
to the interior of the disk, while the
equator, a line of fixed points in this construction, defines its {\it
boundary}. On the other hand, crosscaps are certainly less
familiar. Still, their simplest occurrence is found in another surface
of Euler character
$\chi=1$, the {\it real projective plane}, obtained from a
sphere identifying antipodal points, as in figure \ref{fig1}. 
One can again take
as a fundamental region the upper hemisphere, but now pairs of
points oppositely located on the equator are identified. In loose
terms, we shall call such a line, responsible for the lack of
orientability of this surface, a crosscap. As can be seen from figure
\ref{fig1}, the end result is a closed non-orientable surface, where
the transport of a pair of axes can reverse their relative orientation.

\begin{figure}
\begin{center}
\epsfbox{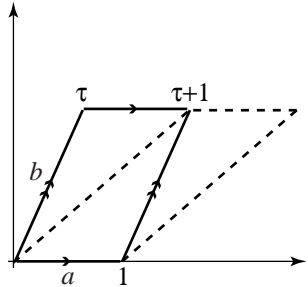}
\end{center}
\caption{The torus as a periodic lattice.}
\label{fig2}
\end{figure}

\begin{figure}
\begin{center}
\epsfbox{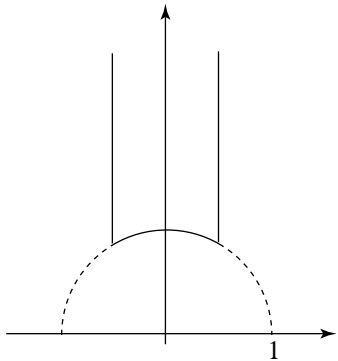}
\end{center}
\caption{Fundamental domain for the torus.}
\label{fig3}
\end{figure}

In general, all these surfaces may be dissected and opened on the
plane by a suitable number of cuts, and for surfaces of vanishing Euler
character the plane can be equipped with a Euclidean metric. 
Thus, for instance, two cuts turn
a torus into the parallelogram of figure \ref{fig2}, whose opposite
sides are to be identified as indicated by the arrows. By a suitable
rescaling, one of the sides may be chosen horizontal and of
length one, and thus a single complex number, $\tau=
\tau_1 + i \tau_2$, with positive imaginary part $\tau_2$, usually
called the Teichm\"uller parameter, or modulus for brevity, 
defines the shape or, more
precisely, the complex structure of this surface. 
There is actually a
subtlety, since not all values of $\tau$ in the upper-half
complex plane correspond to inequivalent tori. Rather, all values
related by the
${\rm PSL}(2,\bb{Z})={\rm SL}(2,\bb{Z})/\bb{Z}_2$ modular group, that 
acts on $\tau$ according to
\be
\tau \to \frac{a \tau + b}{c \tau + d} \qquad {\rm with} \qquad a d -
b c = 1 \, , \quad a , b , c , d  \in \bb{Z} \, ,
\ee 
are to be regarded as equivalent.
This group is generated by the two transformations
\be 
T : \tau \to \tau + 1 \, , \qquad S : \tau \to -
\frac{1}{\tau} \, ,
\ee 
that in SL(2,$\bb{Z}$) satisfy the relation
\be 
S^2 = ( S T )^3 \,. \label{relmod}
\ee 
Notice that $T$ redefines the oblique side of the fundamental cell, 
while $S$ interchanges horizontal and oblique sides.
As a result, the independent values
of $\tau$ lie within a fundamental region of the modular group, for
instance within
\be 
{\cal F}=
\{ -{\textstyle{1\over 2}} < \tau_1 \le {\textstyle{1\over 2}} , 
|\tau| \ge 1\}
\label{fundregion}
\ee  
of figure \ref{fig3}. This property and its generalizations
to other surfaces play a crucial r\^ole in the construction of string
models.  

\begin{figure}
\begin{center}
\epsfbox{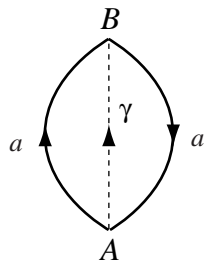}
\end{center}
\caption{A non-contractible loop $\gamma$ and the crosscap.}
\label{fig4}
\end{figure}

In a similar spirit, one can unfold the projective plane into the 
region of figure
\ref{fig4}, where the two sides are again to be identified according
to the arrows, and the additional dashed line
$\gamma$ suffices to reveal a peculiar property of this surface. To
this end, let us imagine to move along 
$\gamma$ from a point $A$ to its opposite image $B$, a closed
path that is clearly not contractible. However, moving $\gamma$ across one
of the two vertical sides of the polygon has the net effect of
reversing its orientation and, as a result, while 
$\gamma$ is non contractible, $\gamma^2$ is, as can be seen
reversing the orientation of one of the two copies.  This illustrates a
familiar result: the fundamental group of the real
projective plane is $\bb{Z}_2$ \cite{stilwell}.

It is simple to extract the fundamental group of a surface
from the corresponding polygon \cite{stilwell}, associating a
generator, or its inverse, to each independent side, according to the
clockwise or counter-clockwise orientation of the corresponding arrows. These
generators are not independent, however, since the interior of the
polygon is clearly contractible, and as a result one has a relation.
For instance, for the torus of figure \ref{fig2} one finds 
$b^{-1} a^{-1} b a = 1$, and the resulting fundamental group is thus
Abelian, since its two generators $a$ and $b$ commute. In a similar
fashion, for the projective disk this leads to the 
condition $a^2 = 1$, so that, as previously stated,  in this case there
is a non-trivial $\bb{Z}_2$ generator.

There are four surfaces with vanishing Euler character. 
Leaving aside the torus, that we have already
discussed, $\chi=0$ can indeed be obtained for three other choices: the {\it
Klein bottle} ($h=0$, $b=0$,
$c=2$), the {\it annulus} ($h=0$, $b=2$, $c=0$) and the {\it M\"obius
strip} ($h=0$, $b=1$, $c=1$). 

\begin{figure}
\begin{center}
\epsfbox{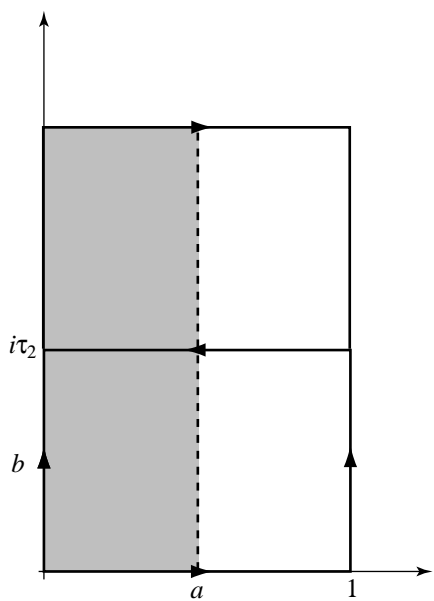}
\end{center}
\caption{Fundamental polygons for the Klein bottle.}
\label{fig5}
\end{figure}

Like the projective disk, the Klein bottle  has the
curious feature of not allowing an embedding in three-dimensional
Euclidean space that is free of self-intersections. Two choices for
the corresponding polygon, together with one for the
doubly-covering torus, are shown
in figure \ref{fig5}. The first polygon, of sides 1 and $i \tau_2$, 
presents two main
differences with respect to the torus of figure \ref{fig2}: the
horizontal sides have opposite orientations, while $\tau$ is now
purely imaginary. The Klein bottle can be obtained from its covering torus,
of Teichm\"uller parameter $2 i \tau_2$, if the lattice translations 
are supplemented by the anticonformal involution 
\be 
z \to 1 - \bar{z} + i { \tau_2} \,, \label{kleininv}
\ee 
where the ``vertical'' time
$\tau_2$
is the ``proper world-sheet time'' elapsed while a {\it closed}
string sweeps it. The second choice of polygon, also
quite interesting, defines an inequivalent ``horizontal'' time.
It is obtained halving the horizontal side while doubling the vertical
one, and thus leaving the area unaltered. The end result has the virtue of
displaying an  equivalent representation of this surface as a tube
terminating at two crosscaps, and the horizontal side is now the ``proper
time'' elapsed while a {\it closed} string propagates between the two
crosscaps. The tube is the interior of the region, whose
horizontal sides have now the same orientation, while the
crosscaps are the two vertical sides, where points differing by
translations by half of their lengths are pair-wise identified. 
It should also be appreciated that, in moving from the
first fundamental polygon to the double cover, the identifications
are governed by eq. (\ref{kleininv}), that has no fixed points
and {\it squares} to the vertical translation
$z \to z + 2 i \tau_2$. Finally, the corresponding relation for the
generators of the fundamental group,
\be 
b^{-1} a^{-1} b a^{-1} = 1 \, ,
\ee 
implies that $a$ and $a^{-1}$ belong to the same conjugacy class,
a result that will have a direct bearing on the ensuing discussion.

\begin{figure}
\begin{center}
\epsfbox{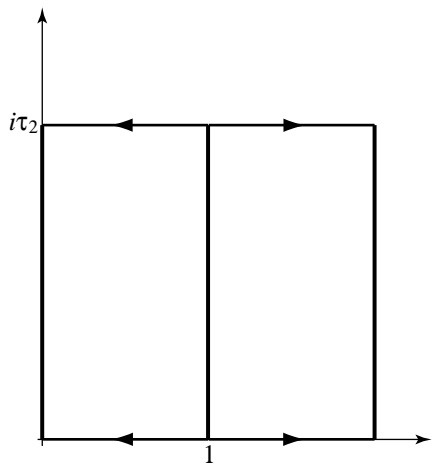}
\end{center}
\caption{Fundamental polygon for the annulus.}
\label{fig6}
\end{figure}

The {\it annulus} is certainly more familiar. Its fundamental polygon 
is displayed in figure \ref{fig6}, together with a polygon for its
doubly-covering torus, obtained by horizontal doubling. In the
original polygon, with vertices at 1 and $i \tau_2$,            
the horizontal sides are identified, while the
vertical ones correspond to the two boundaries. These are
fixed-point sets of the involutions
\be 
z \to - \bar{z} \qquad {\rm and} \qquad z \to 2 - \bar{z} \label{annulusinv}
\ee 
that recover the annulus from the doubly-covering torus. Once
more, $\tau$ is purely imaginary, and $\tau_2$ is now the ``proper time'' 
elapsed while an {\it open} string
sweeps the annulus. One has again a distinct ``horizontal'' choice,
that defines the ``proper time'' elapsed while a {\it
closed} string propagates between the two boundaries.  

\begin{figure}
\begin{center}
\epsfbox{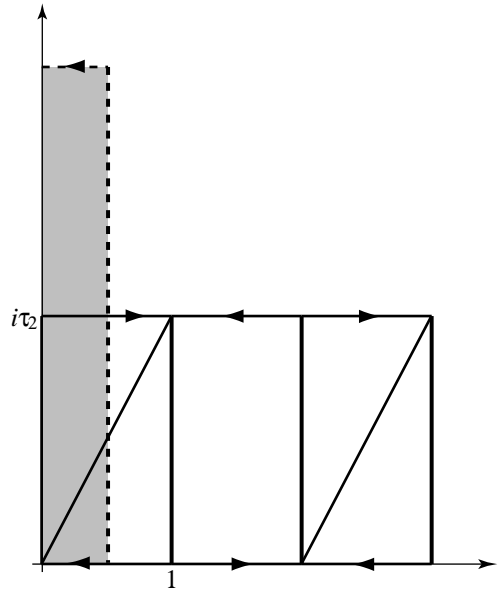}
\end{center}
\caption{Fundamental polygons for the M\"obius strip.}
\label{fig7}
\end{figure}

Finally, the {\it M\"obius strip} corresponds to the polygon in figure
\ref{fig7}, again with vertices at 1 and $i \tau_2$, but 
whose horizontal sides 
have opposite orientations. It
should be appreciated that now the vertical sides describe two
different portions of a single boundary. The parameter $\tau_2$
describes the ``proper time'' elapsed while an {\it open} string sweeps
the M\"obius strip, and one has again the option of choosing a
different fundamental polygon, that displays an 
equivalent representation of the surface as a tube terminating at one
hole and one crosscap. This is simply obtained doubling the
vertical side while halving the horizontal one. One of the two
resulting vertical sides is the single boundary of the M\"obius
strip, while the other, where points are pair-wise identified 
after a vertical translation on account of the involution
\be 
z \to 1 - \bar{z} + i \tau_2 \label{mobiusinv1} \, ,
\ee 
is the crosscap, and the corresponding
horizontal time defines the ``proper time'' elapsed while a {\it
closed} string propagates between the boundary and the crosscap. It
should be appreciated that in this case the polygon obtained doubling the
vertical length defines an annulus, not a torus. A 
doubly-covering torus does exist, of course, but has
the curious feature of having a Teichm\"uller parameter that is {\it
not} purely imaginary. This may be seen  combining the anticonformal
involution of eq. (\ref{mobiusinv1}) with eq.  (\ref{annulusinv}),
that identifies the boundary of the M\"obius strip. Referring to figure
\ref{fig7}, horizontal and skew sides are now consistently
identified, but
\be
\tau = {\textstyle\frac{1}{2}} + {\textstyle{1\over 2}}
i \tau_2 \, , \label{taumobius} 
\ee 
after rescaling to one the length of the horizontal side.

\begin{figure}
\begin{center}
\epsfbox{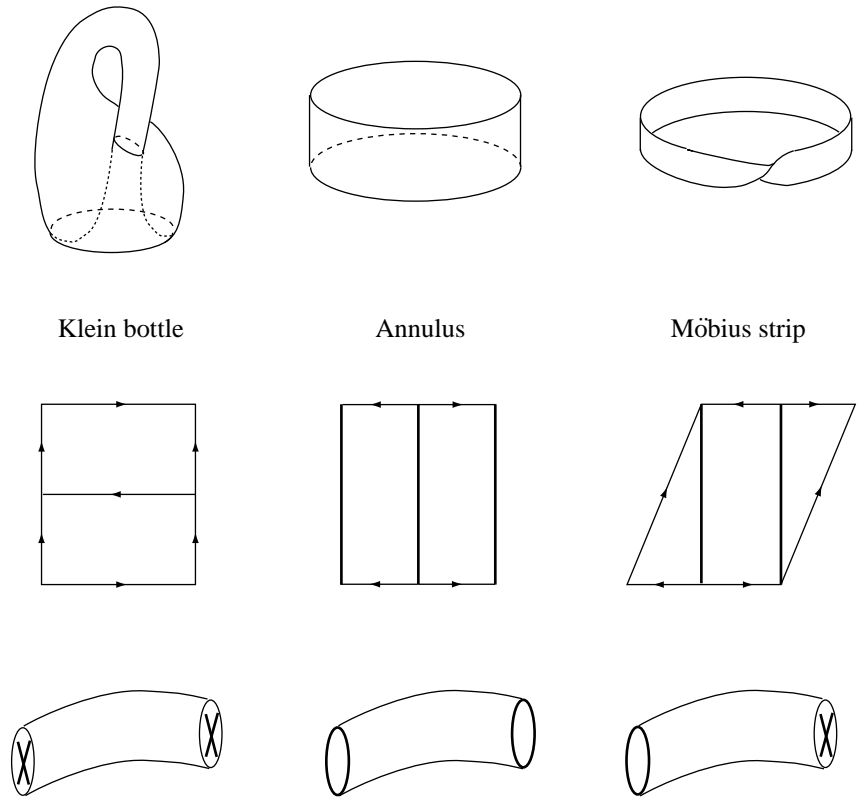}
\end{center}
\caption{Klein bottle, annulus and M\"obius strip.}
\label{figsurfs}
\end{figure}

It is time to summarize these results. Whereas for the torus
one has an infinity of equivalent choices for the ``proper time", that
reflect themselves into the invariance under the modular group
${\rm SL}(2,\bb{Z})$, each of the other three surfaces allows 
two inequivalent canonical choices, na\"{\i}vely related by an
$S$ modular transformation. One of these choices, corresponding to the
``vertical" time, exhibits the propagation of {\it closed} strings in the
Klein bottle and of {\it open} strings in the other two surfaces. 
On the other hand, the ``horizontal" time  
exhibits in all three cases the
propagation of {\it closed} strings between holes and/or crosscaps. There is
actually a technical subtlety, introduced by the doubly-covering torus
of the M\"obius strip, whose Teichm\"uller parameter, given in eq.
(\ref{taumobius}), is not purely imaginary. Since the string integrand
will actually depend on it, one is
effectively implementing the transformation \cite{gpp}
\be 
P : \ \frac{1}{2} + i \frac{\tau_2}{2}  \to \frac{1}{2} + i
\frac{1}{2
\tau_2} \, ,
\ee 
that can be obtained by a sequence of $S$ and $T$ transformations,
as
\be 
P = T S T^2 S  \label{ptransf}
\ee 
and, on account of eq. (\ref{relmod}), satisfies
\be 
P^2 = S^2 = ( S T )^3 \,.
\ee

\begin{figure}
\begin{center}
\epsfxsize=10 truecm
\epsfbox{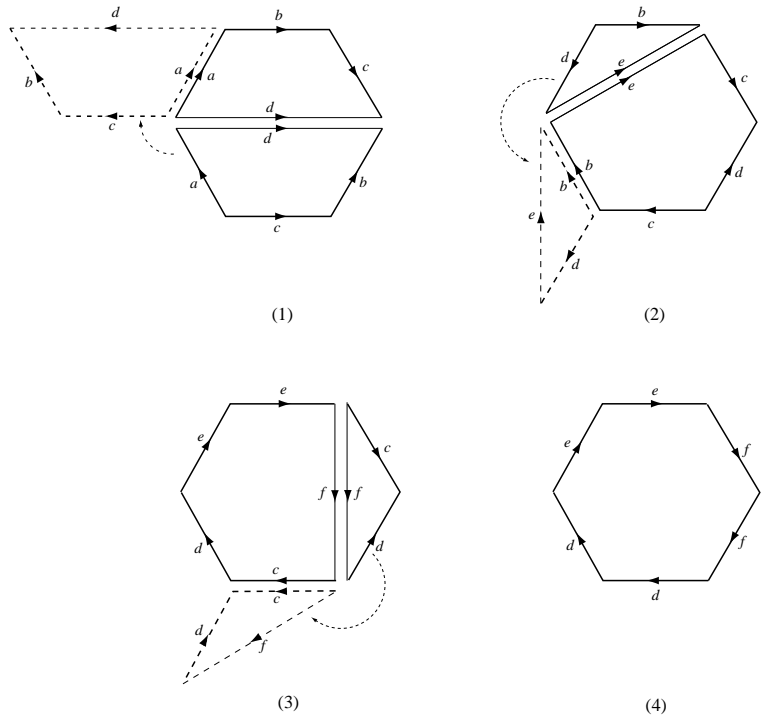}
\end{center}
\caption{An important equivalence:  $3c \equiv h+c$.}
\label{fig8}
\end{figure}

This review, being devoted to the study of string spectra, is
centred on these surfaces of vanishing
Euler character. Still, we would like to conclude
the present discussion showing in some detail an important
topological equivalence between surfaces of higher genera: 
one handle and one crosscap may be replaced
by three crosscaps \cite{stilwell}. This effectively limits the 
Polyakov expansion to
surfaces with arbitrary numbers of handles $h$ and holes
$b$, but with only 0,1 or 2 crosscaps $c$. The simplest setting to
exhibit this equivalence is displayed in figure \ref{fig8}, that
shows a choice of fundamental polygon, a hexagon, for a surface
comprising a crosscap, the sequence of the two $a$ sides, and a handle,
the sequence $b c b^{-1} c^{-1}$. One can now prove the equivalence
performing a series of cuttings and glueings or, equivalently, moving
to different choices for  the fundamental polygon. To this end, let us
begin by introducing a horizontal cut through the centre of the
hexagon, and let $d$ denote the corresponding new pair of sides thus
created. We can then move one of the two resulting trapezia and glue the
two halves $a$ of the crosscap. The new hexagon contains pairs of
sides with clockwise orientations, somewhat reminiscent of the
structure of three crosscaps, albeit still separated from one another.
Two more cuttings and glueings suffice to exhibit three
neighbouring couples. They both remove triangles whose two external
sides have opposite orientations, and then join sides that, in the
hexagon, have like orientations. Thus, referring to the figure, we now
cut out the triangle $b e^{-1} d^{-1}$ in the upper left corner and glue the
two $b$ sides. In the resulting hexagon the two $e$ sides, next to
one another,  define one crosscap. Finally, cutting out the triangle
$c d^{-1} f^{-1}$ and gluing the two resulting $c$ sides fully exhibits the
three crosscaps.

\vskip 12pt
\subsection{Light-cone quantization}

Let us now turn to the quantization of bosonic strings. The starting
point is the action for a set of $D$ world-sheet scalars, identified
with the string coordinates in a $D$-dimensional Minkowski space-time,
coupled to world-sheet gravity \cite{bdhdz}. The corresponding action 
principle is
\footnote[1]{Throughout this paper, space-time metrics have ``mostly
negative" signature, so that in two dimensions
$\eta={\rm diag}(1,-1)$.}
\be 
S = - \frac{1}{4 \pi \alpha^\prime} \int d^2 \xi \sqrt{-g}
g^{\alpha\beta}
\partial_\alpha X^\mu \ \partial_\beta X^\nu \ \eta_{\mu\nu} +
\frac{\langle 
\varphi
\rangle}{4 \pi} 
\int d^2 \xi \sqrt{-g} R \, , \label{bosonicaction}
\ee 
where we have added an Einstein term, that in this case is a topological
invariant, the Euler character of the surface, and a coupling 
$\langle \varphi \rangle$ whose exponential weights the perturbation series.

One can derive rather simply the spectrum of this model, following
\cite{ggrt}. To this end,
one can use the field equations for the background metric, {\it i.e.} the
condition that the energy momentum tensor
\be 
T_{\alpha\beta} = \partial_\alpha X^\mu \partial_\beta X_\mu -
{\textstyle\frac{1}{2}} g_{\alpha\beta} \ \partial^\gamma X^\mu \partial_\gamma
X_\mu  \label{emt}
\ee 
vanish, to express the longitudinal string coordinates in terms of
the transverse ones. The procedure, reminiscent of the usual
light-cone formulation of Electrodynamics, is quite effective since
the string coordinates actually solve
\be
\partial_\alpha \left( \sqrt{-g} g^{\alpha\beta} \partial_\beta \
X^\mu \right) = 0 \, ,
\ee 
that reduces to the standard wave equation
\be
\left( \frac{\partial^2}{\partial \tau^2}  -
\frac{\partial^2}{\partial \sigma^2}
\right) \ X^\mu = 0 \label{wave}
\ee 
if a convenient choice of coordinates
$\xi^\alpha=(\tau,\sigma)$ is used to turn the background metric 
$g_{\alpha\beta}$ to the diagonal form
\be 
g_{\alpha\beta} = \Lambda(\xi) \  \eta_{\alpha\beta} \,.
\ee 
$\Lambda$ then disappears from the classical action
(\ref{bosonicaction}) and, in the critical dimension
$D=26$, that we shall soon recover by a different argument, 
from the functional measure as well \cite{pol}. The longitudinal
string coordinates can be eliminated since, even after this gauge
fixing, the original invariances under Weyl rescalings and
reparametrizations leave behind a residual infinite symmetry that, after a
Euclidean rotation, would correspond to arbitrary analytic and antianalytic
reparametrizations \cite{jacob,bpz,fms,cftrev}. This is the case since
the  string action of 
eq. (\ref{bosonicaction}) effectively 
describes massless free fields, and is thus a simple
instance of a two-dimensional 
conformally invariant model. The infinite dimensional
group of conformal and anticonformal reparametrizations is the basis of 
two-dimensional Conformal Field Theory \cite{bpz,cftrev} that, as we shall 
review briefly in section 6, provides the very rationale for this case, as
well as for more general space-time backgrounds.

Before solving eq. (\ref{wave}) for the simplest case, $D=26$
Minkowski space-time, one need distinguish between two
options. A closed line defines a {\it closed string}, and
in the present, simplest case, calls for the decomposition in periodic
modes \cite{jacob,gsw}
\be 
X^\mu = x^\mu + 2 \alpha^\prime\, p^\mu \ \tau + \frac{i
\sqrt{2\alpha^\prime}}{2}
\ \sum_{n \not= 0} \left( \frac{\alpha^\mu_n}{n} \ e^{-2 i n (\tau -
\sigma)} +
\frac{\tilde{\alpha}^\mu_n}{n}\  e^{-2 i n (\tau + \sigma)} \right)
\, ,
\ee 
consistent with 26-dimensional Lorentz invariance. In a similar fashion, a
segment defines an {\it open string}, and in the present, simplest
case, calls for the Neumann boundary conditions 
$X^\prime =0$ at $\sigma=0,\pi$, and thus for the 
decomposition \cite{jacob,gsw}
\be 
X^\mu = x^\mu + 2 \alpha^\prime\, p^\mu \ \tau + i
\sqrt{2\alpha^\prime}
\ \sum_{n \not= 0} \frac{\alpha^\mu_n}{n} \ e^{- i n \tau} \cos(n
\sigma)  \,. \label{opennmodes}
\ee
 
Using the residual symmetry one can now make a further very
convenient choice for the string coordinates, the light-cone gauge \cite{ggrt}.
Defining $X^{\pm} = (X^0 \pm X^{D-1})/\sqrt{2}$, this corresponds to
eliminating, for both open and closed strings, all oscillations
in the `$+$' direction, so that
\be 
X^+ = x^+ + 2 \alpha^\prime\, p^+ \ \tau \,.
\ee
This condition identifies target-space and world-sheet times, and is the
analogue, in this context, of the condition $A^+=0$ in
Electrodynamics. One can then use the constraints to eliminate $X^-$, 
and indeed eq. (\ref{emt}) results
in the two conditions
\be 
2 \sqrt{2} \, \alpha'\, p^+ \ \partial_{\pm} X^- \ - \ {(\partial_{\pm}
X^i)}^2 = 0 \, ,
\label{constraints}
\ee 
that determine the content of $X^-$ in terms of
the $X^i$'s. The remaining
transverse sums define the transverse Virasoro operators
\be 
L_m = {\textstyle\frac{1}{2}}  \sum_n \ {\alpha}_{m-n}^i {\alpha}_n^i  \,
,
\ee 
and the corresponding $\bar{L}_m$ built out of the
$\tilde{\alpha}$ that, on account of eq. (\ref{constraints}), define
the oscillator modes in the `$-$' direction. The $L_m$ and the
$\bar{L}_m$ are clearly mutually commuting, since they are built out
of independent oscillator modes, and only $L_0$ and
$\bar{L}_0$ need proper normal ordering. Furthermore, both the
$L_m$ and the $\bar{L}_m$ satisfy the
Virasoro algebra with central charge $c=D-2$:
\be 
[ L_m , L_n ] = (m - n) L_{m+n} + \frac{D-2}{12} \ m(m^2 - 1)
\delta_{m+n,0} \,.
\ee
 
The zero modes of eq. (\ref{constraints}) define  the
mass-shell conditions for physical states, and for the closed string one
thus obtains the two conditions
\be 
2 p^+ p^-  = \frac{4}{\alpha'} \left( L_0 - \frac{D-2}{24} \right) =
\frac{4}{\alpha'} \left(\bar{L}_0 - \frac{D-2}{24} \right)
\ee 
or, equivalently,
\be 
2 p^+ p^- = \frac{2}{\alpha'} \left( L_0 + \bar{L}_0 - \frac{D-2}{12}
\right) \, , \label{closedmass2}
\ee 
together with the ``level-matching'' condition $L_0 = \bar{L}_0$
for physical states. The constant term may be
justified from the normal ordering of the Virasoro operators $L_0$
and $\bar{L}_0$,
identifying the corresponding divergent sums over 
zero-point energies with a particular value of the Riemann $\zeta$
function, $\zeta(-1)= -\frac{1}{12}$ \cite{bn}. This result is
a special case of the class of relations
\be
\zeta_\alpha(-1,x) = \sum_{n=1}^\infty \, (n+\alpha) \, e^{-(n+\alpha)x} \ \to \
\zeta_\alpha(-1,0^+) = - \frac{6\alpha(\alpha-1)+1}{12} \label{zriem}
\ee
as $x \to 0^+$, aside from a divergent term, that provide a convenient
way to recover the vacuum shifts
compatible with the Lorentz symmetry, in agreement with the proper 
study of the Lorentz algebra, as in \cite{gsw}. 

In a similar fashion, for the open string, that has only one type of
oscillator modes, one obtains the single mass-shell condition
\be 
2 p^+ p^- = \frac{1}{\alpha'} \left( L_0 - \frac{D-2}{24} \right) \, ,
\label{openmass2}
\ee
where the growth rate, or Regge slope, 
is effectively $\frac{1}{4}$ of the corresponding
one in eq. (\ref{closedmass2}). The 
masses of the string excitations are obtained extracting from $L_0$ and
$\bar{L}_0$ the contributions of transverse momenta, using
\be
L_0 = \frac{\alpha'}{4} \, p^i p^i + N \ , \qquad 
\bar{L}_0 = \frac{\alpha'}{4} \, p^i p^i + \bar N
\ee
for the closed string, and
\be
L_0 = {\alpha'} \, p^i p^i + N \ ,
\ee
for the open string,
with $N$ and $\bar N$ the (normal ordered) number operators that
count the oscillator excitations. Thus, for the closed string
\be 
M^2 = \frac{2}{\alpha'} \left( N + \bar N - \frac{D-2}{12}
\right) \, , \label{closedmass}
\ee 
while for the open string
\be 
M^2 = \frac{1}{\alpha'} \left( N - \frac{D-2}{24} \right) \, ,
\label{openmass}
\ee
but we should warn the reader that, in the
following sections, we shall often be somewhat cavalier in
distinguishing between $L_0$ and $\bar{L}_0$ and the corresponding
number operators. This will make
our expressions very similar to corresponding ones of interest for
Boundary Conformal Field Theory, but hopefully it will cause no confusion,
since it should be clear 
from the outset that momenta along non-compact directions of 
space-time should always be removed from $M^2$ operators.

The particle spectra corresponding to eqs. (\ref{closedmass}) and
(\ref{openmass}) now reveal the r\^ole of the dimensionality of
space-time, since only for $D=26$ are the first excited states 
massless.  For the closed string
$\alpha^i_{-1} \tilde{\alpha}^j_{-1} |0 \tilde 0 \rangle$  describe
the transverse modes of a two-tensor, while for the open string
$\alpha^i_{-1} |0 \rangle$ describe the
transverse modes of a vector. In both cases the longitudinal
components are missing, and thus a Lorentz invariant  spectrum calls
for this ``critical'' dimension. The massless closed spectrum 
then describes a
metric fluctuation $h_{\mu\nu}$, an antisymmetric two-tensor
$B_{\mu\nu}$ and a scalar mode, $\varphi$, usually called the dilaton,
whose vacuum value $\langle \varphi \rangle$, already met in 
(\ref{dilatoncouplingconst}) and (\ref{bosonicaction}), 
weights the perturbative expansion. 
Furthermore, the open and closed spectra contain tachyonic 
modes that, to date, despite much recent progress, are not yet
fully under control \cite{tachyon}.

The open spectrum presents additional subtleties brought about by
the presence of the two ends, that can carry non-dynamical degrees of
freedom, the charges of an internal symmetry group \cite{cp}, to
which we now turn.


\vskip 12pt
\subsection{Chan-Paton groups and ``quarks" at the ends of strings}

One basic feature of open-string amplitudes for identical external
bosons is their cyclic symmetry, and traces of group-valued matrices
$\Lambda^a$ allow a natural generalization that clearly respects this
important property. Indeed, following Chan and Paton \cite{cp}, one can
define ``dressed" $n$-point amplitudes of the type 
\be 
A(1, \dots ,n) \ {\rm tr} ( \Lambda^{a_1} \ldots \Lambda^{a_n} )
\, ,
\label{cpfactors}
\ee  
where $A(1, \ldots , n)$ denotes the ``bare'' amplitude obtained
by standard open-string rules \cite{jacob,schrev,gsw}. This procedure
introduces non-Abelian gauge symmetry in String Theory, but the
modified amplitudes should also be consistent with unitarity, and in particular
all tree amplitudes should factorize at intermediate poles 
consistently with the internal quantum numbers of
the string states. This is certainly possible if the matrices form a 
complete set \cite{cp}, since in this case, at an intermediate pole of mass
$M_I$, where
\be 
A(1,..,m,..,n) \sim A(1,..,m,I)\frac{1}{p_I^2 - M_I^2 + i 
\epsilon} A(I,m+1,..,n) \, ,
\ee 
one can also split the group trace according to
\be 
{\rm tr} ( \Lambda^{a_1}.. \Lambda^{a_m}.. \Lambda^{a_n} ) \sim
\sum_{a_I} {\rm tr} ( \Lambda^{a_1}.. \Lambda^{a_m} \Lambda^{a_I} ) \
{\rm tr} ( \Lambda^{a_I} \Lambda^{a_{m+1}}.. \Lambda^{a_n} ) \,.
\label{compl14}
\ee

\begin{figure}
\begin{center}
\epsfbox{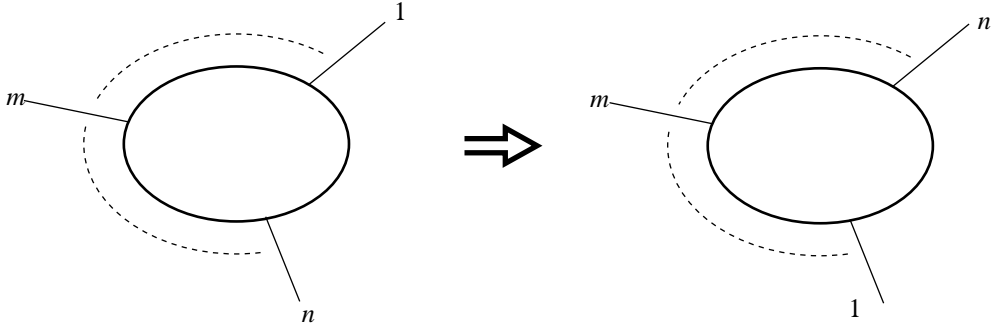}
\end{center}
\caption{Flipping a tree-level amplitude.}
\label{fig9}
\end{figure}

Actually, the amplitudes $A(1,\ldots ,n)$ for the bosonic
string are {\it not} all
independent \cite{cp2}: pairs connected by world-sheet parity are in fact
proportional to one another, and a closer scrutiny reveals that
\be
A(1, \dots ,n) = (-1)^{\sum_i (\alpha' M_i^2 +1)}
A(n, \ldots ,1) \,. \label{wssym}
\ee 
This crucial property may be justified noticing that
$A(n, \dots, 1)$ can be deformed into an amplitude
with {\it flipped} external legs ordered in the sequence $(1,
\ldots, n)$, as in figure \ref{fig9}, while on each
external leg the flip $\Omega$ induces a world-sheet parity reflection, that
results in a corresponding sign $(-1)^{\alpha' M_i^2 +1}$.
This sign can be simply traced to the nontrivial effect of the
world-sheet reflection $\sigma
\to \pi - \sigma$ on the oscillator modes, that must transform as
$\alpha_k \to (-1)^k \alpha_k$, as can be seen from eq.
(\ref{opennmodes}).

This ``flip'' symmetry has strong implications, since the four amplitudes 
\ba
& &A( 1, \dots , m, m+1,\ldots , n) \, , \qquad A( m,\ldots , 1,
m+1,\ldots , n) 
\, , \nonumber \\
& &A( 1,\ldots , m, n,\ldots , m+1) \quad {\rm and} \quad A(
m,\ldots , 1, n,\ldots , m+1)
\ea 
all contribute to the same intermediate pole,  and the condition
(\ref{compl14}) may be correspondingly relaxed. To be definite, let us
consider the factorization of a four-vector amplitude at a
vector pole in the $(1,2)$ channel. 
Eq. (\ref{wssym}) suffices to show that
all three-point amplitudes are proportional,
being related either by cyclic symmetry or by world-sheet parity, and
thus eq. (\ref{compl14}) relaxes into the weaker condition
\ba 
& &{\rm tr} ( \Lambda_o^{a_1} \Lambda_o^{a_2} \Lambda_o^{a_3}
\Lambda_o^{a_4} ) - {\rm tr} ( \Lambda_o^{a_2} \Lambda_o^{a_1} \Lambda_o^{a_3}
\Lambda_o^{a_4} ) 
 - {\rm tr} ( \Lambda_o^{a_1} \Lambda_o^{a_2} \Lambda_o^{a_4} \Lambda_o^{a_3} )
\nonumber \\ & &+ {\rm tr} ( \Lambda^{a_2} \Lambda^{a_1} \Lambda^{a_4}
\Lambda^{a_3} )  = {\rm tr} ( [\Lambda_o^{a_1}, \Lambda_o^{a_2}]
[\Lambda_o^{a_3} ,
\Lambda_o^{a_4}] )
\nonumber \\ & &\sim
\sum_{a_I} {\rm tr} ( [\Lambda_o^{a_1}, \Lambda_o^{a_2}] \Lambda_o^{a_I} )
{\rm tr} (
\Lambda_o^{a_I} [\Lambda_o^{a_3}, \Lambda_o^{a_4}] ) \,, \label{complv14}
\ea 
where the intermediate-state matrices belong to the {\it odd} levels.
The same relative signs are present for poles at all odd mass
levels of the open spectrum\footnote[2]{For the open bosonic string, as
we have seen in the previous section, the vector originates from the
first excited odd level, while the tachyon originates from an even
level, the ground state.}, while even mass levels lead to the condition
\ba 
& &{\rm tr} ( \Lambda_o^{a_1} \Lambda_o^{a_2} \Lambda_o^{a_3}
\Lambda_o^{a_4} ) + {\rm tr} ( \Lambda_o^{a_2} \Lambda_o^{a_1} \Lambda_o^{a_3}
\Lambda_o^{a_4} ) 
 + {\rm tr} ( \Lambda_o^{a_1} \Lambda_o^{a_2} \Lambda_o^{a_4} \Lambda_o^{a_3} )
\nonumber \\ & &+ {\rm tr} ( \Lambda_o^{a_2} \Lambda_o^{a_1} \Lambda_o^{a_4}
\Lambda_o^{a_3} )  = {\rm tr} ( \{\Lambda_o^{a_1}, \Lambda_o^{a_2}\} \{
\Lambda_o^{a_3},
\Lambda_o^{a_4}\} )
\nonumber \\ & &\sim
\sum_{a_I} {\rm tr} ( \{ \Lambda_o^{a_1}, \Lambda_o^{a_2}\} \Lambda_e^{a_I}
) {\rm tr} ( \Lambda_e^{a_I} \{ \Lambda_o^{a_3}, \Lambda_o^{a_4}\} ) \,, 
\label{complt14}
\ea
where the intermediate-state matrices now belong to the {\it even} levels.

Two more cases exhaust all possibilities with four identical external
states:  four even states into one odd or into one even state. 
Summarizing, we have
thus obtained the four conditions
\ba
& &({\rm four \ odd}
\to {\rm odd} ) : 
\nonumber 
\\  & &{\rm tr} ( [\Lambda_o^{a_1}, \Lambda_o^{a_2}] 
[\Lambda_o^{a_3}, 
\Lambda_o^{a_4}] )
\sim
\sum_{a_I} {\rm tr} ( [\Lambda_o^{a_1}, \Lambda_o^{a_2}]
\Lambda_o^{a_I} ) {\rm tr} ( \Lambda_o^{a_I} [\Lambda_o^{a_3},
\Lambda_o^{a_4}] ) \, , \nonumber 
\\
& &({\rm four \ odd}
\to {\rm even} ) : 
\nonumber 
\\  & &{\rm tr} ( \{\Lambda_o^{a_1}, \Lambda_o^{a_2}\} 
\{\Lambda_o^{a_3}, 
\Lambda_o^{a_4}\} )
\sim
\sum_{a_I} {\rm tr} ( \{\Lambda_o^{a_1}, \Lambda_o^{a_2}\}
\Lambda_e^{a_I} ) {\rm tr} ( \Lambda_e^{a_I} \{\Lambda_o^{a_3},
\Lambda_o^{a_4}\} ) \, , \nonumber 
\\
& &({\rm four \ even}
\to {\rm odd} ) : 
\nonumber 
\\  & &{\rm tr} ( [\Lambda_e^{a_1}, \Lambda_e^{a_2}] 
[\Lambda_e^{a_3}, 
\Lambda_e^{a_4}] )
\sim
\sum_{a_I} {\rm tr} ( [\Lambda_e^{a_1}, \Lambda_e^{a_2}]
\Lambda_o^{a_I} ) {\rm tr} ( \Lambda_o^{a_I} [\Lambda_e^{a_3},
\Lambda_e^{a_4}] ) \, , \nonumber 
\\
& &({\rm four \ even}
\to {\rm even} ) : 
\label{fourcases} 
\\  & &{\rm tr} ( \{\Lambda_e^{a_1}, \Lambda_e^{a_2}\} 
\{\Lambda_e^{a_3}, 
\Lambda_e^{a_4}\} )
\sim
\sum_{a_I} {\rm tr} ( \{\Lambda_e^{a_1}, \Lambda_e^{a_2}\}
\Lambda_e^{a_I} ) {\rm tr} ( \Lambda_e^{a_I} \{\Lambda_e^{a_3},
\Lambda_e^{a_4}\} ) \, , \nonumber
\ea 
where the labels $e$ and $o$ anticipate
the freedom of associating different Chan-Paton matrices to the even
and odd mass levels of the open spectrum,
and these imply generalized completeness conditions of the type
\ba 
& &[\Lambda_o,\Lambda_o ] \sim i \, \Lambda_o \, , \quad  [
\Lambda_e,\Lambda_e ]
\sim i \, \Lambda_o
 \, , \quad  [ \Lambda_e,\Lambda_o ] \sim i \, \Lambda_e \, , \nonumber 
\\ & &\{\Lambda_o,\Lambda_o \} \sim \Lambda_e \, , \quad  \{
\Lambda_e,\Lambda_e
\} \sim 
\Lambda_e
 \, , \quad  \{ \Lambda_e,\Lambda_o \} \sim \Lambda_o \,.
\label{sumunoriented}
\ea 

Once the dynamical parts of the amplitudes are given a proper 
normalization, consistently with the flip condition (\ref{wssym}), one
ought to supplement eqs. (\ref{sumunoriented}) with additional
hermiticity conditions, necessary to guarantee the proper sign of
physical residues. For definiteness, let us imagine to have normalized 
all two-point functions so that the $\Lambda$'s are all hermitian. The
algebraic content of eqs. (\ref{sumunoriented}) is then easier to
appreciate in terms of the two auxiliary sets
\be
\lambda = \{ \Lambda_e \} \, , \qquad \mu = \{ i \Lambda_o \} \, ,
\ee 
since, on account of (\ref{sumunoriented}), the $\lambda$'s and
$\mu$'s may be regarded as basis elements of a {\it real associative
algebra}, a vector space closed under multiplication. 
One is thus led to classify the irreducible
real associative algebras\footnote[3]{In the last Section we shall have more 
to say on the Chan-Paton matrices 
for more general models with different, although apparently identical,
sectors of the spectrum.}. As in the case of 
Lie algebras, the problem
simplifies if one considers the complex extension since,
on account of Wedderburn's theorem \cite{wedd}, the only irreducible solutions 
are then the full matrix algebras ${\rm GL}(n, \bb{C})$. 

Our next task is to recover the original non-complexified form of
the algebra generated by the two sets
$\lambda$ and $\mu$, and to this end we should distinguish two cases.
If the original algebra contains an element that squares to $-1$, it
coincides with its complex extension, and is itself ${\rm
GL}(n,\bb{C})$. In this
case the $\mu$'s are antihermitian generators of ${\rm U}(n)$, while the
$\lambda$'s are the remaining hermitian generators of ${\rm GL}(n,\bb{C})$.
States of even and odd mass levels are now defined by equivalent
$\Lambda_e$ and $\Lambda_o$ matrices, and are thus all valued in the
adjoint representation of ${\rm U}(n)$. On the other hand, if the original
algebra does not contain an element that squares to $-1$, on account
of the first line of eq. (\ref{sumunoriented}), it is a real form of
${\rm GL}(n,\bb{C})$ that, besides being an algebra, is also a Lie algebra. A
corollary of Wedderburn's theorem states that these real forms are 
only ${\rm GL}(n,\bb{R})$
and, in the even, $2n$, case, ${\rm GL}(n,\bb{Q})$. 
${\rm GL}(n,\bb{R})$ defines antisymmetric $\Lambda_o$ matrices that span the
adjoint representation of
${\rm SO}(n)$ and symmetric $\Lambda_e$ matrices corresponding to the
symmetric traceless and singlet representations of ${\rm SO}(n)$. Finally,
${\rm GL}(n,\bb{Q})$ 
defines $\Lambda_o$ matrices that span
the adjoint representation of ${\rm USp}(2n)$ and 
$\Lambda_e$ matrices corresponding to the traceless
antisymmetric and singlet representations of ${\rm USp}(2n)$. The
factorization of higher-point functions leads to additional sets of
conditions, that we shall refrain from writing explicitly. All,
however, are satisfied by these solutions, as can be seen by direct
substitution. One can summarize these results saying that the ends
of an open string are valued in the (anti)fundamental
representations of one of the classical groups ${\rm U}(n)$,
${\rm SO}(n)$ and ${\rm USp}(2n)$. 

It is interesting to recover these gauge groups and the corresponding
representations from the dynamics of additional degrees of
freedom living at the two ends of an open string \cite{marcus86}. 
This can be done
adding an even number $n$ of one-dimensional fermions $\psi_I$, with a
corresponding action
\be 
S = {\textstyle\frac{1}{4}} \ \int_{\partial \Sigma} \ ds \ i \ \eta^{IJ} \ 
\psi_I \frac{ d \psi_J}{ds} \, ,
\label{boundarypsi}
\ee 
where $\partial \Sigma$ denotes the world-sheet boundary
and, for the time being, $\eta$ is a Minkowski-like metric with $t$
time-like and $s$ space-like directions. 
Canonical quantization then results in the Clifford algebra
\be
\{ \psi_I , \psi_J \} =  2 \eta_{IJ} \,  , \label{cliffordpsi}
\ee 
and the two ends of an open string are thus to fill the corresponding 
representation, of dimension
$2^{n/2}$, so that each is now endowed with as many ``colours".

A related result may be obtained from the contribution of a
single empty closed boundary, along which the fermions of eq.
(\ref{boundarypsi}) are naturally antiperiodic.  
For a pair of $\psi$ fields the finite contribution to the resulting
determinant, free of zero modes, is independent of the
length $l$ of the boundary. It may be
conveniently calculated from the corresponding antiperiodic
$\zeta$ function, as
\ba 
{\rm det}\left( - {\textstyle\frac{1}{4}} \ \frac{\partial^2}{\partial s^2}
\right) &=&   
\lim_{p \to 0} \ \exp \left\{ - \frac{d}{dp} \left[
\left(\frac{\pi}{2 l} \right)^{-2p} \ (2^{-2p} - 1) \zeta(2p) \right]
\right\} \nonumber \\
&=& \exp[-2 \zeta(0)
\log 2  ] \, ,
\ea 
where $\zeta$ denotes the Riemann $\zeta$ function and
$\zeta(0)=-\frac{1}{2}$, and the end result is therefore $2$, consistently
with the Clifford algebra (\ref{cliffordpsi}).

In a similar fashion, one can associate internal quantum numbers to
open-string states via corresponding dressings of their vertex
operators. These are to be regarded as bi-spinors $V_{\alpha
\beta}$, and involve corresponding expansions in terms of
the $\psi^I$ fields:
\be 
V_{\alpha\beta} = \delta_{\alpha\beta} +
{(\gamma^I)}_{\alpha\beta} \psi_I + \ldots + {(\gamma^{I_1 \ldots
I_p})}_{\alpha\beta} \psi_{I_1} \ldots
\psi_{I_p} + \ldots  \,.
\label{cpfields}
\ee 
The correlation functions of these vertex operators now include
contributions from the fermions $\psi_I$, whose Green function is a
simple square wave for any closed boundary and, as a result, one can see
that the Chan-Paton factors of eq. (\ref{cpfactors}) can be
recovered from correlators of $\psi$ fields. This setting has
been widely used in \cite{dorn} to derive low-energy open-string
couplings, in the spirit of the $\sigma$-model constructions in
\cite{sigma}.

Taking these fermion fields more seriously, one can actually go a bit
further. To this end, let us anticipate a result to be discussed in
detail in later sections: for the bosonic string, there is a special
gauge group, SO(8192),  that from our previous considerations can be
built with 26 boundary fermions. Let us recall that, as we have
seen, each end of the open string is valued in the spinor
representation of the manifest symmetry group of the action
(\ref{boundarypsi}). For unoriented strings, whose states are eigenstates
of the ``flip'' operator $\Omega$, the bi-spinor field $V$
of eq. (\ref{cpfields}) satisfies a corresponding reality condition.
This can be consistently imposed both in the real,
$s-t=0,2 \ {\rm mod \ }8$, and pseudo-real, $s-t=4,6 \ {\rm mod \ }8$,
cases, since it is imposed simultaneously on both indices of
$V_{\alpha\beta}$. However, the resulting Chan-Paton group is
${\rm SO}(2^{n/2})$ in the first case and
${\rm USp} (2^{n/2})$ in the second. Thus, it is SO(8192) precisely with
$26$ boundary fermions, as many as the string coordinates, and with the
same signature. A related, amusing observation, is that only in this case
the linear divergence, proportional to the length of the boundary,
present in the determinant of the Laplace operator for the string
coordinates, naturally compensates a similar divergence of the
fermion determinant, that we have not seen explicitly
having used the $\zeta$-function method. 
Although this simple setting can
naturally recover classical groups whose order is a power of two,
it is apparently less natural to adapt it to cases where the gauge
groups have a reduced rank.


\vskip 12pt
\subsection{Vacuum amplitudes with zero Euler character}

In Field Theory, one usually does not pay much attention to the
one-loop vacuum amplitude.  This is a function of the masses of the
finite number of fields of a given model, fully determined by the free
spectrum \cite{qft} that, aside from its relation to the cosmological
constant, does not embody important structural information.  On the
other hand, strings describe infinitely many modes, and their
vacuum amplitudes satisfy a number of geometric constraints, that 
in a wide class of models essentially determine the full perturbative
spectrum.

In order to define the vacuum amplitudes for closed and open strings,
it is convenient to start from Field Theory, and in particular from
the simplest case of a scalar mode of mass $M$ in $D$ dimensions, for
which
\be 
S = \int \ d^D x \left(  {\textstyle\frac{1}{2}} \, \partial_\mu \phi \
\partial^\mu
\phi  -
{\textstyle\frac{1}{2}}\, M^2 
\phi^2  \right) \,.
\ee 
After a Euclidean rotation, the path integral
defines the vacuum energy 
$\Gamma$ as
\be 
e^{- \Gamma} = \int [D \phi] e^{-S_E} \sim {\rm
det}^{-\frac{1}{2}} \left( -
\Delta + M^2
\right)  \, ,
\ee 
whose $M$ dependence may be extracted using the identity
\be
\log\left({\rm det}(A)\right) =  - \int_\epsilon^\infty \
\frac{dt}{t} \  {\rm tr}
\left( e^{-tA} \right)
\, ,
\ee 
where $\epsilon$ is an ultraviolet cutoff and $t$ is a Schwinger
parameter. 
In our case, the complete set of momentum eigenstates diagonalizes the
kinetic operator, and
\be
\Gamma =  - \frac{V}{2} \ \int_\epsilon^\infty  \ \frac{dt}{t} \ e^{- t
M^2} \
\int \ \frac{d^D p}{(2 \pi)^D} \ e^{- t p^2} \label{gammawithp}
\, ,
\ee 
where $V$ denotes the volume of space-time.
Performing the Gaussian momentum integral then yields
\be
\Gamma =  - \frac{V}{2(4 \pi)^{D/2}} \ \int_\epsilon^\infty  \
\frac{dt}{t^{D/2+1}} \ e^{- t M^2} 
\, ,
\ee 
while similar steps for a Dirac fermion of mass $M$ in $D$
dimensions would result in
\be
\Gamma = \frac{V 2^{[D/2]}}{2 (4 \pi)^{D/2}} \ \int_\epsilon^\infty \
\frac{dt}{t^{D/2+1}} \ e^{- t M^2} \label{gammawithoutp}
\, ,
\ee
 with an opposite sign, on account of the Grassmann nature of the
fermionic path integral. 
These results can be easily extended to generic Bose or Fermi fields, 
since
$\Gamma$ is only sensitive to their physical modes, and is
proportional to their number. Therefore, in the general case they 
are neatly summarized in the expression
\be
\Gamma_{\rm tot} =  - \frac{V}{2 (4 \pi)^{D/2}} \ \int_\epsilon^\infty \
\frac{dt}{t^{D/2+1}} \ {\rm Str} \left( e^{- t M^2} \right) \, ,
\label{gammatot}
\ee 
where ${\rm Str}$ counts the signed multiplicities of Bose and
Fermi states.

We can now try to
apply eq. (\ref{gammatot}) to the closed bosonic string in the
critical dimension $D=26$, whose spectrum, described at the end of
subsection 2.2, is encoded in
\be 
M^2 = \frac{2}{\alpha'} \left( L_0 + \bar{L}_0 - 2
\right) \, , \label{closedmass26}
\ee 
subject to the constraint $L_0 = \bar{L}_0$.
Substituting (\ref{closedmass26}) in (\ref{gammatot}) then gives
\be
\Gamma_{\rm tot} =  - \frac{V}{2 (4 \pi)^{13}}  \int_\epsilon^\infty 
\frac{dt}{t^{14}} \ {\rm tr}
\ \left( e^{- \frac{2}{\alpha'} (L_0 + \bar{L}_0 - 2)t} \right) \, ,
\label{gammaclosed1}
\ee 
an expression that is not quite correct, since it does not take into
account the ``level-matching'' condition $L_0 = \bar{L}_0$ for the
physical states that, however, can be simply accounted for 
introducing a $\delta$-function constraint in (\ref{gammaclosed1}),
so that
\be
\Gamma_{\rm tot} =  - \frac{V}{2 (4 \pi)^{13}}  \int_{-\, \frac{1}{2}}^{\frac{1}{2}} ds 
\int_\epsilon^\infty 
\frac{dt}{t^{14}} \ {\rm tr} \left(
\ e^{- \frac{2}{\alpha'} (L_0 + \bar{L}_0 - 2)t} e^{2\pi i (L_0 - 
\bar{L}_0 )s}
\right)
\, ,
\label{gammaclosed2}
\ee 
since, from our previous discussion, $L_0 - \bar{L}_0$
has integer eigenvalues. Defining the ``complex'' Schwinger parameter
\be
\tau = \tau_1 + i \tau_2 = s + i \frac{t}{\alpha' \pi} \, ,
\ee 
and letting
\be 
q = e^{2 \pi i \tau} \, , \qquad  \bar{q} = e^{-2 \pi i
\bar{\tau}} \, ,
\ee 
eq. (\ref{gammaclosed2}) takes the more elegant form
\be
\Gamma_{\rm tot} 
=  - \frac{V}{2 (4 \pi^2 \alpha')^{13}} \ \int_{- \, \frac{1}{2}}^{\frac{1}{2}}
d \tau_1 \
\int_\epsilon^\infty \
\frac{d \tau_2}{\tau_2^{14}} \ {\rm tr}\ q^{L_0 - 1}
\bar{q}^{\bar{L}_0 - 1}
\,.
\label{gammaclosed3}
\ee

Actually, at one loop a closed string sweeps a torus, whose
Teichm\"uller parameter is naturally identified with the complex Schwinger
parameter $\tau$ but, as we have seen in subsection 2.1, not all
values of $\tau$ within the strip $\{-\, \frac{1}{2} < \tau_1 \le
\frac{1}{2}, \epsilon <
\tau_2 < \infty\}$ of eq. (\ref{gammaclosed3}) correspond to
distinct tori. Hence, one should restrict the integration domain 
to a fundamental region of the modular group, for instance to 
the region ${\cal F}$ of eq.
(\ref{fundregion}), and the restriction to
${\cal F}$
introduces an effective ultraviolet cutoff, of the order of the string
scale, for all string modes.
After a final rescaling, we are thus led to
an important quantity, the torus amplitude, that defines
the partition function for the closed bosonic string
\be {\cal T} = \int_{\cal F} \ \frac{d^2 \tau}{\tau_2^2} \
\frac{1}{\tau_2^{12}}
\  {\rm tr}\ q^{L_0 - 1} \bar{q}^{\bar{L}_0 - 1}
\,.
\label{Zclosed}
\ee
This type of expression actually determines the vacuum amplitude for any 
model of oriented closed
strings, once the corresponding Virasoro operators $L_0$ and
$\bar{L}_0$ are known.

It is instructive to compute explicitly the torus amplitude 
(\ref{Zclosed}) for the
bosonic string. To this end, we should recall that $L_0$ and
$\bar{L}_0$ are effectively number operators for two infinite sets of
harmonic oscillators. In particular, in terms of conventionally
normalized creation and annihilation operators, for each transverse
space-time dimension
\be  
L_0 = \sum_n \ n \ a^\dagger_n \ a_n \, ,
\ee   
while for each $n$
\be 
{\rm tr} \ q^{n  \, a^\dagger_n a_n} = 1 + q^n + q^{2 n} + \ldots
= \frac{1}{1 - q^n} \, ,
\ee  
and putting all these contributions together for the full
spectrum gives
\be 
{\cal T} = \int_{\cal F} \ \frac{d^2 \tau}{\tau_2^2} \
\frac{1}{\tau_2^{12}} \ \frac{1}{|\eta(\tau)|^{48}} \label{bosonictorus}
\, ,
\ee 
where we have defined the Dedekind $\eta$ function
\be
\eta(\tau) = q^\frac{1}{24} \, \prod_{n=1}^\infty \ (1 - q^n ) \,.
\ee 
The integrand of ${\cal T}$ is indeed invariant
under the modular group, as originally noticed by Shapiro \cite{shapmod}, 
since the measure is invariant
under the two generators $S$ and $T$ while, using the transformations
\cite{jacobi}
\be 
T: \ \eta(\tau + 1 ) = e^\frac{i \pi}{12} \, \eta( \tau ) \, ,
\qquad S:
\ \eta( - 1/\tau  ) =  \sqrt{- i \tau } \  \eta( \tau ) \, ,
\ee 
one can verify that the combination $\tau_2^{1/2} |\eta|^2$ is also
invariant. In other words, modular invariance
holds separately for the contribution of each transverse string
coordinate, independently of their total number, {\it i.e.} independently
of the total central
charge $c$. This is a crucial
property of the conformal field theories that {\it define} the torus 
amplitudes for all consistent models of oriented closed strings.

In the case at hand all string 
states are oscillator excitations of the tachyonic
vacuum, while the factor $\tau_2^{-12}$ can be recovered from the
integral over the continuum of transverse momentum modes, as
\be
(\alpha')^{12} \ \int \ d^{24} p \ e^{- \pi \alpha' \tau_2 p^2} \, ,
\ee
so that the partition function (\ref{bosonictorus}) can be
written in the form
\be 
{\cal T} = (\alpha')^{12} \ \int_{\cal F} \ \frac{d^2 \tau}{\tau_2^2} \
\int \ d^{24} p \ \left|\frac{q^{\frac{\alpha'}{4}p^2}}{\eta(\tau)^{24}}
\right|^2 
\label{bosonictorus2}
\, ,
\ee
that exhibits a continuum 
of distinct ground states with corresponding towers of
excitations. In the language of Conformal Field Theory, each tower is a 
``Verma module" \cite{bpz,cftrev}, while
the squared masses of the ground states are determined by the conformal
weights $h_i$ of the primaries. The content of each Verma
module may be encoded in a corresponding {\it character}
\be
\chi_i(q) = {\rm tr} \left( q^{L_0 - c/24} \right)_i = q^{h_i - c/24} 
\sum_k d_k q^k
\, ,
\label{chis}
\ee
where the
$d_k$ are positive integers that count the multiplicities of the corresponding
excitations, of weights $(h_i+k)$. In terms of these characters, 
a general torus
amplitude would read
\be
{\cal T} = \int_{\cal F} \ \frac{d^2 \tau}{\tau_2^2} \ \sum_{i,j} 
\bar{\chi}_i(\bar{q})
\ X_{ij} \ \chi_{j} (q) \, ,
\ee
with $X$ an integer matrix that counts their signed multiplicities,
as determined by spin-statistics. The 26-dimensional 
bosonic string thus belongs to this type of setting, with the double 
sum over Verma
modules replaced by an integral over the continuum of its transverse
momentum modes, each associated to a Virasoro character
\be
\chi_p (q) = \frac{q^{\frac{\alpha'}{4} p^2}}{\eta(\tau)^{24}}\,.
\ee

We are now ready to meet the first and simplest instance of an 
orientifold or open 
descendant \cite{cargese}, where world-sheet parity is used to
project a closed spectrum. Let us begin by recalling the 
low-lying spectrum of the
closed bosonic string that, as we have seen, starts with a tachyonic
scalar, followed by the massless modes associated to
$\alpha_{-1}^i \tilde{\alpha}_{-1}^j |0 \tilde{0} \rangle$: a
traceless symmetric tensor, a scalar mode, identified with the trace, 
and  an antisymmetric tensor. We would like to stress that these states
and all the higher
excitations have a definite symmetry under the interchange of
left, $\alpha$, and right, $\tilde{\alpha}$, oscillator modes. 
Indeed, both the action and the quantization procedure
used preserve the world-sheet parity $\Omega$, while this 
operation squares to the
identity, and thus splits the whole string spectrum in two subsets of
states, corresponding to its two eigenvalues, $\pm 1$. Na\"{\i}vely, one
could conceive to project the spectrum retaining either of these two
subsets, but string states can scatter, and the product of two odd
states would generate even ones. Hence, in this case
one has the unique option of
retaining only the states invariant under world-sheet parity,
and this eliminates, in particular, the massless
antisymmetric two-tensor. Therefore, after the projection 
the massless level, that in the original model contained $(24)^2$ 
states, contains only $24 \, (24 + 1) / 2$ states.

In order to account for the multiplicities in the projected spectrum,
one is thus to halve the torus contribution and to supplement it with  an
additional term, where left and right modes are effectively
identified.  This is accomplished by the Klein-bottle amplitude, that
describes a vacuum diagram drawn by a closed string undergoing a
reversal of its orientation.  From an operatorial viewpoint, one is
computing a trace over the string states with an insertion of the
world-sheet parity operator $\Omega$:
\be 
{\cal K} = {\textstyle\frac{1}{2}} \ \int_{{\cal F}_{\cal K}} \ \frac{d^2
\tau}{\tau_2^2} \
\frac{1}{\tau_2^{12}}{\rm tr} \left( q^{L_0 - 1} \ 
\bar{q}^{\bar{L}_0 - 1} \ \Omega \right) \,.
\ee 
More explicitly, the inner trace can be written
\be
\sum_{{\rm L},{\rm R}} \langle {\rm L},{\rm R}| \ q^{L_0 - 1} \ 
\bar{q}^{\bar{L}_0 - 1} \ \Omega \ |{\rm L},{\rm R}\rangle 
\ee 
and, after using $\Omega | {\rm L},{\rm R} \rangle  =  | {\rm
R},{\rm L} \rangle $, that, as we have anticipated, is the only available 
choice in this case, and the orthonormality conditions for the
states, reduces to
\be
\sum_{{\rm L},{\rm R}} \langle {\rm L},{\rm R}| \ q^{L_0 - 1} \ 
\bar{q}^{\bar{L}_0 - 1} \ | {\rm R},{\rm L} \rangle   = \sum_{\rm L}
\langle {\rm L},{\rm L}| \ (q \bar{q} )^{L_0 - 1} \  | {\rm L},{\rm L}
\rangle \, ,
\ee 
where the restriction to the diagonal subset $| {\rm L},{\rm L}
\rangle $  has led to the effective identification of $L_0$ and
$\bar{L}_0$.

It should be appreciated that the resulting amplitude depends naturally
on $2 i \tau_2$ that, as we have seen, is the modulus of the
doubly-covering torus. The integration
domain, not fully determined by these considerations, is
necessarily the whole positive imaginary axis of the $\tau$ plane,
since the involution breaks the modular group
to a finite subgroup. In
conclusion, after performing the trace, for the bosonic string one finds
\be 
{\cal K} = {\textstyle\frac{1}{2}} \ \int_0^\infty \ \frac{d
\tau_2}{\tau_2^{14}}
\ \frac{1}{\eta^{24}(2 i \tau_2)} \,. \label{klein1d}
\ee 

It is instructive to compare the $q$ expansions of the integrands of
${\cal T}$ and ${\cal K}$, while retaining in the former only terms
with equal powers of $q$ and $\bar{q}$, that correspond to on-shell
physical states satisfying the level-matching condition. 
Aside from
powers of $\tau_2$, these integrands are
\ba 
{\cal T} &\rightarrow& \left( (q \bar{q})^{-1} + (24)^2 +
\ldots\right) \, , \nonumber \\ 
{\cal K} &\rightarrow& {\textstyle\frac{1}{2}} \left(
(q \bar{q})^{-1} + (24) + \ldots\right) \, ,
\ea 
and therefore the right counting of states in the projected spectrum
is indeed attained halving the torus amplitude ${\cal T}$ 
and adding to it the Klein-bottle amplitude ${\cal K}$.

Following \cite{gsop,marcus86}, let us now use as integration variable 
the modulus $t = 2
\tau_2$ of the double cover of the Klein bottle. The corresponding 
transformation recovers a very
important power of two, that we have already met in the discussion of
the ``quarks'' at the ends of the open string, and indeed, taking into
account the rescaling of the integration measure gives
\be 
{\cal K} = \frac{2^{13}}{2} \ \int_0^\infty \ \frac{d t}{t^{14}}
\ \frac{1}{\eta^{24}(it)}  \,. \label{klein2d}
\ee
In our description of the Klein bottle in subsection 2.1, we have 
emphasized that this surface allows for two distinct natural choices of
``time''. The vertical time, $\tau_2$, enters the operatorial
definition of the trace, and defines the {\it direct-channel} or 
{\it loop} amplitude, while the horizontal
time, $\ell=1/t$, displays
the Klein bottle as a tube terminating at two crosscaps, and defines
the {\it transverse-channel} or {\it tree} amplitude. The corresponding
expression, that we denote by $\tilde{\cal K}$, 
\be
\tilde{\cal K} = \frac{2^{13}}{2} \ \int_0^\infty \ d \ell
\ \frac{1}{\eta^{24}(i\ell)}  \label{klein1t}
\ee 
can be obtained from eq. (\ref{klein2d}) by an $S$ modular
transformation. 

Let us now turn to the annulus amplitude. In this case, the trace is
over the open spectrum and, in order to account for the internal
Chan-Paton symmetry, we associate a multiplicity $N$ to each of the
string ends. As in the previous case, let us begin from the 
direct-channel amplitude, defined in terms of a trace over
open-string states,
\be 
{\cal A} = \frac{N^2}{2} \ \int_0^\infty \ \frac{d
\tau_2}{\tau_2^{14}} \ {\rm tr} ( q^{\frac{1}{2}(L_0 - 1)} ) \, ,
\ee 
where the exponent is now rescaled
as demanded by the different Regge slope of the open spectrum, exhibited
in eq. (\ref{openmass}). Computing the trace as above one finds
\be 
{\cal A} = \frac{N^2}{2} \ \int_0^\infty \ \frac{d
\tau_2}{\tau_2^{14}} \ \frac{1}{\eta^{24}\left( \frac{1}{2}i \tau_2
\right)} \, , \label{ann1d}
\ee 
and once more the amplitude is naturally expressed in terms of the
modulus, now ${1\over 2} i \tau_2$, of the doubly-covering torus. 
The first terms in the expansion of the integrand in powers of $\sqrt{q}$
give
\be
{\cal A} \rightarrow \frac{N^2}{2} \left(
(\sqrt{q})^{-1} + (24) + \ldots \right)  \label{annexp}
\ee
and, as for the Klein bottle, it is convenient to move to the modulus
of the double cover, now $t= \tau_2/2$, as
integration variable, obtaining
\be {\cal A} = \frac{N^2 \ 2^{-13}}{2} \ \int_0^\infty \ \frac{d
t}{t^{14}} \ \frac{1}{\eta^{24}(i t)} \,. \label{ann2d}
\ee 
The other choice of time, $\ell=1/t$, then displays the
annulus as a tube terminating at two holes, and defines the
transverse-channel amplitude. The corresponding expression,
that we denote by $\tilde{\cal A}$,
\be
\tilde{\cal A} = \frac{N^2 \ 2^{-13}}{2} \ \int_0^\infty \ d \ell
\ \frac{1}{\eta^{24}(i\ell)}  \, , \label{ann1t}
\ee 
can be obtained from eq. (\ref{ann2d}) by an $S$ modular
transformation. It should be appreciated that, in this tree channel,
the multiplicity  $N$ of the Chan-Paton charge spaces associated to the 
ends of the open string determines the reflection coefficients for
the closed spectrum in front of the two boundaries. 

The M\"obius strip presents some additional subtleties. This can
be anticipated, since the discussion of the other two
amplitudes suggests that the corresponding integrand should depend on the 
modulus of
the doubly-covering torus. In this case, however, as we have seen in 
subsection
2.1,  this is not purely imaginary but has a fixed real part,
equal to
$\frac{1}{2}$, that introduces relative signs for the oscillator
excitations at the various mass levels. These are precisely the
signs
discussed in the previous subsection, as can be appreciated from
the limiting behaviour of the amplitude for large vertical time, that exhibits
the contributions of intermediate open-string states undergoing a flip of
their orientation.

While the integrand is
obviously real for both ${\cal K}$ and ${\cal A}$, that depend on an
imaginary modulus, the same is  not true for the M\"obius amplitude
${\cal M}$, where
$\tau_1=\frac{1}{2}$. 
In order to write it for generic models, that can
include several
Verma modules with primaries of different weights, it is convenient to
introduce a basis of real ``hatted" characters, defined as
\be
\hat{\chi}_i(i \tau_2 +{\textstyle{1\over 2}}) = 
q^{h_i - c/24} \sum_k (-1)^k d_{(i) k} q^k \, ,
\label{chihats}
\ee
where $q=e^{-2 \pi \tau_2}$, that differ from $\chi_i(i \tau_2 +
\frac{1}{2})$ in the overall
phases
$e^{- i \pi(h_i -c/24)}$. This redefinition affects the modular transformation
$P$ connecting direct and transverse M\"obius amplitudes,
${\cal M}$ and $\tilde{\cal M}$, that now becomes
\be
P = T^{1/2} \ S \ T^2 \ S \ T^{1/2} \,,
\ee
where $T^{1/2}$ is a diagonal matrix, with $T^{1/2}_{ij}= \delta_{ij}\, e^{i\pi(h_i - c/24)}$.
For a generic conformal field theory, using the constraints
\be
S^2 = (S T)^3 = {\cal C} \, ,
\ee
it is simple to show that
\be
P^2 = {\cal C} \, ,
\ee
so that $P$ shares with $S$ the important property of squaring to 
the conjugation matrix ${\cal C}$.
In the last section we shall elaborate on the r\^ole of $P$, and of this 
property in particular, in Boundary Conformal Field Theory.

Returning to the open bosonic string, the M\"obius amplitude finally
takes the form
\be
{\cal M} = \frac{\epsilon\,N}{2} \int_0^\infty \ \frac{d \tau_2}{\tau_2^{14}} \
\frac{1}{\hat{\eta}^{24}({\textstyle{1\over 2}}i \tau_2 + 
{\textstyle{1\over 2}} )} \, ,
\ee
where $\epsilon$, equal to $\pm 1$, is an overall sign,
and its expansion in powers of $\sqrt{q}$ gives
\be
{\cal M} \rightarrow \frac{\epsilon \, N}{2} \left(
(\sqrt{q})^{-1} - (24) + \ldots \right) \,. \label{mobexp}
\ee
Then, from eqs. (\ref{annexp}) and 
(\ref{mobexp}), $\epsilon=+1$ corresponds to a total of $N(N-1)/2$
massless vectors, and thus to an orthogonal gauge group, while
(for even $N$) $\epsilon=-1$ corresponds to a symplectic gauge group.

In this case the transition to the transverse channel requires,
as we have emphasized, the redefinition $\tau_2 \to 1/t$ and the
corresponding $P$ transformation. It is then simple to show that
\be
\hat{\eta}\left( \frac{i}{2 t} + \frac{1}{2} \right) = 
\sqrt{t} \ \hat{\eta}\left( \frac{i t}{2} + \frac{1}{2} \right)
\,, 
\ee
and therefore
\be
\tilde{\cal M} = \frac{\epsilon \, N}{2} \int_0^\infty \ d t \
\frac{1}{\hat{\eta}^{24}({\textstyle{1\over 2}} i t + {\textstyle{1\over 2}} 
)} 
\ee
or, in terms of $\ell = t/2$,
\be
\tilde{\cal M} = 2 \ \frac{\epsilon \, N}{2} \int_0^\infty \ d \ell \
\frac{1}{\hat{\eta}^{24}(i \ell + {\textstyle{1\over 2}} )} \,. \label{mob1t}
\ee
The additional factor of two introduced by the last redefinition is
very important, since it reflects the combinatorics of the vacuum
channel: as we have seen, $\tilde{\cal M}$ may be associated
to a tube with one hole and one crosscap at the
ends, and thus needs precisely a combinatoric factor of two compared to 
$\tilde{\cal K}$ and $\tilde{\cal A}$, while the sign $\epsilon$ is
a relative phase between crosscap and boundary reflection coefficients. 
Finally, the Chan-Paton multiplicity $N$ determines
the reflection coefficient for the closed string in front 
of the single boundary present in the tree channel.

\begin{figure}
\begin{center}
\epsfbox{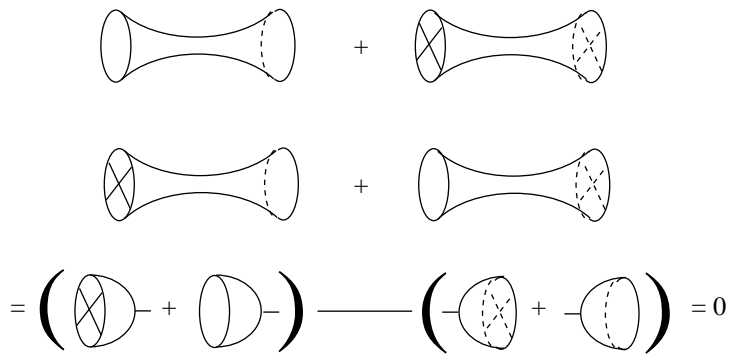}
\end{center}
\caption{Tadpole conditions in orientifold models.}
\label{figtadpoles}
\end{figure}

One can now study the limiting ultraviolet behaviour of the four amplitudes of
vanishing Euler character for small vertical time. As we have seen, the torus 
${\cal T}$
is formally protected by modular invariance, that 
excludes the ultraviolet region from the
integration domain. On the other hand, for
the other three surfaces the integration regions touch the real axis, and 
introduce
corresponding ultraviolet divergences. In order to take a closer look, it is 
convenient
to turn to the transverse channel, where the divergences appear in the 
infrared, or
large $\ell$, limit of eqs. (\ref{klein1t}), (\ref{ann1t}) and (\ref{mob1t}), 
and
clearly originate from the exchange of tachyonic and massless modes. In 
general, a
state of mass $M$ gives a contribution proportional to 
\be
\int_0^\infty \ d \ell \ e^{- M^2 \ \ell} = \frac{1}{M^2} \, , 
\label{limtransv}
\ee
and therefore, although one can formally regulate the tachyonic divergence, 
there is
no way to regulate the massless exchanges. It should be appreciated
that all massive states give sizable 
contributions only for $\ell \le 1/M^2$. Thus, once the massless
term is eliminated, the vertical time ultraviolet region
inherits a natural cutoff of the order of the string scale, precisely
as was the case for oriented closed strings on account of modular invariance.
In this simple model, Lorentz
invariance clearly associates the singular exchange 
to the only massless scalar mode of the closed string,
the dilaton. Moreover, it is simple to convince oneself that, as suggested by
eq. (\ref{limtransv}), the
divergence has a simple origin, clearly exhibited in the factorization limit
$\ell \to \infty$: the propagator $1/(p^2 + M^2)$
diverges 
for a massless state of zero momentum. This is a very important 
point,
since the corresponding residues are actually {\it
finite}, and define two basic building blocks of the theory, the one-point 
functions for
closed-string fields in front of a boundary and in front of a crosscap. 
Since
the former is proportional to the dimension $N$ of the Chan-Paton 
charge space, the two
contributions can cancel one another, leading to a finite amplitude, 
{\it only} for
a single choice of Chan-Paton gauge group. More in detail, the singular terms
of eqs. (\ref{klein1t}), (\ref{ann1t}) and (\ref{mob1t}) group into a 
contribution proportional to
\be
\tilde{\cal K} + \tilde{\cal A} + \tilde{\cal M} \sim { \textstyle 
\frac{1}{2}} \
( 2^{13} + 2^{-13} \ N^2 - 2 \, \epsilon \, N ) = \frac{2^{-13}}{2} \ 
(N - \epsilon \, 2^{13})^2 \, , 
\ee
that clearly vanishes for $N= 2^{13} = 8192$ and $\epsilon=+1$, 
and thus for the Chan-Paton
gauge group SO(8192) \cite{douglas1,weinberg,marcus86,bs88}. This is our
first encounter with a  {\it tadpole condition}. 

While in this model the special choice
of $N$ eliminates a well-defined correction to the low-energy effective field theory, a potential for the dilaton $\varphi$
\be
V \sim (N - \epsilon \, 2^{13}) \, \int \, d^{26} x \ \sqrt{-g} \ e^{- \varphi} \, ,
\ee
with $g$ the space-time metric,
whose functional form is fully determined by general covariance and by
the Euler characters of disk and crosscap, in more 
complicated cases, as we shall soon see, one can similarly
dispose of some {\it inconsistent} contributions, thus eliminating 
corresponding
anomalies in gauge and gravitational currents \cite{pc}. 
In this sense, tadpole 
cancellations provide the very  rationale for the appearance 
of the open sector. 

Let us recall the steps that have led to the SO(8192) model of
unoriented open and closed strings. The direct-channel Klein bottle amplitude 
${\cal K}$
receives contributions only from states of the oriented closed spectrum
built symmetrically out of left and right oscillator modes, and
completes the projection of the closed spectrum to states symmetric
under world-sheet parity. The corresponding transverse channel amplitude 
$\tilde{\cal K}$
receives contributions only from states that can be reflected compatibly with 
26-dimensional Lorentz invariance. It may
be obtained rescaling the integration variable to the modulus of the 
doubly-covering torus and performing an $S$ modular transformation, 
describes the
propagation of the projected closed spectrum on a tube terminating at two
{\it crosscaps}, and is thus quadratic in the corresponding reflection 
coefficients.
In a similar fashion, the transverse-channel annulus amplitude 
$\tilde{\cal A}$ describes
the propagation of the projected closed spectrum on a tube terminating at two
{\it boundaries} compatibly with 26-dimensional Lorentz invariance, 
and is thus quadratic in the
corresponding reflection coefficients, that are proportional to the overall 
Chan-Paton
multiplicity $N$.  The corresponding direct-channel amplitude ${\cal A}$, the
one-loop vacuum amplitude for the open string, may be recovered by a 
rescaling
of the integration variable and an $S^{-1}$ transformation. 
In this picture, $N$
describes the Chan-Paton multiplicity associated to an end of the open 
string.
Finally, the transverse-channel M\"obius amplitude $\tilde{ \cal M}$ describes
the propagation of the projected closed spectrum on a tube terminating at one 
hole
and one crosscap, and as such is a ``geometric mean" of $\tilde{\cal K}$
and $\tilde{\cal A}$, an operation that leaves a sign $\epsilon$
undetermined.  A rescaling of the
integration variable and a  $P^{-1}$ transformation 
turn it into the direct-channel amplitude ${\cal M}$, that
completes the projection of the open spectrum.

In conclusion, leaving the integrations implicit, the open
descendants of the 26-dimensional bosonic string are described by
\ba
{\textstyle\frac{1}{2}} {\cal T} &=& {\textstyle\frac{1}{2}} 
\ \frac{1}{\tau_2^{12} \ |\eta(\tau)|^{48}} \, , 
\nonumber \\
{\cal K} &=& {\textstyle\frac{1}{2}} 
\ \frac{1}{\tau_2^{12} \ \eta(2 i \tau_2)^{24}} \, ,
\ea
that define the projected unoriented {\it closed} spectrum, and by
\ba
{\cal A} &=& \frac{N^2}{2} \ \frac{1}{\tau_2^{12} \ 
\eta({\textstyle{1\over 2}} i \tau_2)^{24}} \,
,
\nonumber
\\ {\cal M} &=& \frac{\epsilon \, N}{2} \ \frac{1}{\tau_2^{12} \ 
\hat{\eta}({\textstyle{1\over 2}} i \tau_2 +{\textstyle{1\over 2}})^{24}} \, ,
\ea
that define the projected unoriented {\it open} spectrum, with $\epsilon=+1$
if one wants to enforce the tadpole condition. In the transverse channel
the last three amplitudes turn into
\ba
\tilde{\cal K} &=& \frac{2^{13}}{2} \ \frac{1}{\eta(i \ell)^{24}} \, ,  
\nonumber \\
\tilde{\cal A} &=& \frac{2^{-13} \ N^2}{2} \ \frac{1}{\eta(i \ell)^{24}}
\, , \nonumber \\ 
\tilde{\cal M} &=& 2 \ \frac{\epsilon \, N}{2} \ \frac{1}{\hat{\eta}
(i \ell+{\textstyle{1\over 2}})^{24}} \,
,
\ea
three closely related expressions that describe the propagation of the 
{\it closed} spectrum on tubes terminating at holes and/or crosscaps.
In the following, all transverse-channel amplitudes will 
be expressed in terms of $\ell$, rather than in terms of the natural
modulus $2 \ell$ of the closed spectrum, while all direct-channel ones will
be expressed in terms of $\tau_2 \sim 1/\ell$, even if this will not
be explicitly stated. For the sake
of brevity, we shall also avoid the use of two different symbols, 
$q$ and $\tilde{q}$, for the exponentials in the two channels.

\newsec{Ten-dimensional superstrings}

We now move on to consider the open descendants of the ten-dimensional
superstrings. After describing how the SO(32) type I model
and a variant with broken supersymmetry 
can be obtained from the ``parent'' type IIB, we turn to
other interesting non-supersym\-metric models
that descend from the tachyonic 0A and 0B strings. 
These have a somewhat richer
structure, and illustrate rather nicely some general features of
the construction.


\vskip 12pt
\subsection{Superstrings in the NSR formulation}

The starting point for our discussion is the supersymmetric
generalization of (\ref{bosonicaction}) \cite{bdhdz,pol}. Leaving aside the
Euler character, the resulting
action,
\ba 
S = &-& \frac{1}{4 \pi \alpha^\prime} \int d^2 \xi \sqrt{-g} \biggl[
g^{\alpha\beta}
\partial_\alpha X^\mu \ \partial_\beta X^\nu \ \eta_{\mu\nu} +
i \bar{\psi^{\mu}} \gamma^{\alpha} \nabla_{\alpha} \psi^{\nu} 
\eta_{\mu\nu} \nonumber \\ &+& i \bar{\chi}_{\alpha} \gamma^{\beta}
\gamma^{\alpha} \psi^{\mu} ( \partial_{\beta} X^{\nu} - 
\frac{i}{4} \bar{\chi}_{\beta} \psi^{\nu} ) \ \eta_{\mu\nu} \biggl]
\, , \label{fermionicaction}
\ea 
also involves two-dimensional Majorana spinors, 
$\psi^{\mu}$, the superpartners of
the $X^\mu$, and 
suitable couplings to the two-dimensional supergravity fields, the zweibein
$e_{\alpha}^a$ and the Majorana gravitino $\chi_{\alpha}$. As for
the bosonic string, these fields may be eliminated by a choice of 
gauge, letting
\be
g_{\alpha\beta}= \Lambda(\xi) \eta_{\alpha\beta} \, ,
\ee
and
\be
\chi_{\alpha} = \gamma_{\alpha} \chi (\xi) \,.
\ee
These conditions reduce (\ref{fermionicaction}) to a free model of
scalars and fermions, described by
\be
 S = - \ \frac{1}{4 \pi \alpha^\prime} \int d^2 \xi \biggl(
\partial^\alpha X^\mu \ \partial_\alpha X^\nu \ \eta_{\mu\nu} +
i \bar{\psi^{\mu}} \gamma^{\alpha} \partial_{\alpha} \psi^{\nu} \
\eta_{\mu\nu} \biggr) 
\ee
and in 
the critical dimension $D=10$ the remaining fields $\Lambda$ and
$\chi$ disappear also from the functional measure \cite{pol}. 
As for the bosonic string,
we shall resort to the light-cone description, 
sufficient to deal with all our subsequent applications to
string spectra. The equations of the two-dimensional supergravity
fields are actually constraints, that set to zero both the
energy-momentum tensor and the Noether current of two-dimensional 
supersymmetry, while the residual super-conformal invariance, left
over after gauge fixing,  may be used to let
\be
X^+ = x^+ + 2 \alpha' p^+ \tau \, , \qquad \psi^+ = 0 \,.
\ee
The constraints then yield the mass-shell conditions for physical 
states and allow one to express $X^-$
and $\psi^-$ in terms of the transverse components $X^i$ and
$\psi^i$.

We should again distinguish between the two cases of {\it closed}
and {\it open} strings. Since the Noether currents of the
space-time Poincar\'e symmetries contain
even powers of the spinors $\psi^\mu$, they are {\it 
periodic} along the string 
both if the spinors $\psi^\mu$ are {\it antiperiodic}
(Neveu-Schwarz, or NS, sector) and if they are
{\it periodic} (Ramond, or R, sector). 
As a result, for a closed string one need distinguish four types of sectors. 
Two, NS-NS and R-R, describe space-time
bosons, while the others, NS-R and R-NS, describe space-time fermions.
On the other hand, the open string has a single set of modes, equivalent
to purely left-moving ones on the double. One, NS, describes 
space-time bosons, while the other, R, describes
space-time fermions \cite{gsw}. In both cases, the perturbative string spectrum
is built acting on the vacuum with the creation 
modes in $X^i$ and
$\psi^i$, while $L_m$ and $\bar{L}_m$  now include contributions 
from both types of oscillators. Thus, in particular,
\be
L_m = {\textstyle \frac{1}{2}} : \sum_m \alpha_{m-n}^i \alpha_n^i : + 
{\textstyle \frac{1}{2}} : \sum_r (r-\frac{m}{2}) \psi_{m-r}^i \psi_r^i : 
+ \delta_{m,0} \Delta \, ,
\ee
where $r$ is half-odd integer in the NS
sector and integer in the R sector. The corresponding
normal-ordering shift $\Delta$ is essentially determined by the simple rule
of eq. (\ref{zriem}): each fermionic
coordinate contributes $-\frac{1}{48}$ in the NS sector and $\frac{1}{24}$ 
in the R sector, while each periodic
boson contributes $-\frac{1}{24}$. As a
result, for each set of modes the total shift in $D$ dimensions, 
induced by $D-2$
transverse bosonic and fermionic coordinates, is $-\frac{1}{16}(D-2)$ in
the NS sector, but vanishes in the R sector.

The NS sector is simpler to describe, since the antiperiodic 
transverse fermions 
$\psi^i$ do not have zero modes, and as a result the corresponding vacuum is
a tachyonic scalar. Its lowest excitation results from the action of
$\psi_{-1/2}^i$: it is a transverse
vector whose squared mass, proportional to $\frac{1}{2} - \frac{1}{16}(D-2)$,
must vanish in a Lorentz-invariant model. As for the bosonic string,
this simple observation suffices to recover the critical dimension, 
$D=10$ in this case, while fixing the level of the
ground state, and
we can now compute ${\rm tr}(q^{L_0})$ resorting to
standard results for the Fermi gas. In the previous section 
we have already obtained the contribution of the bosonic
modes, and for the fermionic oscillators
\be
{\rm tr}\bigl(q^{\sum_r r \psi^{\dag}_{r} \psi_r}\bigr) = 
\prod_r {\rm tr}\bigl(q^{r a^{\dag}_{r} a_r}
\bigr) = \prod_r (1 + q^r)^8 \, ,
\ee
since the Pauli exclusion principle allows at most one fermion in each of
these states. It should be appreciated that this expression actually
applies to both the NS and R sectors, provided $r$ is turned
into an integer in the second case.

Summarizing, in the NS sector
\be
{\rm tr}(q^{L_0}) = \frac{\prod_{m=1}^{\infty} (1 + q^{m-1/2})^8}{
q^{1/2} \prod_{m=1}^{\infty} (1 - q^m)^8} \, , \label{theta00}
\ee
while in the R sector
\be
{\rm tr}(q^{L_0}) = 16 \ \frac{\prod_{m=1}^{\infty} (1 + q^{m})^8}{
\prod_{m=1}^{\infty} (1 - q^m)^8} \,. \label{theta10}
\ee
The factor $q^{1/2}$ is absent in (\ref{theta10}) since, 
as we have seen, the R sector starts with massless modes, while
the overall coefficient reflects the degeneracy of the R vacuum, since the
zero modes of the $\psi^i$, absent in the NS case, imply that 
this carries a 
16-dimensional representation of the SO(8) Clifford algebra
\be
\{ \psi_0^i ,  \psi_0^j \} = 2 \ \delta^{ij} \, ,
\ee
and is thus a space-time spinor, like all its excitations.

Building a sensible spectrum is less straightforward in this
case. The difficulties may be anticipated noting that even and odd numbers of
anticommuting fermion modes have opposite statistics, and the simplest
possibility, realized in the type I superstring, is to 
project out all states created by {\it even} numbers of fermionic oscillators.
This prescription is the original form of the GSO 
projection \cite{gso}, and has the
additional virtue of removing the tachyon. The corresponding projected
NS sector is described by
\be
{\rm tr}\biggl( \frac{(1 - (-1)^F)}{2} q^{L_0} \biggr) = 
\frac{\prod_{m=1}^{\infty} (1 + q^{m-1/2})^8 - 
\prod_{m=1}^{\infty} (1 - q^{m-1/2})^8}{
2 \, q^{1/2} \prod_{m=1}^{\infty} (1 - q^m)^8}  \, , \label{v8part}
\ee
where the insertion of $(-1)^F$, with $F$ the world-sheet fermion number, 
reverses the sign of all contributions associated
with odd numbers of fermionic oscillators.

This expression plays an important r\^ole in the representation
theory of the affine extension of so(8). In order to elucidate this
point, of crucial importance in the following, let us begin
by recalling that the so(8) Lie algebra 
has four conjugacy classes of representations, and
that its level-one affine extension has consequently four integrable
representations. These 
correspond to four sub-lattices of the weight lattice, that 
include the vector, the scalar and the two eight-dimensional
spinors. 
To each of these sub-lattices one can associate a character, and
one of them is directly related to the expression in
(\ref{v8part}) \cite{affine}.

In order to proceed further, let us introduce the Jacobi theta functions
\cite{jacobi}, defined by the Gaussian sums
\be
\vartheta \left[{\textstyle {\alpha \atop \beta}} \right] (z|\tau) = \sum_n \ 
q^{\frac{1}{2} (n+ \alpha)^2} \ e^{2 \pi i (n + \alpha)(z+\beta)}
\ee
or, equivalently, by the infinite products
\ba
\vartheta \left[ {\textstyle {\alpha \atop \beta}} \right] (z|\tau) &=& 
e^{2 i \pi \alpha (z+\beta)} \ q^{\alpha^2/2} \prod_{n=1}^\infty \
( 1 - q^n) \\
& &\times \prod_{n=1}^\infty \ (1 + q^{n + \alpha - 1/2} e^{2 i \pi (z+\beta)} ) 
(1 + q^{n - \alpha - 1/2} e^{-2 i \pi (z+\beta)} ) \nonumber
\,.
\ea
These $\vartheta$ functions have a simple behaviour under $T$ and $S$ modular
transformations:
\ba
& &\vartheta \left[ {\textstyle {\alpha \atop \beta}}
 \right] (z|\tau+1) 
= e^{-i \pi \alpha
(\alpha -1)} \vartheta \left[ {\textstyle {\alpha \atop \beta +\alpha - 1/2}}
\right] 
(z|\tau) \, , \nonumber \\
& &\vartheta \left[ {\textstyle {\alpha \atop \beta}} \right] 
\left(\frac{z}{\tau}\right|\left.-\frac{1}{\tau}\right) = 
(-i \tau)^{1/2} \ e^{2 i \pi \alpha \beta + i \pi z^2/\tau} \   \vartheta \left[  {\textstyle {\beta \atop -\alpha}}
 \right] (z|\tau)
\,.
\label{thetamodular}
\ea

In our case the
fermions $\psi^i$ are periodic or antiperiodic, and it is thus sufficient to 
consider Jacobi theta functions
with vanishing argument $z$, usually referred to as theta-constants, 
with characteristics $\alpha$ and $\beta$ equal to $0$ or $\frac{1}{2}$. 
If $\alpha$ and $\beta$ are both $\frac{1}{2}$ the resulting
expression, usually denoted $\vartheta_1$, vanishes. 
On the other hand, the fourth powers of the
other three combinations, usually denoted $\vartheta_2$, $\vartheta_3$
and $\vartheta_4$, divided by the twelfth power of $\eta$, 
are directly related to the superstring vacuum amplitudes, since
\ba
& &\frac{\vartheta^4 \left[ {\small 1/2} \atop {\small 0} \right] (0|\tau)}
{\eta^{12}(\tau)} = \frac{\vartheta_2^4(0|\tau)}{\eta^{12}(\tau)} =
16 \frac{\prod_{m=1}^{\infty} (1 + q^{m})^8}
{\prod_{m=1}^{\infty} (1 - q^m)^8} \, , \\
& &\frac{\vartheta^4 \left[ {\small 0} \atop {\small 0} \right] (0|\tau)}
{\eta^{12}(\tau)} = \frac{\vartheta_3^4(0|\tau)}{\eta^{12}(\tau)} =
\frac{\prod_{m=1}^{\infty} (1 + q^{m-1/2})^8}{
q^{1/2} \prod_{m=1}^{\infty} (1 - q^m)^8} \, , \\
& &\frac{\vartheta^4 \left[ {\small 0} \atop {\small 1/2} \right] (0|\tau)}
{\eta^{12}(\tau)} = \frac{\vartheta_4^4(0|\tau)}{\eta^{12}(\tau)} =
\frac{\prod_{m=1}^{\infty} (1 - q^{m-1/2})^8}{
q^{1/2} \prod_{m=1}^{\infty} (1 - q^m)^8} \,. \label{vacuumd=10}
\ea

Returning to the so(8) representations, let us define the first two
characters, $O_8$ and $V_8$, as
\ba
O_8 &=& \frac{\vartheta_3^4 + \vartheta_4^4}{2 \eta^4} = 1 + 28 q + \ 
\ldots \, , \label{O8} \\
V_8 &=& \frac{\vartheta_3^4 - \vartheta_4^4}{2 \eta^4} = 8 q^{1/2} + 
64 q^{3/2} + \ \ldots \,.
\label{V8}
\ea
These correspond to an orthogonal decomposition of the NS spectrum, 
where only even or only odd numbers of $\psi$ excitations are retained,
and are thus of primary interest in the construction of string amplitudes.
The $O_8$ character starts at the lowest mass level with the 
tachyon and corresponds to
the conjugacy class
of the singlet in the weight lattice. On the other hand, $V_8$ starts
with the massless vector and corresponds to the conjugacy class
of the vector in the weight lattice.
The previous considerations suggest that two more characters
should be
associated to the two spinor classes, both clearly belonging to the
R sector. However, only $\vartheta_2$ is available, since 
$\vartheta_1$ vanishes at the origin. We are thus facing a rather
elementary example of a system where an ambiguity is present, since
four different characters are to be built out of three non-vanishing
$\vartheta$'s. In 
Conformal Field Theory, powerful methods have been devised to 
deal with this type of
problems \cite{simplec}, and the end result is, in general, 
a finer description of
the spectrum, where each sector is associated to an independent character.
The $T$ and $S$ transformations are then represented on the resolved
characters by a pair of unitary matrices, diagonal and symmetric 
respectively, satisfying the constraints
\be
S^2 = (S T)^3 = {\cal C} \,.
\ee
For the ${\rm SO}(2n)$ groups, that have in general the
four conjugacy classes $O_{2n}$, $V_{2n}$, $S_{2n}$ and $C_{2n}$,
\be
{\cal C} = {\rm diag} (1_2,(\sigma_1)^n ) \, ,
\ee
where $1_2$ denotes the $2 \times 2$ identity matrix and $\sigma_1$ denotes the
usual Pauli matrix. Thus, ${\cal C}$ is the identity for all SO($4n$), that
have only self-conjugate representations, but connects the two conjugate 
spinors for all SO($4n+2$). One can also understand the vanishing of
$\vartheta_1$, that can be ascribed to the insertion of the 
chirality matrix $\Gamma_{9}$ in the trace. $\vartheta_1$
has nonetheless a well-defined
behaviour under the modular group, that may be deduced from eq. 
(\ref{thetamodular}) in the limit $z \to 0$, and the conclusion is that
the two R characters
\ba
S_8 &=& \frac{\vartheta_2^4 + \vartheta_1^4}{2 \eta^4} \, , \label{S8} \\
C_8 &=& \frac{\vartheta_2^4 - \vartheta_1^4}{2 \eta^4} \, , \label{C8}
\ea
describe orthogonal portions of the R spectrum that begin,
at zero mass, with the two spinors of opposite chirality. In both
cases, the excitations are projected by \cite{gso,gsw}
\be
{\textstyle\frac{1}{2}} 
\left(1 + \Gamma_{9} \ (-1)^{: \sum_{n=1}^{\infty} \psi^i_{-n} \psi^i_{n}: }
\right) 
\,,
\ee
that has proper (anti)commutation relations with the superstring
fields $X$ and $\psi$, so that the massive modes of the
$S_8$ and $C_8$ sectors actually involve states of 
both chiralities, as needed to describe massive spinors.

The famous {\it aequatio identica satis abstrusa} of Jacobi \cite{jacobi},
\be
\vartheta_3^4 - \vartheta_4^4 - \vartheta_2^4 = 0 \, ,
\ee
then implies that the full superstring spectrum
built from an eight-dimensional vector and an 
eight-dimensional Majorana-Weyl spinor, the degrees of 
freedom of ten-dimensional 
supersymmetric Yang-Mills, contains equal numbers of Bose and
Fermi excitations at all mass levels, as originally recognized by
Gliozzi, Scherk and Olive \cite{gso}.

The modular transformations in eq. (\ref{thetamodular})
determine the $T$ and $S$ matrices for the four
characters of all so($2n$) algebras, that may be defined as
\ba
O_{2n} &=& \frac{\vartheta_3^n + \vartheta_4^n}{2 \eta^n} \, , \label{O2n} \\
V_{2n} &=& \frac{\vartheta_3^n - \vartheta_4^n}{2 \eta^n} \, , \label{V2n} \\
S_{2n} &=& \frac{\vartheta_2^n + i^{-n} \vartheta_1^n}{2 \eta^n} \, , 
\label{S2n} \\
C_{2n} &=& \frac{\vartheta_2^n - i^{-n} \vartheta_1^n}{2 \eta^n} \, , 
\label{C2n}
\ea
a natural generalization of eqs. (\ref{O8}), (\ref{V8}), (\ref{S8})
and (\ref{C8}), and it is then simple to show that, in all cases
\be
T = e^{ -i n \pi/12} \ {\rm diag} \left( 1,-1,e^{i n \pi/4},e^{i n \pi/4} 
\right) \, ,
\ee
and
\ba
S = \frac{1}{2} \ \left( \begin{array}{rrrr}
1 & 1 & 1 & 1 \\
1 & 1 & -1 & -1 \\
1 & -1 & i^{-n} & -i^{-n} \\
1 & -1 & -i^{-n} & i^{-n} 
\end{array} \right) \,.
\label{s2nmat}
\ea

Taking into account the eight transverse bosonic coordinates, the actual 
superstring vacuum amplitudes may then be
built from the four so(8) characters 
divided by $\tau_2^4 \eta^8(\tau)$, and on the four combinations
\be
\frac{O_8}{\tau_2^4 \eta^8} \, , \quad 
\frac{V_8}{\tau_2^4 \eta^8} \, , \quad \frac{S_8}{\tau_2^4
\eta^8} \, , \quad \frac{C_8}{\tau_2^4 \eta^8}  
\ee
the $T$ matrix acts as
\be
T = \ {\rm diag}(-1,1,1,1)
 \,.
\ee
Consequently, the $P$ matrix also takes a very simple form in this case,
\be
P =  \ {\rm diag}(-1,1,1,1)
 \, , \label{po2n}
\ee
and actually coincides with $T$, up to the usual effect on the powers of
$\tau_2$, that disappear in the transverse channel. 
On the other hand, for the general case of
so($2n$)
\ba
P = \ \left( \begin{array}{rrrr}
c & s & 0 & 0 \\
s & -c & 0 & 0 \\
0 & 0 & \zeta c & i \zeta s \\
0 & 0 & i \zeta s & \zeta c 
\end{array} \right) \, , \label{pradisimato2n}
\ea
where $c=\cos(n \pi/4)$, $s=\sin(n \pi/4)$ and $\zeta = e^{- i n \pi/4}$.

We can now use the constraint of modular invariance to build
consistent ten-dimensional spectra of oriented closed strings. 
The corresponding (integrands for the) torus amplitudes will be of the form
\be
{\cal T} = \chi^{\dag} X \chi \, ,
\ee
where the matrix $X$ defines the GSO projection and satisfies
the two constraints of modular invariance
\be
S^{\dag} X S = X \, , \qquad T^{\dag} X T = X \,.
\ee
Furthermore, $X$ is to describe a single graviton and is to respect the 
spin-statistics relation, so that bosons and fermions must contribute with
opposite signs to ${\cal T}$, a result that can also be recovered from the
factorization of two-loop amplitudes \cite{abk}. 
It is then simple to see that only four
distinct torus amplitudes exist, that correspond to the type IIA and 
type IIB superstrings, described by
\ba
{\cal T}_{{\rm IIA}} &=& (\bar{V}_8 - \bar{S}_8) (V_8 - C_8 ) \, , 
\nonumber \\
{\cal T}_{{\rm IIB}} &=& |V_8 - S_8 |^2 \, , \label{IIAB}
\ea
and to the two non-supersymmetric 0A and 0B models \cite{dhsw}, described by
\ba
{\cal T}_{{\rm 0A}} &=& |O_8|^2 + |V_8|^2 + \bar{S}_8 C_8 + 
\bar{C}_8 S_8 \, , \nonumber \\
{\cal T}_{{\rm 0B}} &=& |O_8|^2 + |V_8|^2 + |S_8|^2 + |C_8|^2 \,.
\label{0AB}
\ea

It is instructive to summarize the low-lying spectra of 
these theories. 
The type II superstrings have no tachyons, and their massless
modes arrange themselves in the multiplets of the type IIA and type IIB  
ten-dimensional supergravities.
Both include, in the NS-NS sector, the graviton $g_{\mu\nu}$, an antisymmetric
tensor $B_{\mu\nu}$ and a dilaton $\varphi$. Moreover, both contain a
pair of gravitinos and a corresponding pair of spinors, in the NS-R and
R-NS sectors. In the IIA string the two pairs contain 
fields of opposite chiralities, while in the IIB string 
both gravitinos are left-handed and both spinors are
right-handed. Finally, in the R-R sector type IIA contains an
Abelian vector $A_{\mu}$ and a three-form potential $C_{\mu\nu\rho}$, while
type IIB contains an additional scalar, an additional antisymmetric
two-tensor and a four-form potential $A^{+}_{\mu\nu\rho\sigma}$ with a
self-dual field strength. The type IIB spectrum, although chiral, is free
of gravitational anomalies \cite{alvgaum}. On the other hand, 
the 0A and 0B strings
do not contain any space-time fermions, while 
their NS-NS sectors comprise two
sub-sectors, related to the $O_8$ and $V_8$ characters, so that the former adds
a tachyon to the low-lying NS-NS states of the previous models. 
Finally,
for the 0A theory the R-R states are two copies of those of type IIA, 
{\it i.e.} a pair of Abelian vectors and a pair of three-forms, 
while for the 0B theory they are a pair of scalars, a pair of two-forms and
a full, unconstrained, four-form. These two additional spectra are clearly not
chiral, and are thus free of gravitational anomalies.
 
It should be appreciated that for all these solutions the
interactions respect the choice of GSO projection. This
condition may be formalized introducing the {\it fusion rules} between
the four families $[O_8]$, $[V_8]$, $[S_8]$ and $[C_8]$, that identify
the types of chiral operators that would emerge from all possible
interactions (technically, from operator products), and demanding
closure for both left-moving and right-moving excitations.
A proper account of the ghost structure would show that,
for space-time characters, 
$[V_8]$ is actually the identity of the fusion algebra, 
and appears in the square of 
all the other families \cite{bert}. All fusion rules are neatly encoded in 
the fusion-rule coefficients ${\cal N}_{ij}{}^{k}$, that can also be recovered 
from the $S$ matrix for the
space-time characters $O_8$, $V_8$, $-S_8$ and $-C_8$. Notice the
crucial sign, that reflects the relation between spin and
statistics and leads to
\ba
S' = \frac{1}{2} \ \left( \begin{array}{rrrr}
1 & 1 & -1 & -1 \\
1 & 1 & 1 & 1 \\
-1 & 1 & 1 & -1 \\
-1 & 1 & -1 & 1 
\end{array} \right) \, ,
\ea
with the result of interchanging the r\^oles of $O_8$ and $V_8$. 
The Verlinde formula \cite{verlinde}
\be
{\cal N}_i = S' \Lambda_i 
S^{' \dagger} \, ,
\ee
with 
\be
(\Lambda_i)_{jk}  = \delta_{jk} \frac{S'_{ij}}{S'_{1j}}
\ee
determines the fusion-rule coefficients $\left({\cal N}_i\right)_j{}^k$, and
may be used to verify these statements.

Summarizing, four ten-dimensional models of oriented closed strings, 
whose spectra are
encoded in the partition functions of eqs. (\ref{IIAB}) and (\ref{0AB}),
can be obtained via consistent GSO projections from the ten-dimensional
superstring action. 
The last three are particularly interesting, since they share with our
original example, the bosonic string, the property of being 
{\it symmetric}
under the interchange of left and right modes. In the next subsection we
shall describe how to associate open and unoriented spectra to the
type IIB model, thus recovering the type I SO(32) superstring and
a non-supersymmetric variant.

\vskip 12pt
\subsection{The type I superstring: SO(32) {\it vs} USp(32)}

The SO(32) superstring contains a single sector,
corresponding to the (super)character $(V_8 - S_8 )$, and is thus rather
simple to build. We can just repeat the steps followed for the 
bosonic string in the
previous section and write, displaying once more both the full integrands
and the modular integrals,
\ba
{\cal K} &=& \frac{1}{2} \int_0^{\infty} \ \frac{ d \tau_2}{\tau_2^6} \ 
\frac{ (V_8 - S_8 ) ( 2 i \tau_2)}{\eta^8(2 i \tau_2)} \, , \\
{\cal A} &=& \frac{N^2}{2} \int_0^{\infty} \ \frac{ d \tau_2}{\tau_2^6} \ 
\frac{ (V_8 - S_8 ) ( {\textstyle{1\over 2}}
i \tau_2 )}{\eta^8( {\textstyle{1\over 2}} i \tau_2)} \, , \\
{\cal M} &=& \frac{\epsilon N}{2} \int_0^{\infty} \ 
\frac{ d \tau_2}{\tau_2^6} \ 
\frac{ (\hat{V}_8 - \hat{S}_8 ) ( {\textstyle{1\over 2}}
i \tau_2 + {\textstyle{1\over 2}})}{\hat{\eta}^8(
{\textstyle{1\over 2}} i \tau_2 + {\textstyle{1\over 2}})} \, , 
\ea
where, as in section 2, $\epsilon$ is a sign.
The Klein-bottle projection symmetrizes the NS-NS sector, thus eliminating
from the massless spectrum the two-form, and antisymmetrizes the R-R 
sector, thus
eliminating the second scalar and the self-dual four-form.
Since the complete projection leaves only one combination of each pair of 
fermion modes, the resulting massless spectrum corresponds to 
the minimal ${\cal N}=(1,0)$ ten-dimensional
supergravity, and comprises a graviton, a two-form, now from the R-R sector,
a dilaton, a left-handed gravitino and a right-handed spinor.
In a similar fashion, the massless open sector is a (1,0) super Yang-Mills
multiplet for the group SO($N$)  if $\epsilon=-1$ or USp($N$) if 
$\epsilon=+1$.

Proceeding as in the previous section, one can also write the
corresponding transverse-channel amplitudes
\ba
\tilde{\cal K} &=& \frac{2^5}{2} \int_0^{\infty} \ d \ell  \ 
\frac{ (V_8 - S_8 ) (i \ell )}{\eta^8(i \ell)} \, , \\
\tilde{\cal A} &=& \frac{2^{-5} \ N^2}{2} \int_0^{\infty} \ 
d \ell \ 
\frac{ (V_8 - S_8 ) ( i \ell )}{\eta^8( i \ell )} \, , \\
\tilde{\cal M} &=& 2 \ \frac{\epsilon N}{2} \int_0^{\infty} \ d \ell  \ 
\frac{ (\hat{V}_8 - \hat{S}_8 ) ( i \ell + {\textstyle{1\over 2}})}
{\hat{\eta}^8(i \ell + {\textstyle{1\over 2}})} \, ,
\ea
and the tadpole condition
\be
\frac{2^5}{2} + \frac{2^{-5} \ N^2}{2} + 2 \ \frac{\epsilon N}{2} =
\frac{2^{-5}}{2} \ ( N + 32 \epsilon )^2 = 0 \, , \label{gsso32}
\ee
that applies to both the NS-NS and R-R sectors, selects uniquely the 
SO(32) gauge group ($N = 32$, $\epsilon=-1$).

This cancellation can be given a suggestive space-time interpretation:
the world-sheet boundaries traced by the ends of the open strings are mapped
to extended objects, D9 branes, that fill the whole of space-time, while 
the crosscaps are mapped to a corresponding non-dynamical object, the
orientifold O9 plane. In general, 
both D$p$ branes and O$p$ planes have tensions and
carry R-R charges with respect to $(p+1)$-form potentials $C_{p+1}$  
\cite{pol95}. 
For D-branes, tension and charge are both positive while, as
we shall soon see, two types of O-planes can be present
in perturbative type I vacua: those with negative tension and negative charge,
here denoted
O$_+$ planes, and those with positive tension and positive charge, here
denoted 
O$_-$ planes \cite{wittor}. 

In addition, there are of course 
D-antibranes and O-antiplanes, in the following often
called for brevity $\overline{\rm D}$-branes and $\overline{\rm O}$-planes, 
with identical tensions
and opposite R-R charges. If these results are combined with non-perturbative
string dualities, a rich zoo of similar extended objects
emerges, with very interesting properties \cite{hananykol}. 
Let us stress that the NS-NS and R-R 
tadpole conditions are conceptually quite different and play quite
distinct r\^oles: while the latter reflect the need for overall charge
neutrality, consistently with the Gauss law for $C_{p+1}$ if its Faraday
lines are confined to a compact space,
and are related to space-time anomalies \cite{pc}, the 
former, as we have seen in the previous subsection, 
give rise to a dilaton-dependent correction to 
the vacuum energy
that, in principle, can well be non-vanishing. This will be loosely referred
to as a dilaton tadpole. That the peculiar $(-{1\over 2},-{3\over 2})$ 
ghost picture
could produce non-derivative R-R couplings, consistently with the emergence
of  zero-momentum tadpoles, when boundaries or crosscaps
are present, was first pointed out in \cite{bps}, while the detailed
coupling was analyzed in detail in \cite{lerdarr}. In space-time
language \cite{pol95}, these couplings reflect the R-R charge
of the branes and orientifolds present in the models.
Notice that our conventions for the O-planes, summarized in table 
\ref{tabplanes}, where $T$ and $Q$ denote their tensions and R-R charges,
are as in \cite{wittor} and in our previous
papers, but are opposite to those in \cite{hananykol}.
\begin{table}
\caption{Conventions for O-planes.}
\label{tabplanes}
\begin{indented}
\lineup
\item[]\begin{tabular}{@{}llr}
\br
$\quad T$ & $Q$ & Type$\quad$ \\
\mr
$\quad<0\qquad $ & $<0\qquad$ & ${\rm O}_+\quad$
\\
$\quad>0\qquad$ & $>0\qquad$ & ${\rm O}_-\quad$
\\
$\quad<0\qquad$ & $>0\qquad$ & $\overline{\rm O}_+\quad$
\\
$\quad>0\qquad$ & $<0\qquad$ & $\overline{\rm O}_-\quad$
\\
\br
\end{tabular}
\end{indented}
\end{table}

In the type I SO(32)
superstring, NS-NS and R-R tadpoles, related by supersymmetry, cancel
at the same time, and therefore this vacuum configuration
involves D9 branes and corresponding O$9_+$ planes. However, it is
simple to generalize this model to a non-supersymmetric
configuration with a residual dilaton tadpole. 
To this end, let us
begin by assigning different reflection coefficients to the NS-NS and
R-R states flowing in $\tilde{\cal A}$ and $\tilde{\cal M}$, so that
\ba
\tilde{\cal A} &=& \frac{2^{-5}}{2} \int  
d \ell \ 
\frac{ (n_+ + n_-)^2 \, V_8 - (n_+-n_-)^2 \, S_8 }{\eta^8} \, , \\
\tilde{\cal M} &=& \frac{2}{2} \int d \ell  \ 
\frac{ \epsilon_{{\rm NS}} \, (n_+ + n_- ) \, \hat{V}_8 - \epsilon_{{\rm R}} 
\, (n_+ - n_-) \, \hat{S}_8 }{\hat{\eta}^8} \,,
\ea
while the corresponding direct-channel expressions become
\ba
\!\!{\cal A} \!\! &=& \!\! {\textstyle\frac{1}{2}} \int_0^\infty \,
\frac{d t}{t^6 \eta^8} 
\left[ (n_+^2 + n_-^2) \, ( V_8 - S_8) + 2 n_+ n_- \, 
(O_8 - C_8) \right]  \, , \\
\!\!{\cal M} \!\! &=& \!\! {\textstyle\frac{1}{2}} 
\int_0^\infty \,  \frac{d t}{t^6 \hat{\eta}^8} 
\left[ \epsilon_{{\rm NS}} \, (n_+ + n_- ) \, \hat{V}_8 - \epsilon_{{\rm R}} 
\, (n_+ - n_-) \, \hat{S}_8 \right] \,.
\ea
There are now two types of Chan-Paton charges, of
multiplicities $n_+$ and $n_-$, and two distinct sectors: strings
with like charges at their ends have the standard GSO projection, that involves
the vector and the $S$ spinor, while the
additional modes with unlike charges have the opposite projection,
that involves the tachyon and the $C$ spinor, as stressed
in \cite{senba,2benot2b}.

The two tadpole conditions
\ba
 \hbox{ NS-NS:}& \quad & 32 +  \epsilon_{{\rm NS}} \, (n_+ + n_- ) = 0 \, , 
\\
 \hbox{ R-R:}& \quad & 32 + \epsilon_{{\rm R}} \, (n_+ - n_-) = 0 \, ,
\ea
have a solution for the multiplicities, 
$\epsilon_{\rm NS}= 
\epsilon_{\rm R} = -1$,
$n_+=32$, $n_-=0$, that corresponds again to the
SO(32) superstring, but relaxing the NS-NS tadpole allows an
infinity of solutions, with, say, $\epsilon_{{\rm R}} =
\epsilon_{\rm NS} = -1$ and
$n_+ - n_- = 32$. From the transverse amplitudes one can read the
relative values of tensions and charges for the D-branes and
O-planes now present. In particular, from the R-R couplings one can
easily see that $n_+$ and $n_-$ count, respectively, the numbers of
D9 branes and D9 antibranes, while the two choices 
$\epsilon_{{\rm NS}} = \pm 1$ 
correspond to O$_\mp$ planes. $\tilde{\cal M}$
encodes rather neatly this last property, since its NS contribution
is sensitive to the relative signs of D-brane and
O-plane tensions. 
All configurations with 
$n_+ \neq 0$ and $n_- \neq 0$ have tachyonic instabilities, 
that reflect the mutual attraction of branes and antibranes \cite{senba}. 
Some definite progress has been made recently
in connection with these ``brane'' tachyons \cite{tachyon}, while
the current understanding
of ``bulk'' tachyons, present for instance in the ``parent'' 0A and 0B
models of subsection 3.1, is still far more primitive.
For $n_+=0$ there is also a new possibility, a non-supersymmetric
model, discovered in \cite{sugimoto}, with $\epsilon_{\rm NS}=
\epsilon_{\rm R}=+1$ and $n_-=32$,
involving D9 antibranes and
${\rm O}9_-$ planes. The resulting gauge group is USp(32), while the massless
spinors are still in the antisymmetric representation.

The emergence of a tree-level dilaton potential induced by 
the relaxed NS-NS tadpole condition, na\"{\i}vely incompatible
with bulk supersymmetry, is actually crucial in order to couple sensibly 
a non-supersymmetric open-string spectrum to a supersymmetric bulk:
on the branes supersymmetry is realized non-linearly, as in 
\cite{samwess}, and the 
dilaton tadpole is the leading term in the expansion of the
Volkov-Akulov action for the goldstino, the singlet in
the $496 = 495+1$ decomposition for the antisymmetric 
two-tensor of USp(32) \cite{dm1}.
The presence of the dilaton tadpole is incompatible with a maximally
symmetric Minkowski space, but the theory admits a background with
manifest SO(1,8) symmetry resulting from a warping of the ninth dimension
\cite{dmwarp}. The fate of this non-supersymmetric
vacuum and of its lower-dimensional analogues still
deserves a fuller investigation since, although tachyons are not present,
there is a net attraction between ${\rm O}_-$-planes and 
$\overline{\rm D}$-branes.

\vskip 12pt
\subsection{Open descendants of the 0A model}

We now turn to a richer class of non-supersymmetric models \cite{bs}. In order
to lighten the notation, from now on we shall mostly leave implicit all
arguments in the amplitudes, while also omitting all measure
factors and the contributions of non-compact bosonic coordinates.

Starting from the torus amplitude of eq. (\ref{0AB}), 
we can thus write the unique Klein bottle projection
\be
{\cal K} = {\textstyle \frac{1}{2}} \left( O_8 + V_8 \right)
 \, ,
\ee
that yields the transverse-channel amplitude
\be
\tilde{\cal K} = \frac{2^5}{2} \left( O_8 + V_8 \right)
 \,.
\ee
This is nicely consistent with the fact that only the NS-NS sectors can reflect
in a Lorentz-invariant fashion at a crosscap. Indeed,  
this operation turns each
of the four self-conjugate SO(8) characters into itself, and this is only
compatible with the propagation of the NS-NS modes, that
appear diagonally in ${\cal T}$. This Klein-bottle amplitude eliminates from
the low-lying closed spectrum the antisymmetric
two-tensor, leaving a tachyon, a graviton
and a dilaton, together with a vector and a three form, linear combinations
of those present in the R-R sectors of the original 0A model.

In a similar fashion, the most general Lorentz invariant
transverse-channel annulus may be written in the form
\be
\tilde{\cal A} = \frac{2^{-5}}{2} \left[ (n_B + n_F)^2 V_8 + (n_B - n_F)^2 O_8
\right] \, , 
\ee
and after an $S$ modular transformation and a rescaling of the modulus
becomes
\be
{\cal A} = {\textstyle \frac{1}{2}} (n_B^2 +n_F^2)
 (O_8 + V_8) - n_B n_F (S_8 + C_8 ) \,. \label{annulus0a}
\ee
Notice that the NS states carry pairs of like
Chan-Paton charges, and are consequently projected by the M\"obius 
amplitude, while the R states carry Chan-Paton charges of two different
types.

Finally, the transverse-channel M\"obius amplitude may be deduced from 
$\tilde{\cal K}$ and $\tilde{\cal A}$, taking geometric means of the
reflection coefficients of the individual characters, and reads
\be
\tilde{\cal M} = - {\textstyle \frac{2}{2}} \left[ (n_B + n_F) \hat{V}_8 +
(n_B - n_F) \hat{O}_8 \right] \, ,
\ee
so that, after a rescaling of the modulus and a $P$ transformation,
\be
{\cal M} = - {\textstyle \frac{1}{2}} \left[ (n_B + n_F) \hat{V}_8 -
(n_B - n_F) \hat{O}_8 \right] \,.
\ee

Enforcing the dilaton tadpole condition fixes the total Chan-Paton
multiplicity, so that $n_B + n_F = 32$, and the
result is the family of gauge groups
${\rm SO}(n_B) \times {\rm SO}(n_F)$. For all configurations with 
$n_F \neq 0$ the SO(32) gauge group is broken to a subgroup and, as 
is usually the case in String Theory,
this brings about new sectors. Thus, 
aside from the gauge vectors, the low-lying open spectrum generally includes 
tachyons in the $(\frac{n_B(n_B+1)}{2},1)$ and
$(1,\frac{n_F(n_F-1)}{2})$ representations and Majorana fermions in the
bi-fundamen\-tal $(n_B,n_F)$ representation. Moreover, as in the previous 
section, one has the option of relaxing the dilaton tadpole, and even of 
reversing the M\"obius projection, at the cost of a non-vanishing contribution 
to the vacuum energy, obtaining a pair of symplectic gauge groups
with spinors in the bi-fundamental representation.
It should be appreciated that no
R-R modes flow in the transverse channels, and that this spectrum is 
correspondingly not chiral and thus free of anomalies.
We shall return in the last section to the structure of the Chan-Paton
matrices of this model.
\vskip 12pt
\subsection{Open descendants of the 0B model}

This case is far richer \cite{bs,susy95}: it leads to several types 
of open descendants, all with chiral spectra, and shows, in a relatively 
simple setting, that the
Klein-bottle projection is in general {\it not unique} \cite{fps,pss,pss2}.
Let us begin by considering the simplest choice,
\be
{\cal K}_1 = {\textstyle\frac{1}{2}} 
\left( O_8 + V_8 - S_8 - C_8 \right) \, ,
\ee
that again symmetrizes the NS-NS sectors and antisymmetrizes the R-R ones.
The resulting low-lying modes comprise a tachyon, the metric tensor, the
dilaton, and a pair of R-R two-forms. The corresponding transverse-channel
amplitude
\be
\tilde{\cal K}_1 = \frac{2^6}{2} V_8 
\ee
is rather simple, and includes a single contribution,
while the transverse-channel annulus amplitude is
\ba
\tilde{\cal A}_1 &=& \frac{2^{-6}}{2} \biggl[
(n_o + n_v + n_s + n_c)^2 V_8 + (n_o + n_v - n_s - n_c)^2 O_8 \nonumber \\  
& & -  (-n_o + n_v + n_s - n_c)^2 S_8 - (-n_o + n_v - n_s + n_c)^2 C_8 
\biggr]
\ea
and, after a rescaling of the modulus and an $S$ transformation, gives
\ba
{\cal A}_1 &=& {\textstyle\frac{1}{2}} \left( n_o^2 + n_v^2 + n_s^2 + n_c^2
\right) V_8 + (n_o n_v + n_s n_c)
O_8
\nonumber \\
& & - (n_v n_s + n_o n_c) S_8 - (n_v n_c + n_o n_s) C_8 \,.
\label{annulusa1}
\ea
This expression reflects an important observation of
Cardy \cite{cardy2}, that plays a prominent r\^ole in Boundary Conformal Field
Theory. In this context, it can be summarized saying that in a model whose
sectors all flow in the transverse channel, as is the case for our 0B
examples, there is a one-to-one correspondence between allowed
boundary conditions and bulk sectors, and with boundary conditions $i$ and
$j$ the sectors flowing in ${\cal A}$ are determined, in terms of the 
fusion-rule coefficients ${\cal N}_{ij}^k$, by the sum
${\cal N}_{ij}^k \, \chi_k$.
We shall return to these important issues in the last section.
Finally,
the transverse-channel M\"obius amplitude, fully determined by $\tilde{\cal 
K}_1$ and $\tilde{\cal A}_1$, is
\be
\tilde{\cal M}_1 = - {\textstyle\frac{2}{2}} 
(n_o + n_v + n_s + n_c) \hat{V}_8 \, ,
\ee
and thus, after a rescaling of the modulus and a
$P$ transformation, one finds
\be
{\cal M}_1 = - {\textstyle\frac{1}{2}} 
(n_o + n_v + n_s + n_c) \hat{V}_8 \,.
\ee

There are now tadpole conditions for $V_8$, $S_8$ and $C_8$, the three
sectors that contain massless modes,
\ba
n_o + n_v + n_s + n_c &=& 64 \, ,\nonumber \\
n_o - n_v - n_s + n_c &=& 0 \, ,\nonumber \\
n_o - n_v + n_s - n_c &=& 0 \, , \label{obtadpoles}
\ea
so that $n_o=n_v$ and $n_s=n_c$, and the resulting gauge group is
${\rm SO}(n_o) \times {\rm SO}(n_v) \times {\rm SO}(n_s) \times {\rm SO}(n_c)$.
Aside from the
gauge vectors, the low-lying excitations comprise tachyons and 
fermions, all in different bi-fundamental representations and, 
as in the previous cases, one has also the options of relaxing the 
dilaton tadpole and of reversing the M\"obius contribution.
Notice that the open spectrum in eq. (\ref{annulusa1}) 
is {\it chiral}, since the $S_8$ and $C_8$ sectors are valued in 
different representations of the gauge group, but it is simple to 
see that the homogeneous 
R-R tadpole conditions eliminate all irreducible gauge anomalies.
A simple way to convince oneself of this important property is
by counting, say, the net number of fermions in the fundamental of the $n_o$
gauge group, that is $n_c - n_s$ and vanishes on account of the
R-R tadpoles in (\ref{obtadpoles}). More details on the anomaly
structure of these models may be found in \cite{susy95,schwitt}.

There are actually two more classes of models, associated to two
additional, inequivalent, choices of ${\cal K}$ \cite{susy95}:
\ba
{\cal K}_2 &=& {\textstyle\frac{1}{2}} 
\left( O_8 + V_8 + S_8 + C_8 \right) \, ,
\nonumber \\
{\cal K}_3 &=& {\textstyle\frac{1}{2}} 
\left( - O_8 + V_8 + S_8 - C_8 \right) \,,
\ea
and the corresponding transverse-channel amplitudes are
\ba
\tilde{\cal K}_2 &=& \frac{2^6}{2} O_8  \, ,
\nonumber \\
\tilde{\cal K}_3 &=& - \frac{2^6}{2} C_8 \,.
\ea
The first choice results in a low-lying spectrum
without a massless two-form, since, aside from the tachyon, one is left with
the graviton, the dilaton, a R-R scalar and an unconstrained four-form.
The last case is particularly interesting: the projected closed
spectrum does not contain tachyons, and is {\it chiral}, since it contains
a self-dual four-form,
left over from the {\it symmetrization} of the $|S_8|^2$ sector. 
The corresponding ${\cal A}$ and ${\cal M}$ amplitudes 
may be simply recovered ``fusing"
the various terms in the first model with $O_8$ and $-C_8$, respectively,
a procedure that we shall discuss further in the last section.

Thus, for model 2
\ba
{\cal A}_2 &=& {\textstyle\frac{1}{2}} \left( n_o^2 + n_v^2 + 
n_s^2 + n_c^2 \right) O_8 + (n_o n_v + n_s n_c)
V_8
\nonumber \\
& &- (n_v n_s + n_o n_c) C_8 - (n_v n_c + n_o n_s) S_8 
\ea
and
\be
{\cal M}_2 = \mp {\textstyle \frac{1}{2}} 
(n_o + n_v - n_s - n_c) \hat{O}_8 \, ,
\ee
while the corresponding transverse-channel amplitudes are
\ba
\tilde{\cal A}_2 &=& \frac{2^{-6}}{2} \biggl[
(n_o + n_v + n_s + n_c)^2 V_8 + (n_o + n_v - n_s - n_c)^2 O_8 \nonumber \\  
& & + (n_o - n_v + n_s - n_c)^2 C_8 + (n_o - n_v - n_s + n_c)^2 S_8 \biggr]
\ea
and
\be
\tilde{\cal M}_2 = \pm {\textstyle\frac{2}{2}} 
(n_o + n_v - n_s - n_c) \hat{O}_8 \, .
\ee
Since $V_8$ does not appear in ${\cal M}$, we must reinterpret
the charges in terms of unitary groups, letting $n_o= n_b$, $n_v = \bar{n}_b$,
$n_s= n_f$, $n_c = \bar{n}_f$, so that
\ba
{\cal A}_2 &=& {\textstyle\frac{1}{2}} \left(
n_b^2 + \bar{n}_b^2 + n_f^2 + \bar{n}_f^2 \right) O_8 + 
(n_b \bar{n}_b + n_f \bar{n}_f) V_8
\nonumber \\
& &- (\bar{n}_b n_f + n_b \bar{n}_f) C_8 - (n_b n_f + \bar{n}_b \bar{n}_f) S_8 
\, ,
\ea
while the $S_8$ and $C_8$ contributions to $\tilde{\cal A}_2$  vanish
if $n_i = \bar n _i$.
As in the previous case, the model has a chiral spectrum with
no net excess of chiral fermions, as demanded by the projected non-chiral
closed spectrum, that does not contribute to the gravitational anomaly. 
Aside from the ${\rm U}(n_b) \times {\rm 
U}(n_f)$
gauge bosons, the low-lying excitations comprise tachyons in symmetric
and antisymmetric
representations and chiral fermions in several bi-fundamental combinations.

Finally, for model 3 we start from
\ba
{\cal A}_3 &=& - {\textstyle\frac{1}{2}} \left(
n_o^2 + n_v^2 + n_s^2 + n_c^2 \right) C_8 - 
(n_o n_v + n_s n_c) S_8
\nonumber \\
& &+ (n_v n_s + n_o n_c) V_8 + (n_v n_c + n_o n_s) O_8 
\ea
and
\be
{\cal M}_3 = {\textstyle\frac{1}{2}} (n_o - n_v - n_s + n_c) \hat{C}_8 \, ,
\ee
and the corresponding transverse-channel amplitudes are then
\ba
\tilde{\cal A}_3 &=& \frac{2^{-6}}{2} \biggl[
(n_o + n_v + n_s + n_c)^2 V_8 - (n_o + n_v - n_s - n_c)^2 O_8 \nonumber \\  
& &- (n_o - n_v - n_s + n_c)^2 C_8 + (n_o - n_v + n_s - n_c)^2 S_8 \biggr]
\ea
and
\be
\tilde{\cal M}_3 = {\textstyle\frac{2}{2}} 
(n_o - n_v - n_s + n_c) \hat{C}_8 \,.
\ee

Since $V_8$ does not appear in ${\cal M}$, we must again reinterpret
the charges in terms of unitary groups, letting $n_v= n$, $n_s = \bar{n}$,
$n_o= m$, $n_c = \bar{m}$, so that
\ba
{\cal A}_3 &=& - {\textstyle\frac{1}{2}} 
\left(n^2 + \bar{n}^2 + m^2 + \bar{m}^2 \right) C_8 + 
(n \bar{n} + m \bar{m}) V_8
\nonumber \\
& &- (m n + \bar{m} \bar{n}) S_8 + (m \bar{n}  + \bar{m} n) O_8 
\ea
and
\be
{\cal M}_3 = {\textstyle\frac{1}{2}} (m + \bar{m} - n - \bar{n} ) \hat{C}_8 
\,.
\ee
The tadpole conditions now fix $m = 32 + n$, and in particular the choice
$n=0$ eliminates all tachyons also from the open spectrum. The resulting
model, usually termed $0^\prime{\rm B}$ in the literature, contains a
net number of chiral fermions, precisely as needed to cancel gravitational
anomalies \cite{susy95}. One can actually see that the gauge group is 
effectively ${\rm SU}(32)$, since the ${\rm U}(1)$
factor is anomalous and the corresponding gauge vector disappears
from the low-energy spectrum, by a mechanism similar to that discussed in
\cite{wito32,dsw} for four-dimensional models.

All these models can be given a geometric interpretation in terms of 
suitable collections of D-branes and O-planes. This is
easier for the open descendants of the 0B model, since in this case there
is a one-to-one correspondence between boundaries and sectors
of the open spectrum. In Conformal Field Theory, this situation corresponds
to the Cardy case \cite{cardy2}, that is indeed the simplest 
possibility. On the other hand,
the 0A model is effectively more complicated, since its boundaries are 
combinations of these, and as a result its branes are uncharged combinations of
the 0B ones. The relation between charged and uncharged D-branes is 
discussed further in subsection 5.12.
Returning to the 0B case, one should first observe that,
since the R-R sector is doubled, both D-branes and O-planes now 
carry a pair of R-R charges, a property to be contrasted with the type I case.
In particular, there are two types of D-branes, D$9^{(1)}$ and D$9^{(2)}$, both
with positive tension but with charges $(+,+)$ and $(+,-)$ with respect
to the 10-forms associated to $S_8$ and $C_8$, as 
first stressed in \cite{kt}. 
On the other hand, the additional option of reversing the tension 
allows one to define {\it four} types of orientifold planes, 
identified by the signs of their
couplings to the $(O_8,V_8,-S_8,-C_8)$ sectors:
O$9_{\pm}^{(1)}$, with couplings $(\mp,\mp,\mp,\mp)$,   
O$9_{\pm}^{(2)}$, with couplings $(\pm,\mp,\mp,\pm)$.
Moreover,
as usual, there are corresponding $\overline{\rm D}$-branes and 
$\overline{\rm O}$-planes, with identical tensions and opposite R-R charges.

A closer look at the transverse amplitudes allows one to identify the
types of objects involved in these models. Thus, from $\tilde{\cal K}$
and $\tilde{\cal M}$,
one can argue that combinations of
four types of these O-planes are involved in each of the 
0B orientifolds, so that
\ba
\tilde{\cal K}_1 &\to& {\rm O}9_{\pm}^{(1)} \, \oplus \,
{\rm O}9_{\pm}^{(2)} \, \oplus \, \overline{{\rm O}9}_{\pm}^{(1)} \, 
\oplus \, \overline{{\rm O}9}_{\pm}^{(2)} \, , \\
\tilde{\cal K}_2 &\to& {\rm O}9_{\mp}^{(1)} \, \oplus \,
{\rm O}9_{\pm}^{(2)} \, \oplus \, \overline{{\rm O}9}_{\mp}^{(1)} \, 
\oplus \, \overline{{\rm O}9}_{\pm}^{(2)} \, , \\
\tilde{\cal K}_3 &\to& {\rm O}9_{\mp}^{(1)} \, \oplus \,
{{\rm O}9}_{\pm}^{(2)} \, \oplus \, 
\overline{{\rm O}9}_{\pm}^{(1)} \, 
\oplus \, \overline{{\rm O}9}_{\mp}^{(2)} \, ,
\ea
where the double choices for the signs reflect the possibility of
reversing the M\"obius projection consistently with the cancellation of
R-R tadpoles. For the sake of comparison, the orientifold planes involved
in type I constructions may also be formally regarded as bound states 
of pairs of
these objects, with vanishing $C_8$ charge, and with 
tension and $S_8$ charges that are $\sqrt{2}$ times larger than
for ${\rm O}9_{\pm}^{(1)}$ and ${\rm O}9_{\pm}^{(2)}$, 
as can be seen from the normalization of the amplitudes. 

Finally, from $\tilde{\cal A}_1$
one can also identify rather neatly the types of D9 branes involved in
this class of descendants, where
\be
n_v \, \rightarrow \, {\rm D}9^{(1)} \, , \quad 
n_o \, \rightarrow \, \overline{{\rm D}9}^{(1)} \, , \quad 
n_s \, \rightarrow \, {\rm D}9^{(2)} \, , \quad 
n_c \, \rightarrow \, \overline{{\rm D}9}^{(2)} \,,
\ee
where all (anti)branes are subject to the orientifold projection.
This is rather standard and simple to understand, since in
this canonical case all types of branes are fixed by the
$\Omega$ projection. The other cases are more peculiar, and thus
more interesting. Let us thus begin by considering
the second model, where the reversed Klein-bottle projection of the
R-R states may be ascribed to the replacement of the standard $\Omega$
by $\Omega_2 = \Omega (-1)^{F_{\rm L}}$ \cite{bfl1}, 
with $F_{\rm L}$
the space-time fermion number for the left-moving modes.
Turning to the transverse Klein-bottle amplitude via an $S$ transformation,
one can deduce the effect of $\Omega_2$ on $\tilde{\cal M}$,
that is now determined by $\Omega (-1)^{f_{\rm L}}$, where $f_{\rm L}$ denotes
the {\it world-sheet} fermion number. Then,
after a $P$ transformation, one can read the new flip operator for the
open spectrum, $\Omega (-1)^{f}$, where $f$ denotes the world-sheet fermion
number for the open sector. This acts on the Chan-Paton
charge-space as
\ba
 \left( \begin{array}{c}
n_v \\
n_o \\
n_s \\
n_c 
\end{array} \right) \to  \left( \begin{array}{cccc}
0 & 1 & 0 & 0 \\
-1 & 0 & 0 & 0 \\
0 & 0 & 0 & 1 \\
0 & 0  & -1 & 0 
\end{array} \right) \left( \begin{array}{c}
n_v \\
n_o \\
n_s \\
n_c 
\end{array} \right) \,.
\ea
The new ``eigencharges'' that lead to a direct-channel annulus 
amplitude compatible with spin-statistics are then
\ba
n_b &={\displaystyle \frac{n_o e^{i\pi/4} + n_v e^{-i\pi/4}}{\sqrt{2}}} \, , 
\qquad 
n_f &= \frac{n_s e^{i\pi/4} + n_c e^{-i\pi/4}}{\sqrt{2}} \, , \nonumber \\
\bar{n}_b &= {\displaystyle
\frac{n_o e^{-i\pi/4} + n_v e^{i\pi/4}}{\sqrt{2}}} \, , \qquad 
\bar{n}_f &= \frac{n_s e^{-i\pi/4} + n_c e^{i\pi/4}}{\sqrt{2}} \,. 
\ea
This implies a similar redefinition for the R-R ten-forms, and therefore
the branes of this model are ``complex'' superpositions of those
present in the ``parent'' 0B model.

The third model presents similar features. Here one starts from $\Omega_3 =
\Omega (-1)^{f_{\rm L}}$ \cite{bfl1} 
for the closed spectrum, and the same sequence of $S$
and $P$ transformations determines its action, $\Omega (-1)^{F}$, on the Chan-Paton charge space. As a result, the ``eigencharges'' for this case are
\ba
n &={\displaystyle \frac{n_v e^{i\pi/4} + n_c e^{-i\pi/4}}{\sqrt{2}}} \, , 
\qquad 
m &= \frac{n_o e^{i\pi/4} + n_s e^{-i\pi/4}}{\sqrt{2}} \, , \nonumber \\
\bar{n} &= {\displaystyle
\frac{n_v e^{-i\pi/4} + n_c e^{i\pi/4}}{\sqrt{2}}} \, , \qquad 
\bar{m} &= \frac{n_o e^{-i\pi/4} + n_s e^{i\pi/4}}{\sqrt{2}} \,. 
\ea


\newsec{Toroidal compactification}

Toroidal compactifications display several interesting new features in a
relatively simple context. In this section we begin by considering
the compactification of the type IIB superstring on a circle and
describe the new type of deformation allowed in its open
descendant, the type I string. This corresponds to the
introduction of continuous Wilson lines, that break the gauge
group while preserving its overall rank. T-dualities turn these
momentum-space shifts into coordinate-space displacements of D8-branes,
on which the ends of open strings terminate. We then move on to
higher-dimensional tori, where a new type of phenomenon can occur: 
discrete deformations can give rise to a rank reduction of the 
Chan-Paton gauge group, while also allowing to continuously connect
orthogonal and symplectic gauge groups.
\vskip 12pt
\subsection{The one-dimensional torus} 

Let us begin by describing
the compactification on a circle of radius $R$. The closed spectrum
presents a long-appreciated surprise: in addition to  the usual
Kaluza-Klein momentum modes, quantized in terms of
an integer $m$ and familiar from point-particle theories, it
includes an infinity of topologically distinct sectors,
associated to closed strings wrapped
$n$ times around the circle \cite{cs,narain}. Given this interpretation, the
integer $n$ is usually called a ``winding number". This is
neatly summarized in the expansion
\be 
X = x + 2 \alpha' \, \frac{m}{R} \ \tau +  2 \, n \, R \,
\sigma + {\rm (oscillators)} \, ,
\ee 
but the structure of the zero modes is often better emphasized
defining the two combinations
\be 
X_{\rm L,R} = {\textstyle\frac{1}{2}} x +  \alpha' p_{L,R} (\tau \mp \sigma)
+ {\rm (oscillators)_{L,R}} \, ,
\ee 
where the two chiral components $p_{\rm L}$ and $p_{\rm R}$
associated to the compact coordinate are defined as
\be 
p_{\rm L,R} = \frac{m}{R} \pm \frac{n R}{\alpha'}  \,. \label{onedimmom}
\ee 
These chiral momenta play a crucial r\^ole in the definition of the
corresponding torus amplitude since, if one of the non-compact
coordinates of a critical string is replaced with a compact
one, the continuous integration over internal momenta is
replaced by a lattice sum, according to
\be
\frac{1}{\sqrt{\tau_2} \eta(\tau) \eta(\bar{\tau})} \to \sum_{m,n}
\frac{ q^{\alpha' p_{\rm L}^2/4} \; \bar{q}^{\alpha' p_{\rm R}^2/4}}{\eta(\tau)
\eta(\bar{\tau})} \,.
\ee 

The lattice sum displays a remarkable symmetry under the
interchange of the two apparently unrelated quantum numbers $m$
and $n$, provided this is accompanied by a corresponding
inversion of the radius, $R \to \alpha'/R$. This is
the simplest instance of a T-duality \cite{cs} and, out of all
models previously discussed, it is actually a
symmetry only for bosonic strings. On the
other hand this operation, a parity transformation
on right-moving world-sheet modes, flips the chirality of the
corresponding GSO projection, mapping the
type IIA and type IIB strings into one another, and relates in
a similar fashion the SO(32) and
${\rm E}_8 \times{\rm E}_8 $ heterotic models \cite{ginsp}. For open
strings the situation is even more subtle, since T-duality
affects the boundary conditions at the string endpoints 
\cite{dlp,horava2,green}. This issue will be discussed further 
in the following. 

Returning to the type IIB model, and confining once more our
attention to the fermion modes and to the contribution
of the internal circle, the partition function is
\be 
{\cal T} = |V_8 - S_8 |^2 \sum_{m,n} \frac{ q^{\alpha'
p_{\rm L}^2/4} \; \bar{q}^{\alpha' 
p_{\rm R}^2/4}}{\eta(\tau) \eta(\bar{\tau})}
\,.
\ee
In order to construct the open descendants \cite{bps}, we should begin by
introducing a Klein-bottle projection. For a generic internal
radius, the only states allowed in the Klein-bottle are those
with $p_{\rm L}=p_{\rm R}$ or, equivalently, with vanishing winding number $n$.
This is easy to see considering a generic vertex operator
\be 
V = e^{i( p_{\rm L} X_{\rm L} + p_{\rm R} X_{\rm R} )} \, ,
\ee 
for which the interchange of $X_{\rm L}$ and $X_{\rm R}$ is clearly
equivalent to the interchange of $p_{\rm L}$ and $p_{\rm R}$, so that
only the states with $n=0$ are fixed and flow 
in the Klein bottle to complete the projection. Therefore
\be 
{\cal K} = {\textstyle\frac{1}{2}} \, (V_8 - S_8) (q^2) \sum_{m}
\frac{ q^{\alpha' m^2/2 R^2}}{\eta(2 i \tau_2)} \, , \label{k1s}
\ee 
but actually this is not the only possible choice for $\Omega$
in this case: one has indeed the option of assigning different
$\Omega$  eigenvalues to different lattice states, provided
this is compatible with the nature of the string interactions 
\cite{fps,pss,pss2}. This
leaves the interesting alternative \cite{dp,gepner}
\be 
{\cal K}' = {\textstyle\frac{1}{2}}\, (V_8 - S_8) (q^2) \sum_{m} \
(-1)^m \ \frac{ q^{\alpha' m^2/2 R^2}}{\eta(2 i \tau_2)} \, ,
\label{k1ns}
\ee 
equivalent to combining the original world-sheet projection
with a  shift along the circle by half of its length. 

These two choices have vastly different effects, as can be seen
turning  eqs. (\ref{k1s}) and (\ref{k1ns}) into the
corresponding transverse-channel amplitudes
\ba
\tilde{\cal K} &=& \frac{2^5}{2} \frac{R}{\sqrt{\alpha'}} \ 
(V_8 - S_8) (i \ell) \sum_{n} \frac{ q^{n^2
R^2/\alpha'}}{\eta(i \ell)} \, ,
\\
\tilde{\cal K}' &=& \frac{2^5}{2} \frac{R}{\sqrt{\alpha'}} \ 
(V_8 - S_8) (i \ell) \sum_{n} \frac{ q^{(n+1/2)^2
R^2/\alpha'}}{\eta(i \ell)}
\, .
\ea 
Here $q = e^{-2\pi \ell}$, and we have used the Poisson summation
formula
\be
\sum_{\{n_i\} \in \sbb{Z}} \, e^{-\pi \, n^{\rm T}\, A \,n \ + \ 2 \, i 
\, \pi \, b^{\rm T} \, n} \ = \
\frac{1}{\sqrt{{\rm det}(A)}} \, \sum_{\{m_i\} \in \sbb{Z}}
 \, e^{-\pi \, (m - b)^{\rm T} \, A^{-1} \, (m - b)} \, , \label{poissonsum}
\ee
where in general $A$ is a $d \times d$ square matrix and $m$ and $n$
are $d$-dimensional vectors,
to connect direct and transverse channels. In the second case the R-R
tadpole is lifted in mass: the theory does not need an open
sector, and can thus be restricted to only unoriented closed strings!

Returning to the standard projection, we can now proceed to
introduce the open sector. The simplest choice, 
\ba 
{\cal A} &=& {\textstyle\frac{1}{2}} \, N^2 \, (V_8 - S_8) 
({\textstyle{1\over 2}} i\tau_2) 
\sum_{m} \ \frac{ q^{\alpha' m^2/2 R^2}}{
\eta({\textstyle{1\over 2}} i \tau_2)} \, ,
\\ {\cal M} &=& - {\textstyle \frac{1}{2}} \, N \, (\hat{V}_8 - \hat{S}_8)
({\textstyle{1\over 2}} i \tau_2 + {\textstyle{1\over 2}}) \sum_{m} \
\frac{ q^{\alpha' m^2/2 R^2}}{\hat{\eta}
({\textstyle{1\over 2}} i \tau_2+{\textstyle{1\over 2}})} \, ,
\ea 
does not affect the gauge group, that is still 
SO(32) even after the compactification, and the corresponding
transverse-channel amplitudes
\ba
\tilde{\cal A} &=& \frac{2^5}{2}\, N^2 \, \frac{R}{\sqrt{\alpha'}}
\ (V_8 - S_8) \ (i \ell) \sum_{n} \
\frac{ q^{n^2 R^2/ 4 \alpha'}}{\eta(i \ell)} \, , \\
\tilde{\cal M} &=& - {\textstyle\frac{2}{2}} \, N \, \frac{R}{\sqrt{\alpha'}} \
(\hat{V}_8 -
\hat{S}_8) (i \ell+ {\textstyle\frac{1}{2}}) \sum_{n} \
\frac{ q^{n^2 R^2/\alpha'}}{\hat{\eta}(i \ell+ {\textstyle\frac{1}{2}}
)} \, ,
\ea 
are also simple generalizations of the ten-dimensional
ones discussed in the previous section. 

The open sector actually allows an interesting type of
continuous deformation, obtained inserting
constant Wilson lines \cite{bps} along boundaries or,
equivalently, translating the momenta of open-string states 
according to their charges. In the transverse channel, this
deformation corresponds to altering by phases the reflection
coefficients of the closed-string modes, while preserving their
structure of perfect squares. It is instructive to consider 
a simple instance, corresponding to the breaking of ${\rm
SO}(32)$ to
${\rm U}(M) \times {\rm SO}(N)$, with $2 M + N =32$.  In this
case, one is actually splitting the 32 charges in the
fundamental representation of SO(32) into three sets.
The first two comprise $M$ charges, and for the charges in the first
set the momentum quantum number $m$ is shifted to $m+a$, while for
those in the second set it is shifted to $m-a$. Finally, for the charges
in the last set the momentum is unaffected. The deformed annulus
amplitude is then
\ba 
{\cal A} &=&  (V_8 - S_8) ({\textstyle\frac{1}{2}}i \tau_2) \sum_{m} \
\biggl\{
\biggl( M \bar{M} + {\textstyle\frac{1}{2}} \, N^2 \biggr)  
\frac{q^{\alpha'
m^2/2 R^2}}{\eta({\textstyle\frac{1}{2}} i \tau_2)} \nonumber \\ & &+ M N \
\frac{q^{\alpha' (m+a)^2/2 R^2}}{\eta(
{\textstyle\frac{1}{2}} i \tau_2)} +  \bar{M} N \
\frac{q^{\alpha' (m-a)^2/2 R^2}}{\eta({\textstyle\frac{1}{2}}i \tau_2)} 
\nonumber \\
& &+
{\textstyle\frac{1}{2}} \, M^2 \ \frac{q^{\alpha' (m+2a)^2/2
R^2}}{\eta({\textstyle\frac{1}{2}} i \tau_2)} +  
{\textstyle\frac{1}{2}}\, \bar{M}^2 
\ \frac{q^{\alpha'
(m-2a)^2/2 R^2}}{\eta({\textstyle\frac{1}{2}} i \tau_2)} \biggr\} \, , 
\ea 
while the corresponding M\"obius amplitude is
\ba 
{\cal M} &=& - (\hat{V}_8 - \hat{S}_8) 
({\textstyle\frac{1}{2}} i \tau_2 + {\textstyle\frac{1}{2}}) 
\sum_{m} \ \biggl\{
 {\textstyle\frac{1}{2}}\, N  \frac{q^{\alpha' m^2/2
R^2}}{\hat{\eta}(
{\textstyle\frac{1}{2}} i \tau_2+ {\textstyle\frac{1}{2}})} 
\nonumber \\ 
& &+ 
{\textstyle\frac{1}{2}} \, M \ \frac{q^{\alpha' (m+2a)^2/2
R^2}}{\hat{\eta}
({\textstyle\frac{1}{2}}i \tau_2+{\textstyle\frac{1}{2}})} 
+  {\textstyle\frac{1}{2}} \, \bar{M} \
\frac{q^{\alpha' (m-2a)^2/2 R^2}}{\hat{\eta}
({\textstyle\frac{1}{2}} i \tau_2+{\textstyle\frac{1}{2}})}
\biggr\} \,. 
\ea

The structure of this important deformation may be better
appreciated considering the corresponding transverse-channel
amplitudes
\ba
\tilde{\cal A} &=& \frac{2^{-5}}{2} \frac{R}{\sqrt{\alpha'}} \
(V_8 - S_8) (i \ell)
\nonumber \\ & &\times
 \sum_{n} \ 
\frac{ q^{n^2 R^2/ 4 \alpha'}}{\eta(i \ell)} \ \left( N + M
\, e^{2 i \pi a n} +  
\bar{M} \, e^{- 2 i \pi a n}    \right)^2 \, , \label{aawil}
\ea
\ba
\tilde{\cal M} &=& - \frac{2}{2} \frac{R}{\sqrt{\alpha'}} \
(\hat{V}_8 - \hat{S}_8)  (i \ell + {\textstyle\frac{1}{2}})
\nonumber \\ & &\times
 \sum_{n} \ 
\frac{ q^{n^2 R^2/ \alpha'}}{\eta(i \ell+ {\textstyle\frac{1}{2}})} 
\ \left( N + M\, 
e^{4 i \pi a n} +  
\bar{M}\, e^{- 4 i \pi a n}    \right) \, , \label{amwil}
\ea 
where the phases are easily associated to the Wilson lines
of the internal components of the gauge vectors. These
expressions display an amusing phenomenon for $a=\frac{1}{2}$: in this
case the two ``complex" charges $M$ and $\bar{M}$ have
identical reflection coefficients in $\tilde{\cal A}$ and
$\tilde{\cal M}$, so that the  direct-channel amplitudes only
depend on their sum and, as a result, the ${\rm U}(M)$ gauge group is
enhanced to ${\rm SO}(2M)$. This procedure can thus generate,
via continuous deformations, all rank-preserving breakings
${\rm SO}(32) \to {\rm SO}(2M) \times {\rm SO}(32 - 2M)$. 

\begin{figure}
\begin{center}
\epsfbox{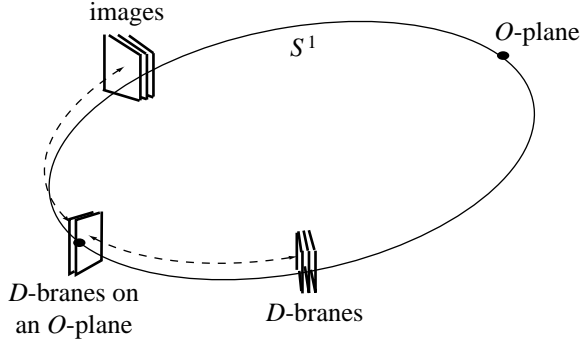}
\end{center}
\caption{Moving a brane away from an orientifold point.}
\label{figtoroidal}
\end{figure}

Chan-Paton symmetry-breaking has an alternative interpretation 
in terms of D-brane displacements in the dual coordinate space 
\cite{horava2,pcj}. 
After a T-duality in the compact dimension, that turns the Neumann boundary
condition of the corresponding coordinate into a Dirichlet one, 
the model lives in a
segment $S^1/\bb{Z}_2$, with a pair of O8 planes at its ends.
In this description, D8 branes sitting at the two endpoints lead to orthogonal 
gauge groups, while D8 branes in the 
interior lead to unitary gauge groups. The previous results can 
thus be recovered moving the branes from one fixed point to the interior and 
then to the second fixed point, as shown in figure \ref{figtoroidal}.
These phenomena  have a natural counterpart in the low-energy
effective field theory, where adjoint scalars can acquire
vacuum expectation values in the Cartan sub-algebra of the gauge group,
compatibly with the vanishing of the corresponding potential,
that in the maximally supersymmetric Yang-Mills theory involves
only their commutators, and thus preserving supersymmetry. Open
strings apparently allow another possibility, 
breakings via parity-like elements of ${\rm
O}(32)$, that as such are not contained in ${\rm SO}(32)$, nor in
the ${\rm Spin}(32)/\bb{Z}_2$ gauge group of the dual heterotic string. 
An example of this type is the breaking ${\rm SO}(32) \to {\rm SO}(17) 
\times {\rm SO}(15)$, that does not preserve the overall 
rank \cite{bps}, but the duality with the ${\rm Spin}(32)/\bb{Z}_2$
heterotic string suggests that breakings of this type would lead to 
inconsistencies in a full non-perturbative treatment \cite{dgg}.

Before leaving the one-dimensional case, it is worth
taking a closer look at these amplitudes, noting that
$\tilde{\cal K}$ and $\tilde{\cal M}$
involve {\it only} even windings, while $\tilde{\cal A}$ involves {\it both}
even and odd ones. This, however, is not demanded by the 
reflective conditions at the ends of the tube in the tree channel, 
that for the two cases are
\ba
{\rm boundary:}& \quad & \frac{\partial}{\partial \tau}\, X_L(\tau - \sigma) = - \, \frac{\partial}{\partial \tau}\, X_R(\tau + \sigma) \, , 
\nonumber \\
{\rm crosscap:}& \quad & 
\frac{\partial}{\partial \tau}\, X_L(\tau - \sigma) = - \, \frac{\partial}{\partial \tau}\,X_R(\tau + \sigma + {\textstyle\frac{1}{2}}\pi) 
\, ,
\ea
and thus have the same effect on zero modes. Still, the presence
of different
lattice sums has a sizable consequence: in the $R \to 0$ limit 
all winding modes collapse to zero mass, and
the resulting odd-level tadpoles in $\tilde{\cal A}$ are apparently unmatched. 
This problem can actually be cured introducing suitable Wilson lines, and to
this end let us reconsider eq. (\ref{aawil}) for the special case 
$a=\frac{1}{2}$.
In the T-dual picture, as we have seen, this corresponds to placing
the branes at the two O8 planes, and
\be
\tilde{\cal A} \sim  
(V_8 - S_8) \sum_{n} 
\frac{ \left[ N_1 + N_2 (-1)^n
  \right]^2\; q^{n^2 R^2/ 4 \alpha'}}{\eta}   \, . 
\ee
Therefore the choice $N_1=N_2$, corresponding to the gauge group
${\rm SO}(16) \times {\rm SO}(16)$ removes all odd windings
from $\tilde{\cal A}$ and solves the problem of the collapsing 
tadpoles \cite{pw}.

There is actually a neater way to understand this result. To this end, let us
rewrite $\tilde{\cal K}$ and $\tilde{\cal M}$ in the form \cite{emrev}
\ba
\tilde{\cal K} &\sim&  16^2 \; 
(V_8 - S_8)  \sum_{n}  \frac{ \left[1 + (-1)^n\right]^2 \; q^{n^2
R^2/4 \alpha'}}{\eta} \, ,  \\
\tilde{\cal M} &\sim& - 2 \times 16 \;
(\hat{V}_8 -
\hat{S}_8)  \sum_{n} 
\frac{ \left[N_1 + N_2 (-1)^n \right]\left[1 + (-1)^n \right]\; 
q^{n^2 R^2/4\alpha'}}{\hat{\eta}} \,. \nonumber
\ea 
We can now collect all the zero-mode contributions to the three 
transverse-channel amplitudes in the form
\be
\sum_n  q^{n^2 R^2/4 \alpha'} \; \left[ (16 - N_1) + 
(16 - N_2) (-1)^n \right]^2 \, , \label{tadpolecirclelocal}
\ee
that shows very clearly how the gauge group ${\rm SO}(16) \times {\rm SO}(16)$
has the unique virtue of eliminating the tadpoles of {\it all} winding modes.
In the T-dual picture, this configuration corresponds to saturating
tadpoles {\it locally} at the two ${\rm O}_+$ planes, since the
cancellation continues to hold in the limit of very large dual radius,
when branes away from the orientifolds would move to an infinite distance
from them. The phases present in (\ref{tadpolecirclelocal}) and in the
previous amplitudes are indeed Fourier coefficients, that reflect rather
clearly the relative positions of these objects on the dual circle.

Repeating the same exercise for the Klein-bottle
projection of eq. (\ref{k1ns}), 
\be
\tilde{\cal K} \sim  16^2 \; 
(V_8 - S_8)  \sum_{n}  \frac{ \left[1 - (-1)^n\right]^2 \; q^{n^2
R^2/4 \alpha'}}{\eta} \, ,
\ee
it is easy to see that in this case the T-dual interval has one 
${\rm O}_+$ plane and one
${\rm O}_-$ plane at its ends. One can  again cancel locally the
R-R tadpoles, but now placing 16 branes at the ${\rm O}_+$ plane and 16 
antibranes at the ${\rm O}_-$ plane. This configuration, however, 
breaks supersymmetry, has
a dilaton tadpole, and is expected to be unstable, as a result
of the net force between branes and antibranes.

We can also reconsider the non-supersymmetric 
ten-dimensional USp(32) model \cite{sugimoto} and study its compactification
to $D=9$. We have already seen in subsection 3.2 how in ten 
dimensions the simultaneous
presence of branes and antibranes, corresponding to nonzero values
for both $n_+$ and $n_-$, results in the appearance of tachyonic modes.
In the circle reduction, if the radius $R$ is sufficiently small, 
the tachyonic modes can be lifted by
suitable Wilson lines. For instance, a discrete Wilson line, that in the
T-dual picture would correspond to placing the (anti)branes at the
two ${\rm O}_-$-planes, would result in the open spectrum
\ba
{\cal A}  &=&  {\textstyle\frac{1}{2}} \sum_{m} \,
\left[ (n_+^2 + n_-^2) \, ( V_8 - S_8) \frac{ q^{\alpha' m^2/ 2 R^2}}{\eta} 
\right. 
\nonumber \\ & & \left. + 2 \, n_+ n_- \, 
(O_8 - C_8) \frac{ q^{\alpha' (m+1/2)^2/ 2 R^2}}{\eta} \right]  \, , \\
{\cal M} &=& {\textstyle\frac{1}{2}} \sum_{m} \,
\left[ (n_+ + n_- ) \, \hat{V}_8 + 
(n_- - n_+) \, \hat{S}_8 \right] \frac{ q^{\alpha' m^2/ 2 R^2}}{\hat\eta} \, ,
\ea
with $n_- - n_+ = 32$,
where for the massless modes supersymmetry is again broken by the 
M\"obius projection or, equivalently, by the presence of O$_-$, rather
than O$_+$, planes. One would therefore expect that supersymmetry
be recovered, at least for the massless modes, if a small continuous 
Wilson line $a$ were introduced in the non-tachyonic model with
$n_+=0$. In the T-dual picture, this deformation would
correspond to moving the antibranes slightly away from the ${\rm O}_-$-planes
and, indeed, in this case a generic Wilson line would break 
the original group, USp(32), to U(16), while recovering a global
supersymmetry for the massless brane modes. The corresponding
amplitudes
\ba
{\cal A} &=&  {\textstyle\frac{1}{2}} ( V_8 - S_8) 
\sum_{m} \,
\left[ 2  M \bar M  \; 
\frac{ q^{\alpha' m^2/ 2 R^2}}{\eta} \right. \nonumber \\
& & \left. + 
 M^2  \; \frac{ q^{\alpha' (m+2a)^2/ 2 R^2}}{\eta} 
+ 
 \bar{M}^2 \; \frac{ q^{\alpha' (m-2a)^2/ 2 R^2}}{\eta} \right] \, ,
\\
{\cal M} &=&  {\textstyle\frac{1}{2}} ( \hat{V}_8 + \hat{S}_8) 
\sum_{m} \ \left[
  M \; \frac{ q^{\alpha' (m+2a)^2/ 2 R^2}}{\hat{\eta}}  +
  \bar{M} \; \frac{ q^{\alpha' (m-2a)^2/ 2 R^2}}{\hat{\eta}} \right] \, ,
\ea
display very clearly the phenomenon:
the massless modes, unaffected by the M\"obius projection, now
fill entire vector multiplets.

\vskip 12pt
\subsection{Higher-dimensional tori}

The compactification on higher-dimensional tori affords even
more interesting possibilities. Indeed, aside from richer breaking
patterns resulting from continuous Wilson lines, this
case allows a peculiar discrete
deformation related to the NS-NS antisymmetric tensor $B_{ab}$ \cite{bps,
ssethi,bianchitor,wittor} \footnote{Recent work on compactifications
with quantized backgrounds focusing on mathematical
aspects of the problem can be found in \cite{kuej} and references
therein.}.
We would like to stress that in the type I superstring the
fluctuations of this field are removed by the  Klein-bottle
projection, but one can nonetheless introduce quantized 
$B_{ab}$ backgrounds compatible with the symmetry of the
type IIB spectrum under the interchange of left and right
modes, and then analyse the properties of the corresponding
open descendants.

Let us begin by defining, as in \cite{narain}, the
generalizations of $p_{\rm L}$ and $p_{\rm R}$ 
to a $d$-dimensional torus $T^d$:
\ba 
p_{{\rm L},a} = m_a + \frac{1}{\alpha'} ( g_{ab} - B_{ab} ) \, n^b
\, , \label{pll} 
\\ 
p_{{\rm R},a} = m_a - \frac{1}{\alpha'} ( g_{ab} + B_{ab}
) \, n^b \,. \label{plr}
\ea 
Here we are slightly changing our notation with
respect to the one-dimensional case and to \cite{bps}, since we are 
scaling an inverse vielbein out of these
momenta, so that the resulting expressions
contain only the internal metric. For
instance, in this notation the torus amplitude reads
\be 
{\cal T} = |V_8 - S_8|^2 \ \sum_{m,n}
\frac{q^{\frac{\alpha'}{4} p_{\rm L}^T g^{-1} p_{\rm L}} \
\bar{q}^{\frac{\alpha'}{4} p_{\rm R}^T g^{-1} p_{\rm R}}}{|\eta(\tau)|^{2d}}
\,.
\ee

In order to construct the descendants of this compactification,
we should make sure that the symmetry $\Omega$ under left-right
interchange be still present. It is not hard to convince
oneself that this is not possible for generic values of
$g_{ab}$ and
$B_{ab}$, while one should insist on allowing generic metric
deformations, since the corresponding moduli are certainly
in the projected spectrum. Hence, the condition that for
any pair $(m_a,n^a)$ there exist another pair $(m^\prime_b,
n^{\prime \, b})$ such that
\be 
m_a + \frac{1}{\alpha'} ( g_{ab} - B_{ab} ) \, n^b = m'_a -
\frac{1}{\alpha'} ( g_{ab} + B_{ab} ) \, {n'}^b
\ee 
is to be understood as a constraint on $B_{ab}$, and
implies that
\be
\frac{2}{\alpha'} \, B_{ab} \in \bb{Z}  \,.
\ee 
As a result, the independent values for its components are
0 and $\frac{\alpha '}{2}$, subject to the condition of antisymmetry.
One can then proceed to compute the Klein-bottle amplitude,
that is not  affected by $B_{ab}$ since it only involves lattice modes
with $p_{\rm L}=p_{\rm R}$, and thus with $n^b=0$,
and for the standard choice of $\Omega$
projection is
\be 
{\cal K} = {\textstyle\frac{1}{2}}\, (V_8 - S_8) (2i\tau_2) \ \sum_m
\frac{q^{\frac{\alpha'}{2} m^T g^{-1} m}} {\eta^d (2 i \tau_2)} \, ,
\ee 
while the corresponding transverse-channel amplitude is
\be
\tilde{\cal K} = \frac{2^5}{2} \sqrt{{\rm det}(g/\alpha')} \
(V_8 - S_8) (i \ell) \ 
\sum_n \frac{q^{\frac{1}{\alpha'} n^T g n}} {\eta^d (i \ell)}
\,.
\ee

The transverse-channel annulus amplitude
is to involve contributions from all closed-string states
that are paired with their conjugates, where the pairing is determined
in this case by the condition of no momentum flow through the ends,
$p_{{\rm L},a} = - p_{{\rm R},a}$, as pertains to standard Neumann 
conditions.  The
resulting expression does not contain $g_{ab}$ and, given the
quantization of $B_{ab}$, becomes a constraint on $n$:
\be
\frac{2}{\alpha '} \, B_{ab} \, n^b = 2 \, m_a \,. \label{babconstr}
\ee 
This may be accounted for introducing a projector  in the
transverse-channel annulus amplitude, that becomes
\be
\tilde{\cal A}^{(r)}  = \frac{2^{r-d-5}}{2} \, \sqrt{{\rm det}
(g/\alpha ') }
\ N^2 \, (V_8 - S_8 )(i\ell) 
\sum_{\epsilon =0,1}
\sum_{n} {q^{{1\over 4 \alpha '} n^{T} g n}
\ e^{{2 i \pi \over \alpha '} n^{T} B \epsilon} \over \eta^d 
(i\ell ) } \, ,
\ee 
where the overall factor depending on the rank $r$ of $B_{ab}$ 
ensures that ${\cal A}$ is properly normalized.

It is worth emphasizing the nature of this expression, since
if one substitutes in the bosonic string action the 
zero modes of eqs. (\ref{pll}) and (\ref{plr}), subject to the 
condition that only windings flow in the tube, only the first 
term is generated. Therefore, it is the constraint (\ref{babconstr}) and 
not the {\it local} 
dynamics that generates the Wess-Zumino phase, as well as the additional
discrete variables $\epsilon_a$.

After an $S$ modular transformation, one then obtains the
direct-channel annulus amplitude
\be
{\cal A}^{(r)} = \frac{2^{r-d}}{2} \ N^2 \, (V_8 - S_8 )
({\textstyle\frac{1}{2}} i \tau_2) 
\sum_{\epsilon =0,1} \sum_{m}  {q^{{\alpha ' \over 2} ( m + 
{1\over \alpha '} B
\epsilon) ^{T} g^{-1} ( m + {1\over \alpha '} B \epsilon )} 
\over \eta^d ({\textstyle\frac{1}{2}} i \tau_2)} \, , \label{adtbab}
\ee
here normalized in such a way that, for any choice of the
antisymmetric tensor $B_{ab}$, $N$ is precisely the 
Chan-Paton multiplicity. This is 
particularly simple to appreciate in the two limiting cases: for $r=d$ 
the massless vectors are all obtained when all $\epsilon_a =0$, while for $r=0$
all $2^d$ values of the $\epsilon_a$ contribute.

The M\"obius amplitudes
\ba
\tilde{\cal M}^{(r)}  &=& - \frac{2 \times 2^{r/2-d/2}}{2} 
\, \sqrt{{\rm det} (g/\alpha ')} \,  
 N\, (\hat V _8 - \hat S _8 ) (i\ell+{\textstyle\frac{1}{2}}) 
\nonumber \\ 
& &\times
\sum_{\epsilon =0,1} \sum_{n} {q^{{1\over \alpha '} n^{T} g n}
\,\, e^{{2 i \pi \over \alpha '}  n^{T} B \epsilon}
\ \gamma_\epsilon \over \hat\eta^d ( i\ell+{\textstyle\frac{1}{2}}) }
\ea 
and
\be
\!\!\!\!\!\!\!\!\!\!{\cal M}^{(r)} = - \frac{2^{r/2 -d/2}}{2}    N  (\hat V _8 -
\hat S _8 ) ({\textstyle\frac{1}{2}} i\tau_2 + {\textstyle\frac{1}{2}}) 
\! \sum_{\epsilon =0,1} \sum_{m} {q^{{\alpha
' \over 2} (m +  {1\over \alpha '} B
\epsilon )^{T} g^{-1} (m + {1\over \alpha '} B \epsilon )} 
\gamma_\epsilon \over \hat\eta^d ({\textstyle\frac{1}{2}}i \tau_2
+ {\textstyle\frac{1}{2}} ) }
\ee
present a new subtlety, since they
involve the additional signs $\gamma_\epsilon$, needed to ensure a
proper normalization of the direct channel compatibly with the
tadpole condition, that now becomes
\be
\frac{2^5}{2} + \frac{2^{r-5}}{2}\;  N^2 + 
\frac{2\times 2^{r/2-d/2}}{2} \; N \sum_{\epsilon =0,1}  \gamma_\epsilon = 0
\, ,
\ee
and is a perfect square only if 
\be
\sum_{\epsilon =0,1}  \gamma_\epsilon = 2^{d/2}\,.
\ee
As a result, some of the $\gamma$'s must be negative. 
It should be appreciated that, in general, the resulting gauge group has a
{\it reduced} rank, since the solution for the Chan-Paton multiplicity $N$ is
\be
N = 2^{5-r/2}
\,.
\ee

The $\gamma$'s play a very important r\^ole also in the direct-channel 
amplitudes, since they allow one to connect orthogonal and symplectic 
groups via
continuous deformations \cite{bps}. In order to illustrate this point, 
it is instructive to consider in some detail the compactification on 
a two-torus, whose  metric $g_{ab}$ and antisymmetric tensor $B_{ab}$ 
can be
conveniently parametrized in terms of the complex structure $X=X_1 + i X_2$
and of the K\"ahler structure \cite{gsw} $Y=Y_1 + i Y_2$ as  
\be
g = \frac{\alpha' Y_2}{X_2} \left( \begin{array}{cc} 1&X_1 \\ X_1&
|X|^2  \end{array} \right) \, , \qquad
B = {\alpha'} \left( \begin{array}{cc} 0&Y_1 \\ -Y_1&
0 \end{array} \right) \, , \label{XY}
\ee
where the independent values for $Y_1$ are 0 and ${1\over 2}$. 
In section 2 we have already met the complex structure of the world-sheet
torus, the ``shape'' of the corresponding parallelogram, and here
$X$ describes in a similar fashion the ``shape'' of the target-space
torus. The imaginary part of the
K\"ahler structure has also a very simple interpretation: up to
a normalization, it is the area of the target-space torus.

For the interesting case $Y_1 = {1\over 2}$, that corresponds to $r=2$, 
the annulus amplitudes are
\ba
\tilde{\cal A}^{(2)} &=& \frac{2^{-5} N^2}{2} \ Y_2 \ (V_8 - S_8)
\nonumber \\
& &\times \sum_{n_1,n_2} \frac{ W_{n_1,n_2} \left[
1 + (-1)^{n_1} + (-1)^{n_2} + (-1)^{n_1+n_2} \right]}{\eta^2} \, ,
\ea
\be
{\cal A}^{(2)} = {\textstyle\frac{1}{2}} \, N^2 \ (V_8 - S_8) \ 
\frac{\left[P_{0,0} + 
P_{-\frac{1}{2},0} + P_{0,\frac{1}{2}} + P_{-\frac{1}{2},\frac{1}{2}}
\right]}{\eta^2} \, ,
\ee
and, in a similar fashion, the M\"obius amplitudes are
\ba
\tilde{\cal M}^{(2)} &=& -{\textstyle\frac{2}{2}}\, N \ Y_2 \ (\hat{V}_8 - \hat{S}_8)
\frac{1}{\hat{\eta}^2} 
\sum_{n_1,n_2} W_{2n_1,2n_2} \nonumber \\
& &\times \left[
\gamma_{0,0}+ (-1)^{n_1}\gamma_{0,1} + (-1)^{n_2}\gamma_{1,0} + 
(-1)^{n_1+n_2} \gamma_{1,1} \right] \, ,
\ea
\ba
{\cal M}^{(2)} &=& -{\textstyle\frac{1}{2}}\, N \ (\hat{V}_8 - \hat{S}_8)
\nonumber \\ & &\times  \frac{\left[\gamma_{0,0} P_{0,0} + 
\gamma_{0,1} P_{-\frac{1}{2},0} + \gamma_{1,0} P_{0,\frac{1}{2}} + 
\gamma_{1,1} P_{-\frac{1}{2},\frac{1}{2}} \right]}{\hat{\eta}^2}
\, ,
\ea
where we have introduced a compact notation for the
winding and momentum sums:
\ba
W_{n_1,n_2} &=& q^{\frac{Y_2}{4 X_2} [ (n_1 + X_1 n_2)^2 + n_2^2 X_2^2
]} \, ,\nonumber \\
P_{\epsilon_1,\epsilon_2} &=&
\sum_{m_1,m_2} q^{\frac{1}{2 X_2 Y_2} [ (m_2 - \epsilon_2 - 
X_1 (m_1 - \epsilon_1))^2 + (m_1 - \epsilon_1)^2
X_2^2 ]} \,.
\ea
One can easily extract the contributions of the
three amplitudes to the R-R tadpole,
\ba
\tilde{\cal K} &\to& - 2^5 \, , \nonumber
\\
\tilde{\cal A}^{(2)} &\to& - 2^{-5} \times N^2 \times 4 \, , \nonumber
\\
\tilde{\cal M}^{(2)}&\to&  2 \times N \times (\gamma_{0,0} + \gamma_{0,1} + 
\gamma_{1,0} + \gamma_{1,1} ) \,,
\ea 
and it is evident that, in order to solve the tadpole condition, one
of the four $\gamma_\epsilon$ is to equal minus one, while the
three others are to equal plus one. As a result, the total
charge is reduced, consistently with the rank reduction for the Chan-Paton
gauge group. Among the four possible choices,
only two lead to different results, depending on the sign of
$\gamma_{0,0}$. If $\gamma_{0,0}=+1$, the massless open-string vector 
is in the adjoint of ${\rm SO} (16)$, 
while if $\gamma_{0,0} =-1$ it is in the adjoint of
${\rm USp} (16)$. 
Notice that a T-duality along both internal coordinates would 
actually alter the
$\Omega$ projection, moving the restriction to
even windings from $\tilde{\cal A}$ to ${\cal K}$ \cite{cab}, as we 
shall see in detail in the next section.

As anticipated, orthogonal and symplectic 
gauge groups can be connected 
via continuous Wilson lines. As an example, let us consider the 
compactification on the two-dimensional torus with a Wilson line in the 
1 direction. For definiteness, let us choose $\gamma_{0,0} =-1$, 
$\gamma_{1,0} =\gamma_{1,1}=\gamma_{0,1} =1$, so that 
the annulus and M\"obius amplitudes in the transverse channel are
\ba
\tilde{\cal A}^{(2)} &=& \frac{2^{-5}}{2} \ Y_2 \ (V_8 - S_8)
\sum_{n_1,n_2} \left(M e^{2 i \pi a n_1} + \bar M 
e^{-2i \pi a n_1} \right)^2 \nonumber \\
& &\times \frac{ W_{n_1,n_2} \left[
1 + (-1)^{n_1} + (-1)^{n_2} + (-1)^{n_1+n_2} \right]}{\eta^2} 
\ea
and
\ba
\tilde{\cal M}^{(2)}&=& {\textstyle\frac{2}{2}} \ Y_2 \ (V_8 - S_8)
\frac{1}{\eta^2} \sum_{n_1,n_2} W_{2n_1,2n_2}
\nonumber
\\ 
& &\times \left(M e^{4 i \pi a n_1} + \bar M 
e^{- 4 i \pi a n_1} \right)
\nonumber \\
&  &\times 
\left[ 1 - (-1)^{n_1} - (-1)^{n_2} - 
(-1)^{n_1+n_2} \right] \, ,
\ea
while the usual rescalings and the $S$ and $P$ modular transformations give 
\ba
{\cal A}^{(2)} &=& {\textstyle\frac{1}{2}} \ (V_8 - S_8)\ 
\frac{1}{\eta^2} \Biggl\{
2M\bar M \left[P_{0,0} + P_{\frac{1}{2},0} + 
P_{0,\frac{1}{2}} + P_{\frac{1}{2},\frac{1}{2}} \right] \nonumber
\\
& &+ M^2\ \left[P_{2a,0} + P_{2a+\frac{1}{2},0} + 
P_{2a,\frac{1}{2}} + P_{2a+\frac{1}{2},\frac{1}{2}} \right] \nonumber
\\
& &+ \bar M^2\ \left[P_{-2a,0} + P_{-2a+\frac{1}{2},0} + 
P_{-2a,\frac{1}{2}} + P_{-2a+\frac{1}{2},\frac{1}{2}} \right] \Biggr\}
\ea
for the annulus amplitude, and
\ba
{\cal M}^{(2)} &=& {\textstyle\frac{1}{2}} 
\ (V_8 - S_8)\frac{1}{\eta^2}
\nonumber \\ \
& &\times  \Biggl\{
M \left[P_{2a,0} - 
P_{2a+\frac{1}{2},0} - P_{2a,\frac{1}{2}} - 
P_{2a+\frac{1}{2},\frac{1}{2}} \right] \nonumber
\\
& &+ \bar M \left[ P_{-2a,0} - 
P_{-2a+\frac{1}{2},0} - P_{-2a,\frac{1}{2}} -
P_{-2a+\frac{1}{2},\frac{1}{2}} \right] \Biggr\}\nonumber
\ea
for the M\"obius amplitude. 
For a generic Wilson line $a$, with $0\leq a<\frac{1}{2}$,
the gauge group is thus U(8), but for
$a=0$ the lattice sums $P_{\pm 2a,0}$ contribute massless modes
to ${\cal A}$ and ${\cal M}$,
and the gauge group enhances to USp(16). On the other hand, 
for $a=\frac{1}{4}$ the lattice sums $P_{\pm 2a+\frac{1}{2},0}$ 
contribute massless modes, and since 
these terms are accompanied  in ${\cal M}$ by additional signs, the 
gauge group 
enhances to SO(16).
The Wilson line is a modulus of the compactified open string, and therefore
orthogonal and symplectic gauge groups are indeed continuously 
connected \cite{bps}.

\vskip 12pt
\subsection{T-duality and discrete moduli}

It is instructive to study a bit further the effect of 
T-duality on toroidally compactified type I
strings. This subject was originally considered in \cite{dlp,horava2,green}, 
and, as we have seen, results in an equivalent, though rather pervasive, 
description of gauge symmetry breaking
by Wilson lines, momentum-space translations, in terms of 
spatial displacements of the
extended objects where the ends of open strings terminate.

In order to appreciate the effect of T-duality on the construction of
open descendants, one can begin by observing, as in subsection 4.1, that
the circle inversion
$R \to {\alpha'}/{R}$ turns quantized momenta $p\sim 1/R$ into windings 
$w \sim R/\alpha'$. These, on the other hand, can only be supported by open
strings whose ends are fixed. Therefore, the T-dual picture of an open
string with Neumann boundary conditions involves a novel type
of string with Dirichlet boundary conditions \cite{dlp,horava2,green,pcj},
and this poses the problem of the identity of the hyper-surfaces where 
the ends live. In fact,
T-duality is a world-sheet duality transformation that interchanges 
$\sigma$ and $\tau$, and thus 
tangential and normal derivatives. At a boundary, where the original Neumann
condition was turning $X_{\rm L}$ into $X_{\rm R}$, the left and right modes of
a closed string, the Dirichlet condition generated by T-duality turns
$X_{\rm L}$ into $-X_{\rm R}$. 
In this sense T-duality acts as a parity transformation,
but, say, limitedly to the right-moving modes, and this 
provides a consistent picture of
the whole construction. To begin with, the direct-channel Klein-bottle
amplitude now involves only winding modes, that are precisely fixed by the
combination of the usual world-sheet parity $\Omega$ with the inversion of
the right-moving modes, $X_{\rm L} \leftrightarrow - X_{\rm R}$. 
In addition, the open-string coordinate
involves an expansion in terms of $\sin(n\sigma)$, 
that can also be associated to the new way of combining left and right modes. 

Open strings thus terminate on extended objects, D-branes, whose dimensions are
altered by T-duality transformations \cite{pcj}. In particular, Neumann 
strings can be thought of
as terminating on D9 branes, that invade the whole ten-dimensional space-time,
and that T-duality along any of the coordinates
turns into D8 branes. In a similar fashion, the ends of a string 
with $p+1$ Neumann
boundary conditions and $9-p$ Dirichlet boundary conditions live on a
D$p$ brane, and, more generally, a T-duality increases or reduces the dimension
of a D-brane according to whether it involves a direction orthogonal or
parallel to it.

An additional feature of D-branes and O-planes that has 
attracted much attention during the
last few years is their tension. Since they both show up in
two genus-${1\over 2}$ world-sheets, the disk and the projective plane, they are
naturally
weighted by a factor $1/g_s$, a rather peculiar feature when compared 
to ordinary
solitons, whose mass scales like $1/g^2$, with $g$ a typical
coupling of a gauge theory \cite{pol95}.

This cursory review of T-duality suffices to bring us to our next topic.
In the previous section we have seen how quantized values of the NS-NS
tensor $B_{ab}$ play an important r\^ole in the vacuum amplitudes for
Neumann strings. We can now describe how
different orientifold projections can result in quantized values for
other moduli of the ``parent'' closed string \cite{ab}.
For simplicity, we shall concentrate on the $T^2$ case,
but similar results apply to higher-dimensional tori.

On $T^2$ one has the two distinct options of performing a T-duality along both 
coordinates or along
a single one. The first, discussed in the previous subsection, can replace
the ${\rm O}9_+$ plane with three ${\rm O}7_+$ and one
${\rm O}7_-$, but $B_{ab}$
remains a discrete deformation. On the other hand, a T-duality along
a single direction combines the $\Omega$ projection with 
the conjugation ${\cal R}:\ Z_{\rm R} \to \bar
Z_{\rm R}$ of the complex right-moving coordinate 
$Z_{\rm R}=(X^7_{\rm R}+i X^8_{\rm R})/\sqrt{2}$
on the target $T^2$, and a simple
analysis of the massless spectrum reveals that in this case the
internal components of the NS-NS antisymmetric tensor
\be
\left( \psi_{-{1\over 2}}^{1} \tilde\psi_{-{1\over 2}}^{2} - 
\psi_{-{1\over 2}}^{2} \tilde\psi_{-{1\over 2}}^{1} \right) |0\tilde 0
\rangle 
\nonumber
\ee
survive the $\Omega {\cal R}$ projection, while the mixed
components of the internal metric
\be
\left( \psi_{-{1\over 2}}^{1} \tilde\psi_{-{1\over 2}}^{2} +
\psi_{-{1\over 2}}^{2} \tilde\psi_{-{1\over 2}}^{1} \right) |0\tilde 0
\rangle 
\nonumber
\ee
do not. Therefore, in this case the antisymmetric
tensor is a continuous modulus of the projected theory, while some
quantization condition should be met by the mixed components of the
metric. Indeed, from the expressions
\ba
p_{\rm L} &=& 
{1\over \sqrt{\alpha ' X_2 Y_2}} \left[ - X m_1 + m_2 + \bar{Y} ( n_1
+ X n_2 ) \right]\, ,
\nonumber 
\\
p_{\rm R} &=& {1\over \sqrt{\alpha ' X_2 Y_2}} \left[ - X m_1 + m_2 + Y ( n_1
+ X n_2 ) \right]\, ,
\nonumber
\ea
for the left and right complex momenta on a $T^2$ with complex 
structure $X$ and 
K\"ahler form $Y$, whose metric and antisymmetric tensor are given
in eq. (\ref{XY}), one can see that requiring the invariance of the
parent theory under 
\be
\Omega {\cal R}:\qquad p_{\rm L} \leftrightarrow \bar p_{\rm R} 
\ee
results in a quantization condition for the real slice of the {\it complex
structure}, so that now 
$2 X_1 \in \bb{Z}$. This is to be compared with the standard case,
discussed previously,
where the quantization condition applied to the real slice 
$B_{ab}$ of the {\it K\"ahler form}.
 
The closed-string states allowed in the transverse-channel amplitudes
$\tilde{\cal A}$ and $\tilde{\cal M}$
are now subject to the constraints 
\be 
2 X_1 m_1 \, , \ 2 X_1 n_2 \in 2 \bb{Z}\, ,
\ee
that, as in the standard case, result in the
insertion of suitable projectors in the annulus and M\"obius
amplitudes.  For half-integer values of $X_1$,
these have the effect of halving the Chan-Paton multiplicities. 
We have thus met a generic
feature of open descendants: non-vanishing, quantized, backgrounds for
the closed-string moduli eliminated by the orientifold projection 
typically reduce the rank of the Chan-Paton gauge group.

In this example, the open descendants with conventional $\Omega$
projection are effectively constructed on
the mirror torus, where the r\^oles of $X$ and $Y$ are interchanged,
 so that now $Y$ and $X$ are the complex and K\"ahler 
structures. This
correspondence is well known from closed strings: on a
$T^2$, T-duality along a one-cycle is indeed the simplest instance
of a mirror symmetry \cite{mirror}.

A rank reduction similar to that induced in open-string models by a
quantized $B_{ab}$ manifests itself also in the 
heterotic models usually termed
CHL strings \cite{chl}, where it originates from
higher-level realizations of the current algebra. 
There is indeed a nice duality correspondence between these two classes
of models, first noticed in \cite{bianchitor,wittor}.

 
\newsec{Orbifold compactification}

Since their first appearance in \cite{dhvw}, toroidal orbifolds
have proved a major source of insight into the structure of
String Theory. In the resulting wide class of exactly solvable models,
strings propagate consistently in singular curved spaces and display 
rich patterns of interesting low-energy spectra.

Orbifolds can be constructed subjecting smooth covering manifolds 
to discrete identifications. These in general leave
some sets of points fixed, 
that as a result support curvature singularities. Whereas in
such singular spaces the dynamics of particles is generally ill-defined, 
{\it modular invariance} fully determines the resulting spectra for 
oriented closed strings. These include an ``untwisted'' sector, 
whose states are subsets of those present in the covering manifolds, 
and additional ``twisted'' sectors
confined to the fixed points.  Typically, the latter also include 
some blow-up moduli, capable of lifting the orbifold 
singularities, thus connecting these spaces to nearby smooth
manifolds.

This section is devoted to orbifold compactifications of models with
open strings. We begin by considering the bosonic string on
the one-dimensional orbifold of
a circle, that allows one to describe the r\^ole of Neumann and Dirichlet
boundary conditions in a relatively simple setting. We then move on to
one-dimensional shift orbifolds of the type-I superstring, the 
simplest models where the breaking of supersymmetry is related
to a continuous parameter, the radius $R$ of a circle. 
Six-dimensional supersymmetric
open-string models display new types of low-energy spectra, that
generally contain several
tensor multiplets, and are thus vastly different from heterotic models.
In orbifold compactifications,
the rank reduction induced by a quantized $B_{ab}$, familiar from the toroidal
case of section 4, is accompanied by a grouping of
the fixed points, that results in the presence of 
several tensor multiplets in the closed 
sector and of several families
of {\it ND} states in the open sector. 
Six-dimensional orbifolds are also the
simplest examples whose consistency can require that supersymmetry be
broken on branes. All these aspects are reviewed in some detail,
before describing the simplest chiral four-dimensional supersymmetric 
model, the open descendant of
the $T^6/\bb{Z}_3$ IIB compactification, whose spectrum 
includes three families of chiral matter. 
``Brane supersymmetry breaking'' provides also
the solution to an interesting class of
otherwise inconsistent four-dimensional models, associated to $\bb{Z}_2 \times
\bb{Z}_2$ orbifolds, that we then review in some detail
before turning to a discussion of magnetic deformations.
We conclude with a cursory view of the D-branes present in the ten-dimensional
strings and in their orientifolds.

\vskip 12pt
\subsection{The one-dimensional orbifold of the bosonic string}

Let us begin by considering the one-dimensional orbifold
of the bosonic string, whose open descendants were first constructed in 
\cite{ps}. The starting point is the
torus amplitude (\ref{bosonictorus}), now projected in order to
retain only states built by operators invariant under the internal 
parity  $X^{25} \to - X^{25}$. This is achieved combining the original
internal lattice operators, with
$P_{\rm L,R}$ defined in (\ref{onedimmom}),
into pairs of definite parity, that can
then be accompanied by even or odd powers of string 
oscillators according to
\ba
\prod_{i,j} \; {\alpha_{-n_i}^{25} \, 
\tilde{\alpha}_{-n_j}^{25}}  \; 
\cos( p_{\rm L} X_{\rm L} + p_{\rm R} X_{\rm R}) 
&  \quad {\rm for} \quad & \#(i + j)
\in 2 \bb{Z} \, , \label{untwope} \\ 
 \prod_{i,j} \;  {\alpha_{-n_i}^{25} \, 
\tilde{\alpha}_{-n_j}^{25}} \; \sin( p_{\rm L} X_{\rm L} + 
p_{\rm R} X_{\rm R})
&  \quad {\rm for} \quad & \#(i + j) 
\in 2 \bb{Z} +1  \, , \label{untwopo}
\ea
since $(-1)^{\#(i + j)}$, with ${\#(i + j)}$ the total number of oscillators
in the vertices, is the overall parity of these oscillator modes.
Hence, away from the origin of the lattice, the operators that
survive the projection are simply half of those allowed in the covering torus,
while an additional contribution is needed to count properly the 
remaining ones, that have $p_{\rm L}=0=p_{\rm R}$. All this
is neatly encoded in the ``untwisted'' partition function
\be
{\cal T}_{\rm (u)} = {\textstyle\frac{1}{2}} \; \sum_{m,n}
\frac{ q^{\alpha' p_{\rm L}^2/4} \; \bar{q}^{\alpha' p_{\rm R}^2/4}}{\eta(\tau)
\eta(\bar{\tau})} + {\textstyle\frac{1}{2}} \; 
\left| \frac{2 \eta}{\vartheta_2}  \right| \, ,
\ee
where, for the sake of brevity, we are omitting the factors
$1/(\sqrt{\tau_2} \eta \bar{\eta})$ associated to the remaining
non-compact transverse coordinates.
The second contribution is not invariant under the $S$ modular
transformation,
while its $S$-transform is not invariant under the $T$ transformation. Still,
combining all these terms gives a modular invariant partition function,
that describes the propagation of the closed 
bosonic string on the segment $S^1/\bb{Z}_2$:
\be
{\cal T} = {\textstyle\frac{1}{2}} \; \sum_{m,n}
\frac{ q^{\alpha' p_L^2/4} \; \bar{q}^{\alpha' p_R^2/4}}{\eta(\tau)
\eta(\bar{\tau})} + {\textstyle\frac{1}{2}} \; 
\left| \frac{2 \eta}{\vartheta_2}  \right| 
+ {\textstyle\frac{2}{2}} \; \left| \frac{ \eta}{\vartheta_4}  \right| +
{\textstyle\frac{2}{2}} \; \left| \frac{ \eta}{\vartheta_3}  \right| \,.
\ee
In the last two terms, that define the ``twisted'' sector, 
the overall factor of two is fixed by
the modular invariance of ${\cal T}$, and accounts
for the two sectors living at the orbifold fixed
points.

As in previous examples, the open descendants are built halving ${\cal T}$
and adding the corresponding Klein-bottle amplitude. Following \cite{ps},
we begin by confining our attention to the simplest choice,
\be
{\cal K} = {\textstyle\frac{1}{4}} \;  \sum_{m}
\frac{ q^{\frac{\alpha'}{2} \left(\frac{m}{R}\right)^2}}{\eta(2i \tau_2)} +
{\textstyle\frac{1}{4}} \; \sum_{n}
\frac{ q^{\frac{1}{2\alpha'} ( n R )^2}}{\eta(2 i \tau_2)}  +
{\textstyle\frac{2}{2}}  \sqrt{ \frac{ \eta}{\vartheta_4}} \,. \label{korb1}
\ee
In order to justify this expression, let us begin by noting that there
are {\it two}
subsets of the untwisted cosine-operators in (\ref{untwope}) that
are fixed under the 
involution $\Omega$, and thus contribute to ${\cal K}$. Aside 
from the terms with zero winding, that
we already met in the previous section, the even nature of the cosine vertices
also fixes the operators with zero momentum, whose argument
is {\it odd} under $\Omega$. Finally, the last term in eq. (\ref{korb1}) 
is associated to the twisted states, and has the proper multiplicity to
account for the two fixed points.
Notice that this expression does not involve the two functions $\vartheta_2$
and $\vartheta_3$, that would reflect an antiperiodic behaviour under
vertical transport in the doubly-covering torus. 
This can be clearly seen from the geometry of the
Klein bottle in fig. 5: in this example 
the orbifold involution relating pairs of image points
is $\bb{Z}_2$ valued, and for consistency must square 
to a {\it periodic} vertical translation on the covering torus.

An equivalent description of this spectrum
exhibits its orthogonal decomposition in 
different sectors \cite{pradrev}. It is rather effective and explicit for the
oscillator excitations at the origin of the lattice, for which one
can define the four combinations
\be
\phi_{+\pm} = {\textstyle\frac{1}{2}} \left( \; \frac{1}{\eta} \pm 
\sqrt{\frac{2\eta}{\vartheta_2}} \; \right) \, , \qquad 
\phi_{-\pm} = {\textstyle\frac{1}{2}} \left( \; 
\sqrt{\frac{\eta}{\vartheta_4}}  \pm 
\sqrt{\frac{\eta}{\vartheta_3}} \; \right) \, ,
\ee
and in terms of these expressions the torus 
partition function becomes
\be
{\cal T} = |\phi_{++}|^2 + |\phi_{+-}|^2 + 2 |\phi_{-+}|^2 + 2 |\phi_{--}|^2 
+  {\textstyle\frac{1}{2}} \; \sum {}' \, .
\ee
The primed sum refers to the operators associated to points of the
lattice away from the origin that, as we have seen, combine in pairs, 
and whose $\bb{Z}_2$ symmetrization is properly accounted for by the overall factor ${1\over 2}$. 
Actually, each of the twisted terms is
not a full specification of the corresponding operators, since we are unable
to distinguish in ${\cal T}$ pairs of operators that belong to different
fixed points. From the conformal field theory viewpoint, we are thus 
facing an ambiguity, but the structure 
is  nonetheless evident, and indeed the Klein-bottle 
amplitude reads
\be
{\cal K} = {\textstyle\frac{1}{2}}\; 
\left[ \phi_{++} + \phi_{+-} + 2 \phi_{-+} + 
2 \phi_{--} 
+  {\textstyle\frac{1}{2}} \; 
( P'_m + W'_n ) \right] \, , \label{korb2}
\ee
where the complete momentum and winding sums present in ${\cal K}$
and in the annulus amplitude ${\cal A}$ that we shall introduce shortly 
are
\be
P_m(q^\gamma ) = \sum_m \; \frac{q^{\frac{\alpha'}{2} 
\left( \frac{m}{R} \right)^2}}{\eta(q^\gamma)} \, , \qquad 
W_n(q^\gamma) = \sum_n \; \frac{q^{\frac{(n R)^2}{2 \alpha'} }}
{\eta(q^\gamma)} \, ,
\ee  
with $\gamma=2$ for ${\cal K}$ and $\gamma=1/2$ for ${\cal A}$,
while the ``primed'' sums lack the terms at the origin of the lattices.
Eq. (\ref{korb2}) reflects precisely our previous comments, and may thus
be regarded as an alternative justification for (\ref{korb1}).

We can now turn ${\cal K}$ to the transverse channel, by an $S$ modular
transformation. Taking into account the contributions of the remaining 
non-compact coordinates, in terms of the modulus of the double cover
the result is
\be
\tilde{\cal K} = \frac{2^{13}}{4} \; \left[
 v \; W_{2n} +  \frac{1}{v}\; P_{2n}
+ 2 \sqrt{\frac{2 \eta}{\vartheta_2}} \right] \, ,
\ee
where, in order to lighten the notation, we have introduced the dimensionless
radius, $v = R/\sqrt{\alpha '}$.
In terms of the $\phi$'s this expression becomes
\ba
\tilde{\cal K} &=& \frac{2^{13}}{2} \; \left[ \left( \sqrt{v} 
+ \frac{1}{\sqrt{v}} \right)^2 \phi_{++} \right. \nonumber \\
& & + \left.
\left( \sqrt{v} - \frac{1}{\sqrt{v}} \right)^2 \phi_{+-} 
+ 
v\; W'_{2n} +  \frac{1}{v} \; P'_{2n} \right] \, , \label{tckba}
\ea
where the zero modes exhibit very clearly the familiar structure of
perfect squares for the reflection coefficients.

The open sector is quite interesting, since on this 
$\bb{Z}_2$ orbifold it 
allows for the simultaneous presence of open strings with different 
boundary conditions. Aside from the standard
strings with Neumann conditions at their two ends, NN strings,
additional ones, DD and ND, with Dirichlet-Dirichlet and mixed 
Neumann-Dirichlet boundary conditions are present, together
with their $\bb{Z}_2$ orbifold projections. 
These different types of strings have by now a 
geometrical interpretation in terms of bosonic
D25 and D24 branes, while the annulus amplitude reads
\ba
{\cal A} &=& {\textstyle{1\over 4}} 
\left[ N^2 \; P_m + D^2 \; W_n +
\left( R_{N}^{2} + R_{D}^{2} \right) 
\sqrt{2 \eta \over \vartheta_2} \right. \nonumber \\
& & + 2\,  \left. N\, D \; \sqrt{\eta \over \vartheta_4}+ 
2\, R_N \, R_D \; \sqrt{\eta \over \vartheta_3} \right] \, .
\ea
Here $N$ and $D$ count the overall numbers of D25 and D24 branes, 
while $R_N$ and $R_D$ encode the $\bb{Z}_2$ orbifold action on the 
corresponding Chan-Paton charges. The first term involves a sum over 
Kaluza-Klein momenta, and thus refers to NN strings, whose 
ends live on D25 branes.
On the other hand, the winding contribution refers to DD strings, whose ends
are free to move in the 24 non-compact dimensions, parallel to the D24 branes,
but are fixed in the compact one. For simplicity we have only considered 
DD strings at a single fixed point of the
$S^1 /\bb{Z}_2$ orbifold, but more general 
configurations, where the D24 branes are distributed between the
fixed points or moved to the bulk of the compact space, are also
possible. We shall return to this option in the
following sections. Finally, open strings with mixed ND boundary conditions,
{\it i.e.} stretched between D25 and D24 branes, 
have half-integer mode expansions and an additional factor of
two that reflects the two orientations of their endpoints.

The M\"obius amplitude receives contributions only from
NN and DD strings, and reads
\be
{\cal M} = {\textstyle{1 \over 4}} \, \epsilon \, 
\left[ N\, \hat P_m + D\, \hat W_n +
(N+D) \sqrt{2 \hat \eta \over \hat \vartheta_2} \right] \, ,
\ee
where $\epsilon$ is an overall sign.
It can be recovered, as usual, from the corresponding transverse-channel 
amplitude $\tilde{\cal M}$, determined by the factorization of 
$\tilde{\cal K}$ and $\tilde{\cal A}$. 

In superstring vacua the new type of 
boundary condition is generally not optional, but 
is demanded by R-R tadpole cancellations, and thus by space-time
anomalies \cite{pc}.  
A closer look at $\tilde{\cal K}$ 
reveals also in this case the existence of {\it two} types of tadpoles.
These can be associated to
the standard bosonic O25 planes, whose contribution scales proportionally
to the length of 
the circle, consistently with the fact that they invade the whole
internal space, and to new bosonic O24 planes, 
whose contribution scales inversely with it, consistently with the
fact that they are localized at the two fixed points. Thus, both
D25 and D24 branes would be needed in this case if one insisted on
cancelling all
tadpoles, as can be seen from the transverse-channel open-string amplitudes 
\ba
\tilde{\cal A} &=& {2^{-13} \over 4} \Biggl[ N^2 \, v
\, W_n + {D^2 \over v} \, P_n + 2\, N\, D \sqrt{2\eta \over
\vartheta_2}
\nonumber
\\
& &
+ 2\left( R^{2}_{N} + R^{2}_{D} \right) \sqrt{\eta \over \vartheta_4}
+ 2\,\sqrt{2} \, R_N \, R_D \, \sqrt{\eta \over \vartheta_3} \Biggr]
\ea
and
\be
\tilde{\cal M} = {\textstyle{2\over 4}} \, \epsilon\, \left[ N \, v
\, W_{2n} + {D \over v}\, P_{2m} + (N+D) 
\sqrt{2\hat\eta \over \hat\vartheta_2} \right] \,,
\ee
that in terms of the $\phi$'s would become
\ba
\tilde{\cal A} &=& {2^{-13}\over 4} \Biggl\{ \left( N \sqrt{v } 
+ {D \over \sqrt{v}} \right)^2 \, \phi_{++}
\nonumber \\
& & 
+ \left( N \sqrt{v} - {D \over \sqrt{v}} \right)^2 \, \phi_{+-}
\nonumber
\\
& & + 2\,\left[ \left( {R_N \over \sqrt{2}} + R_D \right)^2 + \left( 
{R_N \over \sqrt{2}} \right)^2 \right] \, \phi_{-+} 
\nonumber \\ & &
+ 2\,\left[ \left( {R_N \over \sqrt{2}} - R_D \right)^2 + \left( 
{R_N \over \sqrt{2}} \right)^2 \right] \, \phi_{--} \Biggr\} + \ldots
\label{tcaa}
\ea
and
\ba
\tilde{\cal M} &=& {\textstyle{2\over 4}} \, \epsilon \, \Biggl[ 
\left( \sqrt{v} + {1 \over \sqrt{v}} \right)
\left( N \sqrt{v} + {D \over \sqrt{v}} 
\right)\, \hat\phi_{++} \nonumber
\\
& & + \left( \sqrt{v} - {1 \over \sqrt{v}} \right)
\left( N \sqrt{v} - {D  \over \sqrt{v}} 
\right)\, \hat\phi_{+-} \Biggr] + \ldots \,.
\ea

\noindent
These expressions match precisely the Klein-bottle amplitude (\ref{tckba}),
while the tadpole conditions would lead to
\ba
N &= - \, 8192\, \epsilon \,, \qquad \quad R_N &= 0 \, ,
\nonumber
\\
D &= - \, 8192\, \epsilon \,, \qquad \quad R_D &= 0 \, ,
\label{tado}
\ea
thus also requiring that  $\epsilon = -1$.

We would like to stress that the amplitude (\ref{tcaa}) reveals rather 
neatly the geometry of the
D-brane configuration: not only do $N$ and $D$ count the overall numbers 
of D25 and 
D24 branes, but the twisted terms clearly display that only 
one of the two fixed points accommodates all the D24 branes. 

To conclude our description of the $S^1/\bb{Z}_2$ orbifold, we now turn to 
the open spectrum, rewriting also
${\cal A}$ and ${\cal M}$ in terms of the $\phi$'s.
The result is
\ba
{\cal A}_0 &\sim& {\textstyle{1\over 4}} \Bigl\{ 
\left[ (N^2 + R_{N}^{2} ) + (D^2 + R_{D}^{2} ) \right] \, \phi_{++}
\nonumber \\
& & + \left[ (N^2 - R_{N}^{2}) + (D^2 - R_{D}^{2})\right]\, \phi_{+-}
\nonumber \\
& & + 2 \left( N\, D + R_N \, R_D \right) \phi_{-+} + 2 
\left( N\, D + R_N \, R_D \right) \phi_{--} \Bigr\} 
\ea
for the annulus amplitude, and 
\be
{\cal M}_0 \sim {\textstyle{1\over 2}} \, \epsilon \, (N+D) \, \phi_{++} 
\ee
for the M\"obius amplitude. A consistent particle interpretation of
the direct-channel amplitudes calls for a regular action of the 
$\bb{Z}_2$ orbifold group on the charge space,
and thus for a parametrization in terms of ``real'' Chan-Paton multiplicities,
so that
\ba
N &= n_1 + n_2 \, , \qquad\quad R_N &= n_1 - n_2 \, ,
\nonumber \\
D &= d_1 + d_2 \, , \qquad\quad R_D &= d_1 - d_2 \, ,
\ea
Enforcing the tadpole conditions (\ref{tado}), this would result in 
a gauge group comprising two copies of ${\rm SO} (4096) 
\times {\rm SO} (4096)$, associated
to D25 and D24 branes respectively, with tachyons in symmetric 
representations and scalars in bi-fundamentals.

\vskip 12pt
\subsection{The one-dimensional shift-orbifold}

We now turn to orbifold compactifications of the type I superstring
where the target-space coordinates 
are not identified under reflections, as in the previous case, but
under discrete, fractional, shifts of the lattice basis vectors or,
more generally, under the combined action of shifts and internal
symmetries.
This combined action is particularly interesting, since it can
implement in String Theory \cite{kounnas} the Scherk-Schwarz
mechanism \cite{ss} to attain the breaking of supersymmetry.
In the simplest case of circle 
compactification, this allows higher-dimensional modes that are
periodic only up to an internal symmetry transformation.
The Kaluza-Klein momenta of the various fields are thus shifted 
proportionally to their charges, with the consequent possibility of
introducing mass differences between bosons and fermions.

In Field Theory, with only Kaluza-Klein excitations available,
the Scherk-Schwarz mechanism can only result from shifts of internal
momenta. On the other hand, String Theory offers more possibilities, 
since one has also the option of affecting the windings. 
For oriented closed 
superstrings these two deformations, related by T-duality,
describe essentially the same phenomenon. After 
orientifolding, however, they lead to completely different results. 
As in \cite{ads1},
we shall refer to these two mechanisms as Scherk-Schwarz and M-theory 
breaking, since the second can actually be related via string
dualities to conventional Scherk-Schwarz
deformations along the eleventh coordinate. In the latter 
case we shall also meet an interesting phenomenon,
``brane supersymmetry'', where this appropriate term was
actually coined in \cite{kaktye}: 
the low-lying excitations of a brane immersed in
a non-supersymmetric bulk {\it can be} supersymmetric. This phenomenon is
generic and, as we shall see, admits a neat geometrical interpretation. 
Orientifolds of this type with partial breaking of supersymmetry 
were first discussed in \cite{adds}, while 
in more complicated models even entire towers of brane 
excitations can be supersymmetric \cite{bg}. 
\vskip 12pt
\subsection{Momentum shifts: Scherk-Schwarz supersymmetry breaking}

Let us begin by describing the effect of momentum deformations \cite{bd,ads1}. 
As we shall see, in this case the open sector will involve branes
that fill the compactified dimension, and are thus
affected by
the Scherk-Schwarz deformation. These models can be realized 
via freely-acting orbifolds,
projecting the IIB superstring with the $\bb{Z}_2$ 
generator $(-1)^F \, \delta$, where $F=F_{{\rm L}} + F_{{\rm R}}$ is 
the total space-time fermion number and $\delta$ is the shift $ x^9 \to x^9+
\pi R$ along the ninth spatial dimension, a circle of radius $R$. 
The resulting torus partition function is
\ba
{\cal T}_{{\rm KK}} &=& 
{\textstyle{1\over 2}} \Biggl[ |V_8 - S_8 |^2 \; \Lambda_{m,n} 
+ |V_8 + S_8 |^2 \; (-1)^m \Lambda_{m,n}
\nonumber \\
& & + |O_8 - C_8 |^2 \; \Lambda_{m,n+{1\over 2}} +
|O_8 + C_8 |^2 \; (-1)^m \Lambda_{m,n+{1\over 2}} \Biggr] \, ,
\ea
where, for brevity, we have let
\be
\Lambda_{m+a,n+b} = 
{q^{{\alpha ' \over 4} \left( {(m+a)\over R} + 
{(n+b)R \over \alpha '}\right)^2}\; \bar q ^{{\alpha ' \over 4} \left(
{(m+a)\over R} - {(n+b)R \over \alpha '} \right)^2} \over \eta (q) \, 
\eta (\bar q)} \, ,
\ee
while leaving all lattice sums implicit.
Expanding the various terms in ${\cal T}_{{\rm KK}}$ then yields
the orthogonal decomposition of the spectrum
\ba
{\cal T}_{{\rm KK}} &=& (V_8 \bar V_8 + S_8 \bar S_8 ) \Lambda_{2m,n}
+ (O_8 \bar O_8 + C_8 \bar C_8) \Lambda_{2m,n+{1\over 2}} \nonumber \\
& & - (V_8 \bar S_8 + S_8 \bar V_8 ) \Lambda_{2m+1,n}
- (O_8 \bar C_8 + C_8 \bar O_8 ) \Lambda_{2m+1,n+{1\over 2}} \, .
\label{torshift}
\ea
Notice that the torus
amplitude (\ref{torshift}) develops a tachyonic instability for
$R \sim \sqrt{\alpha'}$, while for $R \to \infty$ the
standard supersymmetric IIB string is formally recovered. These
properties are shared by the descendants that we are about to describe,
and therefore we shall implicitly restrict our analysis to values
of $R$ inside the region of stability. 

The Klein bottle amplitude completes the 
projection of the closed sector, and thus receives contributions
from all modes mapped  onto themselves by $\Omega$. 
The relevant lattice states, defined by the condition
$p_{{\rm L}} = p_{{\rm R}}$, have zero winding 
number, and therefore  the resulting amplitude is
\be
{\cal K}_{{\rm KK}} = {\textstyle{1\over 2}} (V_8 - S_8) \; P_{2m} \, ,
\ee
while the corresponding transverse-channel amplitude is
\be
\tilde{\cal K}_{{\rm KK}} 
= {2^5 \over 4} \, v  \, (V_8 - S_8)\;
W_{n} \, ,
\ee
where $v={R\over {\sqrt{\alpha '}}}$.

In a similar fashion, the transverse-channel annulus amplitude is
determined restricting the diagonal portion of the spectrum in
${\cal T}_{{\rm KK}}$ 
to the zero-momentum sector, $m=0$. Thus, the only contributions 
allowed in the $\tilde{\cal A}$  come from $V_8$ and $S_8$ with integer 
windings and from $O_8$ and $C_8$ with half-integer ones.
As a result, one can naturally introduce four different types
of Chan-Paton charges, obtaining 
\ba
\tilde{\cal A}_{{\rm KK}}
&=& {2^{-5}\over 4} \, v \,
 \Bigl\{
\bigl[ (n_1 + n_2 + n_3 + n_4 )^2 V_8 \nonumber \\
& &- (n_1 + n_2 - n_3 - n_4 )^2 S_8 \bigr] \, W_n 
\\
& & + \bigl[ (n_1 - n_2 + n_3 - n_4 )^2 O_8 \nonumber \\
& &-
 (n_1 - n_2 - n_3 + n_4 )^2 C_8 \bigr] \, W_{n+{1\over 2}}
\Bigr\} \,. \nonumber
\ea
The relative signs of the various contributions of the 
closed spectrum to $\tilde{\cal A}$ then reveal that $n_1$ and $n_2$ count the 
D9 branes, while $n_3$ and $n_4$ count the 
D9 antibranes.

Finally, the characters common to $\tilde{\cal K}_{{\rm KK}}$ and to 
$\tilde{\cal A}_{{\rm KK}}$ determine the transverse-channel M\"obius 
amplitude
\ba
\tilde{\cal M}_{{\rm KK}} &=& - \, \frac{v}{2} \, 
\Bigl[ (n_1 + n_2 + n_3 + n_4 ) \, \hat V _8 \; W_{n} \nonumber \\
& & - (n_1 + n_2 - n_3 - n_4 ) \, \hat S _8 \; (-1)^n W_{n} 
\Bigr] \,.
\ea
Whereas the massless contributions are fully fixed by the tadpole
conditions
\ba
\hbox{{\rm NS-NS:}} & \quad n_1 + n_2 + n_3 + n_4 =& 32 \  ,
\nonumber \\
\hbox{{\rm R-R:}} & \quad n_1 + n_2 - n_3 - n_4 =& 32\, , 
\label{KKtad}
\ea
the signs of the massive contributions are to be appropriately chosen 
in order that the direct-channel amplitudes
\ba
{\cal A}_{{\rm KK}} &=& 
{\textstyle{1\over 2}} (n_1^2 + n_2^2 + n_3^2 + n_4^2 )
\left[ V_8\, P_{2m} - S_8 \, P_{2m+1} \right] \nonumber \\
& & 
+ (n_1 n_2 + n_3 n_4 ) \left[ V_8 \, P_{2m+1} - S_8 \, P_{2m} \right]
\nonumber \\
& &+ (n_1 n_3 + n_2 n_4 ) \left[ O_8 \, P_{2m} - C_8 \, P_{2m+1} \right]
\nonumber \\
& & +  (n_1 n_4 + n_2 n_3 ) \left[ O_8 \, P_{2m+1} - C_8 P_{2m} \right] 
\ea
and
\ba
{\cal M}_{{\rm KK}} 
&=& - {\textstyle{1\over 2}} (n_1 + n_2 + n_3 + n_4 )
\, \hat V _8 \, P_{2m} \nonumber \\
& &+ {\textstyle{1\over 2}}(n_1 + n_2 - n_3 - n_4 )\, \hat S _8 \, 
P_{2m+1} 
\ea
have a consistent particle interpretation. 

The R-R tadpole conditions fix the {\it net} number of branes in the model.
If the NS-NS tadpoles are also enforced, no  
antibranes are allowed ($n_3 = n_4 =0$). The resulting spectrum, 
free of tachyons, has then an ${\rm SO} (n_1) \times
{\rm SO} (32-n_1)$ gauge group, with spinors in the
bi-fundamental. This partial breaking of the gauge symmetry can be
ascribed to Wilson lines in the original SO(32) gauge group, and can
be generalized by the methods of section 4 to further breakings.

\vskip 12pt
\subsection{Winding shifts: M-theory breaking and ``brane supersymmetry''}

As we have emphasized, in String Theory one has the additional
option of introducing winding shifts. A T-duality 
can turn these into more conventional 
momentum shifts, but only at the price of turning the type IIB string into 
type IIA, so that the dimension
of the branes is correspondingly affected. Intuitively, one would
then expect that the resulting momentum
shifts, {\it orthogonal} to the D8 branes, be ineffective on their
excitations.
This phenomenon, usually referred to as ``brane supersymmetry'', can 
be nicely illustrated by the following simple nine-dimensional example,
where, however, it is only present for the massless modes.

The starting point is now the partition function for the type IIB superstring
with winding shifts along a circle of radius $R$, 
\ba
{\cal T}_{{\rm W}} &=& ( V_8 \bar V_8 + S_8 \bar S_8 ) 
\Lambda_{m,2n} + ( O_8 \bar O_8 + C_8 \bar C_8 ) 
\Lambda_{m+{1\over 2},2n} \nonumber \\
& & - (V_8 \bar S_8 + S_8 \bar V_8)
\Lambda_{m,2n+1} - (O_8 \bar C_8 + C_8 \bar O_8 )
\Lambda_{m+{1\over 2},2n+1} \, ,
\label{torwshift}
\ea
where again  for $R \sim \sqrt{\alpha'}$ a tachyonic instability appears, 
while now the supersymmetric spectrum is formally recovered for $R \to 0$.
These properties are shared by the descendants that we are about to describe,
and therefore we shall implicitly restrict our analysis to values
of $R$ inside the region of stability. This, however, is beyond the
domain of applicability of field theory considerations and, as we shall
see, some surprises are in store.

In this case, new states contribute to the direct-channel Klein 
bottle amplitude
\be
{\cal K}_{{\rm W}} = {\textstyle{1\over 2}} (V_8 - S_8 ) \; P_{m}
+ {\textstyle{1\over 2}} (O_8 - C_8 ) \; P_{m+\frac{1}{2}} \, ,
\ee
while in the corresponding transverse-channel amplitude
\be
\tilde{\cal K}_{{\rm W}} = {2^5 \over 2}\, 2 \,  v \,
\left( V_8\; W_{4n} - S_8 \; W_{4n+2} \right) \, ,
\ee
where $v = \frac{R}{\sqrt{\alpha'}}$,
the only massless contribution originates from the NS-NS character $V_8$. 
As a result, these winding-shift orientifolds involve 
O9 and $\overline{\rm O} 9$ 
planes, whose overall R-R charge indeed vanishes.

On the other hand, the transverse-channel annulus amplitude can only
accommodate $V_8$ and $S_8$, that have zero-momentum lattice modes, and
can be written in the form
\ba
\tilde{\cal A}_{{\rm W}} &=& {2^{-5} \over 2} \, 2 \, v \,
\Bigl\{ \Bigl[ (n_1+n_2+n_3+n_4)^2 \; V_8 \nonumber \\
& & - \; (n_1+n_2-n_3-n_4)^2 \; S_8 \Bigl] 
W_{4n} \nonumber
\\
& & + \Bigl[ (n_1-n_2+n_3-n_4)^2 \; V_8 \nonumber \\
& & - \; (n_1-n_2-n_3+n_4)^2 \; S_8 
\Bigr]  W_{4n+2} \Bigr\} \, ,
\,.
\ea
where for later convenience we have distinguished four types of contributions.
As usual, from the relative sign of the Chan-Paton multiplicities in the
coefficient of $S_8$ one can see that $n_1$ and $n_2$ count the D9 branes,
while $n_3$ and $n_4$ count the D9 antibranes.

From $\tilde{\cal K}_{{\rm W}}$ and $\tilde{\cal A}_{{\rm W}}$ one can 
derive as usual the transverse-channel M\"obius amplitude
\ba
\tilde{\cal M}_{{\rm W}} &=& - 2 \, v \,  \Bigl[
(n_1+n_2+n_3+n_4) \, \hat V_8\, W_{4n} \nonumber \\
& & - (n_1-n_2-n_3+n_4) \, \hat S_8\, 
W_{4n+2} \Bigr]
\ea
and then extract the tadpole conditions
\be
n_1+n_2+n_3+n_4 = 32\, , \qquad \qquad n_1+n_2=n_3+n_4 \,.
\ee
However, in the limit $R \to 0$, that as we have seen
is well within the stability region, this model develops the
additional tadpoles
\be
n_1+n_3=n_2+ n_4 \, ,\qquad\qquad   n_1-n_2-n_3+ n_4 = 32 \, ,
\ee
arising from the sectors with shifted winding sums $W_{4n+2}$,
whose states collapse to zero mass. This is the analogue, in this
context, of the phenomenon stressed in \cite{pw} and reviewed
in subsection 4.1, and
enforcing all these conditions leads to the unique solution
\be
n_1 = 16 = n_4 \, , \qquad\qquad n_2 = 0 = n_3 \, ,
\ee
Finally, $S$ and $P$ modular transformations yield the direct-channel 
open string amplitudes
\ba
{\cal A}_{{\rm W}} &=& {\textstyle{1\over 2}} (n_1^2 + n_4^2)\,  
(V_8 - S_8 )\; 
(P_{m} +  P_{m+\frac{1}{2}}) \nonumber \\
&& + n_1 n_4 \, (O_8 - C_8) \; 
(P_{m+ \frac{1}{4}} + P_{m+ \frac{3}{4}})
\ea
and
\be
{\cal M}_{{\rm W}} = - {\textstyle{1\over 2}} 
 (n_1+n_4) \left[ (\hat V _8  - \hat S_8)\,  P_{m} + 
(\hat V _8  + \hat S_8)\, P_{m+ \frac{1}{2}} \right]  \,.
\ee
The resulting spectrum is indeed supersymmetric at the massless level,
where it contains a vector multiplet for the gauge group ${\rm SO}(16) \times
{\rm SO} (16)$. On the other hand, the massive excitations 
are not supersymmetric,
as a result of the different M\"obius projections of Bose and Fermi modes, as
well as of the presence of the $O_8$ and $C_8$ sectors, and therefore the
breaking will be transmitted to the massless modes via
radiative corrections.

The peculiar result for the gauge group can actually be given an
interesting interpretation in terms of M theory \cite{mtheory}. Namely, by a
T-duality one can turn these winding shifts into momentum shifts
in a direction orthogonal to the branes, that can be identified with
the eleventh dimension of M theory. We are thus facing 
Scherk-Schwarz breakings in the Ho\v{r}ava-Witten scenario \cite{hw},
that here have a perturbative description. 
These interesting issues are further discussed in \cite{ads1},
while a field theory construction along these lines may be 
found in \cite{horss}.

\vskip 12pt
\subsection{Comment: Scherk-Schwarz and orbifold bases}

In the previous two subsections, we have seen how two
different freely-acting orbifolds of the circle can induce the breaking of
supersymmetry via momentum or winding shifts, and we
have also referred to the first possibility as a conventional
Scherk-Schwarz deformation. 
While correct in spirit, however, this definition does not
correspond to the common use of the term in Field Theory,
since the canonical
Scherk-Schwarz deformation for a circle would lead to
periodic bosons and antiperiodic fermions, a choice manifestly
compatible with any low-energy effective field theory, where
fermions only enter via their bilinears. On the other hand,
from eq. (\ref{torshift}), rewritten more explicitly as
\ba
{\cal T}_{{\rm KK}} &=& (V_8 \bar V_8 + S_8 \bar S_8 ) \Lambda_{2m,n}(R)
+ (O_8 \bar O_8 + C_8 \bar C_8) \Lambda_{2m,n+{1\over 2}}(R) \\
& & - (V_8 \bar S_8 + S_8 \bar V_8 ) \Lambda_{2m+1,n}(R)
- (O_8 \bar C_8 + C_8 \bar O_8 ) \Lambda_{2m +1,n+{1\over 2}}(R) \,,
\nonumber 
\ea
it is clear that bosons and fermions have {\it even} and {\it odd}
momenta in the orbifold, but it is simple to relate the 
two settings: the conventional Scherk-Schwarz basis of Field Theory 
can be recovered letting $R_{\rm SS} = {1\over
2}R$, so that
\ba
{\cal T}_{{\rm SS}} &=& (V_8 \bar V_8 + S_8 \bar S_8 ) \Lambda_{m,2n}
(R_{\rm SS})
+ (O_8 \bar O_8 + C_8 \bar C_8) \Lambda_{m,2n+1}(R_{\rm SS})  \\
& & - (V_8 \bar S_8 + S_8 \bar V_8 ) \Lambda_{m +{1\over 2},2n}(R_{\rm SS})
- (O_8 \bar C_8 + C_8 \bar O_8 ) \Lambda_{m +{1\over 2},2n+1}(R_{\rm
SS}) \, , \nonumber
\ea
where bosons and fermions have indeed the correct momentum quantum numbers.

Similar considerations apply to the orientifolds, where the 
Scherk-Schwarz basis illuminates the geometry of the configurations. 
Thus, for instance, in the M-theory breaking model, letting now
$R = \tilde{R}_{\rm SS}/2$, from 
\be
\tilde{\cal K}_{\rm W} = {2^5 \over 2}\,{\tilde{R}_{\rm SS}\over 
\sqrt{\alpha '}}\,
\left[ V_8\; W_{2n} (\tilde{R}_{\rm SS}) - S_8 \; W_{2n+1} 
(\tilde{R}_{\rm SS})\right] \, 
\ee
and
\be
\tilde{\cal A}_{\rm W} = {2^{-5} \over 2}  {\tilde{R}_{\rm SS}
\over \sqrt{\alpha '}}
\left[ (n_1+ (-1)^n n_4)^2 \, V_8  
-  (n_1-(-1)^n n_4)^2 \, S_8 \right] W_{n}(\tilde{R}_{\rm SS})
\ee
one can see that in the T-dual picture the two fixed points accommodate
an O8 plane together with a stack of $n_1$ D8 branes, and an 
O8 antiplane with a stack of $n_4$ D8 antibranes, respectively.
The corresponding open-string spectrum also takes a simpler form
in the Scherk-Schwarz basis, and
is described by
\be
{\cal A}_{{\rm W}} = {\textstyle{1\over 2}} (n_1^2 + n_4^2)\,  
(V_8 - S_8 )\; P_{m} (\tilde{R}_{\rm SS})  + n_1 n_4 \, (O_8 - C_8) \; 
P_{m+ \frac{1}{2}}  (\tilde{R}_{\rm SS})
\ee
and
\be
{\cal M}_{{\rm W}} = - {\textstyle{1\over 2}}  
 (n_1+n_4) \left[\hat V _8  - \hat S_8 (-1)^m\right]  P_{m} (\tilde{R}_{\rm SS})   \,.
\ee

\vskip 12pt
\subsection{Supersymmetric six-dimensional $T^4 /\bb{Z}_2$ orbifolds}

We now turn to the open descendants of the $T^4/\bb{Z}_2$ 
compactification of the type-IIB superstring. In this case
the $\bb{Z}_2$ action on the bosonic coordinates, described in 
subsection 5.1, has to be supplemented by a corresponding prescription for 
the fermionic modes. To this end, it is convenient to recall the
${\rm SO}(4) \times {\rm SO}(4)$ decomposition of the SO(8)
characters
\ba
V_8 &= V_4 O_4 + O_4 V_4 \, , \qquad O_8 &= O_4 O_4 + V_4 V_4 \, ,
\nonumber \\
S_8 &= C_4 C_4 + S_4 S_4 \, , \qquad C_8 &= S_4 C_4 + C_4 S_4 \, ,
\ea
where the first SO(4) factor refers to the transverse space-time
directions and the second to the internal
ones. World-sheet supersymmetry demands that 
the $\bb{Z}_2$ actions on bosonic and fermionic coordinates 
be properly correlated \cite{dhvw,4d}, and this can be achieved if one assigns
positive eigenvalues to the internal $O_4$ and $C_4$
and negative ones to the internal $V_4$ and $S_4$.
The action on the fermionic coordinates and the results of 
subsection 5.1 for the bosonic string determine completely the
modular invariant torus amplitude
\ba
{\cal T} &=& {\textstyle\frac{1}{2}} \Biggl[ |Q_o + Q_v|^2 
\sum_{m,n} {q^{{\alpha' \over 4} p_{{\rm L}}^{{\rm T}} g^{-1} p_{{\rm L}}}
\bar q^{{\alpha' \over 4} p_{{\rm R}}^{{\rm T}} g^{-1} p_{{\rm R}}} 
\over \eta^4 \bar \eta ^4 }
 + |Q_o - Q_v |^2 \left| {2 \eta \over 
\vartheta_2} \right|^4 
\nonumber \\
& & + 16 |Q_s + Q_c |^2 \left| {\eta \over \vartheta_4}\right|^4
+ 16 |Q_s - Q_c |^2 \left| {\eta \over \vartheta_3} \right|^4 \Biggr]\, ,
\ea
where the left and right momenta are as in (\ref{plr}) with 
vanishing $B_{ab}$, 
and where the multiplicity of the twisted contributions reflects the
number of fixed points. In writing this expression, we have also 
introduced the supersymmetric combinations of characters
\ba
& Q_o = V_4 O_4 - C_4 C_4 \, , \qquad \quad 
& Q_v = O_4 V_4 - S_4 S_4 \, ,
\nonumber \\
& Q_s = O_4 C_4 - S_4 O_4 \, , \qquad \quad 
& Q_c = V_4 S_4 - C_4 V_4 \, ,
\ea
that are eigenvectors of the $\bb{Z}_2$ generator \cite{bs,bs2}.

The partition function clearly encodes the massless string excitations,
that can be identified using the standard ${\rm SO} (4)
\sim {\rm SU} (2) \times {\rm SU} (2)$ decompositions. For instance
\ba
V_4 \times \bar V_4 &=& (2,2) \times (2,2) = (3,3) + (3,1) 
+ (1,3) + (1,1) \, ,
\nonumber \\ 
C_4 \times \bar C_4 &=& (2,1) \times (2,1) = (3,1) + (1,1) \, ,
\nonumber \\
S_4 \times \bar S_4 &=& (1,2) \times (1,2) = (1,3) + (1,1) \, ,
\nonumber \\
V_4 \times \bar C_4 &=& (2,2) \times (2,1) = (3,2) + (1,2) \, ,
\nonumber \\
V_4 \times \bar S_4 &=& (2,2) \times (1,2) = (2,3) + (2,1) \,.
\ea
Hence, $|Q_o|^2$ describes the ${\cal N}=(2,0)$ gravitational multiplet,
that contains the metric (3,3), five self-dual two-forms (3,1)
and two left-handed gravitinos, each described by a pair of (3,2), 
together with a tensor multiplet, 
that contains an antiself-dual two-form (1,3), five scalars (1,1)
and two right-handed spinors, each described by a pair of (1,2).
In fact, six-dimensional fermions 
are conveniently described as Sp(2) doublets of Majorana-Weyl (2,1) or (1,2) 
spinors \cite{romans}. Altogether, the massless spectrum comprises
the ${\cal N}=(2,0)$ gravitational multiplet and 21 tensor multiplets, 
the unique
six-dimensional anomaly-free spectrum with this supersymmetry \cite{alvgaum},
and this result reflects the well-known geometrical interpretation
of the $T^4/\bb{Z}_2$ orbifold as a singular
point in the moduli space of the K3 surface \cite{aspin}. 

As usual, the construction of the open descendants begins with the
Klein-bottle amplitude. The standard choice,
\ba
{\cal K} &=& {\textstyle \frac{1}{4}} \Biggl[ (Q_o + Q_v) \left( 
\sum_m {q^{{\alpha'\over 2} m^{{\rm T}} g^{-1} m} \over \eta^4} + 
\sum_n {q^{{1\over 2\alpha'} n^{{\rm T}} g n} \over \eta^4} 
\right) \nonumber \\
& & + 2 \times 16 (Q_s + Q_c) \left( 
\frac{\eta}{\vartheta_4} \right)^2 \Biggr] \, ,
\label{kt4s}
\ea
yields a projected closed spectrum comprising the ${\cal N}= (1,0)$
gravitational multiplet (the graviton, one self-dual two-form and
one left-handed gravitino), a single tensor multiplet (one antiself-dual
two-form, one scalar and one right-handed spinor) and 20 hyper multiplets
(four scalars and one right-handed spinor), 16 of which originate from
the fixed points of the orbifold. The corresponding transverse-channel
amplitude
\ba
\tilde{\cal K} &=& \frac{2^5}{4} \Biggl[ (Q_o + Q_v) \left( 
v_4 \sum_n {q^{{1\over \alpha'} n^{{\rm T}} g n} \over \eta^4} 
+ \frac{1}{v_4} \sum_m {q^{\alpha' m^{{\rm T}} g^{-1} m} \over \eta^4}
\right) \nonumber \\
& &  +
2 (Q_o - Q_v)  \left( 
\frac{2 \eta}{\vartheta_2} \right)^2 \Biggr] \, ,
\ea
where $v_4=\sqrt{{\rm det} g /(\alpha')^4}$ is proportional to the 
internal volume, determines the massless tadpole contributions
\be
\tilde{\cal K}_0 = \frac{2^5}{4} \left[ Q_o \left( 
\sqrt{v_4} + \frac{1}{\sqrt{v_4}} \right)^2  +
Q_v \left( 
\sqrt{v_4} - \frac{1}{\sqrt{v_4}} \right)^2
 \right] \,. \label{kt4st}
\ee
From this expression one can see that the usual O9 planes are supplemented
with additional O5 ones, with standard negative values for
tension and R-R charge. Referring for simplicity to the tensions, 
the two NS-NS contributions to $\tilde{\cal K}_0$ associated to $Q_o$ and $Q_v$
are indeed the derivatives of 
\be
\Delta S \sim - \sqrt{v_4} \, \int d^6 x \, \sqrt{-g} \, e^{-\varphi_6} -
\frac{1}{\sqrt{v_4}} \, \int d^6 x \, \sqrt{-g} \, e^{-\varphi_6} 
\label{oplanecoupls}
\ee
with respect to the deviations of the
six-dimensional dilaton $\varphi_6$ and of the internal volume $v_4$ 
around their background values, defined via
\be
\varphi_6 \to \varphi_6 + \delta \varphi_6 \, , \qquad
\sqrt{v_4} \to \left( 1 + \delta h \right) \, \sqrt{v_4} \, .
\ee
The meaning of (\ref{oplanecoupls})  is perhaps more transparent in terms of 
the ten-dimensional dilaton, related to $\varphi_6$ by
\be
 v_4 \, e^{-2 \varphi_{10}} = e^{-2 \varphi_{6}} \, ,
\ee
as demanded by the compactification of the Einstein term in the string
frame, since it is then clear that the two terms in
\be
\Delta S \sim - v_4 \, \int d^6 x \, \sqrt{-g} \, e^{-\varphi_{10}} -
 \, \int d^6 x \, \sqrt{-g} \, e^{-\varphi_{10}} \, ,
\ee
refer to the O9 and O5 planes, respectively, and determine 
precisely their relative tensions.

Actually, in this orbifold (\ref{kt4s}) is not the only allowed 
choice for ${\cal K}$. Other interesting choices are 
\ba
{\cal K} &=& {\textstyle \frac{1}{4}} \Biggl[ (Q_o + Q_v) \left( 
\sum_m (-1)^m \, {q^{{\alpha'\over 2} m^{{\rm T}} g^{-1} m} \over \eta^4}
+ \sum_n (-1)^n \, 
{q^{{1\over 2\alpha'} n^{{\rm T}} g n} \over \eta^4}
\right) 
\nonumber \\
& & + 2 \times (8-8) (Q_s + Q_c) 
\left( 
\frac{\eta}{\vartheta_4} \right)^2 \Biggr] \, ,
\label{kt4c}
\ea
where $(-1)^m$ and $(-1)^n$ indicate symbolically a variety of
options available for introducing alternating signs in one or
more tori. All these choices result in identical massless anomaly-free 
${\cal N}= (1,0)$ 
{\it closed} spectra comprising, together with the 
gravitational
multiplet, nine tensor multiplets, eight of which originate from the
twisted sector, and twelve hyper multiplets, eight of which originate from the
twisted sector \cite{poltens}. Together with the toroidal model of 
eq. (\ref{k1ns}), these are notable examples of supersymmetric
orientifolds that are consistent {\it without} 
open strings. The corresponding transverse channel amplitude,
that we write symbolically
\ba
\tilde{\cal K} &=& \frac{2^5}{4} (Q_o + Q_v) 
\Biggl( v_4 \sum_n {q^{{1\over \alpha'} (n+{1\over 2})^{{\rm T}} g 
(n+{1\over2})} \over \eta^4} \nonumber \\
& & +
\frac{1}{v_4} \sum_m {q^{\alpha' (m+{1\over 2})^{{\rm T}} 
g^{-1} (m+{1\over 2})} \over \eta^4} \, ,
\Biggr)
\ea
has only massive contributions, so that indeed no massless tadpoles are
generated. 
A third consistent choice, 
\ba
{\cal K} &=& {\textstyle \frac{1}{4}} \Biggl[ (Q_o + Q_v) \left( 
\sum_m {q^{{\alpha'\over 2} m^{{\rm T}} g^{-1} m} \over \eta^4}
+ \sum_n {q^{{1\over 2\alpha'} n^{{\rm T}} g n} \over \eta^4}
\right) \nonumber \\ & &  - 2 \times 16 (Q_s + Q_c) 
\left( 
\frac{\eta}{\vartheta_4} \right)^2 \Biggr] \, ,
\label{kt4bsb}
\ea
has the peculiar feature of leading to a non-supersymmetric open sector
\cite{bsb,cab,au,bmp,aadds,abg}, and will be 
described in detail in subsection 5.8.

We now turn to the open sector associated to the standard Klein-bottle 
amplitude of eq. (\ref{kt4s}). The simplest choice corresponds to 
introducing only branes sitting at a single fixed point and
no Wilson lines, and is described by
\ba
{\cal A} &=& {\textstyle \frac{1}{4}} \Biggl[ (Q_o + Q_v) \left( 
N^2 \sum_m {q^{{\alpha ' \over 2} m^{{\rm T}} g^{-1} m} \over \eta^4}
+ D^2 \sum_n {q^{{1\over 2\alpha '} n^{{\rm T}} g n} \over \eta^4}
\right) 
\nonumber \\
& & + 
\left(R_N^2 + R_D^2 \right) (Q_o - Q_v) \left( {2\eta \over 
\vartheta_2}\right)^2 \label{annz2orb}
\\
& & + 2 N D \, (Q_s + Q_c ) \left( {\eta \over \vartheta_4}\right)^2
+ 2 R_N R_D \, (Q_s - Q_c ) \left( {\eta \over \vartheta_3}\right)^2
\Biggr] \, , \nonumber
\ea
where, as in subsection 5.1, $N$ and $D$ count the multiplicities of the
string ends with Neumann and Dirichlet boundary conditions, and, as in
\cite{ps},
$R_N$ and $R_D$ define the orbifold action on the Chan-Paton charges. In the
present examples, these are associated to the D9 and D5 branes that must 
be present in order to cancel the R-R tadpoles introduced by $\tilde{\cal K}$.
From the corresponding transverse-channel amplitude
\ba
\tilde {\cal A} &=& \frac{2^{-5}}{4} \Biggl[ (Q_o + Q_v) \left( 
N^2 v_4 \sum_n {q^{{1\over 4\alpha '} n^{{\rm T}} g n} \over \eta^4}
 + \frac{D^2}{v_4} \sum_m {q^{{\alpha ' \over 4} m^{{\rm T}} g^{-1} m} 
\over \eta^4} \right) \nonumber \\ 
& & + 
2 N D \, (Q_o - Q_v ) \left( {2 \eta \over \vartheta_2}\right)^2
\nonumber \\
& & + 16 \left(R_N^2 + R_D^2 \right) (Q_s + Q_c) \left( {\eta \over 
\vartheta_4}\right)^2 \nonumber \\
& & - 2 \times 4 R_N R_D \, (Q_s - Q_c ) 
\left( {\eta \over \vartheta_3}\right)^2
\Biggr] \, ,
\ea
one can then extract the tadpole contributions
\ba
\tilde {\cal A}_0 &=& \frac{2^{-5}}{4} \Biggl\{ Q_o \left( 
N \sqrt{v_4} + \frac{D}{\sqrt{v_4}} \right)^2 +
Q_v \left( N \sqrt{v_4} - \frac{D}{\sqrt{v_4}} \right)^2
\nonumber \\
& & + Q_s \left[ 15 R_N^2 + \left(R_N - 4 R_D \right)^2 \right] 
\nonumber \\ & & +
Q_c \left[ 15 R_N^2 + \left(R_N + 4 R_D \right)^2 \right]
\Biggr\} \,,
\ea
and, as usual, this expression contains several interesting informations.
From the untwisted terms, one can see that the Chan-Paton
multiplicities $N$ and $D$ determine indeed
the overall numbers of D9 and D5 branes, while
these individual terms match precisely the corresponding O9 and O5
contributions in eq. (\ref{kt4st}), a fact often
overlooked in the literature \cite{gp}. The additional terms related to
the exchange of twisted closed-string modes are also quite interesting,
since they neatly encode the distribution of the branes among 
the fixed points.
In this case, with all D5 branes at the same fixed point,
these tadpole terms account precisely for the 15 fixed points seen only
by the space-filling D9 branes, as well as for the single additional
fixed point where also D5 branes are present \cite{tovunp}. 

It is instructive to compare these results with a more general case,
where the D5 branes are distributed over the 16 fixed points, whose
coordinates are denoted concisely by $x$. The direct-channel 
amplitude now reads
\ba
{\cal A} &=& {\textstyle \frac{1}{4}} \Biggl[ (Q_o + Q_v) \Biggl( 
N^2 \sum_m {q^{{\alpha ' \over 2} m^{{\rm T}} g^{-1} m} \over \eta^4}
\nonumber \\ & & 
+ \sum_{i,j=1}^{16}D_i D_j \sum_n {q^{{1\over 2\alpha '} 
(n + x_i - x_j)^{{\rm T}} g (n + x_i - x_j)} \over \eta^4}
\Biggr) 
\nonumber \\ & & + 
\left(R_N^2 + \sum_{i=1}^{16} R_{D,i}^2 \right) (Q_o - Q_v) 
\left( {2\eta \over \vartheta_2}\right)^2
 \\
& & + 2 N \sum_{i=1}^{16} D_i \, (Q_s + Q_c ) 
\left( {\eta \over \vartheta_4}\right)^2 \nonumber \\
& & + 2 R_N \sum_{i=1}^{16} R_{D,i} \, (Q_s - Q_c ) 
\left( {\eta \over \vartheta_3}\right)^2
\Biggr] \, , \nonumber
\ea
while the corresponding tadpole contributions
\ba
\tilde {\cal A}_0 &=& \frac{2^{-5}}{4} \Biggl[ Q_o \left( 
N \sqrt{v_4} + \sum_{i=1}^{16} \frac{D_i}{\sqrt{v_4}} \right)^2 +
Q_v \left( N \sqrt{v_4} - \sum_{i=1}^{16} \frac{D_i}{\sqrt{v_4}} \right)^2
\nonumber \\
& & + Q_s \sum_{i=1}^{16} \left(R_N - 4 R_{D,i} \right)^2 +
Q_c \sum_{i=1}^{16} \left(R_N + 4 R_{D,i} \right)^2
\Biggr] 
\ea
reflect again the distribution of the D5 branes among the fixed points.

One can actually consider a more general situation, where pairs of
image D5 branes are moved away from the fixed points, to generic
positions denoted concisely by $y$, as first shown in \cite{gp}. 
The main novelty
is that the $R_D$ terms are absent for the pairs of displaced branes. 
This reflects
the fact that the projection interchanges the images in each 
pair,  consistently with the structure of the conformal
field theory, and this more general configuration thus results in
the annulus amplitude 
\ba
{\cal A} &=& {\textstyle\frac{1}{4}} \Biggl[ (Q_o + Q_v) \Biggl( 
N^2 \sum_m {q^{{\alpha ' \over 2} m^{{\rm T}} g^{-1} m} \over \eta^4}
\nonumber \\
& &  
+ \sum_{i,j=1}^{16} D_i D_j \sum_n {q^{{1\over 2\alpha '} 
(n + x_i - x_j)^{{\rm T}} g (n + x_i - x_j)} \over \eta^4}
\nonumber \\
& & + \sum_{i=1}^{16}\sum_{k=1}^{2p} D_i D_k \sum_n 
{q^{{1\over 2\alpha '} 
(n + x_i - y_k)^{{\rm T}} g (n + x_i - y_k)} \over \eta^4} 
\nonumber \\ & & +
\sum_{k,l=1}^{2p} D_k D_l \sum_n {q^{{1\over 2\alpha '} 
(n + y_k - y_l)^{{\rm T}} g (n + y_k - y_l)} \over \eta^4}
\Biggr) 
\nonumber \\ 
& & + \left(R_N^2 + \sum_{i=1}^{16} R_{D,i}^2 \right) (Q_o - Q_v) 
\left( {2\eta \over \vartheta_2}\right)^2 \nonumber \\
& & 
+ 2 N \left( \sum_{i=1}^{16} D_i+\sum_{k=1}^{2p} D_k \right) \, 
(Q_s + Q_c ) \left( {\eta \over \vartheta_4}\right)^2
\nonumber \\
& & + 2 R_N \sum_{i=1}^{16} R_{D,i} \, (Q_s - Q_c ) 
\left( {\eta \over \vartheta_3}\right)^2
\Biggr] \, , 
\ea
where the indices $i,j$ refer to the D5 branes at the 16 fixed points
$x$, while the indices $k,l$ refer to the $p$ image pairs of D5 
branes away from the fixed points, at generic positions $y$.

In this case the tadpole contributions may be read from
\ba
\tilde{\cal A}_0 &=& {2^{-5}\over 4} \Biggl\{ 
Q_o \left[
N \sqrt{v_4} + {1\over \sqrt{v_4}} 
\left( \sum_{i=1}^{16} D_i + \sum_{k=1}^{2p} 
D_k \right) \right]^2 
\nonumber \\
& &  +
Q_v \left[ N \sqrt{v_4} - {1\over \sqrt{v_4}} \left(\sum_{i=1}^{16} D_i  +
\sum_{k=1}^{2p} D_k \right) \right]^2
\nonumber \\
& & + Q_s \sum_{i=1}^{16} \left(R_N - 4 R_{D,i} \right)^2 +
Q_c \sum_{i=1}^{16} \left(R_N + 4 R_{D,i} \right)^2
\Biggr\} \, ,
\ea
and, while the untwisted exchanges are sensitive to all branes, the twisted 
ones feel only the branes that touch the fixed points, consistently with
the fact that twisted closed-strings states are confined to them.
In general, some of these can be ``fractional branes'' \cite{douglas3}, 
peculiar 
branes stuck at the fixed points that are responsible for the generalized 
Green-Schwarz couplings of \cite{as92,fabio}. 
While they are not present in this
model, for a reason that will soon be evident, we shall meet them 
in the next subsections.

The transverse-channel M\"obius amplitude for this more general 
brane configuration reads
\ba
\tilde{\cal M} &=& -{\textstyle{2\over 4}} \Biggl[ 
(\hat Q_o + \hat Q_v ) \Biggl( 
N v_4 \sum_n {q^{{1\over \alpha'} n^{{\rm T}} g n} \over \hat\eta^4}
\nonumber \\ & & 
+ \sum_{i=1}^{16} {D_i\over v_4} \sum_m 
{q^{\alpha ' m^{{\rm T}} g^{-1} m} \; \over \hat\eta^4}
\nonumber \\
& & +  \sum_{k=1}^{2p} {D_k \over v_4}
\sum_m 
{q^{\alpha ' m^{{\rm T}} g^{-1} m} \; e^{4i \pi m^{{\rm T}} y_k} \over 
\hat\eta^4}
\Biggr)\nonumber \\ & &  + 
\left( N + \sum_{i=1}^{16} D_i + \sum_{k=1}^{2p} D_k \right) 
(\hat Q_o - \hat Q_v ) \left( {2\hat \eta \over \hat \vartheta_2}\right)^2
\Biggr] \, ,
\ea
and a $P$ transformation can now be used to determine the 
M\"obius projection of the open spectrum. To this end, it is important to 
notice that the $P$ transformations of the SO(4) characters given in 
subsection 3.1 
imply that $\hat Q _o$ and $\hat Q _v$ are simply interchanged,
and one is thus led to
\ba
{\cal M} &=& -{\textstyle{1\over 4}} \Biggl[ (\hat Q _o + \hat Q _v ) \Biggl(
N \sum_m {q^{{\alpha ' \over 2} m^{{\rm T}} g^{-1} m} \over \hat\eta^4}
+ \sum_{i=1}^{16} D_i \sum_n {q^{{1\over 2\alpha '} n^{{\rm T}} g n} \over
\hat\eta^4} 
\nonumber \\ & &  
+ \sum_{k=1}^{2p} D_k \sum_n {q^{{1\over 2\alpha '} 
(n+2y_k)^{{\rm T}} g (n+2 y_k) } \over
\hat\eta^4}
\Biggr) 
\nonumber \\
& & - \left( N + \sum_{i=1}^{16} D_i +\sum_{k=1}^{2p} D_k \right) 
(\hat Q _o - \hat Q _v )
\left( {2\hat\eta\over \hat\vartheta_2} \right)^2 \Biggr] \,.
\label{mt4g}
\ea
We should stress that eq. (\ref{mt4g}) implies an important
property: while the D5 branes at the fixed points lead
to unitary gauge groups whose rank is determined by their total number, 
the remaining D5 branes away from the fixed points lead to symplectic gauge
groups \cite{witsmall,gp} whose rank is determined by the number of 
displaced pairs.
The difference with respect to the toroidal case, where orthogonal groups 
naturally appear, is directly implied by the 
$P$ matrix for the SO(4) characters.
The proper parametrization for the Chan-Paton multiplicities
\ba
N &= n+ \bar n \, , 
\qquad \quad R_N &= i (n- \bar n) \, ,
\nonumber \\
D_i &= d_i + \bar d_i \, , 
\qquad \quad 
R_D &= i (d_i - \bar d_i ) \, ,
\nonumber \\
D_k &= d_k \, , \ (k=1,...,p) \, , \ &{\rm with} \ D_k \equiv D_{2p+1-k}
\ea
identifies the family of gauge groups \cite{gp}
\be
G_{{\rm CP}} = {\rm U} (n) \times \prod_{i=1}^{16} {\rm U} (d_i) 
\times \prod_{k=1}^{p} {\rm USp} (d_k ) \, ,
\ee
and the untwisted tadpole conditions 
\be
n= 16\, , \qquad \quad \sum_{i=1}^{16} d_i + 2 \sum_{k=1}^{p} d_k = 16
\ee
fix the total rank of the $N$ and $D$ factors, while the twisted ones
are identically satisfied, given the numerical coincidence of the
``complex'' Chan-Paton multiplicities $n$ and $d_i$ with their conjugates.
The structure of the $R$ coefficients reflects the absence in this
model of ``fractional'' branes carrying ``twisted'' R-R charges.

The massless spectrum can be simply extracted from ${\cal A}$ and
${\cal M}$, and 
aside from the ${\cal N} = (1,0)$ gauge multiplets
(one vector and one left-handed spinor), it includes hyper multiplets in
antisymmetric representations for the unitary gauge groups, in symmetric
representations for the symplectic
groups, and in bi-fundamental representations. 
The ND sector presents a further subtlety, since $Q_s$ actually
describes only one half of a hyper multiplet, but always presents 
itself in pairs of conjugate representations or in individual
pseudo-real representations, so that in the end 
only full hyper multiplets are consistently obtained \cite{witsmall}. 
The simplest configuration, with all D5 branes at the
same fixed point, leads to the gauge group ${\rm U} (16)_9 \times 
{\rm U} (16)_5$, where the subscripts refer to the D9 and D5 branes,
and the massless spectrum is  neatly encoded in
\ba
{\cal A}_0 &=& ( n \bar{n} + d \bar{d} ) Q_0 + {\textstyle{1\over 2}} ( 
n^2 + \bar{n}^2
+ d^2 + \bar{d}^2 ) Q_v 
+ (n \bar{d} + \bar{n} d) Q_s \nonumber \\
{\cal M}_0 &=& -  {\textstyle{1\over 2}} (n + \bar{n} + d + \bar{d}) 
\hat{Q}_v \, ,
\ea
and thus contains charged hyper multiplets in the 
$(120 + \overline{120} ,1)$ and $(1, 120 + \overline{120})$, together with
ND states that arrange themselves into
complete hyper multiplets in the $(16,\overline{16})$. This spectrum,
first derived in \cite{bs2} and later recovered in \cite{gp},
is free of all irreducible gravitational and gauge anomalies as a
result of tadpole cancellation \cite{pc}, while additional,
reducible non-abelian anomalies are disposed of by a conventional 
Green-Schwarz 
mechanism involving a single two-form, whose self-dual and antiself-dual
parts originate from the gravitational multiplet and from the single
untwisted tensor multiplet present in the model \cite{gs,six}. 

\vskip 12pt
\subsection{Introducing a quantized $B_{ab}$}

We now turn to discuss the effect of a quantized $B_{ab}$ on orbifold
compactifications \cite{kab,cab}. As we shall see, this results in 
a rich class of 
six-dimensional models, where the antisymmetric NS-NS two-tensor  
not only induces the rank reduction 
of the Chan-Paton gauge group already met in toroidal models \cite{bps}, 
but also affects the projected closed 
spectrum, that can actually contain variable numbers of $(1,0)$
tensor multiplets. Although these phenomena
emerged very early in the study of rational compactifications \cite{bs,bs2}, 
they are spelled out in a clearer fashion by the irrational 
analysis. The rational construction, however, 
has the additional virtue of exhibiting some important 
features of Boundary Conformal Field Theory, while also allowing naturally
the construction of additional classes of models with partly frozen
geometric moduli, and will be discussed 
in some detail in section 6.

Let us begin by stressing that the twisted sector of the $T^4/\bb{Z}_2$
orbifold comprises sixteen independent sub-sectors of states confined to the
sixteen fixed points, as can be seen quite clearly from the massless
contributions to ${\cal T}$, 
\be
{\cal T}_{0} = |Q_o |^2 + |Q_v|^2 + 16 \left( |Q_s|^2 + |Q_c|^2 
\right) \,.
\ee
In the usual case, as discussed in the previous subsection, 
the Klein-bottle projection treats
the sixteen fixed points symmetrically, with the end result that 
each of them contributes a $(1,0)$ hyper multiplet to the projected 
spectrum. On the other hand, in the presence of a
quantized $B_{ab}$ not all fixed points have the same $\Omega$-eigenvalue.
A similar phenomenon would also be present in a dual formulation
of the two-dimensional
toroidal model of subsection 4.2 in terms of D7 branes and O7 planes,
and is reflected in the DD terms that we shall soon meet: for a $T^2$
three of the four O7 planes would be conventional ${\rm O}_+$, with
negative tension and R-R charge, while the fourth would be an
${\rm O}_-$, with positive tension and R-R charge \cite{wittor}. In
this formulation, it is
the simultaneous presence of ${\rm O}_+$ and ${\rm O}_-$ that lowers the
background R-R charge, therefore reducing 
the rank of the Chan-Paton gauge group carried by D-branes.
As we shall see shortly, in the $T^4/\bb{Z}_2$ orbifold, there are a
few more possibilities, and
for a generic $B_{ab}$ of rank $r$ the numbers
of ${\rm O}5_+$ and ${\rm O}5_-$ planes are \cite{kab,cab}
\be
n_\pm = 2^3 (1  \pm 2^{-r/2} ) \,. \label{omeig}
\ee
This result, nicely determined by the structure of
the two-dimensional conformal field theory, is crucial to
obtain a consistent transverse Klein-bottle amplitude.

A related observation is that in orbifolds
twisted sectors live at fixed points,
that for this $T^4/\bb{Z}_2$ example coincide with the 
O5 planes. The Klein-bottle amplitude
thus results from the combined action of world-sheet parity on the 
closed-string states and on the fixed points, and for the low-lying 
modes reads
\be
{\cal K}^{(r)}_{0} \sim  {\textstyle{1\over 2}} \left[
Q_o  + Q_v  + (n_+ - n_-) \left( Q_s  + Q_c
\right) \right] \,.
\ee
One can now easily extract the massless spectrum that, aside from
the ${\cal N} =(1,0)$ gravitational multiplet, comprises 
the universal tensor multiplet and four hyper multiplets
from the untwisted sector, together with $n_+$ hyper multiplets and $n_-$
tensor multiplets from the twisted sector. Taking into account the
untwisted contributions, the allowed total numbers of tensor multiplets
are thus $n_T=1$ if $r=0$, $n_T= 5$ for $r=2$ and $n_T=7 $ for $r=4$.
These are precisely the combinations found in rational models of
this type in  \cite{bs,bs2,gepner}. In addition, as we shall also see
in subsection 6.2, the fixed points are effectively grouped into multiplets.

The full Klein-bottle amplitude can now be computed including the
contributions of massive states. As we have seen, both momentum and 
winding lattices contribute to ${\cal K}$, but in this case 
the latter have to satisfy the  constraint (\ref{babconstr}), already 
met in the construction of $\tilde{\cal A}$ for the toroidal models
with a quantized $B_{ab}$. As a result, the sum over winding states
involves a projector, so that the full amplitude reads
\ba
{\cal K}^{(r)} &=& {\textstyle{1\over 4}} (Q_o + Q_v) \, 
\left[ \sum_m {q^{{\alpha ' \over 2} m^{{\rm T}} g^{-1} m } \over
\eta ^4} + 2^{-4} \sum_{\epsilon =0,1} \sum_n 
{q^{{1\over 2\alpha '} n^{{\rm T}} g n} \, e^{{2 i \pi
\over \alpha '} n^{{\rm T}} B \epsilon} \over \eta ^4} \right]
\nonumber \\
& & + {2^{4-r/2} \over 2} \, (Q_s + Q_c) \left( {\eta \over 
\vartheta_{4}} \right)^2  \,, \label{korbbab}
\ea
where the overall coefficient in front of the winding sum ensures that the 
graviton sector is properly normalized. Notice that the two lattice sums
are related by four T-dualities and indeed, as anticipated in subsection 4.2,
the second contains a projector determined by $B_{ab}$ \cite{kab,cab}. 

An $S$ modular transformation determines the 
transverse channel Klein-bottle amplitude
\ba
\tilde{\cal K}^{(r)} &=& {2^5 \over 4} (Q_o + Q_v )
\Biggl[ v_4 \, \sum_n {( e^{-2 \pi\ell})^{{1\over \alpha '}
n^{{\rm T}} g n} \over \eta ^4 }  
\nonumber \\
& & +
{2^{-4}\over v_4} \, \sum_{\epsilon=0,1} \sum_m { ( e^{-2\pi
\ell})^ {\alpha ' (m + {1\over \alpha '} B \epsilon )^{{\rm T}} g^{-1}
(m + {1\over \alpha '} B\epsilon)} \over \eta ^4 }
\Biggr] 
\nonumber \\
& & + {2^{5-r/2}\over 2} (Q_o - Q_v) \left( {2\eta \over \vartheta_{2}}
\right)^2 \,,
\label{ktildebaborb}
\ea
where $2^{4-r}$ independent choices for the vector $\epsilon$ result in
massless contributions to $\tilde{\cal K}(r)$ and, as in 
the previous subsection,
$v_4 = \sqrt{{\rm det} (g/\alpha')}$
is proportional to the internal volume.
Extracting the leading contributions to the tadpoles, 
one can see that with these multiplicities 
all coefficients in
\be
\tilde{\cal K}^{(r)}_{0} = {2^5 \over 4} \left[ 
Q_o \left( \sqrt{v_4} + {2^{-r/2} \over
\sqrt{v_4}} \right)^2 + 
Q_v  \left( \sqrt{v_4} - {2^{-r/2} \over
\sqrt{v_4}} \right)^2 \right]  \label{kt0baborb}
\ee
are perfect squares,
a familiar fact for two-dimensional Conformal Field
Theory in the presence of boundaries and/or crosscaps.

Before turning to the open sector, let us pause to comment
on the effect of the NS-NS antisymmetric tensor on the twisted closed
sector. Although it is evident that a non-vanishing $B_{ab}$ modifies
the lattice sum, it is
less obvious that it should also alter the structure of the twisted
sector, that does not depend on the moduli defining
size and shape of the lattice, and {\it a priori} carries
no information on the $B_{ab}$ background. Still, 
the ``rule of perfect squares'' determines this result in an
unambiguous fashion, consistently with the fact that the presence of 
$B_{ab}$ reverts the $\Omega$-projection of some of
 the fixed points, interchanging the corresponding ${\rm O}_+$ and
${\rm O}_-$ planes. Let us stress that,
in this way, one can easily obtain the correct
parametrization for the Chan-Paton multiplicities, even without 
appealing to
a geometrical picture of the orbifold model. This is indeed how
unusual spectra with several tensor multiplets were originally 
discovered, in the rational models of  \cite{bs,bs2}, but these
techniques are of interest also in more complicated cases, for instance 
in asymmetric orbifolds \cite{bg} or in genuinely curved backgrounds
\cite{bwis}. In addition, they can yield rather simply peculiar
configurations with frozen geometric moduli, for instance the model
of \cite{gepner} with {\it no} tensor multiplets.

The same procedure
can be applied to the annulus and M{\"o}bius amplitudes
that, in the transverse channel, have to satisfy similar constraints.
From the torus amplitude and
from our knowledge of the structure of the fixed points, 
the massless contributions to the annulus amplitude are
\ba
\tilde{\cal A}^{(r)}_{0} &=& {2^{-5}\over 4} \Biggl\{
Q_o  \left( 2^{r/2} \sqrt{v_4} \, 
N + {1 \over \sqrt{v_4}} \sum_{i=1}^{16/2^r} D^i \right)^2 
\nonumber \\
& & + Q_v  \left( 2^{r/2} \sqrt{v_4} \, 
N - {1 \over \sqrt{v_4}} \sum_{i=1}^{16/2^r} D^i  \right)^2 
\nonumber \\
& & + 2^r\,  \sum_{i=1}^{16/2^r} \Biggl[ 
Q_s \left( R_N   -  4 \times {2^{-r/2}} \, R
^{i}_{D} \right)^2 
\nonumber \\
& & + Q_c  \left( 
R_N  + 4 \times {2^{-r/2}} \, R ^{i}_{D} \right)^2
\Biggr] \Biggr\} \,, \label{at0baborb}
\ea
where we have already related the
boundary-to-boundary reflection coefficients to the Chan-Paton
multiplicities, while stressing that there are $16/2^r$ independent
contributions from the fixed points. Both the peculiar 
structure of these twisted exchanges and the grouping of the fixed points
are clearly spelled out by the complete annulus
amplitudes, where $N$ and $D$ count the numbers of D9 and D5
branes, while $R_N$ and $R_D$ describe the corresponding
orbifold projections \cite{cab}. Including the contributions of
momentum and winding modes, one thus obtains
\ba
\tilde{\cal A}^{(r)} &=& {2^{-5}\over 4} \Biggl\{  (Q_o + Q_v )
\Biggl[ 2^{r-4} v_4 \,\, N^2 \sum_{\epsilon =0,1} \sum_n
{ (e^{-2\pi\ell})^{{1\over 4\alpha '} n^{{\rm T}} g n} 
e^{{2 i\pi \over \alpha '} n^{{\rm T}} B\epsilon} \over \eta ^4} 
\nonumber \\
& & + {1\over v_4} \sum_{i,j=1}^{16/2^r} D^i D^j
\sum_m {(e^{-2\pi\ell})^{{\alpha '\over 4} m^{{\rm T}} g^{-1} m}
e^{2i\pi m^{{\rm T}} (x^i - x^j)} \over \eta ^4} \Biggr] 
\nonumber \\
& & + 2\times 2^{r/2} (Q_o - Q_v) \left( {2 \eta \over 
\vartheta_{2}}\right)^2 \,
\sum_{i=1}^{16/2^r} N D^i 
\nonumber \\
& & + 16 (Q_s + Q_c) \left({\eta \over \vartheta_{4}}\right)^{2}
\left[ R_{N}^{2} + \sum_{i=1}^{16/2^r} (R_{D}^{i})^2 \right]
\nonumber \\
& & -8\times 2^{r/2} (Q_s -Q_c) \left({\eta\over\vartheta_{3}}\right)^{2} 
\sum_{i=1}^{16/2^r} R_N R_{D}^{i} \Biggr\} \ ,
\label{atildebaborb}
\ea
and then, in the direct channel
\ba
{\cal A}^{(r)} &=& {\textstyle{1\over 4}} \Biggl\{  (Q_o + Q_v)
\Biggl[ 2^{r-4} N^2 \sum_{\epsilon =0,1} \sum_m
{q^{{\alpha '\over 2} (m + {1\over \alpha '} B\epsilon )^{{\rm T}}
g^{-1} (m + {1\over \alpha '} B\epsilon )} \over \eta ^4} 
\nonumber \\
& &
+ \sum_{i,j=1}^{16/2^r} D^i D^j \sum_n 
{q^{{1\over 2\alpha '} (n + x^i - x^j)^{{\rm T}} g (n +
x^i - x^j)} \over \eta ^4 } \Biggr] 
\nonumber \\
& & + (Q_o - Q_v) \left( {2\eta \over \vartheta_{2}}\right)^{2} 
\left[ R^{2}_{N} + \sum_{i=1}^{2^{4-r}} (R_{D}^{i})^2 \right] 
\nonumber \\
& & + 2 \times 2^{r/2} (Q_s +Q_c) \left({\eta \over \vartheta_{4}}\right)^{2}\,
\sum_{i=1}^{16/2^r} N D^i 
\nonumber \\
& &+2 \times 2^{r/2} (Q_s -Q_c) \left({\eta\over \vartheta_{3}}\right)^{2}\,
\sum_{i=1}^{16/2^r}
R_N R_{D}^{i} \Biggr\} \,.
\ea
From this expression one can clearly see that the ND open-string
states related to the twisted sector acquire multiplicities 
determined by the rank of
the NS-NS antisymmetric tensor \cite{kab,cab}, while the fixed points
group correspondingly into multiplets. Once more, this 
non-trivial feature 
emerges naturally from the familiar condition that the boundary-to-boundary 
reflection coefficients involve perfect squares.

To conclude the construction of the open descendants, one has to add
the M{\"o}bius amplitude. From (\ref{kt0baborb}) 
and (\ref{at0baborb}), one can deduce the terms at the origin 
of the lattices
\ba
\tilde{\cal M}^{(r)}_{0} &=& - {\textstyle{2\over 4}}
\Biggl[ 
\hat Q _o \left( \sqrt{v_4}
+ {2^{-r/2}\over \sqrt{v_4}}\right) 
\left( 2^{r/2} \sqrt{v_4} \, N + {1\over
\sqrt{v_4}} \sum_{i=1}^{16/2^r} D^i \right) 
\nonumber \\
& & + \hat Q_v \left( \sqrt{v_4}
- {2^{-r/2}\over \sqrt{v_4}}\right) 
\left( 2^{r/2} \sqrt{v_4} \, N - {1\over
\sqrt{v_4}} \sum_{i=1}^{16/2^r} D^i \right) \Biggr]
\ea
that, together with corresponding massive lattice modes, determine
\ba
\tilde{\cal M}^{(r)} &=& - {\textstyle{2\over 4}} \Biggl\{  (\hat Q_o +
\hat Q_v ) \Biggl[
2^{(r-4)/2}v_4 \, N \sum_{\epsilon =0,1} \sum_n
{(e^{-2\pi\ell})^{{1\over \alpha '} n^{{\rm T}} g n}
e^{{2i\pi\over\alpha '} n^{{\rm T}} B\epsilon} \gamma_\epsilon
\over \hat\eta ^4 } 
\nonumber \\
& &+{2^{-2}\over v_4} \sum_{i=1}^{16/2^r} D^i
\sum_{\epsilon = 0,1}\sum_m {(e^{-2\pi\ell})^{\alpha ' (m +
{1\over \alpha '} B \epsilon)^{{\rm T}} g^{-1} (m + {1\over \alpha '}
B \epsilon)} \tilde\gamma_\epsilon \over \hat\eta ^4} \Biggr] 
\nonumber \\
& & + (\hat Q_o - \hat Q_v ) \left(
{2\hat\eta \over \hat\vartheta_{2}}\right)^{2} 
\left( N + \sum_{i=1}^{16/2^4} D^i \right) \Biggr\} \,.
\ea
Notice that this expression is somewhat more complicated than
$\tilde{\cal K}$ and $\tilde{\cal A}$, since both its momentum and 
winding sums depend on $B_{ab}$, but in a way perfectly 
compatible with the closed spectrum. This reflects
the factorization constraints that relate $\tilde{\cal M}$ to
the other amplitudes, and the consistency is ensured by the doubling of the
momentum and winding quantum numbers present in $\tilde{\cal M}$,
but another feature is worth stressing.
Namely, the M{\"o}bius amplitude involves the signs $\gamma_\epsilon$,
related as in the toroidal case to the D9 branes \cite{bps}, together with 
the additional signs
$\tilde\gamma_\epsilon$  related to the D5 branes \cite{cab}, 
all needed to ensure the correct normalization of the various
contributions. 

A $P$ modular
transformation then gives the direct-channel M{\"o}bius amplitude
\ba
{\cal M}^{(r)} &=& -{\textstyle{1\over 4}} \Biggl\{ (\hat Q_o + \hat
Q_v ) \Biggl[ 2^{(r-4)/2} N \sum_{\epsilon =0,1} \sum_m
{q^{{\alpha ' \over 2} (m + {1\over \alpha '} B\epsilon )^{{\rm T}}
g^{-1} (m + {1\over \alpha '} B\epsilon)} \gamma_\epsilon \over
\hat\eta ^4 } 
\nonumber \\
& & + 2^{-2} \sum_{i=1}^{16/2^r} D^i \sum_{\epsilon
=0,1} \sum_n {q^{{1\over 2\alpha '} n^{{\rm T}} g n} e^{{2
i\pi\over \alpha '} n^{{\rm T}} B\epsilon}\tilde\gamma_\epsilon \over
\hat\eta ^4 }\Biggr] 
\nonumber \\
& & - (\hat Q_o - \hat Q_v ) \left( {2\hat\eta\over \hat\vartheta_{2}}
\right)^{2} \left( N + \sum_{i=1}^{16/2^r} D^i \right)\Biggr\} \ ,
\ea
that completes the construction of the open descendants, where
the signs $\gamma_\epsilon$ and $\tilde{\gamma}_\epsilon$ 
are to satisfy the constraints
\be
\sum_{\epsilon=0,1} \gamma_\epsilon = 4 \, , \qquad \qquad \sum_
{\epsilon = 0,1 \, \in {\rm Ker}(B)} 
\tilde{\gamma}_\epsilon = 2^{(4-r)/2} \, ,
\label{transversegamma}
\ee
that associate proper tadpole contributions to the transverse channel,
and the additional constraints
\be
\sum_{\epsilon=0,1 } 
\tilde{\gamma}_\epsilon =  4 \, \xi \, , \qquad \qquad  
\sum_{\epsilon = 0,1 \, \in {\rm Ker}(B)} 
{\gamma}_\epsilon = 2^{(4-r)/2} \, \xi 
   \, , \label{directgamma}
\ee
that guarantee a proper particle interpretation
for the direct-channel amplitudes. The restrictions to ${\rm Ker}(B)$ identify 
the independent values of $\epsilon$ such that $B \epsilon = 0$ (mod 2), 
that result 
in massless contributions in the two channels.
Finally, as in the previous subsection, the consistency of the $R_{N,D}$ 
breaking terms, that must be both real or both imaginary in a real
${\cal A}^{(r)}$, allows at most a common sign 
choice $\xi = \pm 1$ for the two terms in (\ref{directgamma}).

We are now ready to extract the tadpole conditions for these 
models. From the untwisted sector one obtains
\be
\sqrt{v_4} \, \left( 2^5 - 2^{r/2}\, 
N \right) 
\,
\pm \, {1\over \sqrt{v_4}} \left( 2^{5-r/2} -
\sum_{i=1}^{16/2^r}
D^i \right) = 0 \ ,
\ee
while the twisted sector yields the additional conditions
\be
R_N - 4 \times 2^{-r/2} \,
R_{D}^{i} = 0 \,, \qquad \quad {\rm for}\quad
i = 1 , \ldots , 16/2^r \,.
\ee

We have already described the 
basic features of the massless closed 
sector for this class of models, that comprises the ${\cal N} =(1,0)$ 
supergravity multiplet coupled to $1+n_-$ tensor multiplets and $4+n_+$
hyper multiplets. The corresponding 
massless open spectrum can be obtained, as usual, expanding
the amplitudes ${\cal A}$ 
and ${\cal M}$ to lowest order in $q$, and the result is
\ba
{\cal A}^{(r)}_{0} &\sim & {\textstyle{1\over 4}}
\biggl\{ N^2 + R_{N}^{2}
+\sum_{i=1}^{16/2^r} \left[ (D^i)^2 +
(R_{D}^{i} )^2 \right]
\biggr\} \, Q_o  
\nonumber \\
& &+ {\textstyle{1\over 4}}
\biggl\{ N^2 - R_{N}^{2} +\sum_{i=1}^{16/2^r} 
\left[ (D^i)^2 - ( R_{D}^{i} )^2 \right]
\biggr\}\, Q_v 
\nonumber \\
& & + {2^{r/2} \over 2}\,\sum_{i=1}^{16/2^r} \left( N \, D^i +
R_N \, R_{D}^{i} \right) Q_s  
\ea
for the annulus amplitude, and
\ba
{\cal M}^{(r)}_{0} &\sim &  -  {\textstyle{1\over 4}} \, \xi \,
(\hat Q_o + \hat Q_v) \left( N
+ \sum_{i=1}^{16/2^r} D^i \right) 
\nonumber \\
& & + {\textstyle{1\over 4}}
(\hat Q_o - \hat Q_v) \left(
N + \sum_{i=1}^{16/2^r} D^i \right)
\ea
for the M{\"o}bius amplitude. One must still introduce an explicit
parametrization of $N$, $D$, $R_N$ and $R_D$ in terms of Chan-Paton
multiplicities, but this is fully determined by the condition that
the resulting direct-channel amplitudes admit a proper
particle interpretation or, 
equivalently, by the condition that the M{\"o}bius amplitude provide 
the correct 
symmetrization of the annulus. 

The sign $\xi$ present in ${\cal M}_0$ is
the counterpart, in these
irrational models, of the discrete Wilson lines of \cite{bs2}.
A positive $\xi$ corresponds to a projective realization of the $\bb{Z}_2$ 
orbifold group on the Chan-Paton charges, since at the massless
level the M{\"o}bius amplitude sees only untwisted hyper multiplets, and,
as a
result, the gauge group is unitary. One is thus led to the following
parametrization in terms of complex Chan-Paton multiplicities:
\ba
N &=& n+\bar n \,,
\nonumber \\
D^i &=& d^i+\bar d^i \,,
\ea
\ba
R_N &=& i\, (n-\bar n) \,,
\nonumber \\
R_{D}^{i} &=& i\, (d^i-\bar d^i) \,,
\ea
consistent with the well known result for the $T^4 /\bb{Z}_2$ orbifold
with vanishing $B_{ab}$, for which the sign $\xi$ is
uniquely fixed by the tadpole
conditions. For instance, with a single $d_i$ 
the massless spectra comprise non-Abelian
vector multiplets for the gauge group 
\be
{\rm U} (2^{4-r/2})_{9} 
\times {\rm U} (2^{4-r/2})_{5} \, ,
\ee
where the subscripts refer to the D9 and D5 branes,
and additional charged hyper multiplets in the representations 
\be
(A + \bar{A}; 1) + (1; A)+ \bar{A} + 2^{r/2} \, (F;\overline{F}) \,,
\ee
where $F$ and $A$ denote the fundamental and the two-index
antisymmetric representation, and are neatly encoded in
\ba
{\cal A}^{(r)}_0 &=& ( n \bar{n} + d \bar{d} ) Q_0 + {\textstyle{1\over 2}} ( 
n^2 + \bar{n}^2
+ d^2 + \bar{d}^2 ) Q_v 
+ 2^{r/2} (n \bar{d} + \bar{n} d) Q_s \nonumber \\
{\cal M}_0 &=& - {\textstyle{1\over 2}} (n + \bar{n} + d + \bar{d}) \hat{Q}_v \, ,
\ea
where $Q_o$ describes a vector multiplet, $Q_v$ describes a hyper multiplet
and $Q_s$ describes one half of a hyper multiplet.
\begin{table}
\caption{Some massless spectra for 
$T^4/\bb{Z}_2$ models with a rank-$r$ $B_{ab}$ ($\xi =+1$).}
\label{tabz21}
\begin{indented}
\lineup
\item[]\begin{tabular}{@{}lllll}
\br
$r$ & $n_{T}^{cl}$ & $n_{H}^{cl}$ & gauge group & charged matter \\
\mr
0 & 1 & 20 & ${\rm U} (16)_{9} \times {\rm U} (16)_{5}$ &  $
(120+\overline{120};1) + (1;120+\overline{120}) + (16;\overline{16})$ 
\\
2 & 5 & 16 & ${\rm U} (8)_{9} \times {\rm U} (8)_{5}$ & $
(28+\overline{28};1)+ (1;28+\overline{28})+ 2\, (8;\overline{8})$
\\
4 & 7 & 14 & ${\rm U} (4)_{9} \times {\rm U} (4)_{5}$ & $
(6+\overline{6};1)+ (1;6+\overline{6})+ 4\, (4;\overline{4})$ \\
\br
\end{tabular}
\end{indented}
\end{table}

The second option, $\xi=- 1$, calls instead for
the real Chan-Paton multiplicities
\ba
N &=& n_1 + n_2 \, ,\nonumber \\
D^i &=& d_1^i + d_2^i  \, ,
\ea
\ba
R_N &=& n_1-n_2 \,,
\nonumber \\
R_{D}^{i} &=& d_1^i- d_2^i \, ,
\ea
and for a single $d^i$ leads to the massless spectra
\ba
{\cal A}^{(r)}_0 &=& {\textstyle{1\over 2}} ( 
n_1^2 + n_2^2
+ d_1^2 + d_2^2 ) Q_o +  ( n_1 n_2 + d_1 d_2 ) Q_v
+ 2^{r/2} (n_1 d_1 + n_2 d_2) Q_s \nonumber \\
{\cal M}_0 &=& {\textstyle{1\over 2}} (n_1 + n_2 + d_1 + d_2) \hat{Q}_o \, ,
\ea
where $Q_o$ describes a vector multiplet, $Q_v$ describes a hyper multiplet
and $Q_s$ describes one half of a hyper multiplet,
with symplectic gauge groups and $n_1=n_2$, $d_1=d_2$ on account
of the twisted tadpole conditions. As a result, fractional branes 
\cite{douglas3} are now generically present at the fixed points
coinciding with ${\rm O}_-$ planes, while the resulting 
twisted two-forms take part in a generalized Green-Schwarz 
mechanism \cite{as92,fabio}. This
is neatly reflected in the $R$ coefficients for the individual group factors, 
proportional to $n_{1,2}$ and $d_{1,2}$, that are no more identically 
vanishing \cite{cab}. On the contrary, in the previous case complex 
charges were present and these couplings vanished identically, due to the 
numerical coincidence of the multiplicities for the individual unitary 
gauge groups with their conjugates.
Tables \ref{tabz21} and \ref{tabz22}, where  $n_{T}^{cl}$ and $n_{H}^{cl}$
denote the numbers of tensor and hyper multiplets from the projected closed
sectors, summarize the massless spectra for 
the simplest choices allowed for $r=0,2,4$ and $\xi =\pm 1$.
As in the toroidal case, continuous Wilson lines can be used to
connect unitary and symplectic gauge groups.
For simplicity, in these examples we have confined 
all the D5 branes to the same
fixed point but, as we have seen, in general one could
place them at generic positions in the internal space.

\begin{table}
\caption{Some massless spectra for 
$T^4/\bb{Z}_2$ models with a rank-$r$ $B_{ab}$ ($\xi=-1$).}
\label{tabz22}
{\small
\begin{tabular}{@{}lllll}
\br
$r$ & $n_{T}^{cl}$ & $n_{H}^{cl}$ & gauge group & charged matter \\
\mr
2 & 5 & 16 & ${\rm USp} (8)^{2}_{9} \times {\rm USp} (8)^{2}_{5}$ & $
(8,8;1,1)+ (1,1;8,8) + (8,1;8,1) + (1,8;1,8)$
\\
4 & 7 & 14 & ${\rm USp} (4)^{2}_{9} \times {\rm USp} (4)^{2}_{5}$ & 
$(4,4;1,1)+(1,1;4,4) +2\, (4,1;4,1) + 2\, (1,4;1,4)$ \\
\br
\end{tabular}
}
\end{table}

\vskip 12pt
\subsection{Brane supersymmetry breaking}

In our discussion of ten-dimensional models, we already met a 
rather surprising phenomenon: a
projected closed sector with a residual amount of
supersymmetry can be tied to an open sector where, to lowest order, 
supersymmetry is broken at the string scale \cite{sugimoto}. 
In that case, the phenomenon was ascribed
to the replacement of the conventional ${\rm O9}_+$ plane with an ${\rm O9}_-$
one, with the end result that the R-R tadpole cancellation  required
antibranes and a consequent breaking of supersymmetry. In 
lower-dimensional models with $\bb{Z}_2$ projections, the simultaneous 
presence of O9 and O5 planes offers additional possibilities. The first
option, directly related to the ten-dimensional example, would be to reverse
simultaneously tensions and charges of both O9 and O5 planes. This choice,
consistent with the standard Klein-bottle projection, would
not alter the supersymmetric closed spectrum, but 
the reversed R-R charges would call for the introduction
of antibranes, with the end result that supersymmetry would be 
broken in the whole open sector. Models with $\bb{Z}_2$ projections, however,
offer an additional possibility \cite{bsb}: one can reverse tension
and charge of only one type of orientifold plane, say the O5. This
induces a different Klein-bottle projection in the twisted closed
sector, and requires the introduction of D5 antibranes,
where the supersymmetry preserved by the D9 branes
is thus broken at the string scale. The origin
of the breaking is simple to understand: branes and antibranes break
two different halves of the original supersymmetry, and therefore
when they are simultaneously present no residual supersymmetry is
left.

We can now present a relatively simple six-dimensional $T^4/\bb{Z}_2$
model where this mechanism is at work \cite{bsb}. As we anticipated, 
the simultaneous 
presence of ${\rm O}9_+$ and ${\rm O}5_-$ planes translates in
a different Klein-bottle projection
\be 
{\cal K} = {\textstyle \frac{1}{4}} \left[ ( Q_o + Q_v ) ( P_m + W_n ) - 2
\times 16 ( Q_s + Q_c ){\left(\frac{\eta}{\vartheta_4}\right)}^2 \right]
\ , 
\label{bsb1}
\ee  
where the twisted NS-NS sectors are antisymmetrized,
while the corresponding R-R ones are symmetrized. As a result,
the projected closed spectrum, that still has (1,0) supersymmetry, comprises
seventeen tensor multiplets and four hyper multiplets.
In the corresponding transverse-channel amplitude
the terms from the origin of the lattice sums,
\be
\tilde{\cal K}_0 = \frac{2^5}{4} \left[ Q_o \left( \sqrt{v_4}  -
\frac{1}{\sqrt{v_4}}\right)^2 + Q_v \left( \sqrt{v_4}  +
\frac{1}{\sqrt{v_4}}\right)^2 \right]  \label{bsb3} \ ,
\ee  
whose coefficients are as usual perfect squares, display
rather clearly the relative signs of tensions and R-R charges for
the O-planes if compared to eq. (\ref{kt4st}).

The corresponding annulus amplitude 
\ba 
{\cal A} &=& {\textstyle \frac{1}{4}} 
\Biggl[(Q_o + Q_v) ( N^2 P_m  + D^2 W_n ) \nonumber \\
&& + (R_N^2 + R_D^2) (Q_o - Q_v) {\left(\frac{2
\eta}{\vartheta_2}\right)}^2 \nonumber 
\\
&& + 
2 N D ( O_4 S_4 - C_4 O_4 + V_4 C_4 -
S_4 V_4) {\left(\frac{\eta}{\vartheta_4}\right)}^2 
\nonumber
\\ 
&& + 2 R_N R_D ( - O_4 S_4 - C_4 O_4 + V_4 C_4 +
S_4 V_4 ){\left(\frac{
\eta}{\vartheta_3}\right)}^2 \Biggr]  \label{bsb4}
\ea  
involves D9 branes and, for simplicity, a single set of D5 antibranes,
needed to compensate the R-R charge of the orientifold planes, and indeed
the GSO projection for the ND strings is reversed with respect to the
standard supersymmetric case of eq. (\ref{annz2orb}), as stressed
in \cite{senba}. This is neatly reflected in
the structure of the untwisted massless contributions to the 
transverse-channel amplitude 
\ba
\tilde{\cal A}_0 &\sim&  (V_4 O_4 - S_4 S_4) \left( N
\sqrt{v_4}  +
\frac{D}{\sqrt{v_4}}\right)^2 \nonumber \\
&& +  (O_4 V_4 - C_4 C_4) \left( N\sqrt{v_4}  -
\frac{D}{\sqrt{v_4}}\right)^2  \,. \label{bsb5} 
\ea
The tension, encoded in the dilaton coupling, can be
read from the $V_4 O_4$ character, and is positive for both types of
branes that, however, have opposite R-R charges, as can be seen
from the coefficient of the $C_4 C_4$ character.

Finally, the contributions to the M{\"o}bius amplitude from 
the origin of the lattices 
\ba
\tilde{\cal M}_0 &=& - {\textstyle \frac{1}{2}} \Biggl[ \hat{V}_4
\hat{O}_4 \left( \sqrt{v_4}  -
\frac{1}{\sqrt{v_4}}\right) \left( N \sqrt{v_4}  +
\frac{D}{\sqrt{v_4}}\right) \nonumber \\
&& + \hat{O}_4
\hat{V}_4 \left( \sqrt{v_4}  +
\frac{1}{\sqrt{v_4}}\right) \left( N \sqrt{v_4}  -
\frac{D}{\sqrt{v_4}}\right) \nonumber \\ 
&& -\hat{C}_4 \hat{C}_4 \left(
\sqrt{v_4}  -
\frac{1}{\sqrt{v_4}}\right) \left( N \sqrt{v_4}  -
\frac{D}{\sqrt{v_4}}\right) \nonumber \\
&& - \hat{S}_4 \hat{S}_4 \left( \sqrt{v_4}  +
\frac{1}{\sqrt{v_4}}\right) \left( N \sqrt{v_4}  +
\frac{D}{\sqrt{v_4}}\right)
\Biggr] \, , \label{bsb6}
\ea  
can be easily obtained combining
${\tilde {\cal K}_0}$ and ${\tilde {\cal A}_0}$, and allow one to
reconstruct the full M{\"o}bius amplitude
\ba  
{\cal M} &=& - {\textstyle \frac{1}{4}} \Biggl[ N  ( \hat{O}_4
\hat{V}_4  + \hat{V}_4 \hat{O}_4  - \hat{S}_4 \hat{S}_4 - \hat{C}_4
\hat{C}_4 ) P_m \nonumber \\
&& -  D ( \hat{O}_4
\hat{V}_4  + \hat{V}_4 \hat{O}_4  + \hat{S}_4 \hat{S}_4 + \hat{C}_4
\hat{C}_4 ) W_n \nonumber \\ 
&& - N( 
\hat{O}_4 \hat{V}_4 - \hat{V}_4 \hat{O}_4 - \hat{S}_4 \hat{S}_4
+ \hat{C}_4 \hat{C}_4 )\left(
{2{\hat{\eta}}\over{\hat{\vartheta}}_2}\right)^2  \nonumber \\
&& + D( \hat{O}_4
\hat{V}_4 - \hat{V}_4 \hat{O}_4 + \hat{S}_4 \hat{S}_4
- \hat{C}_4 \hat{C}_4)\left(
{2{\hat{\eta}}\over{\hat{\vartheta}}_2}\right)^2  \Biggr] \,.
\label{bsb7} 
\ea  

Since the vector multiplet flows in ${\cal M}$, one is led to 
introduce real Chan-Paton multiplicities, so that
\ba  N&=n_1+ n_2 \, , \qquad D&=d_1+ d_2 \, , \nonumber \\
R_N&=n_1- n_2 \, , \qquad R_D&=d_1- d_2 \, , \label{bsb8}
\ea  
and the resulting massless spectrum is summarized in
\ba  
{\cal A}_0 + {\cal M}_0 &=& \frac{n_1(n_1-1) + n_2(n_2-1) +
d_1(d_1+1) +  d_2(d_2+1) }{2} \ V_4 O_4 \nonumber \\
 &&- \frac{n_1(n_1-1) + n_2(n_2-1) + d_1(d_1-1) + d_2(d_2-1) }{2} \ C_4
C_4 \nonumber \\ &&+ (n_1 n_2 + d_1 d_2 ) ( O_4 V_4 - S_4 S_4 ) + (
n_1 d_2 + n_2 d_1 ) \ O_4 S_4 \nonumber \\
&& - (n_1 d_1 + n_2 d_2 ) \ C_4 O_4 \,. 
\label{bsb9}
\ea  

The R-R tadpole conditions $N=D=32,R_N=R_D=0$
($n_1=n_2=d_1=d_2=16$) determine the gauge group $[ {\rm SO}(16) \times
{\rm SO}(16) ]_9 \times  [ {\rm USp}(16) \times {\rm USp}(16) ]_5$,
where the subscripts refer to D9 and $\overline{\rm D}5$ branes.
The NN spectrum is supersymmetric, and comprises the (1,0)
vector multiplet for the ${\rm SO}(16) \times {\rm SO}(16)$ gauge group and a
hyper multiplet in the representation $(16,16,1,1)$. 
On the other hand, the DD spectrum is
not supersymmetric, and contains, aside from the gauge vectors of $[
{\rm USp}(16)
\times {\rm USp}(16) ]$, quartets of scalars in the $(1,1,16,16)$,
right-handed Weyl fermions in the $(1,1,120,1)$ and in the $(1,1,1,120)$, 
and left-handed Weyl fermions in the
$(1,1,16,16)$. Finally, the ND sector, also non-supersymmetric,
comprises doublets of scalars in the $(16,1,1,16)$ and in
the $(1,16,16,1)$, together with additional symplectic
Majorana-Weyl fermions in the $(16,1,16,1)$ and $(1,16,1,16)$. 
These Majorana-Weyl fermions, already met in the previous subsections, 
are a peculiar feature of
six-dimensional space-time, where the fundamental Weyl fermion, a
pseudo-real spinor of
${\rm SU}^*(4)$, can be subjected to an additional Majorana condition,  if
this is supplemented by the conjugation in a pseudo-real representation
\cite{romans,witsmall}. In this case, this is indeed possible, since the ND
fermions are valued in the fundamental representation of ${\rm USp}(16)$.

Notice that the $\overline{\rm D} 5$ 
spectrum reflects the results already emerged in
the discussion of the ten-dimensional USp(32) model. Namely, all bosonic
and fermionic modes affected by the M\"obius projection are in different
representations, while the remaining NN and DD matter in bi-fundamental
representations fills complete hyper multiplets. The novelty here
is the ND sector, where supersymmetry is broken due to the reversed GSO
projection resulting from brane-antibrane exchanges. As in the ten-dimensional
model of \cite{sugimoto}, the open spectrum contains singlet spinors that
play a key r\^ole in the low-energy couplings discussed in \cite{dm1}.

Even in this case one can introduce a quantized $B_{ab}$ \cite{cab},
and the resulting models now contain 17, 13 or 11 tensor multiplets,
according to whether the rank $r$ of $B_{ab}$ is 0, 2 or 4, the allowed
values for $T^4$. The ranks of the resulting gauge groups are correspondingly
reduced by the familiar factors $2^{r/2}$, the ND sector occurs in
multiple families and, as in the supersymmetric 
case, one can connect orthogonal or symplectic gauge groups to unitary ones
by a choice of the $\xi$
coefficient in ${\cal M}$, the irrational
counterpart of the ``discrete Wilson lines'' of \cite{bs2}, as in
subsection 5.7.

As is typically the case for non-supersymmetric models, a dilaton potential,
here localized on the $\overline{\rm D} 5$ branes, is generated. This can be 
easily deduced from the transverse-channel amplitudes, that
in general encode the one-point functions of bulk fields on branes
and orientifold planes, and in this case the uncancelled tadpoles 
\be  
\left[ (N-32){\sqrt v_4}+{D+32 \over{\sqrt
v_4}}\right]^2 \! V_4O_4 +\left[ (N-32){\sqrt v_4}-{D+32 \over{\sqrt
v_4}}\right]^2 \! O_4V_4
 \label{bsb10}
\ee 
are associated to the characters $V_4O_4$ and $O_4V_4$, and thus
to the deviations of the six-dimensional dilaton $\varphi_6$ and
of the internal volume $v_4$ with respect to their background 
values. Proceeding as in subsection 5.6,
factorization and the R-R tadpole conditions $N=32=D$
determine the residual potential, that in the string frame reads
\be  
V_{\rm eff}=c{e^{-\varphi_6}\over{\sqrt v}}=ce^{-\varphi_{10}} ={c\over
g_{\rm YM}^2}\, , \label{bsb11}
\ee  
where we have also expressed this result in terms of
$\varphi_{10}$, the ten-dimensional dilaton, that determines the
Yang-Mills coupling $g_{\rm YM}$ on the $\overline{\rm D} 5$ branes, and where
$c$ is a {\it positive}
numerical  constant. The potential (\ref{bsb11}) is indeed localized on
the $\overline{\rm D} 5$'s, and is clearly positive. This can be understood 
noticing that
the negative O9 plane contribution to the vacuum energy
exactly cancels against that of the D9 branes for $N=32$, and this fixes
the sign of the $\overline{\rm D} 5$ and ${\rm O}5_+$ contributions,  
both positive, consistently with the interpretation of this mechanism as
global supersymmetry breaking. The potential (\ref{bsb11}) has the usual
runaway behaviour, as expected by general arguments.

As in the higher-dimensional examples, one can actually enrich this
configuration adding brane-antibrane pairs \cite{au,aadds}. 
These, however, are expected to lead to instabilities,
reflected by the generic presence of tachyonic modes.
In some cases one can have some control on the fate of these unstable 
systems, also attaining some understanding of the resulting configurations
\cite{abg}.
      
\vskip 12pt
\subsection{Chiral asymmetry with three generations in four-dimensional models}

The simplest four-dimensional type I vacuum can
be obtained starting from the IIB compactification on $T^6/\bb{Z}_3$ 
\cite{abpss}, and the resulting spectrum, with ${\cal N}=1$ supersymmetry, 
has the interesting feature of containing three generations of
chiral matter.

The $\bb{Z}_3$ projection has the natural action  
\be
Z^k \sim \omega \, Z^k \,, \qquad  {\rm with} \ \ \omega = e^{{2 i \pi} \over
3}\ \ {\rm and} \ \ k=1,2,3\,, \label{z31}
\ee
on the complex coordinates of the internal $T^6=T^2 \times T^2 \times
T^2$, where each $T^2$ corresponds to a hexagonal lattice with metric
\cite{dhvw}
\be   
g_{ab} =  {R^2\over 3} \left( \matrix{ 2 & 1 \cr 1 & 2
\cr}\right) \, ,
\label{z32}
\ee
and results in three fixed points in each $T^2$, for a 
total of 27. As a result, the Hodge numbers of the
corresponding Calabi-Yau manifold are $h_{1,1}= 36$ and $h_{1,2}=0$,
and indeed the resulting massless spectrum of the IIB superstring,
with ${\cal N}=2$ supersymmetry, comprises a total of 37 
hyper multiplets\footnote[7]{The R-R scalars are 
actually two-forms, so that the matter is better described
in terms of tensor multiplets.} \cite{chsw}.
This is neatly encoded in the torus amplitude 
\ba
{\cal T} &=& {\textstyle{1 \over 3}} \Biggl[ \Xi_{0,0}(q) \Xi_{0,0}(\bar q) 
\Lambda_{6,6} +\sum_{\lambda = \pm 1} 
\Xi_{0,\lambda}(q) \
\Xi_{0,-\lambda}(\bar q)
\nonumber \\
& & + \sum_{\rho = \pm 1}  
\sum_{\lambda = 0,\pm 1} 
\Xi_{\rho,\lambda}(q) \Xi_{-\rho,-\lambda}(\bar q) \Biggr] \, ,
\label{z33}
\ea  
where, as in the preceding subsections, 
 we have not displayed the contributions of the
transverse space-time coordinates, 
$\Lambda_{6,6}$ denotes the usual Narain lattice sum for the internal
$T^6$ and
\ba
\Xi_{0,\lambda}(q) &=& \left( {{ A_0 \chi_0 + \omega^{\lambda} A_+ \chi_- + 
{\bar\omega}^{\lambda} A_- \chi_+ } \over {{H_{0,\lambda}}^3}} \right) (q) 
\, ,
\nonumber \\
\Xi_{+,\lambda}(q) &=& \left( {{ A_0 \chi_+ + 
\omega^{\lambda} A_+ \chi_0 +  {\bar\omega}^{\lambda} A_- \chi_- } \over
{{H_{+,\lambda}}^3}} \right) (q) \, ,
\nonumber \\
\Xi_{-,\lambda}(q) &=& \left( {{ A_0 \chi_- + \omega^{\lambda} A_- 
\chi_0 +  {\bar\omega}^{\lambda} A_+ \chi_+} \over {{H_{-,\lambda}}^3}}
\right)(q) 
\,.
\label{z34}
\ea 

The projection
of the untwisted bosons involves the combinations of $\vartheta$ 
and $\eta$ functions
\be 
H_{0,{\lambda}}(q)  =  q^{1\over {12}}
\prod_{n=1}^{\infty}(1-\omega^{\lambda} q^n)  (1 - {\bar \omega}^{\lambda} 
q^n) 
\,,
\label{z35}
\ee  
with  $\lambda = {0,\pm 1}$, while the contributions of the 
twisted bosons may be similarly expressed in terms of
\be  
H_{+,{\lambda}}(q) = H_{-,{- \lambda}}(q) = 
\frac{1}{\sqrt{3}}\, {q^{-{1\over{36}}}}
\prod_{n=0}^{\infty}(1 - \omega^{\lambda} q^{n+{1\over 3}}) (1 -  {\bar 
\omega}^{\lambda} q^{n+{2\over3}}) \,.
\label{z36}
\ee

Notice the slight change of notation with respect to the previous 
subsections:
here the $\Xi$ combine the contributions of
world-sheet fermions and bosons, while the multiplicities in the
twisted $H$ account for the 27 fixed points.

The contributions of the world-sheet fermions $\psi^k$, encoded in the 
combinations of $A$ and $\chi$ characters, are to be properly correlated
to those of the world-sheet bosons $Z^k$ in order to preserve  
${\cal N}=2$ space-time supersymmetry. The transverse SO(8) thus
breaks to ${\rm SO}(2) \times {\rm SU}(3) \times {\rm U}(1)$,
and standard group theory branchings determine the decomposition
\be  
V_8 - S_8 = A_0 \, \chi_0 + A_+ \, \chi_- + A_- \, \chi_+  \, ,
\label{z37}
\ee
where we have introduced the level-one SU(3)
characters
$\{\chi_0,\chi_{+},\chi_{-} \}$, of conformal weights $\{0,\frac{1}{3},
\frac{1}{3} \}$, and 
the supersymmetric characters
\ba 
A_0 &=& V_2 \, \xi_0 + O_2 \, \xi_6 - S_2 \, \xi_{-3} - C_2 \, \xi_3 \, ,
\nonumber \\  
A_{+} &=& V_2 \, \xi_4 + O_2 \, \xi_{-2} - S_2 \, \xi_1 - C_2 \, \xi_{-5} \, ,
\nonumber \\  
A_{-} &=& V_2 \, \xi_{-4} + O_2 \, \xi_2 - S_2 \, \xi_5 - C_2 \, \xi_{-1} 
\, ,
\label{z38}
\ea 
of conformal weights $\{\frac{1}{2},\frac{1}{6},\frac{1}{6} \}$.
These, in their turn, are combinations of the four level-one SO(2)
characters defined in subsection 3.1 and of the 12 characters 
$\xi_m$ $(m=-5, \ldots ,6)$, of conformal weight 
$h_m={{m^2} \over {24}}$, of the ${\cal N}=2$ super-conformal model 
with $c=1$, that can be realized by a free boson on 
the rational circle of radius $\sqrt{6\alpha'}$ \cite{cftrev}.
At the massless level, $A_0 \chi_0$ contains an ${\cal N}=1$
vector multiplet,  $A_+ \chi_-$ contains {\it three} copies of a
real scalar and of the positive-helicity component of a Weyl
spinor, while $A_+ \chi_-$ contains  {\it three}  copies of
a real scalar 
and  of the negative helicity component of a Weyl spinor. 
Together, the last two
characters thus describe a triplet of Wess-Zumino multiplets 
of four-dimensional
${\cal N}=1$ supersymmetry, while a chiral spectrum
results if they are valued in different representations.
Standard properties of $\vartheta$ functions and Poisson summations,
as in eq. (\ref{poissonsum}), determine the $S$ and $P$ matrices
\be
S_{\chi} = \frac{1}{\sqrt{3}} \left(  \begin{array}{ccc}
1&1&1\\1&\omega&\bar{\omega}\\1&\bar\omega&\omega
\end{array} \right) \, , \qquad
P_{\chi} = \frac{1}{\sqrt{3}} \left(  \begin{array}{ccc}
1&-1&-1\\-1&\bar\omega&\omega\\-1&\omega&\bar\omega\end{array} \right) \, ,
\label{z39}
\ee
\be
S_{A} = \frac{1}{\sqrt{3}} \left(  \begin{array}{ccc}
1&1&1\\1&\bar\omega&\omega\\1&\omega&\bar\omega
\end{array}  
\right) \, , \qquad 
P_{A} = \frac{1}{\sqrt{3}} \left(  \begin{array}{ccc}
1&-1&-1\\-1&\omega&\bar\omega\\-1&\bar\omega&\omega
\end{array}  
\right) \,.
\label{z310}
\ee

In constructing the open descendants, one starts as usual
by halving the torus amplitude (\ref{z33}). Since the $\bb{Z}_3$ 
action of the target space twist is left-right symmetric, the torus
amplitude if off-diagonal and only the
graviton orbit contributes to the Klein-bottle amplitude, for all
others appear off-diagonally in ${\cal T}$. Moreover
\be  
{\cal K} = {\textstyle{1 \over 6}} \left[  
\Xi_{0,0} P_6 + \Xi_{0,+} + \Xi_{0,-}  \right] 
\label{z311}
\ee  
contains only the conventional momentum
lattice since, for generic values of $R$, the condition
${p_{\rm L}}^{\omega}=p_{\rm R}$ does not have non-trivial solutions
while, in contrast with the previous $\bb{Z}_2$ examples,
the Klein-bottle now includes two projections. 
Only O9 planes are thus present, and therefore a supersymmetric
open sector can only involve D9 branes, while $\tilde{\cal K}$
includes twisted contributions.  The massless states in the 
projected closed
spectrum comprise the ${\cal N} =1$ supergravity multiplet, 
10 linear multiplets from the 
untwisted sector and 27 additional ones from the twisted sectors.

The description of the open sector starts with the annulus amplitude, that 
for this $\bb{Z}_3$ model reads
\ba   
{\cal A} &=& {\textstyle{1\over 6}} \biggl[
(n+m + \bar{m})^2 \, \Xi_{0,0} P_6
+ (n+
\omega m + \bar \omega \bar m )^2\,
 \Xi_{0,+}
\nonumber \\
&& + (n + \bar\omega m + 
\omega \bar m )^2 \, \Xi_{0,-} \biggr] \, ,
\label{z312}
\ea   
where $P_6$ denotes the internal momentum sum and, as usual,
$n,m$ and $\bar m$ are Chan-Paton multiplicities.
The M\"obius amplitude involves the real ``hatted'' characters
\ba
\hat\Xi_{0,\lambda} &=& \left( { \hat{A}_0 \hat\chi_0 + 
\omega^{\lambda} \hat{A}_+\hat\chi_- +   {\bar\omega}^{\lambda} \hat{A}_-
\hat\chi_+ \over \hat H^{3}_{0,\lambda}} \right) \,,
\nonumber \\
\hat\Xi_{\lambda,0} &=& \left( {\hat{A}_0 \hat\chi_{\lambda} + 
\hat{A}_{\lambda}\hat\chi_0 -  
\hat{A}_{-\lambda}\hat\chi_{-\lambda}  \over \hat H^{3}_{\lambda,0}}
\right) \, ,
\label{z313}
\ea  
where the choice of signs defines a flip operator for
open strings that ensures the compatibility of direct and transverse M\"obius
channels, related by a $P$ transformation, that maps ${\hat{\Xi}}_{0,0}$
to ${\hat{\Xi}}_{0,0}$ and 
${\hat{\Xi}}_{0,\pm 1}$ to  $- \, {\hat{\Xi}}_{\mp 1,0}$.   
One can then verify that
\ba  
{\cal M} &=& -{\textstyle{1\over 6}} \biggl[
(n+m+\bar m ) \, \hat \Xi_{0,0} 
 P_6 + (n+\bar\omega m +\omega\bar m ) 
\, \hat \Xi_{0,+} 
\nonumber \\
& & + (n +\omega m + \bar\omega \bar m ) 
\, \hat \Xi_{0,-}  \biggr]
\label{z314}
\ea 
completes the open sector of the spectrum, while $\tilde{{\cal K}}$, 
$\tilde{{\cal A}}$ and $\tilde{{\cal M}}$ are compatible with factorization
and lead to the tadpole conditions 
\ba   
& & n+m+\bar m = 32 \,,
\nonumber \\  
& & n - {\textstyle{1 \over 2}} (m +\bar m ) = -4  \, ,
\label{z315}
\ea   
originating from untwisted and twisted exchanges, respectively. 

The massless open spectrum can be read from 
\ba
{\cal A}_0 + {\cal M}_0 &=& \left[ {\textstyle{1\over 2}} n(n-1) + m \bar m 
\right] A_0 \chi_0 
+ \left[ n\bar m + {\textstyle{1\over 2}} m (m -1) \right] A_+ \chi_-
\nonumber \\
& & + \left[ nm + {\textstyle{1\over 2}} \bar m (\bar m -1) \right] 
A_- \chi_+\,,
\label{z316} 
\ea
and is characterized by an ${\rm SO}(8) \times {\rm U}(12)$ gauge group,
with {\it three} generations of chiral matter in the representations
$(8,\overline{12})$ and $(1 , 66)$, while tadpole cancellation
guarantees that the anomalies are confined to the
U(1) factor, whose gauge boson acquires a mass by the mechanism of
\cite{wito32,dsw}.

Even in this model one can introduce a quantized $B_{ab}$ in the
internal $T^6$, whose rank $r$ can now be 0,2,4 or 6 \cite{cab}.
As we have seen, the Klein-bottle amplitude involves only
O9 planes, and therefore is not affected by the 
background field. The open sector,
however, presents some subtleties. Whereas the annulus amplitude 
\ba   
{\cal A}^{(r)} &=& {\textstyle{1\over 6}} \biggl[
(n+m + \bar{m})^2 \, \Xi_{0,0} \, 2^{r-6} \sum_{\epsilon =0,1}
P_6 (B, \epsilon ) \label{z317} \\
& & + (n+ \omega m + \bar \omega \bar m )^2\,
 \Xi_{0,+} + (n + \bar\omega m + 
\omega \bar m )^2 \, \Xi_{0,-} \biggr] \, ,
\nonumber
\ea   
has the structure familiar from the toroidal case, 
and thus involves a shifted momentum sum as in (\ref{adtbab}), that we denote 
concisely by $P_6 (B,\epsilon)$, the M\"obius amplitude 
\ba
{\cal M}^{(r)} &=& -{\textstyle{1\over 6}} \biggl[
(n+m+\bar m ) \, \hat \Xi_{0,0} \, 2^{(r-6)/2} 
\sum_{\epsilon=0,1} P_6 (B,\epsilon ) \gamma_\epsilon 
\label{z318} \\
& & + 
\delta_+ \, (n+\bar\omega m +\omega\bar m ) 
\, \hat \Xi_{0,+}  
+ \delta_- \, (n +\omega m + \bar\omega \bar m ) 
\, \hat \Xi_{0,-} \biggr]
\nonumber
\ea
involves the additional signs $\delta_\pm$, that 
are to be equal for the reality of ${\cal M}$  and
draw their origin from the factorization constraints and from the 
twisted contributions to $\tilde{\cal K}$. These signs
play a crucial 
r\^ole in allowing integer solutions to the tadpole conditions, that now read
\ba
& & n + m+ \bar m = 2^{5-r/2} \,,
\nonumber \\ 
& & n -{\textstyle{1\over 2}} (m + \bar m) = -4 \, \delta_\pm \,,
\label{z319}
\ea
and thus require that $\delta_\pm$ be $(-1)^{r/2}$. As a result, orthogonal and
symplectic factors alternate in the allowed gauge groups, as do
antisymmetric and 
symmetric matter representations, depending on the rank $r$.
The matter representations can be clearly read from (\ref{z317}) 
and (\ref{z318}), using the proper analogues of (\ref{transversegamma})
and (\ref{directgamma}),
\be
\sum_{\epsilon=0,1} \gamma_\epsilon = 8 \, , \qquad \qquad \sum_
{\epsilon = 0,1 \, \in {\rm Ker}(B)} 
{\gamma}_\epsilon = 2^{(6-r)/2} \, \xi \, ,
\label{gammaz3}
\ee
where $\xi$, the overall sign ambiguity that in subsection 5.7 was
allowing the choice of unitary or symplectic gauge groups, must here be
equal to $\delta_\pm$ in order to obtain a consistent M\"obius projection.
One then obtains
\ba
{\cal A}_0^{(r)} + {\cal M}_0^{(r)}  &=& \left[ {\textstyle{1\over 2}} 
n(n-(-1)^{r/2}) + m \bar m 
\right] A_0 \chi_0 \nonumber \\
&& + \left[ n\bar m + {\textstyle{1\over 2}} m (m -(-1)^{r/2}) \right] A_+ \chi_- \nonumber \\
& & 
+ \left[ nm + {\textstyle{1\over 2}} \bar m (\bar m -(-1)^{r/2}) \right] 
A_- \chi_+\,,
\label{z3160} 
\ea
and the massless spectra for all these models 
are summarized in table \ref{z3tab}.
The U(1) factors are anomalous, and the corresponding gauge fields 
acquire a mass by the mechanism of
\cite{wito32,dsw}. More comments on the low-energy structure of this model,
that has also a perturbative heterotic dual, can be found 
in \cite{abpss,bfpss,kakudual,abd}.

\begin{table}
\caption{Some massless spectra for $T^6/\bb{Z}_3$ models with a rank-$r$ $B_{ab}$.}
\label{z3tab}
\begin{indented}
\lineup
\item[]\begin{tabular}{@{}lll}
\br
$r$ & gauge group & chiral matter (always in three families)\\
\mr
0 & ${\rm SO} (8) \times {\rm U} (12)$ & $(8,\overline{12} ) + (1,66)$
\\
2 & ${\rm USp} (8) \times {\rm U} (4)$ & $(8,\overline{4} ) + (1,10)$
\\
4 & U(4) & 6
\\
6 & USp(4) & --
\\
\br 
\end{tabular}
\end{indented} 
\end{table}

\vskip 12pt 
\subsection{Discrete torsion in four-dimensional models}

Whereas for $\bb{Z}_M$ orbifolds the closed spectra are fully
determined by the modular invariance of the one-loop torus amplitude, in
other classes of models
ambiguities can be present in the projections of twisted sectors. 
These reflect themselves in the freedom of associating suitable
phases, usually termed {\it discrete torsion},
to disconnected modular orbits \cite{vafatorsion}. The simplest instance of
this phenomenon presents itself in the $T^6 /\bb{Z}_2 \times \bb{Z}_2$ 
orbifold, and has interesting consequences for the 
corresponding orientifolds. Most notably, in some cases consistency 
demands that supersymmetry be broken in the open sector \cite{aadds}.

\begin{figure}
\begin{center}
\epsfbox{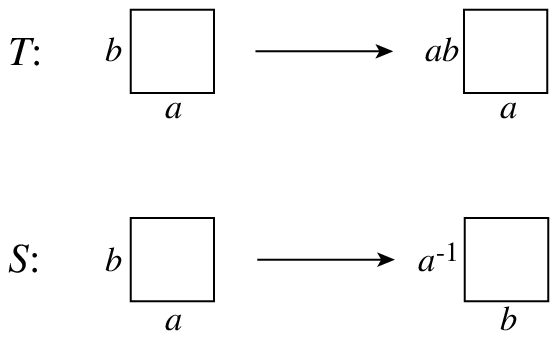}
\end{center}
\caption{Modular transformations for orbifold models.}
\label{figmt}
\end{figure}

\begin{figure}
\begin{center}
\epsfbox{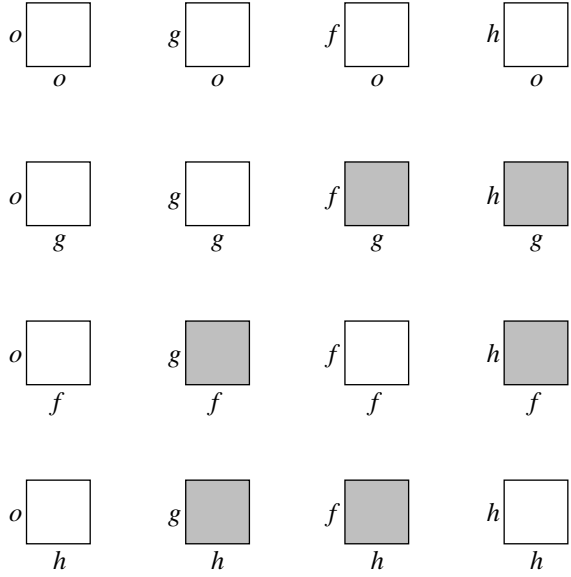}
\end{center}
\caption{Modular orbits for the $\sbb{Z}_2 \times \sbb{Z}_2$ orbifold.}
\label{figmo}
\end{figure}

Aside from the identity, that we shall denote by $o$, the $\bb{Z}_2
\times \bb{Z}_2$ orbifold group contains three other
elements acting on the internal $T^6 = T^2 \times T^2 \times T^2$ as
\be
g: (+,-,-) \, , \qquad f: (-,+,-) \, , \qquad h: (-,-,+) \, ,
\label{dta4}
\ee
that give rise to three independent twisted
sectors, confined to corresponding fixed tori. 
The modular transformations of the Jacobi
theta-functions in eq. (\ref{thetamodular}) and the corresponding properties 
of the individual
orbifold amplitudes in figure \ref{figmt} show rather clearly
that the sixteen blocks of this $\bb{Z}_2 \times
\bb{Z}_2$ orbifold do not belong to a single modular orbit. For
instance, there is no way to generate the $(g,f)$ amplitude
from the untwisted ones, and actually
the white and grey boxes in 
figure \ref{figmo} are associated to two independent orbits, 
both closed under the
action of the modular group. Consequently, one has the freedom to 
modify the
projections on the twisted sectors associating to the disconnected
orbit a phase, in this case a pure sign, consistently with the order
of the orbifold-group generators \cite{vafatorsion}, and as a result
the IIB string on the $\bb{Z}_2 \times \bb{Z}_2$ orbifold
admits the two inequivalent torus amplitudes  
\ba
{\cal T} &=& {\textstyle{1 \over 4}} \Biggl\{ 
|T_{oo}|^2 \Lambda_1 \Lambda_2 \Lambda_3
+|T_{og}|^2  \Lambda_1 
\left|{4\eta^2 \over \vartheta_2^2}\right|^2 \label{z2z2a1} \\
&&+ |T_{of}|^2 \Lambda_2 
\left|{4\eta^2 \over \vartheta_2^2}\right|^2+
|T_{oh}|^2 \Lambda_3 
\left|{4\eta^2 \over \vartheta_2^2}\right|^2 
\nonumber \\
& & + |T_{go}|^2 \Lambda_1
\left|{4\eta^2 \over \vartheta_4^2}\right|^2+ |T_{gg}|^2  \Lambda_1 
\left|{4\eta^2 \over \vartheta_3^2}\right|^2 
+ |T_{fo}|^2 
\Lambda_2 \left|{4 \eta^2 \over \vartheta_4^2}\right|^2 
\nonumber \\
& & 
+ |T_{ff}|^2 \Lambda_2 \left|{4 \eta^2 \over \vartheta_3^2}\right|^2  
+ |T_{ho}|^2 \Lambda_3
\left|{4\eta^2 \over \vartheta_4^2}\right|^2 +|T_{hh}|^2 
\Lambda_3 \left|{4\eta^2 \over \vartheta_3^2}\right|^2 
\nonumber \\
& & + \epsilon \left( |T_{gh}|^2 + |T_{gf}|^2 + |T_{fg}|^2 + |T_{fh}|^2 
+ |T_{hg}|^2
+ |T_{hf}|^2 \right) \left|{8\eta^3 \over \vartheta_2 \vartheta_3 
\vartheta_4} \right|^2 
\Biggr\} \, , 
\nonumber
\ea
where the $\Lambda_k$'s are lattice sums for the three internal 
two-tori and the choices $\epsilon = \mp 1$ identify the models with and 
without discrete torsion. Here we have expressed the torus amplitude in
terms of the 16 quantities $(k=o,g,h,f)$
\ba
T_{ko} &=&  \tau_{ko} +  \tau_{kg} + \tau_{kh} + \tau_{kf} \, , \qquad
T_{kg} =  \tau_{ko} +  \tau_{kg} - \tau_{kh} - \tau_{kf} \, , 
\nonumber \\
T_{kh} &=&  \tau_{ko} -  \tau_{kg} + \tau_{kh} - \tau_{kf} \, , \qquad
T_{kf} =  \tau_{ko} -  \tau_{kg} - \tau_{kh} + \tau_{kf} \, ,
\label{z2z2a2}
\ea
where the $16$ $\bb{Z}_2 \times \bb{Z}_2$ characters $\tau_{kl}$, 
combinations of products of level-one so(2) characters, are displayed
in table \ref{z2z2tab1}.
The low-energy spectra of the two models are quite different: 
{\it with} discrete torsion, {\it i.e.} if $\epsilon = -1$, one has 
${\cal N}=2$ supergravity coupled to 52 hyper multiplets and 
3 vector multiplets, while {\it without} discrete torsion, {\it i.e.}
if $\epsilon = +1$, one has again ${\cal N}=2$ supergravity, but with 4 
hyper multiplets and 51 vector 
multiplets. These two choices describe orbifold limits
of mirror Calabi-Yau 
manifolds with Hodge numbers (51,3) and (3,51), and, from the
conformal field theory viewpoint, the former corresponds to the
charge-conjugation modular invariant while  the latter
corresponds to the diagonal one.

\begin{table}
\caption{Space-time characters for the \\ supersymmetric $\sbb{Z}_2 \times
\sbb{Z}_2$ model.}
\label{z2z2tab1}
\begin{indented}
\lineup
\item[]\begin{tabular}{@{}l} 
\br
$\tau_{oo} = V_2O_2O_2O_2+O_2V_2V_2V_2-S_2S_2S_2S_2-C_2C_2C_2C_2 $ 
\\
$\tau_{og} = O_2V_2O_2O_2+V_2O_2V_2V_2-C_2C_2S_2S_2-S_2S_2C_2C_2 $ 
\\
$\tau_{oh} = O_2O_2O_2V_2+V_2V_2V_2O_2-C_2S_2S_2C_2-S_2C_2C_2S_2 $
\\
$\tau_{of} = O_2O_2V_2O_2+V_2V_2O_2V_2-C_2S_2C_2S_2-S_2C_2S_2C_2 $ 
\\
$\tau_{go} = V_2O_2S_2C_2+O_2V_2C_2S_2-S_2S_2V_2O_2-C_2C_2O_2V_2 $ 
\\
$\tau_{gg} = O_2V_2S_2C_2+V_2O_2C_2S_2-S_2S_2O_2V_2-C_2C_2V_2O_2 $ 
\\
$\tau_{gh} = O_2O_2S_2S_2+V_2V_2C_2C_2-C_2S_2V_2V_2-S_2C_2O_2O_2 $ 
\\
$\tau_{gf} = O_2O_2C_2C_2+V_2V_2S_2S_2-S_2C_2V_2V_2-C_2S_2O_2O_2 $ 
\\
$\tau_{ho} = V_2S_2C_2O_2+O_2C_2S_2V_2-C_2O_2V_2C_2-S_2V_2O_2S_2 $ 
\\
$\tau_{hg} = O_2C_2C_2O_2+V_2S_2S_2V_2-C_2O_2O_2S_2-S_2V_2V_2C_2 $ 
\\
$\tau_{hh} = O_2S_2C_2V_2+V_2C_2S_2O_2-S_2O_2V_2S_2-C_2V_2O_2C_2 $ 
\\
$\tau_{hf} = O_2S_2S_2O_2+V_2C_2C_2V_2-C_2V_2V_2S_2-S_2O_2O_2C_2 $ 
\\
$\tau_{fo} = V_2S_2O_2C_2+O_2C_2V_2S_2-S_2V_2S_2O_2-C_2O_2C_2V_2 $ 
\\
$\tau_{fg} = O_2C_2O_2C_2+V_2S_2V_2S_2-C_2O_2S_2O_2-S_2V_2C_2V_2 $ 
\\
$\tau_{fh} = O_2S_2O_2S_2+V_2C_2V_2C_2-C_2V_2S_2V_2-S_2O_2C_2O_2 $ 
\\
$\tau_{ff} = O_2S_2V_2C_2+V_2C_2O_2S_2-C_2V_2C_2O_2-S_2O_2S_2V_2 $
\\
\br
\end{tabular}
\end{indented}
\end{table}

The $\Omega$ projections for the two classes of models
can be implemented by the Klein-bottle amplitudes
\ba
{\cal K} &=& {\textstyle\frac{1}{8}} 
\Biggl\{ ( P_1 P_2 P_3 + P_1 W_2 W_3 + W_1 P_2
W_3 + W_1 W_2 P_3 ) T_{oo} 
\nonumber 
\\ & & +  
2 \times 16 \Bigl[\epsilon_1 (P_1 + \epsilon W_1 ) T_{go} 
+  \epsilon_2 (P_2 + \epsilon W_2 ) T_{fo}
\nonumber \\
& & + \epsilon_3 (P_3 + \epsilon W_3 ) T_{ho} \Bigr] 
\left( \frac{\eta}{\vartheta_4} \right)^2 \Biggr\} \, ,
\label{z2z2a5}
\ea
where, as usual, $P_k$ and $W_k$ denote the restrictions of the lattice sums 
$\Lambda_k$ to their momentum and winding sub-lattices.
Discrete torsion has a neat effect on $(P_k+\epsilon W_k)$: 
as anticipated, if $\epsilon= +1$ the massless 
twisted contributions are diagonal combinations of the $\tau_{kl}$, and
appear in the Klein bottle, while  
if $\epsilon = -1$ they are off-diagonal 
combinations, and thus do not contribute to it.  Consistently with 
the crosscap constraint of \cite{fps,pss,pss2}, (\ref{z2z2a5}) can 
actually accommodate three additional 
signs $\epsilon_k$ that, however, are not independent, but are 
to satisfy the constraint
\be
\epsilon_1 \, \epsilon_2 \, \epsilon_3 \, = \, \epsilon \,.
\label{z2z2a6}
\ee
One can write this amplitude more compactly as
\be
{\cal K} = {\textstyle\frac{1}{8}} 
\left[ (P_1 P_2 P_3 + {\textstyle\frac{1}{2}} P_k W_l W_m) T_{oo}
+ 2 \times 16 \epsilon_k (P_k + \epsilon W_k ) T_{ko}
\left( \frac{\eta}{\vartheta_4} \right)^2  
\right] \, ,
\label{z2z2a7}
\ee
where for the sake of brevity we have introduced a convenient 
shorthand notation, so that {\it summations} over 
repeated indices and {\it symmetrizations} over 
distinct indices are left implicit, with $k$, $l$ and $m$ 
taking the values $(1,2,3)$. An $S$ 
transformation then turns this expression into the 
corresponding transverse-channel amplitude
\ba
\tilde{\cal K} &=& \frac{2^5}{8} \Biggl[ \left( v_1v_2v_3 W^e_1 W^e_2 
W^e_3 + \frac{v_k}{2v_l v_m} W_k^e P_l^e P_m^e  \right) T_{oo} 
\nonumber \\
& & + 2 \epsilon_k \, \left( v_k W_k^e + \epsilon \frac{P_k^e}{v_k}
\right ) \, T_{ok}
\left( \frac{2\eta}{\vartheta_2} \right)^2 \Biggr] \, ,
\label{z2z2a8}
\ea
where the superscript $e$ denotes the restriction of the
lattice sums to their even terms and the $v_k$'s are proportional to the
dimensionless volumes of the
three internal tori. At the origin of the lattices, the constraint
(\ref{z2z2a6}) leads to
\ba
\tilde{\cal K}_0 &=& \frac{2^5}{8} \Biggl\{ \left( \sqrt{v_1v_2v_3} +
 \epsilon_1 \sqrt{\frac{v_1}{v_2 v_3}} + 
\epsilon_2 \sqrt{\frac{v_2}{v_1 v_3}}  + \epsilon_3
\sqrt{\frac{v_3}{v_1 v_2}} \right)^2 \tau_{oo} 
\nonumber \\ 
& &+  \left( \sqrt{v_1v_2v_3} +
 \epsilon_1 \sqrt{\frac{v_1}{v_2 v_3}} - 
\epsilon_2 \sqrt{\frac{v_2}{v_1 v_3}}  - \epsilon_3
\sqrt{\frac{v_3}{v_1 v_2}} \right)^2 \tau_{og} 
\nonumber \\
& &+ \left( \sqrt{v_1v_2v_3} -
 \epsilon_1 \sqrt{\frac{v_1}{v_2 v_3}} + 
\epsilon_2 \sqrt{\frac{v_2}{v_1 v_3}}  - \epsilon_3
\sqrt{\frac{v_3}{v_1 v_2}} \right)^2 \tau_{of} 
\nonumber \\
&&+ \left( \sqrt{v_1v_2v_3} -
 \epsilon_1 \sqrt{\frac{v_1}{v_2 v_3}} - 
\epsilon_2 \sqrt{\frac{v_2}{v_1 v_3}}  + \epsilon_3
\sqrt{\frac{v_3}{v_1 v_2}} \right)^2 \tau_{oh} 
\Biggr\} \, ,
\label{z2z2a9}
\ea 
whose coefficients are as usual perfect squares,
that shows clearly how moving from $\epsilon_k=1$ to
$\epsilon_k=-1$ {\it reverses} both the tension and the charge of
the O$5_k$ orientifold plane, thus trading an O$5_{k,+}$ for
an O$5_{k,-}$. While
manifestly compatible with the usual positivity requirements, this reversal
clearly affects the tadpole conditions, that, as in subsection 5.8,
require the introduction of antibranes. We are thus facing another,
more intricate manifestation of ``brane supersymmetry breaking''.
In this respect, it should be
appreciated that, according to (\ref{z2z2a6}), discrete torsion implies the
reversal of tension and charge for an odd number of O5 planes.
Therefore, the allowed $\epsilon_k$'s identify
four classes of models. If $\epsilon=+1$, the choice 
$(\epsilon_1,\epsilon_2,\epsilon_3)=(+,+,+)$ recovers the model 
discussed in \cite{erice,bl}, with $48$ chiral 
multiplets from the closed twisted sectors, while the choice $(+,-,-)$ 
gives a model with $16$ chiral multiplets and $32$ vector multiplets from the 
twisted sectors. 
On the other hand, for $\epsilon = -1$ the two choices $(+,+,-)$ 
and $(-,-,-)$ yield identical massless twisted spectra, with 
$48$ chiral multiplets. 

In order to describe the open sector, it is 
convenient to introduce a compact notation, defining
\be
\tilde{T}_{kl}^{(\zeta)} = T_{kl}^{\rm NS} - \zeta 
T_{kl}^{\rm R} \, , \qquad \zeta = \pm 1 \,,
\label{z2z2a10}
\ee
where $T_{kl}^{\rm NS}$ and $T_{kl}^{\rm R}$ denote the NS and R parts of
the usual supersymmetric $\bb{Z}_2 \times \bb{Z}_2$ characters.
Whereas transverse and direct supersymmetric
annulus amplitudes involve the identical sets of characters 
$\tilde{T}_{kl}^{(+)}$ and ${T}_{kl}^{(+)}$, simply denoted by 
$T_{kl}$ in the following, in the presence of
brane supersymmetry breaking the transverse amplitudes involve
the $\tilde{T}_{kl}^{(-)}$, 
with reversed R-R contributions, that are mapped into new characters
by  the $S$ modular transformation.
As a result, the terms in ${\cal A}$ describing open strings 
stretched between branes and antibranes contain the new combinations
$T^{(-)}_{kl}$, obtained from the $T^{(+)}_{kl}$ 
interchanging $O_2$ with $V_2$ and $S_2$ with $C_2$ in the last 
three factors, as summarized in table \ref{z2z2tab2}. 

\begin{table}
\caption{Space-time characters for the \\ non-supersymmetric $\sbb{Z}_2 \times
\sbb{Z}_2$ model.}
\label{z2z2tab2}
\begin{indented}
\lineup
\item[]\begin{tabular}{@{}l} 
\br
$\tau^{(-)}_{oo} = O_2O_2O_2O_2+V_2V_2V_2V_2-C_2S_2S_2S_2-S_2C_2C_2C_2 $ 
\\
$\tau^{(-)}_{og} = V_2V_2O_2O_2+O_2O_2V_2V_2-S_2C_2S_2S_2-C_2S_2C_2C_2 $
\\
$\tau^{(-)}_{oh} = V_2O_2O_2V_2+O_2V_2V_2O_2-S_2S_2S_2C_2-C_2C_2C_2S_2 $
\\
$\tau^{(-)}_{of} = V_2O_2V_2O_2+O_2V_2O_2V_2-S_2S_2C_2S_2-C_2C_2S_2C_2 $
\\
$\tau^{(-)}_{go} = O_2O_2S_2C_2+V_2V_2C_2S_2-C_2S_2V_2O_2-S_2C_2O_2V_2 $
\\
$\tau^{(-)}_{gg} = V_2V_2S_2C_2+O_2O_2C_2S_2-C_2S_2O_2V_2-S_2C_2V_2O_2 $
\\
$\tau^{(-)}_{gh} = V_2O_2S_2S_2+O_2V_2C_2C_2-S_2S_2V_2V_2-C_2C_2O_2O_2 $
\\
$\tau^{(-)}_{gf} = V_2O_2C_2C_2+O_2V_2S_2S_2-C_2C_2V_2V_2-S_2S_2O_2O_2 $
\\
$\tau^{(-)}_{ho} = O_2S_2C_2O_2+V_2C_2S_2V_2-S_2O_2V_2C_2-C_2V_2O_2S_2 $
\\
$\tau^{(-)}_{hg} = V_2C_2C_2O_2+O_2S_2S_2V_2-S_2O_2O_2S_2-C_2V_2V_2C_2 $
\\
$\tau^{(-)}_{hh} = V_2S_2C_2V_2+O_2C_2S_2O_2-C_2O_2V_2S_2-S_2V_2O_2C_2 $
\\
$\tau^{(-)}_{hf} = V_2S_2S_2O_2+O_2C_2C_2V_2-S_2V_2V_2S_2-C_2O_2O_2C_2 $
\\
$\tau^{(-)}_{fo} = O_2S_2O_2C_2+V_2C_2V_2S_2-C_2V_2S_2O_2-S_2O_2C_2V_2 $
\\
$\tau^{(-)}_{fg} = V_2C_2O_2C_2+O_2S_2V_2S_2-S_2O_2S_2O_2-C_2V_2C_2V_2 $
\\
$\tau^{(-)}_{fh} = V_2S_2O_2S_2+O_2C_2V_2C_2-S_2V_2S_2V_2-C_2O_2C_2O_2 $
\\
$\tau^{(-)}_{ff} = V_2S_2V_2C_2+O_2C_2O_2S_2-S_2V_2C_2O_2-C_2O_2S_2V_2 $
\\
\br
\end{tabular}
\end{indented}
\end{table}

The transverse-channel annulus amplitude is 
\ba
\tilde{{\cal A}} &=& \frac{2^{-5}}{8} \Biggl\{
\left( N_o^2 v_1v_2v_3W_1W_2W_3 + \frac{D_{k;o}^2v_k}{2v_lv_m} W_k P_l P_m 
\right) T_{oo} 
\nonumber 
\\
& & + 4\left[ ( N_k^2 + D^2_{k;k} ) v_kW_k  + D^2_{l \neq k;k} 
\frac{P_k}{v_k}\right] 
T_{ko} \left( \frac{2\eta}{\vartheta_4} \right)^2 
\nonumber \\
& & + 2 N_o D_{k;o} v_kW_k {\tilde T}_{ok}^{( \epsilon_k )}
\left( \frac{2\eta}{\vartheta_2} \right)^2 
+ 2 N_k D_{k;k} v_kW_k {\tilde T}_{kk}^{(\epsilon_k)} 
\left( \frac{2\eta}{\vartheta_3} \right)^2
\nonumber \\ 
& & + 4 N_l D_{k \neq l;l}{\tilde T}_{lk}^{(\epsilon_k)}
\frac{8 \eta^3}{\vartheta_2 \vartheta_3 \vartheta_4} + 
D_{k;o} D_{l;o} \frac{P_m}{v_m} {\tilde T}_{om}^{(\epsilon_k \epsilon_l )}
\left( \frac{2\eta}{\vartheta_2} \right)^2  
\nonumber \\
& & + D_{k;m}D_{l;m} \frac{P_m}{v_m} {\tilde T}_{mm} ^{(\epsilon_k 
\epsilon_l )} 
\left( \frac{2\eta}{\vartheta_3} \right)^2 
+ 4 D_{k;k}D_{l;k} {\tilde T}_{km} ^{(\epsilon_k \epsilon_l )}
\frac{8 \eta^3}{\vartheta_2 \vartheta_3 \vartheta_4} 
 \Biggr\} \, ,
\label{z2z2a14}
\ea
where $N_o$, $D_{g;o}$, $D_{f;o}$ and
$D_{h;o}$ count the numbers of D9 branes and of the three 
sets of D5 or $\overline{\rm D} 5$ branes wrapped around the first, second and 
third torus, denoted for brevity $5_{1,2,3}$ or $\bar{5}_{1,2,3}$
in the following. 
In a similar fashion, $N_k$, $D_{g;k}$, $D_{f;k}$ and $D_{h;k}$ 
$(k=g,f,h)$ parametrize the breakings induced by the three orbifold 
operations $g$, $f$ and $h$. Notice that, as in subsection 5.8,
 the R-R portions of all terms 
describing brane-antibrane exchanges have reversed signs.
The untwisted terms at the origin of the lattice sums rearrange 
themselves into perfect squares, so that
\ba
& & \tilde{\cal A}_0 =
\\
& & \frac{2^{-5}}{8} \Biggl\{ \ \left( N_o \sqrt{v_1v_2v_3} +
 D_{g;o} \sqrt{\frac{v_1}{v_2 v_3}} + 
D_{f;o} \sqrt{\frac{v_2}{v_1 v_3}}  + D_{h;o}
\sqrt{\frac{v_3}{v_1 v_2}} \right)^2 \tau^{\rm NS}_{oo} 
\nonumber \\ 
& &- \left( N_o \sqrt{v_1v_2v_3} +
 \epsilon_1 D_{g;o} \sqrt{\frac{v_1}{v_2 v_3}} + 
\epsilon_2 D_{f;o} \sqrt{\frac{v_2}{v_1 v_3}}  + \epsilon_3
D_{h,o} \sqrt{\frac{v_3}{v_1 v_2}} \right)^2 \tau^{\rm R}_{oo} 
\nonumber\\
& &+  \left( N_o \sqrt{v_1v_2v_3} +
 D_{g;o} \sqrt{\frac{v_1}{v_2 v_3}} - 
D_{f;o} \sqrt{\frac{v_2}{v_1 v_3}}  - D_{h;o}
\sqrt{\frac{v_3}{v_1 v_2}} \right)^2 \tau^{\rm NS}_{og} 
\nonumber \\
& &- \left( N_o\sqrt{v_1v_2v_3} +
\epsilon_1 D_{g;o} \sqrt{\frac{v_1}{v_2 v_3}} -
\epsilon_2 D_{f;o} \sqrt{\frac{v_2}{v_1 v_3}}  - \epsilon_3
D_{h;o} \sqrt{\frac{v_3}{v_1 v_2}} \right)^2 \tau^{\rm R}_{og} 
\nonumber\\
&&+ \left( N_o \sqrt{v_1v_2v_3} -
 D_{g;o} \sqrt{\frac{v_1}{v_2 v_3}} + 
D_{f;o} \sqrt{\frac{v_2}{v_1 v_3}}  - D_{h;o}
\sqrt{\frac{v_3}{v_1 v_2}} \right)^2 \tau^{\rm NS}_{of} 
\nonumber \\ 
& &- \left( N_o \sqrt{v_1v_2v_3} -
 \epsilon_1 D_{g;o} \sqrt{\frac{v_1}{v_2 v_3}} + 
\epsilon_2 D_{f;o} \sqrt{\frac{v_2}{v_1 v_3}}  - \epsilon_3
D_{h;o} \sqrt{\frac{v_3}{v_1 v_2}} \right)^2 \tau^{\rm R}_{of} 
\nonumber\\
& &+ \left( N_o \sqrt{v_1v_2v_3} -
 D_{g;o} \sqrt{\frac{v_1}{v_2 v_3}} - 
D_{f;o} \sqrt{\frac{v_2}{v_1 v_3}}  + D_{h;o}
\sqrt{\frac{v_3}{v_1 v_2}} \right)^2 \tau^{\rm NS}_{oh} 
\nonumber \\
& &- \left( N_o\sqrt{v_1v_2v_3} -
 \epsilon_1 D_{g;o} \sqrt{\frac{v_1}{v_2 v_3}} - 
\epsilon_2 D_{f;o} \sqrt{\frac{v_2}{v_1 v_3}}  + \epsilon_3
D_{h;o} \sqrt{\frac{v_3}{v_1 v_2}} \right)^2 \tau^{\rm R}_{oh} 
 \ \Biggr\} \,. \nonumber
\label{z2z2a15}
\ea

The twisted tadpoles reflect rather neatly the 
geometry of the brane configuration. As usual,
the reflection coefficients are sums of squares associated to 
the various fixed tori, and each square contains 
the projections for the branes that are present, 
with factors $\sqrt{v_i}$ if they are wrapped around them and
$1/\sqrt{v_i}$ if they are localized on them. The relative coefficients 
of these terms, also directly linked to the brane geometry, are given by
\be
\sqrt{\frac{\rm \# \ of \ fixed \ tori}{\rm \# \ of 
\ occupied \ fixed \ tori}} 
\, .
\label{z2z2a15b}
\ee
Thus, for a given twisted sector, the numerator counts the fixed 
tori, while the denominator counts the fixed tori where branes
are actually present.  Moreover, the R portions of the characters 
describing brane-antibrane exchanges have reversed signs even in
these twisted contributions, as expected. For instance, in the 
$g$-twisted sector of the $(+,+,-)$ model, that  contains $\overline{\rm D}5_3$
branes, the reflection coefficients for the massless modes in $\tau_{gh}$ are
\ba
& & \frac{2^{-5}}{8} \Biggl[\left( N_g \sqrt{v_1}-4D_{g;g}\sqrt{v_1}
-2D_{f;g}\frac{1}{\sqrt{v_1}}+2D_{h;g}\frac{1}{\sqrt{v_1}} \right)^2 
+9N_g^2v_1 \nonumber \\
& & +3 \left( N_g\sqrt{v_1}-2D_{f;g}\frac{1}{\sqrt{v_1}} \right)^2 
+ 3\left( N_g\sqrt{v_1}+2D_{h;g}\frac{1}{\sqrt{v_1}} \right)^2 
\Biggr] 
\label{z2z2a15c}
\ea
for the NS-NS portion, and
\ba
& & \frac{2^{-5}}{8} \Biggl[\left(N_g\sqrt{v_1}-4D_{g;g}\sqrt{v_1}
-2D_{f;g}\frac{1}{\sqrt{v_1}}-2D_{h;g}\frac{1}{\sqrt{v_1}} \right)^2  
+9N_g^2v_1 
\nonumber \\
& & +3 \left( N_g\sqrt{v_1}-2D_{f;g}\frac{1}{\sqrt{v_1}} \right)^2 
+ 3 \left( N_g\sqrt{v_1}-2D_{h;g}\frac{1}{\sqrt{v_1}} \right)^2 
\Biggr] 
\label{z2z2a15d}
\ea
for the R-R portion. According to ($\ref{z2z2a15b}$),
the coefficient of $N_g$ is $\sqrt{v_1}$, since the D9's are wrapped 
around all fixed tori, the coefficient of $D_{g;g}$ is $4\sqrt{v_1}$, 
since the D$5_1$'s are only wrapped 
around one fixed torus, while the coefficients of $D_{f;g}$ and  
$D_{h;g}$ are $2/\sqrt{v_1}$, since the D$5_2$'s and $\overline{\rm D}5_3$'s 
are 
confined to four of the fixed tori. Finally, out of the 16 $g$-fixed tori, 
one sees all the branes, three see only the 
D9's and the D$5_2$'s, three see only the D9's and the $\overline{\rm D}5_3$'s
and,
finally,  nine see only the D9's.

The direct-channel annulus amplitude is then
\ba
{\cal A} &=& {\textstyle\frac{1}{8}} \Biggl\{
\left( N_o^2 P_1P_2P_3 + \frac{D_{k;o}^2}{2} P_k W_l W_m \right)
T_{oo} 
\\
& & + 
\left[ ( N_k^2 + D^2_{k;k} ) P_k  + D^2_{l \neq k ;k} W_k\right] T_{ok}
\left( \frac{2\eta}{\vartheta_2} \right)^2 
\nonumber
\\
& & + 2 N_o D_{k;o} P_k T_{ko}^{( \epsilon_k )}
\left( \frac{\eta}{\vartheta_4} \right)^2 
- 2 N_k D_{k;k} P_k T_{kk}^{(\epsilon_k)}
\left( \frac{\eta}{\vartheta_3} \right)^2  
\nonumber \\
& & + 2 i (-1)^{k+l} N_l D_{k \neq l ;l}T_{kl}^{(\epsilon_k)}
\frac{2\eta^3}{\vartheta_2 \vartheta_3 \vartheta_4}
+ D_{k;o} D_{l;o} W_m  T_{mo}^{(\epsilon_k \epsilon_l )}
\left( \frac{\eta}{\vartheta_4} \right)^2   
\nonumber
\\
& & - D_{k;m}D_{l;m} W_m T_{mm} ^{(\epsilon_k \epsilon_l )}
\left( \frac{\eta}{\vartheta_3} \right)^2   
+ 2 i (-1)^{m+k} D_{k;k}D_{l;k} T_{mk}^{(\epsilon_k \epsilon_l)} 
\frac{2\eta^3}{\vartheta_2 \vartheta_3 \vartheta_4} 
\Biggr\} \, , \nonumber 
\label{z2z2a13bis}
\ea
where in the signs $(-1)^{k+l}$ and $(-1)^{m+k}$ the integers 
$k,l,m$ take the values $1,2,3$ for the $g$, $f$, and $h$ generators.
$\tilde{\cal K}$ and  $\tilde{\cal A}$
determine by standard methods the transverse-channel  M{\"o}bius amplitude
\ba
\tilde{{\cal M}} &=& 
- {\textstyle\frac{1}{4}} \Biggl\{ N_o v_1v_2v_3W^e_1 W^e_2 W^e_3 \hat{T}_{oo} 
+ N_o v_kW^e_k \epsilon_k\hat{T}_{ok}
\left( \frac{2 \hat{\eta}}{\hat{\vartheta_2}}\right)^2  
\nonumber \\
& & + \frac{v_k}{2v_lv_m}D_{k;o} W^e_k P^e_l P^e_m  \epsilon_k 
\hat{\tilde T}{}_{oo} ^{(\epsilon_k)} 
\nonumber \\
& & +\left( D_{l;o} \epsilon_k \frac{P^e_m}{v_m} 
\hat{\tilde T}{}_{om}^{(\epsilon_l)} 
+ D_{k;o} v_kW^e_k \hat{\tilde T}{}_{ok}^{(\epsilon_k)} \right)
\left( \frac{2 \hat{\eta}}{\hat{\vartheta_2}}\right)^2  \Biggr\} \, ,
\label{z2z2a16}
\ea
and after a $P$ transformation,
\ba
{\cal M} &=& - {\textstyle\frac{1}{8}} \Biggl\{ N_o P_1 P_2 P_3 \hat{T}_{oo} 
- N_o P_k \epsilon_k
\hat{T}_{ok} \left( \frac{2 \hat{\eta}}{\hat{\vartheta_2}}\right)^2 
+ {\textstyle\frac{1}{2}} D_{k;o} P_k W_l W_m  \epsilon_k 
\hat{\tilde T}{}_{oo} ^{(\epsilon_k)} 
\nonumber 
\\ 
& & - \left(D_{l;o} \epsilon_k W_m \hat{\tilde T}{}_{om}^{(\epsilon_l)} 
+ D_{k;o} P_k \hat{\tilde T}{}_{ok}^{(\epsilon_k)}\right)
\left( \frac{2 \hat{\eta}}{\hat{\vartheta_2}}\right)^2 
\Biggr\} \,.
\label{z2z2a17}
\ea
From the transverse amplitudes one can finally read the R-R tadpole conditions
\ba
& N_o = 32 \, , \qquad & N_g = N_f = N_h = 0 \, , 
\nonumber 
\\
& D_{k;o} = 32 \, , \qquad & D_{k;g} = D_{k;f} = D_{k;h} = 0 \,.
\ea
In the presence of discrete torsion the NS-NS tadpole conditions 
are incompatible
with the R-R ones, and, as in all models featuring brane supersymmetry
breaking, a dilaton tadpole and a potential for
geometric moduli are generated.

We can now conclude with a brief discussion of the massless spectra. 
The models where only one $\epsilon_k$ is negative have discrete torsion 
and contain one $\overline{\rm D}5$. For the D9 and the 
two sets of D5 branes, the gauge groups are ${\rm U}(8)\times {\rm U}(8)$, 
with ${\cal N}=1$ supersymmetry, while for the 
$\overline{\rm D}5$ branes the gauge group is ${\rm USp}(8)^4$, with ${\cal N}=0$. 
Moreover, the 95 and 
$5_k5_l$ strings are supersymmetric, while the $9\bar{5}$ and
$5_k\bar{5}$ strings are not. 
Let us discuss in some detail the case $(\epsilon_1,\epsilon_2,
\epsilon_3) = (+,+,-)$, that contains 
$\overline{\rm D}5$
branes wrapped around the third torus. To this end, let us
parametrize the charges as
\ba
N_o &= o+g+\bar{o}+\bar{g} \, , \hspace{2.4cm} 
& N_g = i(o+g-\bar{o}-\bar{g}) \, ,
\nonumber \\
N_f &= i(o-g-\bar{o}+\bar{g}) \, ,
& N_h = o-g+\bar{o}-\bar{g} \, ,
\nonumber \\
D_{g;o} &= o_1+g_1+\bar{o}_1+\bar{g}_1 \, ,   
& D_{g;g} = i(o_1+g_1-\bar{o}_1-\bar{g}_1) \, ,
\nonumber \\
D_{g;f} &= o_1-g_1+\bar{o}_1-\bar{g}_1 \, , 
& D_{g;h} = -i(o_1-g_1-\bar{o}_1+\bar{g}_1) \, ,
\nonumber \\
D_{f;o} &= o_2+g_2+\bar{o}_2+\bar{g}_2 \, , 
& D_{f;g} = o_2-g_2+\bar{o}_2-\bar{g}_2 \, ,
\nonumber \\
D_{f;f} &= i(o_2+g_2-\bar{o}_2-\bar{g}_2) \, ,  
& D_{f;h} = i(o_2-g_2-\bar{o}_2+\bar{g}_2) \, ,
\nonumber \\
D_{h;o} &= a+b+c+d \, ,  
& D_{h;g} = a+b-c-d \, ,
\nonumber \\
D_{h;f} &= a-b+c-d \, , 
& D_{h;h} = a-b-c+d \, ,
\label{z2z2a19}
\ea
and extract the massless spectrum from the amplitudes at the 
origin of the lattices. As anticipated,
the $99$, $5_15_1$ and $5_25_2$ sectors have ${\cal N}=1$ supersymmetry, 
with ${\rm U}(8) \times {\rm U}(8)$ gauge groups and chiral 
multiplets in the representations
$(8,8)$, $(8,\bar{8})$, $(28,1)$, $(1,28)$ 
and their conjugates. Moreover, as expected, the $95_1$, $95_2$ and $5_15_2$ 
strings are also supersymmetric, and contain chiral multiplets in the 
representations 
\ba
95_1 : \ & & (8,1;1,\bar{8}) \, , \ \ (1,8;\bar{8},1) \, , 
          \ \ (\bar{8},1;8,1) \, , \ \ (1,\bar{8};1,8) \, , 
\nonumber \\
95_2 : \ & & (8,1;1,\bar 8) \, , \ \ (1,\bar 8;\bar 8,1) \, , 
          \ \ (\bar 8,1;8,1) \, , \ \ (1,8;1,8) \, , 
\nonumber \\
5_15_2 : \  & &  (8,1;8,1) \, , \ \ (1,8;\bar{8},1) \, , 
          \ \ (\bar{8},1;1,8) \, , \ \ (1,\bar{8};1,\bar{8}) \,.
\nonumber
\ea 

On the other hand, the strings that end on the antibrane yield
non-supersymmetric spectra, even if the annulus contains supersymmetric 
characters, since bosons and fermions are treated differently by ${\cal M}$. 
Thus, the $\bar{5}_3\bar{5}_3$ excitations have a 
${\rm USp}(8)^4$ gauge group,
with Weyl spinors in the $(28,1,1,1)$ and in three
additional permutations, and chiral multiplets in the 
$(8,8,1,1)$ and in five additional permutations. Finally,
the strings stretching between a brane
and an antibrane have non-supersymmetric spectra, with
Weyl spinors and complex scalars in the representations
\ba
9\bar{5}_3  &{\rm spinors \ :}& \
  (\bar{8},1;8,1,1,1) \, , \ \ (1,\bar{8};1,8,1,1) \, , 
\nonumber \\
& &           \ (1,8;1,1,8,1) \, , \ \ (8,1;1,1,1,8) \, , 
\nonumber \\
           &{\rm scalars \ :}&  \ 
(\bar{8},1;1,8,1,1) \, , \ \ (1,\bar{8};8,1,1,1) \, , 
\nonumber \\
& &           \ (1,8;1,1,1,8) \, , \ \ (8,1;1,1,8,1) \, , 
\nonumber \\
5_1\bar{5}_3 &{\rm \ spinors \ :}& \
  (\bar{8},1;1,1,8,1) \, , \ \ (1,\bar{8};1,1,1,8) \, ,
\nonumber \\ 
& &           \ (1,8;1,8,1,1) \, , \ \ (8,1;8,1,1,1) \, , 
\nonumber \\
           &{\rm scalars \ :}&  \ 
(\bar{8},1;1,1,1,8) \, , \ \ (1,\bar{8};1,1,8,1) \, ,
\nonumber \\ 
& &          \ (1,8;8,1,1,1) \, , \ \ (8,1;1,8,1,1) \, , 
\nonumber \\
5_2\bar{5}_3 &{\rm \ spinors \ :}& \
  (8,1;8,1,1,1) \, , \ \ (1,\bar{8};1,1,1,8) \, ,
\nonumber \\ 
& &           \ (1,8;1,1,8,1) \, , \ \ (\bar{8},1;1,8,1,1) \, , 
\nonumber \\
           &{\rm scalars \ :}&  \ 
(\bar{8},1;1,1,1,8) \, , \ \ (1,\bar{8};1,8,1,1) \, , 
\nonumber \\
& &           \ (1,8;8,1,1,1) \, , \ \ (8,1;1,1,8,1) \,. \nonumber
\ea
The choice $(\epsilon_1,\epsilon_2,\epsilon_3) = (-,-,-)$ 
corresponds again to a model with 
discrete torsion. In this case, however, there are D9 branes and three sets of 
$\overline{\rm D}5$
 branes, while the charges are to be parametrized as
\ba
N_o &= a+b+c+d \, , \hspace{2.4cm} 
& N_g  =  a+b-c-d \, ,
\nonumber \\
N_f &= a-b+c-d \, ,  
& N_h =  a-b-c+d \, , 
\nonumber \\
D_{g;o} &= o_1+g_1+\bar{o}_1+\bar{g}_1 \, ,  
& D_{g;g}  =  o_1-g_1+\bar{o}_1-\bar{g}_1 \, ,
\nonumber \\
D_{g;f} &= i(o_1+g_1-\bar{o}_1-\bar{g}_1) \, ,  
& D_{g;h} = i(o_1-g_1-\bar{o}_1+\bar{g}_1) \, ,
\nonumber \\
D_{f;o} &= o_2+g_2+\bar{o}_2+\bar{g}_2 \, , 
& D_{f;g}  = i(o_2+g_2-\bar{o}_2-\bar{g}_2) \, ,
\nonumber \\
D_{f;f} &= o_2-g_2+\bar{o}_2-\bar{g}_2 \, , 
& D_{f;h} = -i(o_2-g_2-\bar{o}_2+\bar{g}_2) \, ,
\nonumber \\
D_{h;o} &= o_3+g_3+\bar{o}_3+\bar{g}_3 \, , 
& D_{h;g} = i(o_3+g_3-\bar{o}_3-\bar{g}_3) \, ,
\nonumber \\
D_{h;f} &= i(o_3-g_3-\bar{o}_3+\bar{g}_3) \, , 
& D_{h;h} = o_3-g_3+\bar{o}_3-\bar{g}_3 \,.
\label{z2z2a22}
\ea
The D9 branes have ${\cal N}=1$ supersymmetry, but now with gauge 
group ${\rm SO}(8)^4$ and chiral multiplets in 
the $(8,8,1,1)$ and five permutations.
Moreover, each antibrane gives a non-supersymmetric 
spectrum, with gauge group ${\rm U}(8) \times {\rm U}(8)$, chiral 
multiplets in the $(8,8)$, $(8,\bar{8})$ and in their conjugates, spinors in 
the $(28,1)$, $(\overline{28},1)$, 
$(1,28)$, $(1,\overline{28})$ and complex scalars in 
the $(36,1)$, $(\overline{36},1)$, $(1,36)$, $(1,\overline{36})$.
We would like to stress that in this case the would-be gauginos are massless,
since the M{\"o}bius amplitude does not affect the adjoint 
representations of unitary 
groups. Finally, the $\bar{5}_k\bar{5}_l$ sectors give chiral multiplets
in the representations
\ba
\bar{5}_1\bar{5}_2  :\ \ & & (8,1;8,1) \ , \ \ (\bar{8},1;1,8) \ , 
          \ \ (1,8;\bar{8},1) \ , \ \ (1,\bar{8};1,\bar{8}) \ , 
\nonumber \\
\bar{5}_1\bar{5}_3  :\ \ & & (8,1;\bar{8},1) \ , \ \ (\bar{8},1;1,\bar{8}) \ , 
          \ \ (1,8;1,8) \ , \ \ (1,\bar{8};8,1) \ , 
\nonumber \\
\bar{5}_2\bar{5}_3 :\ \ & &   (8,1;\bar{8},1) \ , \ \ (\bar{8},1;1,8) \ , 
          \ \ (1,8;1,\bar{8}) \ , \ \ (1,\bar{8};8,1) \ ,
\nonumber
\ea 
while the non-supersymmetric $9\bar{5}_k$ sectors give Weyl spinors
and complex scalars in the representations
\ba
9\bar{5}_1 &{\it \ {\rm spinors} \ :}& \
  (8,1,1,1;8,1) \, , \ \ (1,8,1,1;\bar{8},1) \, , 
\nonumber \\
& &           \ (1,1,8,1;1,8) \, , \ \ (1,1,1,8;1,\bar{8}) \, , 
\nonumber \\
           &{\it \ {\rm scalars} \ :}&  \ 
(8,1,1,1;1,8) \, , \ \ (1,8,1,1;1,\bar{8}) \, , 
\nonumber \\
& &          \ (1,1,8,1;8,1) \, , \ \ (1,1,1,8;\bar{8},1) \, , 
\nonumber \\
9\bar{5}_2 &{\it \ {\rm spinors} \ :}& \
  (8,1,1,1;8,1) \, , \ \ (1,8,1,1;1,8) \, , 
\nonumber \\
& &           \ (1,1,8,1;\bar{8},1) \, , \ \ (1,1,1,8;1,\bar{8}) \, , 
\nonumber \\
           &{\it \ {\rm scalars} \ :}&  \ 
(8,1,1,1;1,8) \, , \ \ (1,8,1,1;8,1) \, , 
\nonumber \\
& &           \ (1,1,8,1;1,\bar{8}) \, , \ \ (1,1,1,8;\bar{8},1) \, , 
\nonumber \\
9\bar{5}_3 &{\it \ {\rm spinors} \ :}& \
  (8,1,1,1;\bar{8},1) \, , \ \ (1,8,1,1;1,\bar{8}) \, , 
\nonumber \\
& &           \ (1,1,8,1;1,8) \, , \ \ (1,1,1,8;8,1) \, , 
\nonumber \\
           &{\it \ {\rm scalars} \ :}&  \ 
(8,1,1,1;1,\bar{8}) \, , \ \ (1,8,1,1;\bar{8},1) \, , 
 \nonumber \\
& &         \ (1,1,8,1;8,1) \, , \ \ (1,1,1,8;1,8) \,. \nonumber
\ea
All chiral spectra thus obtained are free of non-Abelian anomalies. 

On the other hand, the models without discrete torsion are not chiral. 
The choice $(\epsilon_1,\epsilon_2,\epsilon_3) = (+,+,+)$,
discussed in \cite{erice} and worked out in detail in \cite{bl}, 
leads to a gauge group ${\rm USp}(16)^4$. Another model, without 
discrete torsion but with two $\overline{\rm D}5$
branes, can be obtained
if two of the $\epsilon_k$'s are negative. The D9 and D5 branes give 
orthogonal gauge groups with ${\cal N}=1$ supersymmetry, while the two 
$\overline{\rm D}5$
branes give symplectic gauge groups with broken supersymmetry. 
For instance, with the choice $(+,-,-)$
\ba
{\cal A}_0 &=& {\textstyle\frac{1}{8}} \biggl \{ 
(N_o^2+ D^2_{g;o}+ D^2_{f;o}+ D^2_{h;o})T_{oo}
\nonumber \\ 
& & 
+2N_oD_{g;o}T_{go}+2N_oD_{f;o}T^{(-)}_{fo}+2N_oD_{h;o}T^{(-)}_{ho} 
\nonumber \\
& & + 2D_{g;o}D_{f;o}T^{(-)}_{ho}+2D_{g;o}D_{h;o}T^{(-)}_{fo}
+2D_{f;o}D_{h;o}T_{go} \biggr \}  \, , 
\nonumber \\
{\cal M}_0 &=& - {\textstyle\frac{1}{4}} \biggl\{ 
(N_o +D_{g;o}) (\tau_{oo}-\tau_{og}+\tau_{of}+\tau_{oh})
\nonumber 
\\
& & - (D_{f;o}+D_{h;o}) 
(\tau^{{\rm NS}}_{oo}-\tau^{{\rm NS}}_{og}+
\tau^{{\rm NS}}_{of}+\tau^{{\rm NS}}_{oh})
\nonumber \\
& & - (D_{f;o}+D_{h;o}) 
(\tau^{{\rm R}}_{oo}-\tau^{{\rm R}}_{og}+
\tau^{{\rm R}}_{of}+
\tau^{{\rm R}}_{oh})
\biggr \} \, ,
\label{z2z2a26}
\ea
where no breaking terms are present. After a suitable rescaling
of the charge multiplicities,
the $\overline{\rm D} 5_2$ and the $\overline{\rm D} 5_3$ branes give non-supersymmetric spectra, 
with ${\rm USp}(16)$ gauge groups, spinors in the 136 and 
in three copies of the 120
and scalars in the 120 and in two copies of the 136. 
The 99 and $5_15_1$ sectors
have ${\cal N}=1$ supersymmetry, ${\rm SO} (16)$ gauge groups and 
chiral multiplets in the 136 and in two copies of the 120.
Finally, there are two chiral multiplets in the
$(16,16)$ arising from the $95_1$ and $\bar{5}_2 \bar{5}_3$
sectors and complex scalars and Weyl spinors in bi-fundamental
representations arising from the $9\bar{5}_2$, $9\bar{5}_3$, 
$5_1 \bar{5}_2$ and $5_1 \bar{5}_3$ sectors. More details can
be found in \cite{aadds}.  

\vskip 12pt
\subsection{Magnetic deformations and supersymmetry}

Homogeneous magnetic fields provide an interesting example of a 
non-trivial deformation
compatible with two-dimensional conformal invariance. The study of their
effect on open strings is relatively simple \cite{ftse,acny}, for they 
interact only
with the string ends, affecting the world-sheet
dynamics via boundary terms. They also provide an interesting
way to break supersymmetry in open string models 
\cite{wito32,bachasmag,penta},
an option extensively investigated in \cite{berlinmadrid}.

Let us begin by considering the bosonic string in the presence of
a uniform magnetic field, that can be described by the 
vector potential
\be
A_\mu = - {\textstyle \frac{1}{2}}  F_{\mu\nu} X^{\nu} \,. \label{mag1}
\ee
The variational principle for the world-sheet action
\ba
S &=& - \frac{1}{4 \pi \alpha'} \, \int d\tau \int_0^{\pi} d\sigma
\partial_\alpha X \cdot \partial^\alpha X \nonumber \\
&& -
\left. q_{\rm L} \int d\tau A_\mu \partial_\tau X^{\mu} \right|_{\sigma =0}
-
\left. q_{\rm R} \int d\tau A_\mu \partial_\tau X^{\mu} \right|_{\sigma =\pi}
\,, \label{mag2}
\ea
here written in the conformal gauge for a strip of width $\pi$, yields the 
wave equations
\be
\left( {\partial^2 \over \partial \tau^2} - 
{\partial^2 \over \partial \sigma^2} \right) X^{\mu} =0 \,, \label{mag3}
\ee
together with the boundary conditions
\ba
{1\over 2\pi\alpha '} \, \partial_\sigma X^\mu - q_{\rm L} F^{\mu}{}_{\nu} \,
\partial_\tau X^\nu &=& 0 \,,
\nonumber \\
{1\over 2\pi\alpha '} \, \partial_\sigma X^\mu + q_{\rm R} F^{\mu}{}_{\nu} \,
\partial_\tau X^\nu &=& 0 \,, \label{mag4}
\ea
for $\sigma = 0,\pi$, that interpolate between the Neumann and Dirichlet 
cases. 

Eq. (\ref{mag4}) admits an alternative geometric interpretation in 
terms of rotated
branes \cite{douglas2}. To be specific, let us consider a magnetic 
field $F_{12} =H$ 
in a plane $(X^1 , X^2)$ and perform a T-duality along the $X^2$ 
direction, so that $\partial_\alpha X^2 = \epsilon_{\alpha\beta}\,
\partial^\beta Y^2$ links $X^2$ to the dual coordinate $Y^2$. 
The boundary conditions then become standard
Neumann and Dirichlet ones 
\ba
\partial_\sigma \left( X^1 -  2 \pi \alpha '
q_{\rm L} H \, Y^2 \right) &=& 0 \,,
\nonumber \\
\partial_\tau \left( Y^2 +  2 \pi \alpha ' q_{\rm R} H \,
X^1 \right) &=& 0 \,, \label{mag4bis}
\ea
that indeed identify branes rotated by an angle 
\be
\theta_{{\rm L,R}} = - \tan^{-1} (2 \pi \alpha ' q_{{\rm L,R}} H) \,.
\ee 

It is now convenient to introduce the complex coordinates 
\be
Z = {\textstyle{1\over \sqrt{2}}} (X^1 + i X^2)\,, \qquad
\bar Z = {\textstyle{1\over \sqrt{2}}} (X^1 - i X^2)\,, \label{mag5}
\ee
so that the action becomes
\ba
S &=& {1\over 2\pi\alpha '} \int d \tau \int_{0}^{\pi} d\sigma \,
\partial_\alpha \bar Z \partial^\alpha Z \nonumber \\
& &+ i q_{\rm L} H \left. \int d\tau \bar Z \partial_\tau Z
\right|_{\sigma = 0} 
 + i q_{\rm R} H \left. \int d\tau \bar Z \partial_\tau Z
\right|_{\sigma = \pi} \,, \label{mag6}
\ea
while the boundary conditions (\ref{mag4}) reduce to
\ba
\partial_\sigma Z +i \alpha \, 
\partial_\tau Z &=& 0\,,
\nonumber \\
\partial_\sigma Z -i \beta \,
\partial_\tau Z &=& 0 \,,
\label{mag7}
\ea
with
\be
\alpha = 2 \pi \alpha' q_{\rm L} H \,, \qquad \beta= 2 \pi \alpha' 
q_{\rm R} H \,.\label{mag8}
\ee

The solution of the wave equation with these boundary conditions
depends on the total charge $Q=q_{\rm L}+q_{\rm R}$ 
of the open string. If $Q$ is different from zero, 
the frequencies of the oscillator modes are shifted by \cite{acny}
\be
\zeta = {1\over \pi} (\gamma + \gamma' ) \,,
\qquad {\rm with}
\qquad
\gamma= \tan ^{-1} (\alpha ) \,,\quad \gamma' =\tan^{-1} (\beta ) \, ,
\label{mag9}
\ee
and the mode expansion becomes
\be
Z(\tau,\sigma) = z + i \sqrt{2 \alpha'} 
\left[ \sum_{n=1}^\infty a_n \psi_n(\tau,\sigma) - 
\sum_{m=0}^\infty b^\dag_m \psi_{-m}(\tau,\sigma) \right] \, ,
\label{mag10}
\ee
with
\be
\psi_n(\tau,\sigma) = \frac{1}{\sqrt{|n-\zeta|}} \, \cos\left[
(n - \zeta)\sigma + \gamma \right]\, e^{-i(n-\zeta)\tau} \,. 
\label{mag11}
\ee
The momentum canonically conjugate to $Z$ is now
\be
\bar \Pi(\tau,\sigma) = \frac{1}{2 \pi \alpha'} 
\left\{ \partial_\tau  Z (\tau,\sigma) -
i Z (\tau,\sigma) \left[ \alpha \delta(\sigma) + 
\beta \delta(\pi - \sigma) \right]  
\right\} \, ,\label{mag12}
\ee
and the usual commutation relations
imply that $a_m$ and $b^\dag_m$ are independent Fourier coefficients, while
the zero modes do not commute
\be
[z,\bar z]=\frac{2\pi\alpha'}{\alpha+\beta} \, ,
\label{mag13}
\ee
so that their contribution to the Hamiltonian results in the familiar
spectrum of Landau levels.

The solution is quite different for the ``dipole'' strings, with 
$\beta=-\alpha$, for which $Q=0$. 
In this case, the oscillator frequencies are no
more shifted, while the boundary conditions allow for the presence of
new zero modes, so that one can write \cite{acny}
\ba
Z(\tau,\sigma) &=& \frac{z + \bar{p}\left[ \tau - i\alpha(\sigma - \frac{1}{2}
\pi) \right]}{\sqrt{1 + \alpha^2}} 
\nonumber \\
&&+ i \sqrt{2 \alpha'} \sum_{n=1}^\infty 
\left[ a_n \psi_n(\tau,\sigma) - 
b^\dag_n \psi_{-n}(\tau,\sigma) \right] \,. \
\label{mag14}
\ea
The canonical commutation relations now imply that the $a_m$ and 
$b^\dag_m$ are independent oscillator modes, while the cartesian
components in $z$ and $p$ satisfy
\be
[x_i,x_j] = 0\,, \quad  [p_i,p_j] = 0\,, \quad   
[x_i,p_j] = i \delta_{ij}\,.
\label{mag15}
\ee

The operators $p$ and $\bar{p}$ are actually related to the conserved charges
of the particle orbits in a homogeneous magnetic field, that in 
Classical Electrodynamics define their centre, and only in the limit of
a vanishing magnetic field do they reduce to ordinary momentum components.
This can be simply justified considering the equations for
a particle in a uniform magnetic field
\be
\frac{d {\bf p}}{dt} = \frac{q}{c}\, {\bf v} \times {\bf H} \, ,
\ee
for which
\be
{\bf R} = {\bf p} - \frac{q}{c}\, {\bf r} \times {\bf H} 
\ee
is clearly a conserved quantity. If these arguments are applied to the 
two  open-string ends, the overall conserved quantities
\be
{\bf R}_{\rm tot} = {\bf p}_1 + {\bf p}_2 - 
\frac{q_{\rm L}}{c} \, {\bf r}_1 \times {\bf H}
- \frac{q_{\rm R}}{c} \, {\bf r}_2 \times {\bf H} 
\ee 
are indeed
mutually commuting for a ``dipole'' string, with $q_{\rm L}= - q_{\rm R}$,
since in this case ${\bf R}_{\rm tot}$ involves the total momentum and
the relative coordinate of the string ends.

We now have all the ingredients to compute the partition function of
open strings in the presence of a uniform magnetic field. As we shall
see, the annulus amplitude encodes very interesting
properties of the low-energy interactions. Let 
us begin by considering the simpler case of
bosonic strings \cite{acny}, concentrating on a pair of
coordinates, whose contribution
to the annulus amplitude in the absence of a magnetic field was
\be
{\cal A} = \frac{N^2}{2 \tau_2 \eta^2} \label{mag16}
\ee
in the case of $N$ D25 branes. If the magnetic field affects
only some of them, one can let  $N = N_0 + m + \bar m$, 
where $N_0$ counts the number of neutral branes, while $m$ and $\bar m$ 
count the equal numbers of magnetized
branes with U(1) charges $\pm 1$. The resulting embedding of the magnetized
U(1), clearly consistent with the traceless SO(32) generators,
leads to neutral strings, with multiplicities $N_0^2$ and
$m \bar m$, and to charged ones, with multiplicities $N_0 m$, $N_0 \bar
m$, $m^2$ and $\bar m ^2$. 

Both the uncharged, $N_0^2$, and the
``dipole'', $m\bar m$, strings have unshifted oscillators, 
that give identical contributions to the partition function, 
but they differ crucially in their zero modes. This feature
is easier to exhibit if the system is put momentarily 
in a box of size $R$. Then, from (\ref{mag14}), if
$Z$ is associated to circles of radius $R$, $z$ has a period $2 \pi R 
\sqrt{1+\alpha^2}$ and, consequently, $p$ is quantized in units of 
$1/(R\sqrt{1+\alpha^2})$. Therefore, in the large-volume limit \cite{acny}
\be
\sum_{n_1} \sum_{n_2} \to R^2 (1 +
\alpha^2 ) \int dp_1 \int d p_2 \,, \label{mag17}
\ee
and the neutral strings contribute to the partition function 
\be
{\cal A}_0 \sim \left[ {\textstyle{1\over 2}} N_0^2 +
 m\bar m (1 + (2 \pi \alpha' q H)^2 ) \right]
{1 \over \tau_2 \eta^2}\,, \label{mag18}
\ee
where we have inserted the explicit value for $\alpha$ given in
(\ref{mag8}).

The charged-string contributions differ in two respects: their
modes are shifted and, as a result, their contribution
to the annulus amplitude involves theta-functions with
non-vanishing arguments, while no overall $\tau_2$ factors are present
in the partition function, as befits the absence of zero modes.
Altogether, one obtains
\be
{\cal A}_{\pm} \sim i N_0 ( m + \bar m ) {\eta \over 
q^{{1\over 2}\zeta^2}\vartheta_1 (\zeta
\tau | \tau)} + {\textstyle{1\over 2}}i
(m^2 + \bar m ^2 ) {\eta \over 
q^{2\zeta^2} \vartheta_1 (2 \zeta
\tau | \tau )} \,, \label{mag19}
\ee
where we have made use of the symmetry properties of $\vartheta_1$,
and $\zeta$ is given in (\ref{mag9}).

Up to overall normalizations, from (\ref{mag18}) and (\ref{mag19}) one 
can read the open-string spectrum. However, more interesting results 
can be extracted from the 
transverse-channel amplitude, and after an $S$ modular transformation
\ba
\tilde{\cal A} &\sim& \left[ N_0^2 + 2 m \bar m  
(1 + (2 \pi \alpha' q H)^2 ) \right] {1 \over \tau_2 \eta^2}
\nonumber \\
& & + 2 N_0 ( m + \bar m ) {\eta \over \vartheta_1 (\zeta
| \tau)} + (m^2 + \bar m ^2 ) {\eta \over \vartheta_1 (2 \zeta
| \tau )} \,, \label{mag20}
\ea
where the massless contribution is proportional to
\be
\left[ N_0 + (m+\bar m) \sqrt{ 1 + (2 \pi \alpha'
q H)^2} \right]^2 \,.
\ee
This corresponds to the dilaton tadpole that, as we have seen,
can be linked to the
derivative of the low-energy effective action with respect to the
dilaton. Since these tree-level interactions originate from the 
disk, one can easily associate the charged contributions to  the Lagrangian 
\be
{\cal S}_{\rm {DBI}} \sim 
\int e^{-\varphi} \sqrt{-{\rm det} (g_{\mu\nu} + 2 \pi \alpha ' q
F_{\mu\nu} )} \,,
\ee
recovering the celebrated result that the low-energy
open-string dynamics is governed by the Dirac-Born-Infeld action 
\cite{bidirac,ftse}.

We can now turn to the superstring, and in particular to its
compactification on the $[T^2(H_1)\times T^2(H_2)]/\bb{Z}_2$ orbifold,
while allowing for a pair of uniform  Abelian magnetic
fields $H_1$ and $H_2$ in the two $T^2$ factors \cite{magnetic}. As we
shall see, their simultaneous presence will bring about an interesting 
new effect. 

Let us begin by recalling the Klein-bottle projection
already met in subsection 5.6, in the discussion of ${\cal N}= (1,0)$
supersymmetric vacua with one tensor multiplet and 20 hyper multiplets
from the closed sector, that here we write concisely as
\ba
{\cal K} &=& {\textstyle{1\over 4}} \Biggl\{ (Q_o + Q_v) (0;0) \left[ P_1 P_2 +
W_1 W_2 \right] \nonumber \\
&&+ 16\times 2 (Q_s + Q_c ) (0;0) \left( {\eta \over
\vartheta_4 (0)} \right)^2 \Biggr\} \, , \label{mag16k}
\ea
where $P_i$ and $W_i$ denote momentum and winding sums in the two tori,
and where the six-dimensional $Q$ characters are endowed with
a pair of arguments, anticipating the effect of the magnetic
deformations in the two internal tori. In general
\begin{eqnarray}
Q_o (\eta ; \zeta) &=& V_4 (0) \left[ O_2 (\eta ) O_2 (\zeta ) +
V_2 (\eta ) V_2 (\zeta ) \right] \nonumber\\ 
&&- C_4 (0) \left[ S_2 (\eta ) C_2 (\zeta ) +
C_2 (\eta ) S_2 (\zeta ) \right] \, ,
\nonumber
\\
Q_v (\eta ; \zeta) &=&  O_4 (0) \left[ V_2 (\eta ) O_2 (\zeta ) +
O_2 (\eta ) V_2 (\zeta ) \right] \nonumber \\
&&- S_4 (0) \left[ S_2 (\eta ) 
S_2 (\zeta ) + C_2 (\eta ) C_2 (\zeta ) \right] \, ,
\nonumber
\\
Q_s (\eta ; \zeta ) &=& O_4 (0) \left[ S_2 (\eta ) C_2 (\zeta ) +
C_2 (\eta ) S_2 (\zeta ) \right] \nonumber \\ 
&&- S_4 (0) \left[  O_2 (\eta ) O_2 (\zeta ) +
V_2 (\eta ) V_2 (\zeta ) \right] \, ,
\nonumber
\\
Q_c (\eta ; \zeta) 
&=&  V_4 (0) \left[ S_2 (\eta ) S_2 (\zeta ) +
C_2 (\eta ) C_2 (\zeta ) \right] \nonumber \\
&&- C_4 (0) 
\left[  V_2 (\eta ) O_2 (\zeta ) +
O_2 (\eta ) V_2 (\zeta ) \right] \, , \label{mag17k}
\end{eqnarray}
where the four level-one O($2n$) characters in (\ref{mag17k}) 
are related to the
four Jacobi theta-functions with non-vanishing argument according to
\ba
O_{2n}(\zeta) &=& \frac{1}{2 \eta^n (\tau)} \left[
\vartheta_3^n(\zeta|\tau)
+ \vartheta_4^n(\zeta|\tau)\right] \, , 
\nonumber \\
V_{2n}(\zeta) &=&\frac{1}{2 \eta^n (\tau)} \left[
\vartheta_3^n(\zeta|\tau)
- \vartheta_4^n(\zeta|\tau) \right] \, , \nonumber \\ 
S_{2n}(\zeta) &=& \frac{1}{2 \eta^n (\tau)} \left[
\vartheta_2^n(\zeta|\tau) + i^{-n} \vartheta_1^n(\zeta|\tau)\right] \,
, \nonumber \\
C_{2n}(\zeta) &=& \frac{1}{2 \eta^n (\tau)} \left[
\vartheta_2^n(\zeta|\tau)
- i^{-n} \vartheta_1^n(\zeta|\tau) \right] \,. \label{mag18k}
\ea

In this case, the original open strings carry a unitary gauge group,
and therefore one 
is led to distinguish three types of complex multiplicities:
$(m,\bar{m})$ for the string ends aligned with the magnetic U(1), 
that here we shall take within the D9 gauge group, 
$(n,\bar{n})$ for the remaining D9 ends, and finally $(d,\bar{d})$ for
the D5 ones.
As a result, the annulus amplitude involves several types of open strings:
the dipole strings, with Chan-Paton multiplicity $m \bar{m}$, the
uncharged ones, with multiplicities independent
of $m$ and $\bar{m}$, the singly-charged ones, with multiplicities
linear in $m$ or $\bar{m}$, and finally the doubly-charged ones, with
multiplicities proportional to $m^2$ or $\bar{m}^2$. The
annulus amplitude is then
\begin{eqnarray}
{\cal A} &=& {\textstyle{1\over 4}} \Biggl\{ (Q_o + Q_v)(0;0) \left[
(n+\bar n)^2 P_1 P_2 + (d+\bar d)^2 W_1 W_2 
+ 2 m \bar{m} \tilde P_1 \tilde P_2 \right]
\nonumber
\\
&-& 2 (m+\bar m) (n + \bar{n}) (Q_o + Q_v )(\zeta_1 \tau ; \zeta_2 \tau
) {k_1 \eta \over
\vartheta_1 (\zeta_1 \tau)} {k_2 \eta \over \vartheta_1 (\zeta_2 \tau)} 
\nonumber
\\
&-& ( m^2 + \bar{m}^2 ) (Q_o + Q_v ) (2 \zeta_1 \tau ; 2 \zeta_2 \tau ) 
{2 k_1 \eta \over
\vartheta_1 (2 \zeta_1 \tau)} {2 k_2 \eta \over \vartheta_1 (2 \zeta_2 \tau)} 
\nonumber 
\\
&-& \left[ (n-\bar n)^2 -2 m\bar m + (d-\bar d)^2 \right] (Q_o - Q_v ) (0;0)
\left( {2\eta \over \vartheta_2 (0)}\right)^2 
\nonumber
\\
&-& 2 (m-\bar m) (n - \bar{n}) (Q_o - Q_v ) (\zeta_1 \tau ; \zeta_2 \tau) 
{2\eta \over \vartheta_2
(\zeta_1 \tau)} {2\eta \over \vartheta_2 (\zeta_2 \tau)} 
\nonumber
\\
&-& (m^2 + \bar{m}^2) (Q_o - Q_v ) (2\zeta_1 \tau ; 2\zeta_2 \tau)
{2\eta \over \vartheta_2
(2\zeta_1 \tau)} {2\eta \over \vartheta_2 (2\zeta_2 \tau)} 
\nonumber
\\
&+& 2 (n+\bar n ) (d+\bar d) (Q_s + Q_c) (0;0) \left({\eta \over
\vartheta_4 (0)}\right)^2 
\nonumber
\\
&+& 2 (m + \bar{m})(d+\bar d)(Q_s + Q_c) (\zeta_1 \tau ; \zeta_2 \tau)
{\eta \over \vartheta_4
(\zeta_1 \tau )} {\eta \over \vartheta_4 (\zeta_2 \tau )}
\nonumber
\\
&-& 2 (n-\bar n) (d - \bar d) (Q_s - Q_c )
(0;0) \left( {\eta \over \vartheta_3 (0)}\right)^2 \label{annsusy}
\\
&-& 2  (m - \bar{m})(d-\bar d) (Q_s - Q_c) (\zeta_1 \tau ; \zeta_2 \tau)
 {\eta \over \vartheta_3
(\zeta_1 \tau )} {\eta \over \vartheta_3 (\zeta_2 \tau )} \Biggr\} \, ,
\nonumber
\end{eqnarray}
while the corresponding M\"obius amplitude is
\begin{eqnarray}
{\cal M} &=& -{\textstyle{1\over 4}} \Biggl[ 
(\hat Q_o + \hat Q_v )(0;0) \left[ (n+\bar n) P_1 P_2 + (d+\bar d) W_1
W_2 \right]
\nonumber
\\
&-& ( m + \bar{m}) (\hat Q_o + \hat Q_v ) (2\zeta_1 \tau ; 2\zeta_2 
\tau) {2 k_1
\hat\eta \over \hat \vartheta_1 (2\zeta_1\tau)} {2 k_2
\hat\eta \over \hat \vartheta_1 (2\zeta_2\tau)}
\nonumber
\\
&-& \left( n+ \bar n + d + \bar d \right) (\hat Q_o - \hat Q_v )(0;0) \left(
{2\hat\eta \over \hat \vartheta_2 (0)}\right)^2 \label{mobsusy}
\\
&-& (m + \bar{m}) (\hat Q_o - \hat Q_v ) (2 \zeta_1 \tau ; 2 \zeta_2 \tau )
{2\hat\eta \over \hat\vartheta_2 (2\zeta_1\tau)}
{2\hat\eta \over \hat\vartheta_2 (2\zeta_2\tau)} \Biggr] \,.
\nonumber
\end{eqnarray}
Here the arguments $\zeta_i$  and $2\zeta_i$ 
are associated to strings with one
or two charged ends, and, for brevity, both the imaginary modulus 
$\frac{1}{2} i \tau_2$ of ${\cal A}$ and the complex modulus $
\frac{1}{2} + \frac{1}{2} i \tau_2$ of ${\cal M}$ are denoted
by the same symbol $\tau$, although
the proper ``hatted'' contributions to the M\"obius amplitude are explicitly
indicated. Finally, while $P_i$ and $W_i$ are conventional momentum
and winding sums for the two-tori, a ``tilde'' denotes a sum over
``boosted'' momenta $m_i/R\sqrt{1+ (2 \pi \alpha' q H_{i})^{2}}$,
and terms with
opposite U(1) charges, and thus with opposite $\zeta_i$ arguments, 
have been grouped together, using the symmetries of 
the Jacobi theta-functions.

For generic magnetic fields, the open spectrum is indeed non-supersymmetric
and develops Nielsen-Olesen instabilities \cite{nole},
tachyonic modes induced by the magnetic moments of internal 
Abelian gauge bosons \cite{bachasmag}. 
For instance, small magnetic fields
affect the mass formula for the untwisted string modes according to
\ba
\Delta M^2 &=& \frac{1}{2 \pi \alpha'} \; \sum_{i=1,2} 
\Bigl[  (2 n_i + 1) |2 \pi \alpha ' (q_{\rm L} + q_{\rm R}) H_i| \nonumber
\\ 
&&+ 4 \pi\alpha' (q_{\rm L} + q_{\rm R}) \Sigma_i H_i \Bigr] \, ,
\ea
where the first term originates from the Landau levels and
the second from the magnetic moments of the spins $\Sigma_i$. 
For the internal
components of the vectors, the magnetic moment coupling generally
overrides the zero-point contribution, leading to tachyonic modes,
unless $|H_1|=|H_2|$,
while for spin-$\frac{1}{2}$ modes it can at most compensate it. 
On the other hand, for twisted modes the zero-point contribution
is absent, since ND strings have no Landau levels, but in this case
the low-lying space-time fermions, that originate from the fermionic part 
$S_4 O_4$ of $Q_s$, are scalars in the internal space and
have no magnetic moment couplings. However, their bosonic partners, 
that originate from $O_4 C_4$,
are affected by the magnetic deformations and have mass shifts
$\Delta M^2 \sim \pm (H_1 - H_2)$.  The conclusion is that
if $H_1=H_2$ all tachyonic instabilities
are indeed absent, and actually with this choice
the supersymmetry charge, that belongs to the $C_4 C_4$ sector, 
is also unaffected\footnote[8]{Type II
branes at angles preserving some supersymmetry 
were originally considered in \cite{douglas2}.  After T-dualities, these 
can be related to special choices for the internal 
magnetic fields. Type I toroidal models, however, can not lead to 
supersymmetric vacuum configurations, since the resulting R-R tadpoles would
require the introduction of antibranes.}. A residual supersymmetry is 
thus present for the entire string spectrum, and indeed, 
using Jacobi identities for non-vanishing arguments \cite{jacobi}, 
one can see that for $\zeta_1=\zeta_2$ 
both ${\cal A}$ and ${\cal M}$ vanish identically. Still, the 
resulting supersymmetric models are rather peculiar, 
as can be seen from the deformed tadpole conditions, 
to which we now turn.

Let us begin by examining the untwisted R-R tadpoles. 
For $C_4 S_2 C_2$ one finds
\ba
& & \left[ n+\bar n + m + \bar{m} - 32 +  (2 \pi \alpha 'q)^2  H_1 H_2 (m +
\bar m ) \right] \sqrt{v_1 v_2} 
 \nonumber \\
& & + {1\over \sqrt{v_1 v_2}} \left[ d+\bar d
- 32\right]  = 0 \, , \label{rrutad}
\ea
aside from terms that vanish after identifying the multiplicities
of conjugate representations $(m,\bar{m})$, $(n,\bar{n})$ and $(d,\bar{d})$,
while the additional untwisted R-R tadpole conditions 
from $Q_o$ and $Q_v$ are compatible with (\ref{rrutad}) and do not add further 
constraints. This expression reflects the familiar Wess-Zumino
coupling of D-branes \cite{dgm}, that in this context reduces to
\be
{\cal S}_{{\rm WZ}} \sim \sum_{p,a} 
\int_{{\cal M}_{10}}  e^{q_a F} \wedge C_{p+1}  \, ,
\ee
and therefore the various powers of $H$
couple to R-R forms of different degrees. 
In this class of models the term bilinear in the magnetic fields, the
first that can arise since the group generators are traceless, has a very
neat effect: it charges the D9 brane with respect to the
six-form potential, and as a result one can replace some of the D5 branes
with their blown-up counterparts thus obtained. This process thus
reverses the familiar relation of \cite{witsmall} between small-size
instantons and D5 branes: a fully blown-up instanton, corresponding to
a uniform magnetic field, provides an exact description of
a D5 brane smeared over the internal torus in terms of a magnetized D9 brane.
This can be seen very clearly making use of the Dirac
quantization condition  
\be
2 \pi \alpha' q H_i v_i = k_i \qquad (i=1,2) \,,
\label{dirac}
\ee
that turns (\ref{rrutad}) into
\begin{eqnarray}
& & m+\bar m + n + \bar n = 32 \, ,
\nonumber
\\
& & k_1 k_2 (m + \bar m )  + d + \bar d = 32 \,. \label{urrt}
\end{eqnarray}
Thus, if $k_1 k_2 > 0$ the D9 branes indeed acquire the
R-R charge of $|k_1 k_2|$ D5 branes, while if $k_1 k_2 < 0$ 
they acquire the R-R charge of as many $\overline{\rm D} 5$ 
branes. As an aside,
notice that for the T-dual system eq. (\ref{dirac})
would become
\be
R_{1,i} \tan\theta_i = k_i \tilde R _{2,i}\,,
\ee
the condition that the rotated branes do not fill densely the tori
of radii $R_{1,i}$ and $\tilde R_{2,i} = \alpha' / R_{2,i}$, but
close after $k_i$ wrappings.

The untwisted NS-NS tadpoles exhibit very nicely their relation to the 
Born-Infeld term. For instance, the dilaton tadpole
\ba
& & \left[ n+\bar n + (m + \bar m) \sqrt{\left( 1 +  (2 \pi 
\alpha' q)^2 H_1^2 \right) 
\left( 1 + (2 \pi \alpha 'q) ^2 H_2^2 \right) }  -32 \right] \sqrt{v_1 v_2}
\nonumber \\
& & + {1\over \sqrt{v_1 v_2}} \left[ d +
\bar d - 32 \right] \label{diltad}
\ea
originates from $V_4 O_2 O_2$, and can be clearly linked to
the $\varphi$-derivative of ${\cal S}_{{\rm DBI}}$, computed for this
background. On the other hand, the volume of the first internal torus 
originates from $O_4 V_2 O_2$, and the corresponding tadpole,
\ba
& & \left[ n+\bar n + (m + \bar m) \, {1 - (2\pi\alpha' q H_1)^2 \over 
\sqrt{ 1 + (2\pi\alpha' q H_1)^2 }}\,
\sqrt{ 1 + (2 \pi\alpha' q H_2)^2 } -32 \right] \sqrt{v_1
v_2} \nonumber \\
& & - {1\over \sqrt{v_1 v_2}} \left[ d+\bar
d - 32 \right]  \, , \label{mettad}
\ea
can be linked to the derivative of the Dirac-Born-Infeld action 
with respect to
the corresponding breathing mode $\Lambda$, 
that in the first square root in (\ref{diltad}) would deform 
1 to $\Lambda^2$ and 
$H_1^2$ to $H_1^2/\Lambda^2$. A similar result holds for the
volume of the second torus, with the proper interchange
of $H_1$ and $H_2$, and, for the sake of brevity, in these NS-NS tadpoles
we have omitted all terms
that vanish using the constraint $m = \bar m $. The complete form
of eq.~(\ref{mettad}) is also
rather interesting, since, in contrast with the usual structure
of unoriented string amplitudes, it is {\it not} a perfect square. 
This unusual feature can be ascribed to the behaviour of the 
internal magnetic fields under time
reversal. Indeed, as stressed long ago in \cite{cardy2}, these
transverse-channel amplitudes involve a time-reversal operation ${\cal T}$, 
and are thus of the form $\langle {\cal T} (B) | q^{L_0} | B \rangle$. 
In the present examples, additional signs are introduced by the
magnetic fields, that are odd under time reversal and, as a result, in
deriving from factorization the M\"obius amplitudes of these models, 
it is crucial to add the two contributions $\langle {\cal T} (B) 
| q^{L_0} | C \rangle $ and $\langle {\cal T} (C) | q^{L_0} |
B \rangle $, that are different and effectively eliminate the additional 
terms from the transverse-channel.

Both (\ref{mettad}) and the dilaton tadpole (\ref{diltad}) simplify
drastically in the interesting case $H_1=H_2$ where, using the
Dirac quantization conditions (\ref{dirac}), they become
\be
\left[ n+\bar n + m + \bar m -32 \right] \sqrt{v_1 v_2} 
\mp {1\over \sqrt{v_1 v_2}} \left[ k_1 k_2 (m
+ \bar m) + d +
\bar d - 32 \right] \,.
\ee
Therefore, they both vanish, as they should, in these supersymmetric 
configurations, once the corresponding R-R tadpole conditions 
(\ref{urrt}) are enforced.

The twisted R-R tadpoles
\begin{equation}
15 \left[ {\textstyle{1\over 4}} (m-\bar m + n -\bar n )
\right]^2
+ \left[ {\textstyle{1\over 4}} (m-\bar m + n - \bar n ) - (d-\bar d)
\right]^2 \label{twistedrrmag}
\end{equation}
originate from the $S_4 O_2 O_2$ sector,
whose states are scalars in the internal space. As in the undeformed
model of subsection 5.6, they 
reflect very neatly the distribution of the branes among the sixteen
fixed points, only one of which accommodates D5 branes in our
examples,
are not affected by the magnetic fields, and vanish identically for 
unitary gauge groups.
The corresponding NS-NS tadpoles, originating from the
$O_4 S_2 C_2$ and $O_4 C_2 S_2$ sectors, 
are somewhat more
involved, and after the identification of conjugate multiplicities
are proportional to
\be
{2 \pi \alpha ' q \, (H_1 - H_2 )  \over \sqrt{ (1 + (2 \pi \alpha' q H_1)^2 ) 
(1 + (2\pi \alpha' q H_2)^2 ) }} \,.
\ee
They clearly display new couplings for twisted
NS-NS fields that, as expected,  vanish for $H_1 = H_2$.

We can now describe  some supersymmetric models corresponding to
the special choice $H_1=H_2$. It suffices to confine our
attention to the case $k_1 = k_2 =2$, the minimal
Landau-level degeneracies allowed in this $\bb{Z}_2$ orbifold. 
Although the projected closed spectra of all the resulting models 
are identical, and comprise the ${\cal N}=(1,0)$ 
gravitational multiplet, together with one tensor multiplet and
20 hyper multiplets, the corresponding open spectra are quite different
from the standard one, whose gauge groups have a total rank 32.
Still, they are all free of irreducible gauge and gravitational anomalies,
consistently with the vanishing of all R-R tadpoles \cite{pc}.
The massless open spectra can be read from
\ba 
{\cal A}_0 & = & 
m\bar m \, Q_o (0) + n\bar n \, Q_o (0) + \frac{m^2 + \bar m ^2 }{2} \, Q_v (0)
\nonumber
\\
&& + \left( \frac{k_1 k_2}{2} + 2 \right) (m n + \bar m \bar n ) \,  Q_v
(\zeta \tau) \nonumber
\\
&& + \left( \frac{k_1 k_2}{2} - 2 \right) ( m \bar n + \bar m n )\, 
Q_v (\zeta \tau ) \nonumber
\\
&& + 2 \, (k_1 k_2 + 1 ) \, \frac{n^2 + \bar n ^2 }{2} \, Q_v (\zeta \tau)
\label{dannmzhnb}
\ea
and 
\be
{\cal M}_0 \sim - {\textstyle{1\over 2}} (m+\bar m) \, \hat Q_v (0) -
2 \, ( k_1 k_2 + 1) \, \frac{n+\bar n}{2} \, 
\hat Q_v (\zeta \tau) \, ,
\label{dmoebmzhnb}
\ee
obtained expanding the previous amplitudes for $H_1=H_2$, noting that
\ba
Q_o (0) &\sim V_4 - 2 C_4 \ \ ;
\qquad Q_v (0) &\sim 4 O_4 - 2 S_4 \ ; \nonumber
\\
Q_o (\zeta\tau ) &\sim {\rm massive}\ ;
\qquad
Q_v (\zeta\tau ) &\sim 2 O_4 - S_4 \ . \label{susyh1h2exp}
\ea

A possible solution of the R-R tadpole conditions is
$n=13$, $m=3$, $d=4$, that corresponds to a gauge group of rank 20,
${\rm U}(13)_9 \times {\rm U}(3)_9 \times {\rm U}(4)_5$, with
charged hyper multiplets in the representations $(78 + \overline{78},1,1)$,
in five copies of the $(1,3  + \overline{3},1)$, in one copy of the
$(1,1,6  + \overline{6})$, in four copies of the
$(\overline{13},3,1)$, in one copy of the $(13,1,\overline{4})$ and
in one copy of the  $(1,\overline{3},4)$. 
Alternatively, one can take $n=14$, $m=2$, $d=8$,
obtaining a gauge group of rank 24,
${\rm U}(14)_9 \times {\rm U}(2)_9 \times {\rm U}(8)_5$. The
corresponding matter comprises charged
hyper multiplets in the $(91 + \overline{91},1,1)$, 
in one copy of the $(1,1,28  + \overline{28})$, in four copies of the
$(\overline{14},2,1)$, in one copy of the $(14,1,\overline{8})$,
in one copy of the  $(1,2,8)$, and in five copies of the 
$(1,1 + \overline{1},1)$. On the other hand, the choice
$n=12$, $m=4$, and thus $d=0$, results in a rather unusual supersymmetric 
$\bb{Z}_2$ model {\it without} D5 branes, with a gauge group of rank 16, 
${\rm U}(12)_9 \times {\rm U}(4)_9$, and charged 
hyper multiplets in the representations $(66 + \overline{66},1)$,
in five copies of the $(1,6 + \overline{6})$, and in four copies of
the $(\overline{12},4)$. 
A distinctive feature of these
spectra is that some of the matter occurs in multiple families. 
This peculiar phenomenon is a consequence of the multiplicities of
Landau levels, that in these $\bb{Z}_2$ orbifolds are multiples of two for
each magnetized torus. It should be appreciated that in this class
of models the
rank reduction for the gauge group is not simply by powers of two as
with a quantized antisymmetric tensor \cite{bps,kab,cab}.
These results are summarized in table \ref{magbtabsusynob}.

\begin{table}
\caption{Some supersymmetric massless spectra for $H_1=H_2$ $(n_T=1,n_H=20)$.}
\label{magbtabsusynob}
\begin{indented}
\lineup
\item[]
\begin{tabular}{@{}cccccl} 
\br
$G_{{\rm CP}}$  & ${\rm rank}( G_{{\rm CP}})$ 
& charged hyper multiplets  
\\
\mr
${\rm U} (13)_9 \times {\rm U} (3)_9 \times {\rm U} (4)_5$ & 20 & 
$(78+ \overline{78} ,1,1)+ 5\,(1,3 + \overline{3},1)+(1,\overline{3},4)$
\\
& & $+ (1,1,6+ \overline{6})+ 4\, (\overline{13},3,1)+ (13,1,\overline{4})
$
\\
${\rm U} (14)_9 \times {\rm U} (2)_9 \times {\rm U} (8)_5$ & 24 & 
$(91+ \overline{91} ,1,1)+ 5\,(1,1 + \overline{1},1)+(1,\overline{2},8)$
\\
& & $+ (1,1,28+ \overline{28})+ 4\, (\overline{14},2,1)+ (14,1,\overline{8})
$
\\
${\rm U} (12)_9 \times {\rm U} (4)_9$ & 16 & 
$(66+ \overline{66},1)+ 5\,(1,6 + \overline{6})+ 4\, (\overline{12},4)$
\\
\br
\end{tabular}
\end{indented}
\end{table}

One can also consider similar deformations of the model of \cite{bsb},
reviewed in subsection 5.8,
that has an ${\cal N}=(1,0)$ supersymmetric bulk spectrum with 
17 tensor multiplets
and four hyper multiplets. As we have seen, this alternative projection, 
allowed by the
constraints in \cite{fps,pss,pss2}, introduces ${\rm O}9_{+}$
and ${\rm O}5_{-}$ planes and thus requires, for consistency, an open sector
resulting from the simultaneous presence of D9 branes and D5 antibranes,
with ``brane supersymmetry breaking''. A magnetized torus
can now mimic D5 antibranes provided $H_1 = - H_2$, and one can then build
several non-tachyonic configurations as in the previous case,
but there is a subtlety. The different GSO projections for
strings stretched between a D9 brane and a D5 antibrane associate
the low-lying twisted ND bosons to the characters $O_4 S_2(\zeta_1) 
S_2(\zeta_2)$
and $O_4 C_2(\zeta_1) C_2(\zeta_2)$, and thus now
the choice $H_1 = - H_2$, rather than the previous one $H_1 = H_2$, 
eliminates all tachyons. A simple and 
interesting choice corresponds to a vacuum configuration without D5 
antibranes, where the ${\rm O}5_{-}$ charge is fully saturated by the
magnetized D9 branes. In this blown-up version of the ``brane
supersymmetry breaking'' model of \cite{bsb},
the annulus and M\"obius amplitudes
can be obtained deforming the corresponding ones 
in subsection 5.8, and read
\begin{eqnarray}
{\cal A} &=& {\textstyle{1\over 4}} \Biggl\{ (Q_o + Q_v)(0;0) \left[
(n_1+n_2)^2 P_1 P_2 + 2 m \bar{m} \tilde P_1 \tilde P_2 \right]
\nonumber
\\
&-& 2 (n_1+ n_2) (m + \bar{m}) (Q_o + Q_v )(\zeta_1 \tau ; \zeta_2 \tau
) {k_1 \eta \over
\vartheta_1 (\zeta_1 \tau)} {k_2 \eta \over \vartheta_1 (\zeta_2 \tau)} 
\nonumber
\\
&-& ( m^2 + \bar{m}^2 ) (Q_o + Q_v ) (2 \zeta_1 \tau ; 2 \zeta_2 \tau ) 
{2 k_1 \eta \over
\vartheta_1 (2 \zeta_1 \tau)} {2 k_2 \eta \over \vartheta_1 (2 \zeta_2 \tau)} 
\nonumber 
\\
&+& \left[ (n_1-n_2)^2 + 2 m\bar m \right] (Q_o - Q_v ) (0;0)
\left( {2\eta \over \vartheta_2 (0)}\right)^2 
\nonumber
\\
&+& 2 (n_1-n_2) (m + \bar{m}) (Q_o - Q_v ) (\zeta_1 \tau ; \zeta_2 \tau) 
{2\eta \over \vartheta_2
(\zeta_1 \tau)} {2\eta \over \vartheta_2 (\zeta_2 \tau)} 
\nonumber
\\
&+& (m^2 + \bar{m}^2) (Q_o - Q_v ) (2\zeta_1 \tau ; 2\zeta_2 \tau)
{2\eta \over \vartheta_2
(2\zeta_1 \tau)} {2\eta \over \vartheta_2 (2\zeta_2 \tau)} \Biggr\} \, ,
\label{annsusyb}
\end{eqnarray}
and
\begin{eqnarray}
{\cal M} &=& -{\textstyle{1\over 4}} \Biggl\{ 
(n_1 + n_2) ( \hat Q_o + \hat Q_v )(0;0)   P_1 P_2 
\nonumber
\\
&-& ( m + \bar m ) (\hat Q_o + \hat Q_v ) (2\zeta_1 \tau ; 2\zeta_2 \tau) 
{2 k_1
\hat\eta \over \hat \vartheta_1 (2\zeta_1\tau)} {2 k_2
\hat\eta \over \hat \vartheta_1 (2\zeta_2\tau)}
\nonumber
\\
&+& \left( n_1+ n_2 \right) (\hat Q_o - \hat Q_v )(0;0) \left(
{2\hat\eta \over \hat \vartheta_2 (0)}\right)^2 \label{mobsusyb}
\\
&+& (m + \bar m ) (\hat Q_o - \hat Q_v ) (2 \zeta_1 \tau ; 2 \zeta_2 \tau )
{2\hat\eta \over \hat\vartheta_2 (2\zeta_1\tau)}
{2\hat\eta \over \hat\vartheta_2 (2\zeta_2\tau)} \Biggr\} \,.
\nonumber
\end{eqnarray}
In extracting the massless spectra of this class of models, 
it is important to notice that at the
special point $H_1 = - H_2$ all bosons from $Q_o$
with non-vanishing arguments and all fermions from
$Q_v$ with non-vanishing arguments become massive. 
As a result, the massless fermions arising from strings affected by the
internal magnetic fields have a reversed chirality, precisely as
demanded by the cancellation of all irreducible anomalies. 
For instance, with $|k_1 |= |k_2 | = 2$ one can obtain
a gauge group ${\rm SO} (8) \times {\rm SO} (16) \times {\rm U} (4)$
and, aside from the corresponding ${\cal N}= (1,0)$ vector multiplets,
the massless spectrum contains a hyper multiplet in the representation 
$(8,16,1)$, eight scalars in the $(1,16,4+\overline{4})$, two
left-handed spinors in the $(8,1,4+\overline{4})$, and twelve scalars 
and five left-handed spinors in the $(1,1,6+\overline{6})$. 
Supersymmetry is clearly broken on the magnetized D9 branes, but
the resulting dilaton potential scales with the internal volume 
as the $\overline{\rm D}5$ contribution in the undeformed model of \cite{bsb}.

It is also instructive to extend this construction, 
allowing for quantized values
of the NS-NS antisymmetric tensor $B_{ab}$, whose rank
will be denoted by $r$ \cite{magnetic2} as in the previous 
subsections.
As we have seen, the quantized $B_{ab}$ has a twofold
effect on the Klein-bottle amplitude: it induces a projection in 
the winding lattice, while reversing the $\Omega$ eigenvalues of 
some of the twisted contributions, as in (\ref{korbbab}).

Turning to the open sector, for the sake of brevity we shall again confine
our attention to models without D5 branes, since the other cases can be
easily reconstructed from these results. The quantized $B_{ab}$ has
again a twofold effect on ${\cal A}$: it modifies the momentum lattice and
it endows the contributions related to the Landau levels with additional
$r$-dependent multiplicities, so that
\begin{eqnarray}
{\cal A}^{(r)} &=& {\textstyle{1\over 4}} \Biggl\{ (Q_o + Q_v ) (0;0) \Biggl[
(n+\bar n)^2 2^{r-4} \sum_\epsilon P_{1,\epsilon} (B) P_{2,\epsilon} (B) 
\nonumber
\\
& & \qquad\qquad\qquad + 2 m \bar m
2^{r-4} \sum_\epsilon \tilde P _{1,\epsilon} (B) 
\tilde P _{2,\epsilon} (B) \Biggr]
\nonumber
\\
&& - 2 \cdot 2^r (m+\bar m) (n+\bar n ) (Q_o + Q_v) (\zeta_1 \tau ; 
\zeta_2 \tau)
{k_1 \eta \over \vartheta_1 (\zeta_1 \tau)}
{k_2 \eta \over \vartheta_1 (\zeta_2 \tau)} \nonumber
\\ 
&&- 2^r (m^2 + \bar m^2 ) (Q_o + Q_v) (2 \zeta_1 \tau ; 2 \zeta_2 \tau ) 
{2 k_1 \eta \over \vartheta_1 (2 \zeta_1 \tau)}
{2 k_2 \eta \over \vartheta_1 (2 \zeta_2 \tau)} \nonumber
\\
&&- \left[ (n- \bar n )^2 - 2 m \bar m \right] (Q_o - Q_v ) (0;0)
\left( {2 \eta \over \vartheta_2 (0)}\right)^2 \nonumber
\\
&& - 2 (m-\bar m)(n-\bar n) (Q_o - Q_v) (\zeta_1 \tau ; \zeta_2 \tau ) 
{2 \eta \over \vartheta_2 (\zeta_1 \tau)}
{2 \eta \over \vartheta_2 (\zeta_2 \tau)} \nonumber
\\
&& - (m^2 + \bar m^2) (Q_o - Q_v )(2 \zeta_1 \tau ; 2 \zeta_2 \tau)
{2 \eta \over \vartheta_2 (2\zeta_1 \tau)}
{2 \eta \over \vartheta_2 (2\zeta_2 \tau)} \Biggr\} \,. \label{dannhb}
\end{eqnarray}
Notice that the breaking terms are as in (\ref{annsusy}),
while the corresponding transverse channel tadpoles remain a
perfect square, as the first term in eq. (\ref{twistedrrmag}):
despite their D5 charge, the 
magnetized D9 branes are still spread over the internal torus.

The M\"obius amplitude can now 
be recovered, as usual, after a $P$ transformation,
from the transverse amplitudes $\tilde{\cal K}$ and $\tilde{\cal A}^{(r)}$, and
reads
\ba
{\cal M}^{(r)} &=& - {\textstyle{1\over 4}} \Biggl\{ 
(n+\bar n) (\hat Q_o + \hat Q_v ) (0;0) 2^{(r-4)/2} \sum_\epsilon
\gamma_\epsilon P_{1,\epsilon} (B) P_{2,\epsilon} (B) \nonumber
\\
&& - (n+ \bar n) (\hat Q_o - \hat Q_v) (0;0) \left( {2 \hat \eta \over
\hat \vartheta_2 (0)}\right)^2 \nonumber
\\
&& - 2^{r/2} (m+\bar m) (\hat Q _o + \hat Q_v ) (2 \zeta_1 \tau ; 2 \zeta_2
\tau ) {2 k_1 \hat \eta \over \hat \vartheta_1 (2 \zeta_1 \tau)}
{2 k_2 \hat \eta \over \hat \vartheta_1 (2 \zeta_2 \tau)} \nonumber
\\
&& - (m+\bar m) (\hat Q_o - \hat Q _v ) (2 \zeta_1 \tau ; 2 \zeta_2 \tau ) 
{2 \hat \eta \over \hat \vartheta_2 (2 \zeta_1 \tau)}
{2 \hat \eta \over \hat \vartheta_2 (2 \zeta_2 \tau)}
\Biggr\} \, ,
\label{dmoebhb}
\ea
where, as in subsection 5.7, the $\gamma$'s are signs, required by the
compatibility with the transverse channel, that determine the charge
of the resulting O-planes. The modified R-R tadpoles are then 
\ba 
& & m+\bar m + n + \bar n = 2^{5-r/2} \, , \nonumber
\\
& & k_1 k_2 (m+\bar m) = 2^{5-r} \, , \label{tadpoleshb}
\ea
so that 
the ranks of the gauge groups are reduced as usual, albeit 
here in an asymmetrical fashion. 

The massless spectrum clearly
depends on the signs $\gamma_\epsilon$ in  ${\cal M}$, or, 
equivalently, on the sign $\xi$ of eq. (\ref{directgamma}), that
determine the type of action, regular or projective, of the
orbifold on the gauge group or, equivalently, the
type, real or complex, of Chan-Paton multiplicities present.
 
The more standard choice $\xi = +1$ results in a
projective $\bb{Z}_2$ action, and therefore the
massless annulus and M\"obius amplitudes
\ba 
{\cal A}_0^{(r)} & \sim & {\textstyle{1\over 4}} \Bigl\{
4 m\bar m Q_o (0) + 4 n\bar n Q_o (0) + 2 (m^2 + \bar m ^2 ) Q_v (0)
\nonumber
\\
&& + (2\cdot 2^r \cdot k_1 k_2 + 2 \cdot 4 ) (m n + \bar m \bar n ) Q_v
(\zeta \tau) \nonumber
\\
&& + (2 \cdot 2^r \cdot k_1 k_2 - 2 \cdot 4) ( m \bar n + \bar m n )
Q_v (\zeta \tau ) \nonumber
\\
&& + (4\cdot 2^r \cdot k_1 k_2 + 4 ) (n^2 + \bar n ^2 ) Q_v (\zeta \tau)
\Bigr\} \, ,
\label{dannmzhb}
\ea
and 
\be
{\cal M}_0^{(r)} \sim - {\textstyle{1\over 2}} (m+\bar m) \hat Q_v (0) -
{\textstyle{1\over 2}} (2 \cdot 2^{r/2} \cdot k_1 k_2 + 2) (n+\bar n)
\hat Q_v (\zeta \tau) \, ,
\label{dmoebmzhb}
\ee
involve complex multiplicities $m$ and $n$.

Na\"{\i}vely, these amplitudes would seem  inconsistent since,
as a result of the further multiplicities
related to the rank $r$ of $B_{ab}$, {\it only some} of the string states 
with identical U(1) charges at their ends appear to
contribute to ${\cal M}$. This is actually not the case, and
the solution of this little puzzle follows a pattern that first 
emerged in the SU(2) WZW models \cite{pss,pss2}: ${\cal A}$ and
${\cal M}$ need only be equal modulo 2. In general, the proper group
assignments can be obtained associating to each quadratic 
multiplicity in ${\cal A}$ the sum of a symmetric representation $S$ and 
an antisymmetric representation $A$, and to each linear 
multiplicity in ${\cal M}$ their difference. For instance, 
an overall multiplicity $c_1 m^2 + c_2 m$ in ${\cal A}_0 + {\cal M}_0$
for a given character $\chi$ would lead to
\be
{\cal A}_0 + {\cal M}_0 = \left[ c_1 (S+A) + c_2 (S-A) \right] \chi \,,
\ee
and thus the corresponding spectrum would include $c_1+c_2$ copies of
the two-index symmetric representation and $c_1-c_2$ copies of the 
antisymmetric one.
In the case at hand, using the expansions in eq. (\ref{susyh1h2exp})
one can arrive at the massless spectra
\begin{eqnarray}
& & ( A + \overline{A},1) + {2 \cdot 2^r \cdot k_1 k_2 + 2 \cdot
4 \over 4} \, (m,n) + {2 \cdot 2^r \cdot k_1 k_2 - 2 \cdot
4 \over 4} \, (m,\bar n) \nonumber
\\
& & + \left[ k_1 k_2 \, (2^r + 2^{r/2}) + 2 \right] \, (1, A) +
k_1 k_2 \,(2^r - 2^{r/2}) \, (1, S) \, ,
\end{eqnarray}
with ${\rm U} (m) \times {\rm U} (n)$ gauge groups. 
Altogether, the tadpole conditions admit four inequivalent solutions, 
with proper multiplicities even for $k_{1,2}$ odd.
The corresponding spectra, aside from the universal ${\cal N} =
(1,0)$ gravity multiplet, are summarized in table \ref{magbtab}, where
$n^{cl}_{T}$ and $n^{cl}_{H}$ denote the numbers of tensor and hyper
multiplets from the projected closed sector.

\begin{table}
\caption{Some supersymmetric massless spectra with a rank-$r$ $B_{ab}$
($\xi =+1$).}
\label{magbtab}
\begin{indented}
\lineup
\item[]\begin{tabular}{@{}cccccl} 
\br
$r$ & $(k_1,k_2)$ & $n^{cl}_{T}$ & $n^{cl}_{H}$ & $G_{{\rm CP}}$  & 
charged hyper multiplets  
\\
\mr
2 & (1,1) &  5 & 16 & ${\rm U} (4) \times {\rm U} (4)$ & 
$(6+ \overline{6} ,1)+ 4\,(4,4)$
\\
 & & & & & $+ 8\,(1,6)+ 2\, (1,10)$
\\
2 & (1,2) & 5 & 16 & ${\rm U} (6) \times {\rm U} (2)$ & 
$(15+ \overline{15} ,1_0)+ 6\,(6,2_+)$
\\
 & & & & & $+ 2\, (6,2_-) + 14\, (1,1_{++})+ 4\, (1,3)$
\\
2 & (2,2) & 5 & 16 & ${\rm U} (7) \times {\rm U} (1)$ & 
$(21+ \overline{21} ,1_0)+ 10\,(7,1_+)$
\\
 & & & & & $+ 6\,(7,1_-)+ 8\, (1,1_{++})$
\\
4 & (1,1) & 7 & 14 & ${\rm U} (3) \times {\rm U} (1)$ & 
$(3+ \overline{3} ,1)+ 10\,(3,1_+)$
\\
 & & & & & $+ 6\,(3,1_-) + 12\, (1,1_{++})$
\\
\br
\end{tabular}
\end{indented}
\end{table}

On the other hand, as in the non-magnetized case  \cite{cab} 
of subsection 5.7, the choice $\xi =-1$ 
 induces a regular action of the $\bb{Z}_2$ orbifold on
the charges. The corresponding real multiplicities  
require a different embedding of the magnetic U(1)'s, so that
\begin{eqnarray}
n + m + \bar n + \bar m &\to& n_1 + m + \bar m + n_2 \, , \nonumber
\\
n + m - \bar n - \bar m &\to& n_1 + m + \bar m - n_2 \, ,
\end{eqnarray}
and the direct-channel massless contributions become
\begin{eqnarray}
{\cal A}_0^{(r)} & \sim & {\textstyle{1\over 2}} (n_1^2 + n_2^2 ) Q_o (0) + m \bar
m Q_o (0) + n_1 n_2 Q_v (0) \nonumber
\\
& & + \Bigl\{ {\textstyle{1\over 4}} \left[ 2\cdot 2^r \cdot k_1 k_2 -
2\cdot 4 \right] n_1 (m+\bar m) 
\nonumber
\\
& & + {\textstyle{1\over 4}} \left[ 2\cdot 2^r \cdot k_1 k_2 +
2\cdot 4 \right] n_2 (m+\bar m) \Bigr\} Q_v (\zeta\tau ) \nonumber
\\
& &+ {\textstyle{1\over 2}} \left[ 2 \cdot 2^r \cdot k_1 k_2 - 2\right]
(m^2 + \bar m^2 ) Q_v (\zeta \tau ) \, ,
\label{dannhbn}
\end{eqnarray}
and
\begin{equation} 
{\cal M}_0^{(r)} \sim {\textstyle{1\over 2}} \, (n_1 + n_2 ) \hat Q_o (0)
- {\textstyle{1\over 2}} \, \left[ 2\cdot 2^{r/2}\cdot k_1 k_2 + 2 \right] (m+\bar m) \hat Q_v
(\zeta\tau)   \,.
\label{rmoebhbn}
\end{equation}

For these models the untwisted tadpole conditions  
\begin{eqnarray}
& & n_1 + n_2 + m + \bar m = 2^{5-r/2} \, , \nonumber
\\
& & k_1 k_2 ( m + \bar m) = 2^{5-r} \, ,
\end{eqnarray}
are to be supplemented by the twisted one
\begin{equation}
n_1 + m + \bar m = n_2 \,,
\end{equation}
and, for instance, a possible solution with $r=2$ 
and $k_1 = k_2 =1$ is $n_1 =0$,
$n_2 = 8$ and $m=4$. This yields a massless spectrum with 
a gauge group ${\rm USp} (8) \times {\rm U} (4)$ comprising,
aside from the ${\cal N} =(1,0)$ gravity multiplet, 5 tensor
multiplets, 16 neutral hyper multiplets, and additional
charged hyper multiplets in the representations $4 (8,4) + 6 (1,6)$. 
As in conventional tori \cite{bps} and orbifolds \cite{cab}, 
a continuous Wilson line can actually connect these two classes of 
magnetized vacua.

\vskip 12pt
\subsection{Orientifolds and D-brane spectra}

All the preceding sections have been devoted to the general issue
of associating one or more classes of open descendants to a given
``parent'' closed string. As we have seen, from a space-time viewpoint 
the resulting vacua contain dynamical defects, the D-branes, and 
additional apparently non-dynamical ones, the O-planes. The exact D-brane
and O-plane content depends on the type of compactification,
so that, for instance, the ten-dimensional
type I string contains only D9 branes and O9 planes, while its $T^4/\bb{Z}_2$
reduction also involves D5 branes and O5 planes. Let us stress, once more,
that these orientifolds are to be regarded as
genuine vacuum configurations, where the O-planes somehow account for 
the back-reaction of space-time to the presence of the branes.
Similar methods, however, apply also to a different class
of problems, where D-brane probes inserted in a given background 
do not affect it sizably. This has the flavour of familiar situations 
in Classical Electrodynamics, where one is often interested in the
effect of external fields on small test particles. The result, of course, is no
more a vacuum configuration, but bears nonetheless
an important r\^ole both for the
non-perturbative aspects of String Theory, where the probe branes 
describe solitonic sectors or account for instanton-like
corrections, and for dual descriptions of 
their low-lying excitations \cite{adscft}. Following
Polchinski \cite{pol95}, we thus turn to describe the 
D-brane content of the ten-dimensional strings. 
D-brane charges can generally be 
associated to K-theory classes \cite{ktheory}, that also
give a rationale for their patterns, but this subject is not touched upon 
here, where the D-branes of the ten-dimensional models
are retrieved by direct constructions
adapting the orientifold techniques reviewed so far. All these 
results, first derived to a large extent by other authors 
using a variety of 
different methods \cite{dbranes,dm2}, can be recovered nicely and 
efficiently in this way \cite{dms}.

Let us begin with the simplest case, the BPS (charged) ${\rm
D}p$ branes of the
type IIB string. Their spectra can be simply deduced starting from the
transverse annulus amplitude for the bulk
modes propagating between two of them, 
\be
\tilde{\cal A}_{pp} = 2^{-(p+1)/2} d \bar{d} \ ( V_{p-1} O_{9-p} + 
 O_{p-1} V_{9-p} 
-  S_{p-1} S_{9-p} -  C_{p-1} C_{9-p} ) \, , 
\label{IIBbranestilde}
\ee
where we have decomposed the SO(8) characters with respect to the $(p-1)$
light-cone directions longitudinal to the branes and where the
reflection coefficients are squared absolute values of the
corresponding complex multiplicity $d$. An $S$ modular
transformation then gives
\be
{\cal A}_{pp} = d \bar{d} \ ( V_{p-1} O_{9-p} +  O_{p-1} V_{9-p} 
-  S_{p-1} S_{9-p} -  C_{p-1} C_{9-p} ) \, ,
\label{IIBbranes}
\ee
that encodes the full perturbative spectrum of brane excitations.
At the massless level, this comprises the maximal supersymmetric Yang-Mills
theory with a  U($d$) gauge group, the reduction of the ten-dimensional 
(1,0) model to the $(p+1)$-dimensional brane world-volume.
These D$p$ branes exist for odd $p$, have even-dimensional
world-volumes, and are charged with respect to the even-dimensional R-R $(p+1)$
forms of the type IIB theory. T-duality relates them to the BPS branes
of type IIA, that have odd-dimensional world volumes and couple to its
odd-dimensional R-R forms \cite{pol95}. 

The type I branes are more subtle, since they are defined in the
corresponding D9-O9 background, encoded in the familiar amplitudes
\ba
{\cal K} &=& \frac{1}{2} (V_8 - S_8) \, , \nonumber \\ 
{\cal A}_{99} &=& \frac{N^2}{2} (V_8 - S_8) \, , \nonumber \\ 
{\cal M}_{9} &=& - \frac{N}{2} (\hat{V}_8 - \hat{S}_8) \, , 
\label{i4}
\ea
where $N=32$ on account of tadpole cancellation. The
interaction between the probe branes and the background thus
requires that the ${\rm D}p$-${\rm D}p$ amplitude be accompanied 
by additional ones where
the bulk spectrum propagates between the probe and
the background defects. In the presence of O9 planes, 
the strings become unoriented, carry real Chan-Paton charges, and 
the annulus coefficients become {\it perfect squares}. 

Let us begin by considering the closed-channel amplitudes
\ba
\tilde{\cal A}_{pp} &=& \frac{2^{-(p+1)/2}\, d^2}{2} \ ( V_{p-1} O_{9-p} +
O_{p-1} V_{9-p} -  S_{p-1} S_{9-p} -  C_{p-1} C_{9-p} ) \, , 
\nonumber \\
\tilde{\cal A}_{p9} &=& 2^{-5}\, N d \ ( V_{p-1} O_{9-p} -
O_{p-1} V_{9-p} +  S_{p-1} S_{9-p} -  C_{p-1} C_{9-p} ) \, , 
\nonumber \\
\tilde{\cal M}_{p} &=& - d \ ( \hat{V}_{p-1} \hat{O}_{9-p} -
\hat{O}_{p-1} \hat{V}_{9-p} +  \hat{S}_{p-1} \hat{S}_{9-p} -  
\hat{C}_{p-1} \hat{C}_{9-p} ) \,, 
\label{i5}
\ea
that originate from the D$p$-D$p$, D$p$-D9 and D$p$-O9 exchanges.
Notice that $\tilde{\cal A}_{p9}$ and $\tilde{\cal M}_p$ involve 
relative signs
between the different contributions that break the SO(8) space-time
symmetry, with a crucial consequence for the probe spectrum.
An $S$ transformation gives the annulus amplitudes
\ba
{\cal A}_{pp} &=& \frac{d^2}{2} ( V_{p-1} O_{9-p} +
O_{p-1} V_{9-p} -  S_{p-1} S_{9-p} -  C_{p-1} C_{9-p} ) \, , \nonumber \\
{\cal A}_{p9} &=& N d \Bigl[ (O_{p-1} + V_{p-1})(S_{9-p} + C_{9-p})
- (S_{p-1} + C_{p-1})(O_{9-p} + V_{9-p})
\nonumber \\
& & + e^{-{(9-p)i\pi\over 4}} (O_{p-1} - V_{p-1})(S_{9-p} - C_{9-p})
\nonumber \\
& & - e^{-{(9-p)i\pi\over 4}}
(S_{p-1} - C_{p-1})(O_{9-p} - V_{9-p}) \Bigr]\,, 
\ea
while a $P$ transformation gives
\ba
{\cal M}_{p} &=& - \frac{d}{2} \Biggl[    
\sin \frac{(p-5)\pi}{4} \, ( \hat{O}_{p-1} \hat{O}_{9-p} + \hat{V}_{p-1} 
\hat{V}_{9-p}) 
\nonumber \\
& & + 
\cos \frac{(p-5)\pi}{4} \, ( \hat{O}_{p-1} \hat{V}_{9-p} - \hat{V}_{p-1} 
\hat{O}_{9-p}) 
\nonumber \\
&& - i \sin \frac{(p-5)\pi}{4} ( \hat{C}_{p-1} \hat{S}_{9-p} - 
\hat{S}_{p-1} \hat{C}_{9-p}) 
\nonumber \\
& & - \cos \frac{(p-5)\pi}{4} ( \hat{S}_{p-1} \hat{S}_{9-p} - \hat{C}_{p-1}
\hat{C}_{9-p}) \Biggr]
\,. \label{i6}
\ea 
Notice that ${\cal A}_{p9}$ and ${\cal M}_p$ 
are clearly inconsistent unless $p=1,5,9$, so that
the D9, D5 and D1 branes are the only allowed BPS ones in the SO(32) 
type I string. Moreover, since in these three cases the 
left-over cosines are equal to $\pm 1$, stacks of these D9 and D1
branes yield SO groups, while stacks of D5 branes yield USp 
groups \cite{pol95,witsmall}.

Aside from these BPS branes, the type IIB, type IIA and type I strings
contain additional uncharged non-BPS ones that, as described 
by Sen \cite{senba}, 
can be generated subjecting brane-antibrane pairs,
described in type IIB by
\ba
{\cal A}_{pp} &=& (m \bar{m} + n \bar{n}) ( V_{p-1} O_{9-p} +  O_{p-1}
V_{9-p} - S_{p-1} S_{9-p} - C_{p-1} C_{9-p} ) \label{i7} \\ 
&&+ (m \bar{n} + n \bar{m}) 
(O_{p-1} O_{9-p} +  V_{p-1} V_{9-p}- S_{p-1} C_{9-p} -  C_{p-1} S_{9-p}) 
\, , \nonumber
\ea
to an orbifold operation interchanging them. In the bulk type IIB
theory this corresponds to the action of the left space-time fermion
parity, that effectively flips the left R-R charges, 
turning the original type IIB into type IIA. Hence, one is finally
relating non-BPS branes in type IIA, with even-dimensional
world-volumes, to brane-antibrane pairs in type IIB. Their excitations
can then be simply read from (\ref{i7}), after identifying
$n$ and $m$ with a single complex charge multiplicity $N$, while rescaling 
the amplitudes by an overall factor $\frac{1}{2}$, so that
\ba
{\cal A}_{pp} &=& N \bar N \,  
\left[ ( O_{p-1}+ V_{p-1})( O_{9-p}+ V_{9-p}) \right.
\nonumber \\
& & \left. - ( S_{p-1}+ C_{p-1})( S_{9-p}+ C_{9-p}) \right] \,. 
\label{i8}
\ea
The low-lying excitations of non-BPS $p$-branes thus comprise a vector boson, 
$9-p$ massless scalars, a tachyon and non-chiral fermions, 
all in the adjoint of a unitary gauge group.
Notice that the absence of any GSO projection in the open-string
spectrum implies that these non-BPS branes do not carry any R-R
charge, while their tension is $\sqrt{2}$ times larger than that of
the BPS ones, as can be seen from the corresponding 
transverse-channel amplitude
\be
\tilde{\cal A}_{pp} = 2 \times 2^{-(p+1)/2} N \bar N \left( V_{p-1}
O_{9-p} + O_{p-1} V_{9-p} \right) \,.
\ee

Following the same procedure, one can then easily study systems of different 
branes. For instance, strings stretching between 
$n$ ${\rm D}p$ and $d$ ${\rm D}q$ non-BPS branes, where
$p-q=0 \ {\rm mod} \ 2$ and, for definiteness, $p > q$, have
$q+1$ NN coordinates, $9-p$ DD
coordinates and $p-q$ ND coordinates. The
corresponding annulus amplitudes read
\ba
{\cal A}_{pq} &=& (n {\bar d}+ {\bar n} d) \left[
(O_{q-1} O _{9-p} + V_{q-1} V_{9-p} \right. 
\\
& & + V_{q-1} O_{9-p} + O_{q-1} V_{9-p}) 
(S_{p-q} + C_{p-q}) 
\nonumber \\
& & - \left. (S_{q-1} S_{9-p} + C_{q-1} C_{9-p} + C_{q-1} S_{9-p} 
+ S_{q-1} C_{9-p})  (O_{p-q} + V_{p-q}) \right] \, , 
\nonumber
\ea
and, aside from non-chiral space-time 
massless fermions, the massless spectra contain tachyons for $|p-q| < 4$,
massless scalars for $|p-q|=4$ and only massive bosons for $|p-q| > 4$,
all in bi-fundamentals of ${\rm U}(n) \times {\rm U}(d)$. 
One can similarly write the ${\rm D}p$-${\rm D}q$ amplitude between a 
BPS and a non-BPS brane ($p-q=1 \ {\rm mod} \ 2$),
\ba
{\cal A}_{pq} &=& (n {\bar d}+ {\bar n} d) \ \left[
(O_{q-1} O_{9-p} + V_{q-1} V_{9-p} + V_{q-1} O_{9-p} + O_{q-1}
V_{9-p}) S'_{p-q} \right.  
\nonumber \\
& & \left. - S'_{q-1} (S_{9-p} + C_{9-p}) (O_{p-q} + V_{p-q}) \right] 
\,. \label{i09}  
\ea  
Notice the appearance in ${\cal A}_{pq}$, due to the odd number of 
ND coordinates, of the non-chiral fermion characters of ${\rm SO}(2\ell+1)$
\be
S'_{2\ell + 1}= {1\over \sqrt{2}} \left( {\vartheta_2 \over
\eta}\right)^{\ell +\frac{1}{2}} \,,
\ee
properly normalized in order to give the ground state its 
$2^{\ell}$-fold degeneracy. In the transverse-channel amplitudes
\ba 
\tilde{\cal A}_{pq} &=& \sqrt{2} \times 2^{-(p+1)/2} (n {\bar d}+ {\bar
n} d) \left[ (V_{q-1} O_{9-p} + O_{q-1} V_{9-p} ) O_{p-q} \right.
\nonumber \\
& & \left. - (O_{q-1} O_{9-p} + V_{q-1} V_{9-p}) V_{p-q} \right]
\ea
they are responsible for the $\sqrt{2}$ factor, that indeed
reflects the geometric average of BPS and non-BPS brane 
tensions.

We can now turn to the non-BPS branes of type I that, as we have
already stressed, are immersed in the proper D9 and O9 background.
Stacks of $d$ non-BPS D$p$ branes for $even$ $p$ can be discussed 
applying the orientifold projection to the corresponding non-BPS
branes of the parent type IIB. Since they are uncharged with respect to 
the R-R fields, the $\Omega$ projection acts diagonally on their 
Chan-Paton factors, and therefore one expects orthogonal or 
symplectic gauge groups. The corresponding D$p$-D$p$ annulus
amplitudes are thus
\ba
{\tilde {\cal A}}_{pp} &=& 2^{-(p+1)/2} d^2 \,  (V_{p-1} O_{9-p} +  O_{p-1}
V_{9-p}) \ ,  \label{I1} \\
{\cal {\cal A}}_{pp} &=& {d^2 \over 2} \left[ 
(O_{p-1}+V_{p-1})(O_{9-p}+V_{9-p}) - 2  S'_{p-1} S'_{9-p} \right]
 \ , \nonumber  
\ea
and involve the non-chiral fermion characters $S'$, as pertains to 
odd-dimensional world-volumes.

The D$p$-O9 exchanges are encoded in the M\"obius amplitudes
\ba
{\tilde {\cal M}}_{p} &=& - \sqrt{2}\, d  
({\hat V}_{p-1}{\hat O}_{9-p}- {\hat O}_{p-1}{\hat V}_{9-p}) \,, 
\label{I2} 
\\
{\cal M}_{p} &=&  - {d \over \sqrt{2}} \left[ \sin{(p-5) \pi \over 4}
({\hat O}_{p-1}{\hat O}_{9-p} \right.
+ {\hat V}_{p-1}{\hat V}_{9-p})
\nonumber \\
& & +
\left. \cos{(p-5) \pi \over 4}
({\hat O}_{p-1}{\hat V}_{9-p}\!-\! {\hat V}_{p-1}{\hat O}_{9-p})  
\right] \, , 
\nonumber
\ea   
whose precise normalizations are unambiguously determined by the 
non-BPS tension in (\ref{I1}) and by the BPS O9 tension. Therefore, 
the annulus and M\"obius amplitudes have indeed a correct 
particle interpretation only for
even $p$. Moreover the fermions, absent in ${\cal M}$, enter the
annulus amplitude with a crucial multiplicity 2. This phenomenon is similar
to the one already met in the description of compactifications
on magnetized tori, and actually reflects a general property of
Boundary Conformal Field Theory \cite{fps,pss,pss2}: 
${\cal A}$ and ${\cal M}$ need only 
match modulo 2 for terms quadratic in a given type of multiplicity.
Whenever they do not coincide, the spectrum contains at the same time  
symmetric and antisymmetric representations, that
can be determined as in subsection 5.11. 
Finally, the D$p$-D9 spectrum can
be easily extracted from the annulus amplitudes
\ba
{\cal A}_{p9} = N d \left[ (O_{p-1}+V_{p-1}) S'_{9-p} - 
S'_{p-1} (O_{9-p}+ V_{9-p}) \right] \, ,  \label{I3}
\ea
where $N$, equal to 32, accounts for the background D9 branes.
These expressions summarize the open spectra for the various 
non-BPS D$p$ branes with even $p$ of the type I string, that are
as follows.

\begin{itemize}
\item[] {\it D0 brane:} ${\rm SO}(d)$ Chan-Paton group, tachyons in the 
adjoint, scalars in the symmetric
representation and fermions in the symmetric and antisymmetric
representations. The massless D0-D9 spectrum contains only 
fermions in the $(32,d)$ of ${\rm SO}(32) \times {\rm SO}(d)$. The tachyon
is absent if $d$=1, and therefore a single D particle is
stable, as correctly pointed out in \cite{senba}.
\item[] {\it D2 brane:} ${\rm SO}(d)$ gauge group, tachyons and
scalars in the symmetric representation and fermions in the symmetric 
and antisymmetric representations. The massless D2-D9 spectrum contains only
fermions in the $(32,d)$ of ${\rm SO}(32) \times {\rm SO}(d)$. 
The tachyon cannot be eliminated, and therefore the D2 brane is unstable.
\item[] {\it D4 brane:} ${\rm USp}(d)$ gauge group, tachyons in 
the adjoint representation, scalars in the antisymmetric
representation and fermions in the symmetric and antisymmetric
representations. The massless D4-D9 spectrum contains only 
fermions in the $(32,d)$ of ${\rm SO}(32) \times {\rm USp}(d)$. 
The tachyon cannot be eliminated, and therefore the D4 brane is unstable.
\item[] {\it D6 brane:} ${\rm USp}(d)$ gauge group, tachyon 
and scalars in the antisymmetric repre\-sentation and fermions in 
the symmetric and antisymmetric representations. The D6-D9 spectrum 
contains tachyons and massless fermions in the $(32,d)$ of 
${\rm SO}(32) \times {\rm USp}(d)$, and therefore the D6 brane is unstable.
\item[] {\it D8 brane:} similar to the D0-D0
spectrum above, reduces to it upon dimensional reduction of all spatial 
coordinates. The D8-D9 spectrum contains 
tachyons and massless fermions in the $(32,d)$ of ${\rm SO}(32) \times 
{\rm SO}(d)$, and therefore the D8 brane is unstable.
\end{itemize} 

Type I strings have additional non-BPS D$(-1)$, D3 and D7 branes 
that, however,
have a more peculiar structure, since for these dimensions
$\Omega$ interchanges branes and antibranes in type IIB. 
As a result, stacks of these additional branes 
have unitary gauge groups, while the corresponding annulus amplitudes are
\ba
{\tilde {\cal A}}_{pp} &=& {2^{-(p+1)/2} \over 2} \left[ (d+{\bar d})^2 
(V_{p-1} O_{9-p} +  O_{p-1} V_{9-p}) \right.
\nonumber \\
& & \left. +  (d-{\bar d})^2 
(S_{p-1} S_{9-p}+C_{p-1} C_{9-p}) \right] \, ,
\nonumber \\
{\cal A}_{pp} &=& d {\bar d} \ (O_{p-1} V_{9-p} +V_{p-1} O_{9-p}-
S_{p-1} S_{9-p}- C_{p-1} C_{9-p})  
\nonumber \\
&& + \, {d^2+ {\bar d}^2 \over 2} (O_{p-1} O_{9-p} +V_{p-1} V_{9-p}
- S_{p-1} C_{9-p}- C_{p-1} S_{9-p}) \,. \label{I4}  
\ea
Notice that the R-R coupling in the closed channel
vanishes when conjugate multiplicities are
identified, in agreement with the fact that these 
non-BPS branes are uncharged. As usual, the corresponding closed-channel
M\"obius amplitudes
\be
{\tilde {\cal M}}_{p} = (d+{\bar d})
({\hat O}_{p-1} {\hat V}_{9-p} -{\hat V}_{p-1} {\hat O_{9-p}})
-  (d- {\bar d})
({\hat S}_{p-1} {\hat S}_{9-p} -{\hat C}_{p-1} {\hat C_{9-p}})  
\ee
can be obtained as ``geometric means'' of the probe D$p$-D$p$  and 
background O9-O9 amplitudes, while the corresponding M\"obius projections
\ba
{\cal M}_{p} &=& - \, {d + {\bar d} \over 2} \, \sin{(p-5) \pi \over 4}\,
({\hat O}_{p-1} {\hat O}_{9-p} +{\hat V}_{p-1} {\hat V_{9-p}}) \nonumber \\
&& - \, {d - {\bar d} \over 2} \, e^{i (p-5) \pi \over 4}\,
(-i) \, \sin{(p-5) \pi \over 4} \, ({\hat S}_{p-1} {\hat C}_{9-p}
- {\hat C}_{p-1} {\hat S_{9-p}})  
 \ , \label{I5}  
\ea 
follow after a $P$ transformation.
We have thus found, as anticipated, ${\rm U}(d)$ gauge groups, with $9-p$
scalars and fermions in the adjoint representation, the latter obtained
dimensionally reducing a ten-dimensional Majorana-Weyl fermion to 
the D$p$ brane world-volumes. For the D3 (D7) brane there are also complex 
tachyons in (anti)symmetric representations,
Weyl fermions of positive chirality in the symmetric representation and
Weyl  fermions of negative chirality in the antisymmetric 
representation of the gauge group. Finally, the low-lying D$p$-D9 spectra, 
encoded in 
\ba
{\cal A}_{p9} &=& d N  (O_{p-1} S_{9-p} +V_{p-1} C_{9-p}-
C_{p-1} O_{9-p}- S_{p-1} V_{9-p})   \nonumber \\
&& +  {\bar d} N (O_{p-1} C_{9-p} +V_{p-1} S_{9-p}
- S_{p-1} O_{9-p}- C_{p-1} V_{9-p}) 
 \, , \label{I6}  
\ea
where $N$, equal to 32, is the D9 Chan-Paton multiplicity,
comprise in both cases massless Weyl fermions in 
the $(32,d)$ of ${\rm SO}(32) \times {\rm U}(d)$, and for the D7
branes also complex tachyons in the $(32,d)$
representation. These chiral spectra embody
non-trivial cancellations of irreducible gauge anomalies between
the D$p$-D$p$ and the D$p$-D9 sectors, discussed in more detail in
\cite{dms}. 

The non-BPS branes of the USp(32) string can be easily 
obtained from these interchanging symmetric and 
antisymmetric representations, while also flipping space-time and 
internal chiralities in the D$p$-D9 sector. Notice that a single 
${\rm D}(-1)$ 
brane in the SO(32) string and a single D3 brane in the USp(32)
string are {\it stable}, being free of tachyonic excitations 
\cite{lerda,sugimoto}. More details can be found in \cite{dms}.

One can similarly study the branes of the 0A and 0B
models. The doubling of the R-R sector implies that for odd $p$ 
the 0B theory has two types of stable charged D$p$ branes and the
corresponding antibranes, while the 0A theory has two types of unstable
uncharged ones. Since these two theories are related by odd numbers
of T-dualities, this result also implies that for even $p$ the 0B theory has
two types of uncharged unstable branes, while the 0A has two types of
charged stable ones. Their orientifolds, reviewed in section 3, 
are more interesting in
this respect, since in some cases the corresponding projections remove
all tachyons. This happens both for stacks of charged branes
and for some individual uncharged ones. Their main properties are 
summarized in tables 7-10, while more details can be found in
\cite{dms}. 
Notice that here we always refer to the minimal orientifold
background, that only for the $0^\prime{\rm B}$ model includes an open 
sector, and we consider only D$p$ branes with $p<9$, since the maximal
branes were already described in section 3.
The ``parent'' 0A and 0B branes can be simply recovered from the cases
with all real charges associating to all characters allowed in 
$\tilde{\cal A}$ corresponding complex multiplicities.

\begin{table}
\caption{Branes of the ${\rm 0A}/\Omega$ orientifold.}
\label{0abranes}
\begin{indented}
\lineup
\item[]\begin{tabular}{@{}lccc} 
\br
brane & Chan-Paton group & charged & stable  
\\
\mr
${\rm D}8_1$ & ${\rm SO} (m)$ & yes & yes
\\
${\rm D}8_2$ & ${\rm U} (m)$ & no & for $m=1$
\\
D7 & ${\rm SO} (m) \times {\rm USp} (n)$ & no & for $m=1$, $n=0$
\\
${\rm D}6_1$  & ${\rm USp} (m)$ & yes & yes
\\
${\rm D}6_2$  & ${\rm U} (m)$ & no & no
\\
D5 & ${\rm USp} (m) \times {\rm USp} (n)$ & no & no
\\
${\rm D}4_1$ & ${\rm USp} (m)$ & yes & yes
\\
${\rm D}4_2$ & ${\rm U} (m)$ & no & no
\\
D3 & ${\rm SO} (m) \times {\rm USp} (n)$ & no & no
\\
${\rm D}2_1$ & ${\rm SO} (m)$ & yes & yes
\\
${\rm D}2_2$ & ${\rm U} (m)$ & no & for $m=1$
\\
D1 & ${\rm SO} (m) \times {\rm SO} (n)$ & no & for $m=1$, $n=0$
\\
${\rm D}0_1$ & ${\rm SO} (m)$ & yes & yes
\\
${\rm D}0_2$ & ${\rm U} (m)$ & no & for $m=1$
\\
${\rm D}(-1)$ & ${\rm SO} (m) \times {\rm USp} (n)$ & no & for $m=1$, $n=0$
\\
\br
\end{tabular}
\end{indented}
\end{table}

\begin{table}
\caption{Branes of the ${\rm 0B}/\Omega$ orientifold.}
\label{0b1branes}
\begin{indented}
\lineup
\item[]\begin{tabular}{@{}lccc} 
\br
brane & Chan-Paton group & charged & stable  
\\
\mr
D8 & ${\rm USp} (m) \times {\rm USp} (n)$ & no & no
\\
D7 & ${\rm U} (m) \times {\rm U} (n)$ & no & for $m=1$, $n=1$
\\
D6  & ${\rm SO} (m) \times {\rm SO} (n)$ & no & for $m=1$, $n=1$
\\
D5 & ${\rm USp} (m) \times {\rm USp} (n)$ & yes & yes
\\
D4  & ${\rm SO} (m) \times {\rm SO} (n)$ & no & no
\\
D3 & ${\rm U} (m)\times {\rm U} (n)$ & no & no
\\
D2 & ${\rm USp} (m) \times {\rm USp} (n)$ & no & no
\\
D1 & ${\rm SO} (m) \times {\rm SO} (n)$ & yes & yes
\\
D0 & ${\rm USp} (m) \times {\rm USp} (n)$ & no & no
\\
${\rm D}(-1)$ & ${\rm U} (m)\times {\rm U} (n)$ & no & for $m=1$, $n=1$
\\
\br
\end{tabular}
\end{indented}
\end{table}

\begin{table}
\caption{Branes of the ${\rm 0B}/\Omega_2$ orientifold.}
\label{0b2branes}
\begin{indented}
\lineup
\item[]\begin{tabular}{@{}lccc} 
\br
brane & Chan-Paton group & charged & stable  
\\
\mr
D8 & ${\rm SO} (m) \times {\rm USp} (n)$ & no & no
\\
D7 & ${\rm SO} (m) \times {\rm USp} (n)$ & yes & yes
\\
D6  & ${\rm SO} (m) \times {\rm USp} (n)$ & no & no
\\
D5 & ${\rm U} (m) \times {\rm U} (n)$ & no & for $m=1$, $n=0$
\\
D4  & ${\rm SO} (m) \times {\rm USp} (n)$ & no & no
\\
D3 & ${\rm SO} (m) \times {\rm USp} (n)$ & yes & yes
\\
D2 & ${\rm SO} (m) \times {\rm USp} (n)$ & no & no
\\
D1 & ${\rm U} (m) \times {\rm U} (n)$ & no & for $m=1$, $n=0$
\\
D0 & ${\rm SO} (m) \times {\rm USp} (n)$ & no & no
\\
${\rm D}(-1)$ & ${\rm SO} (m) \times {\rm USp} (n)$ & yes & yes
\\
\br
\end{tabular}
\end{indented}
\end{table}

\begin{table}
\caption{Branes of the ${\rm 0B}/\Omega_3$ orientifold.}
\label{0b3branes}
\begin{indented}
\lineup
\item[]\begin{tabular}{@{}lccc} 
\br
brane & Chan-Paton group & charged & stable  
\\
\mr
D8 & ${\rm U} (m)$ & no & no
\\
D7 & ${\rm U} (m)$ & yes & no
\\
D6  & ${\rm U} (m)$ & no & no
\\
D5 & ${\rm U} (m)$ & yes & yes
\\
D4  & ${\rm U} (m)$ & no & no
\\
D3 & ${\rm U} (m)$ & yes & yes
\\
D2 & ${\rm U} (m)$ & no & no
\\
D1 & ${\rm U} (m)$ & yes & yes
\\
D0 & ${\rm U} (m)$ & no & no
\\
${\rm D}(-1)$ & ${\rm U} (m)$ & yes & yes
\\
\br
\end{tabular}
\end{indented}
\end{table} 
 
\newsec{Boundary conformal field theory, orientifolds and branes}

Conformal Field Theory \cite{bpz,cftrev} lies at the heart 
of the string world-sheet
and of its space-time manifestations \cite{fms}, since
conformal invariance provides the vertex operators for the string modes
and determines the space-time dynamics of the
string excitations \cite{gsw}. Conformal invariance is generally
violated by quantum effects, that in the Virasoro algebra for the
Laurent modes of the energy-momentum tensor 
\vskip 8pt
\be
[ L_m , L_n ] = (m-n) L_{m+n} + \frac{c}{12}\, m(m^2-1)\,\delta_{m+n,0} 
\label{bcft1}
\ee
manifest themselves via the emergence of the central extension.
However, the very consistency of
String Theory demands that it be {\it exact}, since it provides
a measure for the world-sheet moduli, that play the r\^ole
of Schwinger parameters in the string amplitudes \cite{friedlh}. 
In the critical case, 
as we have seen, the central charge $c$ of the light-cone modes is 
fixed to 24 and 12 for bosonic and fermionic strings, corresponding to 
their critical dimensions, 26 and 10, compatible in both cases with
Minkowski backgrounds. Equivalently, in a covariant
formulation \cite{gsw} the total central charge of ghosts and coordinates 
would vanish. On the other hand, away from criticality, 
the Liouville field \cite{pol,liouville} complicates the string 
dynamics, and the resulting models are incompatible with a Minkowski
background already at the sphere level \cite{cave}. In this
review we have thus followed the common trend of restricting the attention
to critical models, but the generic features of Boundary Conformal Field 
Theory that we are about to review are also of interest for the
off-critical case.

In two dimensions, conformal invariance is an infinite symmetry 
\cite{bpz,cftrev}, that
as such can unify infinitely many fields into a single conformal
family. Each family is identified by the corresponding {\it primary field}
$\varphi_{i\bar{\imath}}(z,\bar{z})$, characterized by a pair of
conformal weights $(h_i,\bar{h}_{\bar\imath})$,
and the spectrum of a bulk conformal theory is encoded in  
torus partition functions of the type
\be
{\cal T} = \sum_{i,j} \, \bar{\chi}_i \, X_{ij} \, \chi_j \,,
\label{bcft2}
\ee
a structure that we have met repeatedly in the previous sections.
These partition functions
involve in general an infinite number of families, ordered
by the conformal symmetry or by some extension of it, and the
theory is said ``rational'' when this number is finite \cite{verlinde}. 
In the following we shall restrict our attention to this
case, where $X_{ij}$ is a finite-dimensional matrix of non-negative 
integers, subject to the constraints of modular invariance
\be
S^\dag  \, X  \, S = X \,, \qquad T^\dag  \, X  \, T = X \,.
\label{bcft3}
\ee
Further, we shall implicitly assume that (\ref{bcft2}) defines
a permutation invariant, so that $X_{ij} = \delta_{i,\sigma (j)}$,
where $\sigma$ denotes a permutation of the labels, so that $T$
is unitary and diagonal while $S$ is unitary and symmetric. In this
case the family associated to a field $ \varphi_{i\bar{\imath}}$ 
is completely characterized, say, by its holomorphic label, and can 
well be denoted by $[ \varphi_i ]$. The interactions between
pairs of conformal families, determined by the operator
product coefficients $C_{i\bar{\imath}j\bar{\jmath}}{}^{k\bar{k}}$,
are generally subject to super-selection rules, neatly
encoded in the fusion algebra
\be
[\varphi_i] \times [\varphi_j ] = \sum_k \, {\cal N}_{ij}{}^k \, 
[\varphi_k ] \,.
\label{bcft4}
\ee
The fusion-rule coefficients ${\cal N}_{ij}{}^k$ are non-negative 
integers that count how many times states in the $k$-th
family occur in the fusion of the $i$-th and 
$j$-th ones, and can be retrieved from the $S$ matrix, whose entries
are generally complex numbers, via the Verlinde formula \cite{verlinde}
\be
{\cal N}_{ij}{}^k = \sum_l \, \frac{S_i^l \, S_j^l \, {S^\dag}{}_l^k}{S_1^l} \,. 
\label{bcft5}
\ee
Alternatively, the ${\cal N}_{ij}{}^k$ can be regarded as entries of the
set of matrices $({\cal N}_{i})_j{}^k$ so that, defining the
diagonal matrices
\be
(\lambda_i)_l^m = \frac{S_i^m}{S_1^m}\, \delta_l^m  \,,
\label{bcft6}
\ee
eq. (\ref{bcft5}) can be written in the more compact form
\be
{\cal N}_i = S \, \lambda_i \, S^\dag \,.
\label{bcft7}
\ee
Therefore, the non-negative integer matrices ${\cal N}_i$ are
mutually commuting, since they are obtained from the diagonal $\lambda_i$
by a common unitary transformation, and satisfy the fusion algebra
\be
{\cal N}_i\, {\cal N}_j = \sum_k\, {\cal N}_{ij}{}^k \, {\cal N}_k \,.
\label{bcft8}
\ee

One can actually define an additional set of matrices, ${\cal Y}_i$, 
built from the $P$ matrix that, as we have seen, plays a
key r\^ole in the M\"obius amplitude, as \cite{pss,pss2}
\be
{\cal Y}_i = P \, \lambda_i \, P^\dag \,,
\label{bcft9}
\ee
that are also mutually commuting and satisfy the fusion algebra
\be
{\cal Y}_i\, {\cal Y}_j = \sum_k\, {\cal N}_{ij}{}^k \, {\cal Y}_k \,.
\label{bcft10}
\ee
Differently from the ${\cal N}_i$, however, the elements of these new
matrices are in general {\it signed} integers~\cite{bantgan}.

The proper description of a bulk conformal field theory, as needed
for models of oriented closed strings, rests on a number
of polynomial constraints on the
structure constants \cite{bpz,mseib}. These equations, however, 
contain additional
ingredients, the fusion and braiding matrices, model
dependent and known only in very special cases. 
Thus, while in
principle they could determine completely the data of a conformal
model, in practice their use is rather limited and, in the spirit
of this review, we shall content ourselves with some comments on
the enumeration of complete sets of operators.
For bulk conformal theories, these constraints are related to the
sphere and the torus, as described in \cite{bpz,cardy1,sonoda}.
The consistency of conformally invariant spectra is essentially 
guaranteed by the
modular invariance of the torus amplitude \cite{cardy1}, and classifying 
modular invariants is a far simpler task than studying generic scattering 
amplitudes, to which in any case it is preliminary.

The inclusion of boundaries and crosscaps adds new data to a
conformal theory, and above all new types of fields, that live
on boundaries and, in the string picture, describe open-string
vertices. Furthermore, boundary
conditions must be enforced on bulk fields, and
therefore each boundary carries a label that accounts for the
multiplicity of these choices, while a boundary operator 
$\psi_i^{ab}$ generally mediates between them, and thus carries a
pair $a$ and $b$ of such labels. New boundary data, such as the 
one-point functions $B_i^a$ for bulk fields in front of boundaries,
appear, and new non-linear constraints relate them to the operator-product
coefficients \cite{cardy2,lew}. 
Actually, the constraints in \cite{pss,pss2} can 
be turned into those in \cite{lew}, and are thus 
equivalent to them, as shown in \cite{bppz},
but the resulting equations depend once more on the fusion
and braiding matrices, and are of use only in very special
cases. In string language, the multiplicity of boundaries translates into
the multiplicity of types of Chan-Paton labels, and this poses
the enumerative problem of classifying conformally invariant boundary 
conditions. This is also of crucial importance for String Theory, where
as we have seen it
is equivalent to identifying all possible types of D-branes.
Moreover, non-orientable projections 
bring about new data, the crosscap reflection coefficients $\Gamma_i$,
and a new {\it linear} constraint related to the
Klein-bottle and M\"obius amplitudes, the crosscap constraint of 
\cite{fps,pss,pss2}. Even this constraint can be solved only
in special cases, but it can also be used to obtain 
the structure constants rather efficiently \cite{pss,pss2},
since it is a set of linear relations between them,
while the $\Gamma_i$'s can also be recovered directly from the
Klein-bottle, annulus and M\"obius amplitudes.

If one restricts again the attention to the problem of enumerating
bulk and boundary operators, in the non-orientable case a convenient
algebraic setting emerges, that parallels the construction of orientifolds
reviewed in the previous sections, and in rational models provides a
precise algorithm to determine them. We shall refer to this
algorithm  as the method
of open descendants. Aside from the torus amplitude, this involves
in general the Klein bottle, annulus and M\"obius amplitudes, and we
shall now confine our attention momentarily to this non-orientable case,
in the spirit of what we did in most of the review. 

For a general rational conformal theory, the direct channel
amplitudes can be written in the form
\ba
{\cal K} &=& {\textstyle \frac{1}{2}} \sum_i \, {\cal K}^i \, \chi_i \,, \\
{\cal A} &=& {\textstyle \frac{1}{2}} \sum_{i,a,b} \, {\cal A}^i{}_{ab} 
\, n^a\, n^b\,
\chi_i \,, \\
{\cal M} &=& {\textstyle \frac{1}{2}}\sum_{i,a} \, {\cal M}^i{}_{a} \, n^a\,
\hat{\chi}_i \,,
\label{bcft11}
\ea
in terms of the integer-valued coefficients 
${\cal K}^i$, ${\cal A}_{ab}^i$ and ${\cal M}_{a}^i$.
For brevity, we shall also express the corresponding transverse-channel 
amplitudes in
terms of the boundary $B_i^a$ and crosscap $\Gamma^i$ reflection
coefficients as
\ba
\tilde{\cal K} &=& {\textstyle \frac{1}{2}} \sum_i  ({\Gamma}^i)^2 \,
\chi_i \,, \\
\tilde{\cal A} &=& {\textstyle \frac{1}{2}} \sum_{i} \, \chi_i 
\left( \sum_a B^i_a \, n^a \right)^2
\,, \label{bcft12a} \\
\tilde{\cal M} &=& {\textstyle \frac{1}{2}} \sum_{i}\,
 \hat{\chi}_i \, {\Gamma}^i  \left( \sum_a B^i_a \, n^a \right) \, ,
\label{bcft12}
\ea
although these could in general depend
on complex reflection coefficients and complex charges.
Notice that, in moving from critical string amplitudes to Boundary Conformal
Field Theory,
one looses the factor 2 in $\tilde{\cal M}$, that reflected the
combinatorics of string diagrams, and similar overall factors in
$\tilde{\cal K}$ and $\tilde{\cal A}$, since all these drew their 
origin from the modular integrals, absent in this case. 

\begin{figure}
\begin{center}
\epsfbox{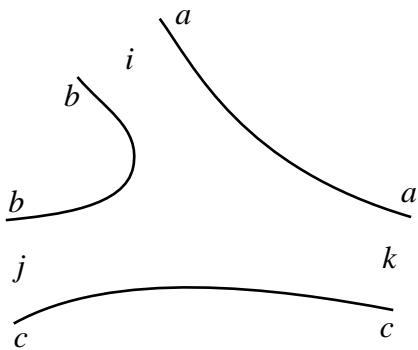}
\end{center}
\caption{Cubic vertex for open strings.}
\label{cubic}
\end{figure}

For the special case $X = {\cal C}$, where
${\cal C}$ denotes the conjugation matrix of the conformal theory,
defined by
\be
S^2 = P^2 = (S\, T)^3 = \cal C \,,
\label{bcft13}
\ee
and provided the boundaries respect the maximal symmetry of the bulk,
Cardy \cite{cardy2} uncovered an important link between 
${\cal A}$ and the fusion-rule
coefficients that, in retrospect, may be regarded as the very 
rationale for the fusion 
algebra (\ref{bcft8}). His argument can be justified by 
the string-inspired picture of the merging of a pair of open strings
at a cubic vertex, as in figure \ref{cubic}, noticing that
if $X = {\cal C}$ {\it all} closed-string sectors can reflect at the
two boundaries, so that in this case 
there are as many independent reflection coefficients, or boundary labels
$(a,b,c)$, as holomorphic bulk labels or, equivalently,
sectors of the bulk spectrum $(i,j,k)$. As a result, in this case 
the number of charge sectors is also equal to the number of bulk sectors,
while on the one hand the fusion-rule coefficients ${\cal N}_{ij}{}^k$ 
have the right structure to count string states in sector 
$i$ with a pair of boundaries labelled by $j$ and $k$, and on the other
hand they fuse together, as in eq. (\ref{bcft8}), like open strings
 should. Moreover, if one 
identifies the annulus coefficients with the ${\cal N}_{ij}{}^k$, writing
\be
{\cal A} = {\textstyle \frac{1}{2}} \sum_{i,j,k} {\cal N}_{ij}{}^k \, 
n^i\, n^j\, \chi_k \,,
\ee
the Verlinde formula guarantees that
\be
\tilde{\cal A} = {\textstyle \frac{1}{2}} \sum_{i} \, \chi_i 
\left( \sum_j \frac{S^i_j}{\sqrt{S^i_1}} n^j \right)^2
\label{bcft14}
\ee
has the proper structure (\ref{bcft12a}), while determining the $B_a^i$ 
in terms of the $S$ matrix. This structure, as first observed in \cite{bs},
is instrumental in allowing consistent M\"obius projections. 

We can actually move a bit further, supplementing ${\cal A}$ with
corresponding Klein-bottle and M\"obius amplitudes. However, as we have
seen in several examples, these choices are not unique in general, and
Cardy's Ansatz in terms of independent charge sectors applies only to 
the canonical choice
\ba
\tilde{\cal K}  & = & {\textstyle{1 \over 2}} \sum_{i} \, 
\chi_i \, {\biggl( {{P_{1i}} \over
{\sqrt{S_{1i}}}} \biggr)}^2 \, , \label{bcft15} \\
\tilde{\cal M} \ & = & \ \pm \, {\textstyle{1 \over 2}} \ \sum_{i,j} \
\hat{\chi}_i \ \biggl({{P_{1i} \ S_{ij} \ n^j}
\over {S_{1i}}} \biggr) \,, \label{bcft16}
\ea
where the crosscap one-point functions can be expressed in
terms of the first line of $P$ in a way reminiscent of Cardy's relation between
the boundary coefficients and the $S$ matrix \cite{pss,pss2} while, turning
to the direct channel,
\be
{\cal K}  =  {\textstyle{1 \over 2}} \ \sum_{i} \, 
{\cal Y}^{i}{}_{1}{}^1 \, \chi_i  \label{bcft17}
\ee
and
\be
 {\cal M }  =    \pm \, {\textstyle{1 \over 2}} \ \sum_{i,j} \,  
{\cal Y}_{j1}{}^i \ n^j \hat{\chi}_i \, , \label{bcft18} 
\ee
are both expressible in terms of the first line of the
${\cal Y}_i$ matrices. These results
admit interesting generalizations to cases where boundaries and 
crosscaps preserve only part of the bulk symmetry \cite{bertgamma},
that correspond to allowing (discrete)
deformations of the types described in the previous section in
the geometries underlying these rational constructions.

\vskip 12pt
\subsection{The ten-dimensional models revisited}

In section 3 we have already met the open descendants of the ten-dimensional
0A and 0B models. We can now revisit them, since they give us an
opportunity to exhibit the structure of Boundary Conformal 
Field Theory in a very simple setting. To this end, one must turn 
to the properly redefined basis, $\{V_8,O_8,-S_8,-C_8\}$, 
already introduced in subsection 3.1, that
accounts for the spin-statistics relation \cite{bert}. Since
these examples are actually critical string models, we can 
well retain the various overall coefficients
introduced by the modular integrals. 
One can also note that in this case the explicit expressions for 
$S$ and $P$ in 
eqs. (\ref{s2nmat}) and (\ref{po2n}) imply that for this class of
models
\be
{\cal Y}_{ij}{}^k = S_{jk} \, \delta_i{}^k \, ,
\label{bcft19}
\ee
while ${\cal C}=1_4$, and therefore all indices can be raised
and lowered at no cost.  The 0B 
descendants associated to ${\cal K}_1$ are the simplest ones in this 
respect, since their annulus 
amplitude is precisely of Cardy type, while ${\cal K}_1$ and
${\cal M}_1$
are precisely as in eqs. (\ref{bcft17}) and (\ref{bcft18}). 
The other descendants
of \cite{susy95} reviewed in section 3 
can then be recovered replacing the fixed indices, equal
to 1 in (\ref{bcft17}) and (\ref{bcft18}), with 2,3 and 4, 
although the last two choices
are connected by an overall parity transformation. The charge 
assignments in ${\cal A}$ change accordingly, in a simple and amusing
fashion, so that
\be
{\cal A}_{(l)} = {\textstyle \frac{1}{2}} \sum_{i,j,k} {\cal N}_{jk}{}^i \, 
n^j\, n^k\, \chi_{[i]\times [l]} \,,
\label{bcft20}
\ee
but this introduces a slight subtlety, since the new Chan-Paton multiplicities
form conjugate pairs, corresponding to the product of two
unitary gauge groups, as we already saw in section 3.

On the other hand, the 0A model is not of Cardy type, and
it is simple to convince oneself
that in this case there are two charges, since only two
bulk sectors, $V_8$ and $O_8$, can flow in the transverse channel.
Thus, at best one can start from \cite{bs,bs2}
\be
\tilde{\cal A} = \frac{2^{-5}}{2}\, \left( \alpha^2 O_8 + \beta^2 V_8 \right)
\,,
\label{bcft21}
\ee
that reverting to the direct channel gives
\be
{\cal A} = {\textstyle \frac{1}{4}}\, \left[ 
(\alpha^2 + \beta^2)\, (O_8 + V_8) - (\beta^2 - \alpha^2)\, (S_8
+ C_8) \right] \,,
\label{bcft22}
\ee
while now the transverse Klein bottle
\be
\tilde{\cal K} = \frac{2^5}{2} ( O_8 + V_8 )
\label{bcft23}
\ee
and the annulus of eq. (\ref{bcft21}) imply that
\be
\tilde{\cal M} = 
 \alpha \hat{O}_8 + \beta \hat V_8 \,,
\label{bcft24}
\ee
and thus, after a $P$ transformation and the usual redefinition of
the measure, that
\be
{\cal M} =  {\textstyle{1\over 2}} \left( -\alpha \,  
\hat{O}_8 + \beta \, \hat{V}_8 \right) \,.
\label{bcft25}
\ee
In section 3 we have already come to this point, and we have actually
completed the construction introducing a parametrization in
terms of two real Chan-Paton multiplicities,
\be
\alpha = n_{B} - n_{F} \,, \qquad \beta = n_{B} + n_{F} \,.
\label{bcft26}
\ee
We now want to see how this result can be retrieved following two
different, albeit equally instructive, routes.

The first derivation is based on the link between the 0B and 0A
models, similar to that between 
the IIB and IIA strings, so that one can recover the latter
as a $(-1)^{F_{\rm L}}$ orbifold of the former, where $F_{\rm L}$ denotes
the left space-time fermion number. This operation reverses all
R-R fields, inducing the interchange of branes and antibranes,
a symmetry only if they occur in equal
numbers, {\it i.e.} if $n_o\equiv n_v$
and $n_s \equiv n_c$. After an overall rescaling, that in
space-time language recovers the correct brane tension,
identifying the former with $n_B$ and the latter with $n_F$ yields the
annulus amplitude in eq. (\ref{annulus0a}) or, equivalently, 
the parametrization in (\ref{bcft26}).  
The corresponding M\"obius amplitude, however, 
cannot be obtained in this way, since now ${\cal K}$ is also
modified, but it can be directly recovered from $\tilde{\cal A}$ and
$\tilde{\cal K}$. Actually, in this particular case
\be
{\cal M}  =   \pm \, {\textstyle {1 \over 2}} \, \sum_{i,j} \, 
\; \frac{{\cal Y}_{1j}{}^i + {\cal Y}_{2j}{}^i  }{2} \; n^j \, \hat{\chi}_i\,,
\label{bcft27}
\ee
where the multiplicities are to be identified as above. This setting is
typical of off-diagonal models where no fixed points are introduced by
the identifications, an additional subtlety nicely illustrated by 
the $D_{\rm odd}$ SU(2) WZW models, that we shall discuss 
in the last subsection.

\begin{figure}
\begin{center}
\epsfbox{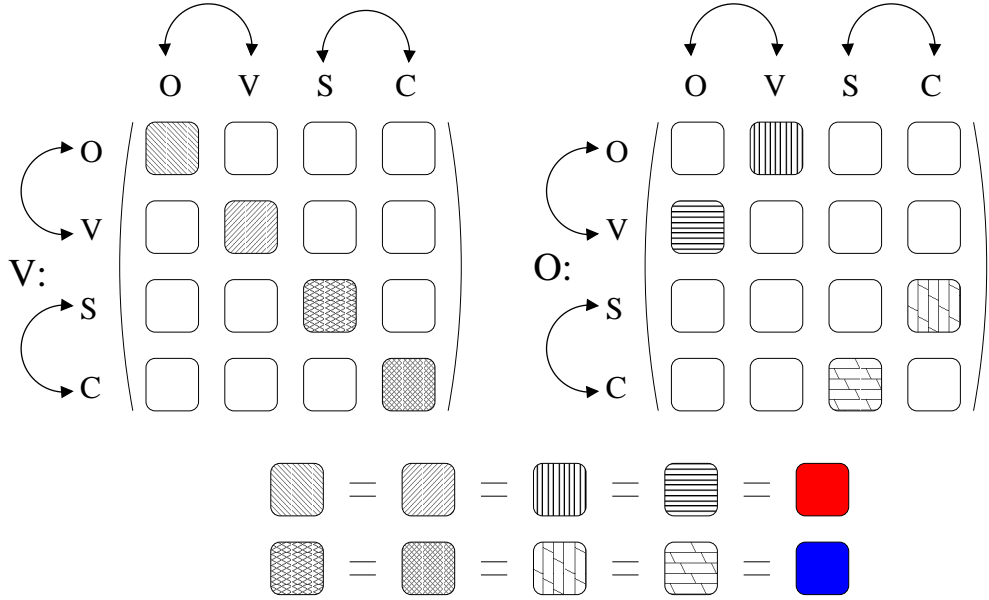}
\end{center}
\caption{Chan-Paton matrices for the 0A orientifold.}
\label{cpmatr0a}
\end{figure}

This derivation has the additional virtue of revealing the precise link
between the Chan-Paton assignments of the two models. For instance,
the matrices for the $V_8$ and $O_8$
sectors of the 0B descendant of eq. (\ref{annulusa1}) are depicted in
figure \ref{cpmatr0a}. As we have
seen, the $(-1)^{F_{\rm L}}$ orbifold brings about the conditions
$n_o \equiv n_v$ and $n_s \equiv n_c$, necessary for a consistent
0A partition function. These effectively identify pairs of
quantum numbers that were originally distinct, but the resulting
Chan-Paton matrices for the $O_8$ and $V_8$ and $S_8$
and $C_8$ sectors of the 0A model are still
{\it distinct}. It is amusing to see how  a close
scrutiny of the partition functions can exhibit this peculiarity, 
first noticed in \cite{senba}. Similar considerations apply
to other models obtained as orbifolds of Cardy-like ones.

The second derivation, on the other hand, is meant
to illustrate in the simplest possible setting a constructive algorithm,
of general applicability in rational models, that uses ${\cal M}$ to
{\it linearize} the constraints on the charge multiplicities. It results in
a set of Diophantine equations that are solved by small signed integers,
and whose solutions can thus be found by trial and error. The basic trick
is to ``turn on one charge at a time'' \cite{pss,pss2}.
{\it All} terms in ${\cal A}$
and ${\cal M}$ are then to be equal modulo 2, as can be seen specializing
any of our previous amplitudes, while ${\cal M}$, being {\it
linear} in the charge multiplicities, allows a superposition of the
independent solutions to the resulting Diophantine constraints.
Our starting point in this case is then \cite{bs,bs2}
\ba
{\cal M} &=& {\textstyle \frac{1}{2}} \, \left( - \alpha \hat O_8 + 
\beta \hat V_8\right) \,,  \nonumber  \\
\tilde{\cal M} &=& \alpha \hat O_8 + \beta \hat V_8 \,,
\label{bcft28}
\ea
where we must express $\alpha$ and $\beta$ in terms of two multiplicities
$n_B$ and $n_F$ as
\ba
\alpha &=&  a_1 n_B + a_2 n_F 
\,, \nonumber \\
\beta &=& b_1 n_B + b_2 n_F \,,
\label{bcft29}
\ea
with the $a_i$'s and $b_i$'s the signed integers to be 
determined. We now proceed
in a slightly different order with respect to the previous cases, and
{\it use} $\tilde{\cal M}$ to derive
\ba
\tilde{\cal A} &=& \frac{2^{-5}}{2} \left[
(a_1 n_B + a_2 n_F)^2 O_8 + (b_1 n_B + b_2 n_F)^2 V_8 \right] \,, \nonumber \\
{\cal A} &=& {\textstyle \frac{1}{4}} \bigl\{
\left[ (a_1 n_B + a_2 n_F)^2 + (b_1 n_B + b_2 n_F)^2\right](O_8+ V_8)
\nonumber \\
&& - \left[ - (a_1 n_B + a_2 n_F)^2 + (b_1 n_B + b_2 n_F)^2\right](S_8+ C_8)
 \bigr\} \,.
\label{bcft30}
\ea
Next we let $n_B=1$ and $n_F=0$, and demand that, within this single-charge
assignment, the contributions in ${\cal A}$ and ${\cal M}$ be equal
modulo two. This gives the conditions
\be
b_1^2 = a_1^2 \,, \qquad {\textstyle \frac{1}{2}}(a_1^2 + b_1^2) = a_1 \ 
{\rm mod} \ 2  \,, \qquad {\textstyle \frac{1}{2}}(a_1^2 + b_1^2) = b_1 \ 
{\rm mod} \ 2 \, ,
\label{bcft31}
\ee
that clearly admit the four solutions $a_1 = b_1 = \pm 1$ 
$a_1 = - b_1 = \pm 1$, two of which can be obtained from the others
by the usual overall reversal of ${\cal M}$, while the choice $n_B=0$,
$n_F=1$ leads again to these solutions. The
general solution can then be obtained
{\it superposing} an independent pair of these,
say $a_1=b_1=b_2=1$, $a_2=-1$, so that, as before
\ba
{\cal A} &=& {\textstyle\frac{1}{2}} 
(n_B^2 + n_F^2) (O_8 + V_8) - n_B n_F (S_8 + C_8) 
\,,  \nonumber  \\
{\cal M} &=& - {\textstyle{1\over 2}} (n_B - n_F) \hat O_8 + 
{\textstyle{1\over 2}} (n_B + n_F) \hat V_8 \,. 
\label{bcft32}
\ea
Notice that the restriction to the orientifold can be simply relaxed, so
that the oriented D9 brane spectrum of the ``parent'' 0A model,
\ba
\tilde{\cal A} &=& \frac{2^{-5}}{2} \left[ |n_B + n_F|^2 V_8 + 
|n_B - n_F|^2 O_8 \right] \,, \nonumber \\ 
{\cal A} &=& (n_B \bar n_B  + n_F \bar n_F) (O_8 + V_8) - 
(n_B \bar n_F  + n_F \bar n_B) (S_8 + C_8) 
\ea
can be recovered complexifying all multiplicities, as in the previous section.
In analogy with our discussion of subsection 5.12, these amplitudes 
also describe uncharged branes of the 0B string.

Although this is a non-diagonal model, the algorithm was somewhat
simplified in this example, since $\tilde{\cal A}$ and
$\tilde{\cal M}$ allow the same types of sectors. When $\tilde{\cal A}$
admits additional sectors, these behave like twisted orbifold projections, 
that split different charge sectors. We shall return to this point in the last
subsection, after a
cursory review of the SU(2) WZW models. This method is quite effective:
it led to the first derivation of the boundary-operator spectrum for the 
$D_{\rm odd}$ series
\cite{pss,pss2}, a result later generalized in \cite{fscl} and 
recovered  in \cite{bppz} by a direct construction based on 
the ADE adjacency matrices.

\vskip 12pt
\subsection{Rational models and tensor multiplets in six dimensions}

It is also instructive to reconsider some lower-dimensional string models
from the viewpoint of Rational Conformal Field Theory. For the sake
of brevity, we shall confine our attention to the simplest class of
six-dimensional  rational orbifolds, that can be obtained as  
$\bb{Z}_2$ orbifolds of the toroidal compactification on the SO(8) lattice.
Our starting point is then
\be
{\cal T} = |V_8 - S_8|^2 \, \left( |O_8|^2 + |V_8|^2 + |S_8|^2 + 
|C_8|^2 \right)
\,. 
\label{bcft33}
\ee
The internal SO(8) partition function corresponds to a lattice whose
metric is given by the $D_4$ Cartan matrix, and that includes

\be
B = \frac{\alpha'}{2} \,
\left( \begin{array}{rrrr}
0 & 1 & 0 & 0 \\    
-1 & 0 & 1 & 1 \\   
0 & -1 & 0 & 0 \\    
0 & -1 & 0 & 0 
\end{array} \right) \, ,
\label{bcft34}
\ee
a quantized $B_{ab}$
of rank $r=2$, determined by the corresponding adjacency matrix.
The results of section 4 thus imply that a toroidal compactification on this 
lattice should give
a Chan-Paton gauge group of rank 8. Indeed the Klein bottle,
annulus and M\"obius amplitudes
\ba
{\cal K} &=& {\textstyle \frac{1}{2}} \, \left(V_8 - S_8 \right)\,
\left( O_8 + V_8 + S_8 + C_8 \right) \, , \nonumber \\
{\cal A} &=& \frac{N^2}{2} \, \left(V_8 - S_8 \right)\, O_8 \, , \nonumber \\
{\cal M} &=&  \frac{N}{2} \, (\hat{V}_8 - \hat{S}_8 )\, 
\hat{O}_8  \, , \label{bcft35}
\ea
and the corresponding transverse-channel amplitudes
\ba
\tilde{\cal K} &=& \frac{2^4}{2} \, \left(V_8 - S_8 \right)\, O_8 
\, , \nonumber \\
\tilde{\cal A} &=& \frac{2^{-4} \, N^2}{2} \, \left(V_8 - S_8 \right)\,
\left( O_8 + V_8 + S_8 + C_8 \right) \, , \nonumber \\
\tilde{\cal M} &=& - N \, (\hat{V}_8 - \hat{S}_8 ) \, \hat{O}_8  
\, , \label{bcft36}
\ea
define a consistent spectrum with a ${\rm USp}(16)$ gauge group.
Although simply implied by the $P$ matrix of eq. (\ref{po2n}) 
for SO(4) characters, that interchanges $\hat O_4$
and $\hat V_4$, the USp(16) gauge
group of this model was in itself a surprise in a
toroidal compactification of the SO(32) superstring \cite{bs,bs2}. In section 4
we have already seen how a quantized $B_{ab}$ is accompanied by
symplectic and orthogonal groups at the end points of
continuous Wilson lines connecting $\gamma$ coefficients of opposite signs, 
as in subsection 4.2. In this rational setting, this peculiar effect manifests
itself as a discrete deformation of the M\"obius amplitude.  
The basic idea \cite{bs2} is that relative
phases between boundaries and crosscaps can alter the lattice
contribution to $\tilde{\cal M}$, so that while the natural choice
would be
\be
\hat{O}_8 = \hat{O}_4 \hat{O}_4 - \hat{V}_4 \hat{V}_4 \, ,
\label{bcft37}
\ee
where the relative sign between the two terms reflects
their different conformal weights, one could also start from
\be
\tilde{\cal M} = - N \, ( \hat{V}_8 - \hat{S}_8 )\, 
( \hat O_4 \hat O_4  + \hat V_4 \hat V_4  ) 
\, .
\label{bcft38}
\ee
The different choice of twist would have a very clear effect in
the direct-channel amplitude
\be
{\cal M}=- \frac{N}{2} \, (\hat{V}_8 - \hat{S}_8 )\, 
( \hat O_4 \hat O_4  + \hat V_4 \hat V_4 ) 
\, , 
\label{bcft39}
\ee
determined by the SO(4) $P$ matrix of eq. (\ref{pradisimato2n}), that
would antisymmetrize the vector, yielding an SO(16) gauge group. These
multiple choices were referred to in \cite{bs2} as ``discrete Wilson
lines'' in ${\cal M}$.

Constructing $\bb{Z}_2$ orbifolds of this model is also quite 
simple and rewarding, for they capture the most striking 
feature of six-dimensional type I
vacua, the generic presence of several tensor 
multiplets \cite{bs,bs2}. All one needs
is to combine the breaking of the space-time characters with a
proper action on the internal ones compatible with the world-sheet
supercurrent, and this is simply achieved in this case 
if the internal SO(8) is
broken to ${\rm SO} (4) \times {\rm SO} (4)$. The resulting models
contain a total of sixteen characters $\chi_i$, all listed in table
\ref{16ch}, whose $S$ and $P$ matrices 
\be{\tiny
S = \frac{1}{4}\, \left( \begin{array}{rrrrrrrrrrrrrrrr}
+ & + & + & + & + & + & + & + & 
+ & + & + & + & + & + & + & + \\
+ & + & - & - & + & + & - & - & 
+ & + & - & - & + & + & - & - \\
+ & - & + & - & + & - & + & - & 
+ & - & + & - & + & - & + & - \\
+ & - & - & + & + & - & - & + & 
+ & - & - & + & + & - & - & + \\
+ & + & + & + & + & + & + & + & 
- & - & - & - & - & - & - & -\\
+ & + & - & - & + & + & - & - & 
- & - & + & + & - & - & + & + \\
+ & - & + & - & + & - & + & - & 
- & + & - & + & - & + & - & + \\
+ & - & - & + & + & - & - & + & 
- & + & + & - & - & + & + & - \\
+ & + & + & + & - & - & - & - & 
+ & + & + & + & - & - & - & - \\
+ & + & - & - & - & - & + & + & 
+ & + & - & - & - & - & + & + \\
+ & - & + & - & - & + & - & + & 
+ & - & + & - & - & + & - & + \\
+ & - & - & + & - & + & + & - & 
+ & - & - & + & - & + & + & - \\
+ & + & + & + & - & - & - & - & 
- & - & - & - & + & + & + & + \\
+ & + & - & - & - & - & + & + & 
- & - & + & + & + & + & - & - \\
+ & - & + & - & - & + & - & + & 
- & + & - & + & + & - & + & - \\
+ & - & - & + & - & + & + & - & 
- & + & + & - & + & - & - & +  
\end{array} \right) \,,}
\label{bcft42}
\ee
\be
P = {\rm diag}(-,+,+,+,-,+,+,+,-,+,+,+,+,-,-,-) \,,
\label{bcft43}
\ee
are, up to an overall normalization for $S$, just collections
of signs. In particular, the diagonal modular invariant
\be
{\cal T} = \sum_{i=1}^{16} \; |\chi_i |^2 
\label{bcft40}
\ee
recovers the unique (2,0) supersymmetric anomaly-free massless spectrum of
\cite{alvgaum}. This comprises the gravitational multiplet and 
21 tensor multiplets: one from $|\chi_1|^2$, four from
$|\chi_5|^2$, and four from each of the twisted terms $|\chi_9|^2$,
$|\chi_{14}|^2$, $|\chi_{15}|^2$ and $|\chi_{16}|^2$.
As in the previous sections,
let us begin by discussing the simplest Klein-bottle projection
\be
{\cal K} = {\textstyle{1\over 2}} \sum_{i=1}^{16} \; \chi_i \,, 
\label{bcft41}
\ee
that determines a (1,0) massless spectrum comprising the gravitational
multiplet, sixteen hyper multiplets, twelve of which originate from the twisted
sector, and five tensor multiplets, four of which originate from the
twisted sector. This was a major surprise of the original construction
in \cite{bs,bs2}, since a 
na\"{\i}ve K3 reduction of the type I superstring would yield
only one antisymmetric two-tensor. This oddity presents itself
since the combination of orientifold and orbifold
projections brings back into the physical spectrum
remnants of the self-dual four-form of type IIB, although this 
field was projected
out in ten-dimensions. As we have seen in the previous section, the
twisted tensors signal the presence of ${\rm O}_-$ planes in the background
\cite{cab}. Notice also how the fixed-point contributions occur in groups of
four, consistently with the discussion of subsection 5.7 for the case
$r=2$. In addition, from the four corresponding (2,0) tensor multiplets
${\cal K}$ extracts one (1,0) tensor multiplet and three (1,0)
hypermultiplets, consistently with the presence of three ${\rm O}_+$ and
one ${\rm O}_-$.

\begin{table}
\caption{Characters for the ${\rm SO} (8) / \sbb{Z}_2$ orbifold. At the
massless level, $\chi_1$ contains a vector multiplet,  $\chi_5$ 
a hyper multiplet, and $\chi_9$, $\chi_{14}$, $\chi_{15}$ and
 $\chi_{16}$ one half of a hyper multiplet each. The remaining
characters contain only massive modes.}
\label{16ch}
\begin{indented}
\lineup
\item[]\begin{tabular}{@{}lll} 
\br
$\chi_1 = Q_o O_4 O_4 + Q_v V_4 V_4$ & & 
$\chi_9\;\, = Q_s S_4 O_4 + Q_c C_4 V_4$
\\
$\chi_2 = Q_o O_4 V_4 + Q_v V_4 O_4$ & & 
$\chi_{10} = Q_s S_4 V_4 + Q_c C_4 O_4$
\\
$\chi_3 = Q_o C_4 C_4 + Q_v S_4 S_4$ & & 
$\chi_{11} = Q_s V_4 C_4 + Q_c O_4 S_4$
\\
$\chi_4 = Q_o C_4 S_4 + Q_v S_4 C_4$ & &
$\chi_{12} = Q_s V_4 S_4 + Q_c O_4 C_4$
\\
$\chi_5 = Q_o V_4 V_4 + Q_v O_4 O_4$ & & 
$\chi_{13} = Q_s C_4 V_4 + Q_c S_4 O_4$
\\
$\chi_6 = Q_o V_4 O_4 + Q_v O_4 V_4$ & &
$\chi_{14} = Q_s C_4 O_4 + Q_c S_4 V_4$
\\
$\chi_7 = Q_o S_4 S_4 + Q_v C_4 C_4$ & & 
$\chi_{15} = Q_s O_4 S_4 + Q_c V_4 C_4$
\\
$\chi_8 = Q_o S_4 C_4 + Q_v C_4 S_4$ & &
$\chi_{16} = Q_s O_4 C_4 + Q_c V_4 S_4$
\\
\br
\end{tabular}
\end{indented}
\end{table}

Notice that (\ref{bcft40}) is a diagonal modular invariant, a
Cardy-like torus amplitude,
and in analogy with the previous ten-dimensional type 0 examples
one can then write
\ba
{\cal A} &=& {\textstyle{1\over 2}}\sum_{i,j,k=1}^{16}\,
{\cal N}_{ij}{}^{k} \, n^i \, n^j \, \chi_k \, ,  \nonumber \\
{\cal M} &=& {\textstyle{1\over 2}}\sum_{i,k=1}^{16}\,
{\cal Y}_{1j}{}^{k} \, n^i \, \hat\chi_k   \,. 
\label{bcft44}
\ea
The simplest solution of the resulting tadpole conditions involves four
types of real multiplicities and results in the gauge group 
${\rm USp}(8) \times {\rm USp}(8) \times {\rm USp}(8) \times {\rm USp}(8)$, 
with hyper multiplets in bi-fundamentals. This is the model where 
the generalized Green-Schwarz
mechanism was first noticed \cite{as92}, since here the reducible
anomaly polynomial does not factorize as in the ten-dimensional SO(32)
superstring \cite{gs}, but can nonetheless be reduced to a sum of 
independent contributions that induce new two-form
couplings in the low-energy model. In this case, the 
discrete Wilson lines of \cite{bs2} can turn symplectic groups 
into unitary ones, 
and one can obtain a similar model with a
${\rm U}(8) \times {\rm U}(8)$ gauge group and hyper multiplets in 
antisymmetric
and bi-fundamental representations. The two choices correspond to the
two different signs for the $\xi$ coefficient
of subsection 5.7, while the two models described are among those in tables 
\ref{tabz22} and \ref{tabz21}.

As for the ten-dimensional models, the $P$ and $T$ matrices are identical
diagonal collections
of signs, and as a result the ${\cal Y}_i$ matrices, also diagonal, 
are related to the $S$ matrix as in (\ref{bcft19}). Further,
for all allowed
Klein bottles, $\tilde{\cal K}$ contains a single character, identified by
the fixed index of ${\cal Y}$ and, as a result,
there are three additional classes of models, according to whether this
corresponds to a massive character, to $\chi_5$ or to one of 
the massless twisted ones.
In the first case, the tadpole conditions imply that the model is consistent
without an open sector, and indeed the corresponding projected
closed spectrum is the only anomaly free one, with nine tensor multiplets
and twelve hyper multiplets \cite{poltens,gepner,dp},
while na\"{\i}vely the second class of models appears to be
inconsistent. We have already come across this type of models,
since this setting is precisely what led to brane supersymmetry
breaking in six dimensions in subsection 5.8. As in that case, we cannot
proceed if we insist on working with the supersymmetric characters in
the table. Rather, in $\tilde{\cal A}$ we should separate the NS and R
contributions and, if this is done, all R-R tadpoles can be cancelled,
while the
resulting anomaly-free massless spectra are accompanied by a dilaton
tadpole. Finally, the third class of models has the peculiar feature of having
twisted tadpoles in $\tilde{\cal K}$, so that their open spectra
are bound to involve collections of fractional branes that are neutral 
with respect to the untwisted R-R charges \cite{abg}.
As an example let us consider the model associated to ${\cal
Y}_9$, whose Klein-bottle amplitude
\be
{\cal K} = {\textstyle{1\over 2}} \sum_{k=1}^{16} 
{\cal Y}^{k}{}_{9}{}^{9} \, \chi_k 
\ee
yields a projected spectrum comprising the
gravitational multiplet, thirteen tensor multiplets 
and eight hyper multiplets. This closed spectrum is
anomalous, consistently with the fact that the transverse-channel Klein
bottle amplitude
\be
\tilde{\cal K} = {2^5 \over 2} \, \chi_9
\ee
develops a non-vanishing R-R tadpole but, differently from the previous
cases, the massless tadpole now corresponds to a
twisted character. As a result, the brane configuration should
involve net numbers of fractional branes with no net  
untwisted R-R charge, and whose twisted charges should
cancel locally the contribution of $\tilde{\cal K}$. 

As for the original ``brane supersymmetry breaking'' model of
\cite{bsb}, the
construction of the open descendants must be slightly modified,
since the direct-channel open-string
amplitudes must include new sectors corresponding to brane-antibrane 
strings, that involve
different GSO projections \cite{abg}. The transverse-channel 
annulus amplitude, however,
can be easily obtained from the Cardy case, if the R portions of the
characters are fused with the NS part corresponding to the
new index, `9' in our case, of the ${\cal Y}$-tensor present in ${\cal K}$,
\be
\tilde{\cal A} = {2^{-5}\over 2} \sum_{i=1}^{16} \epsilon_i \left(
\sum_{j=1}^{16} S_{ij} n^j \right)^2 \,\left( \chi_{i}^{\rm NS} +
\chi^{\rm R}_{[i]\times [9]} \right) \,,
\ee
where the signs $\epsilon_i$, equal to $S_{i9}$ for the model we are
considering, have to be introduced in order to guarantee a consistent
interpretation for the direct-channel amplitudes, and imply that charge
multiplicities
form complex pairs. By standard methods one can then write 
\be
\tilde{\cal M} = - (n_1 + n_5 + n_9 +n_{13})\,\hat\chi_9^{\rm R} -
(n_1 - n_5 + n_9 - n_{13} )\, \hat\chi_9^{\rm NS} \,,
\ee
where we have introduced a minimal set of Chan-Paton multiplicities,
that are to be subjected to the R-R tadpole conditions
\ba
n_1 + n_5 + n_9 + n_{13} &=& 32 \,,
\nonumber
\\
n_1 + n_5 - n_9 - n_{13} &=& 0 \,,
\nonumber
\\
n_1 - n_5 + n_9 - n_{13} &=& 0 \,,
\nonumber 
\\
n_1 - n_5 - n_9 + n_{13} &=& 0 \,.
\ea
Finally, $S$ and $P$ modular transformations and a suitable relabelling of the
multiplicities, 
\be
n_1 =n\,,\quad n_9 = \bar n \,,\quad n_5 =m \,, \quad n_{13} =\bar
m\,,
\ee
give the direct-channel amplitudes
\ba
{\cal A} &=& (n\bar n + m \bar m)\chi_1 + (n\bar m + \bar n m)
\tilde\chi_1
\nonumber
\\
& & +{\textstyle{1\over 2}} (n^2 + \bar n^2 + m^2 + \bar m^2 ) \chi_9
+ (n m + \bar n \bar m ) \tilde\chi _9 \,,
\ea
and 
\be
{\cal M} = (n + \bar n) \, (\chi_{9}^{\rm NS} + \chi_9^{\rm R} ) -
(m + \bar m )\,(\chi_{9}^{\rm NS} - \chi_9^{\rm R} ) \,,
\ee
where, as in section 5, new combinations
\ba
\tilde\chi_1 &=& (V_4 V_4 - C_4 S_4 ) O_4O_4 + (O_4 O_4 - S_4 C_4) V_4
V_4 \,, 
\nonumber
\\
\tilde\chi_9 &=& (O_4 S_4 -S_4 V_4 ) S_4 O_4 + (V_4 C_4 - C_4 O_4 )
C_4 V_4 \,,
\ea
pertain to open strings stretched between branes and
antibranes. Notice the absence of the $\chi_5$ character, that contains
the internal components of the brane gauge field, and
whose presence would
signal the possibility of displacing the branes, a reflection of the fact
that the model indeed contains fractional branes. The tadpole conditions
determine a ${\rm U}(8) \times{\rm U}(8)$ gauge group, and the model is
free of gauge and gravitational anomalies.

One might conclude that,
despite the simultaneous presence of branes and antibranes of the
same type, the model be stable, since no tachyons are present while
apparently the
branes cannot be displaced. The detailed analysis of 
similar orientifolds \cite{abg}, however,
indicates that this should not be the case. In fact, although one is 
considering fractional branes, that cannot be moved away from 
the fixed points, there are still closed-string moduli related to 
the compactification torus. Tilting the $T^4$
alters the distance between brane-antibrane pairs, modifying 
the mass of the
corresponding open-string states and, as a result, tachyons can indeed 
appear for
some values of the geometric moduli. They reflect the stresses on
the background geometry, and indeed these brane configurations can
decay into magnetized (non-)BPS branes \cite{abg}.

\vskip 12pt
\subsection{Examples from WZW models}

These constructions apply directly to more complicated,
interacting rational conformal theories, and we would like to conclude
this review with some examples drawn from \cite{pss,pss2}.
These have the virtue of illustrating several new features in a
relatively simple context, but the
techniques apply with essentially no modifications to more physical, if more
involved, settings, that describe branes in genuinely curved
critical string backgrounds.
For instance, an early, interesting application to four-dimensional
Gepner models can be found in \cite{bwis}.

The characters $\chi_\alpha$ for the level-$k$ SU(2) WZW 
model \cite{gepwit}, with central charge
\be
c = \frac{3 k}{k+2}\,,
\label{bcft200}
\ee
are $k+1$, have isospins $(\alpha - 1)/2$ and conformal weights
\be
h_\alpha = \frac{\alpha^2-1}{4(k+2)} \,.
\label{bcft201}
\ee
The corresponding $S$ and $P$ matrices are
\be
S_{\alpha\beta}  =  \sqrt{2 \over {k+2}} \ \sin \left( {{\pi \alpha\beta} 
\over {k + 2}} \right)
\, ,
\label{bcft202}
\ee
and \cite{bps2,pss}
\be
P_{\alpha\beta}  =  {2 \over {\sqrt{k+2}}} \ \sin \left( {{\pi \alpha\beta} 
\over {2 (k + 2)}} \right)
( E_k E_{\alpha+\beta} + O_k O_{\alpha+\beta} )
\, ,
\label{bcft203}
\ee
where $E$ and $O$ denote even and odd projectors, while
the allowed modular invariants fall in the
ADE classification of Cappelli, Itzykson and Zuber \cite{ciz} and
are summarized in table \ref{cizmi}. 

\begin{table}
\caption{The ADE modular invariants for the SU(2) WZW models.}
\label{cizmi}
\begin{indented}
\lineup
\item[]\begin{tabular}{@{}lcl} 
\br
series & level & modular invariant \\
\mr
$A_{k+1}$ & $k$ & $\sum_{{}^{\alpha=1}_{\alpha\in\ssbb{Z}}}^{k+1} 
|\chi_\alpha|^2$
\\
$D_{2\ell+2}$ & $k=4\ell$ & 
$\sum_{{}^{\alpha=1}_{\alpha\in 2\ssbb{Z}+1}}^{2\ell-1}
|\chi_\alpha + \chi_{4\ell-\alpha}|^2+ 2 |\chi_{2\ell}|^2$
\\
$D_{2\ell+1}$ & $k=4\ell-2$ & $\sum^{4\ell-
1}_{{}^{\alpha=1}_{\alpha\in2\ssbb{Z}+1}}|\chi_\alpha
|^2 + |\chi_{2\ell-1}|^2 + \sum^{2\ell-2}_{{}^{\alpha=2}_{\alpha\in 
2\ssbb{Z}}} (
\bar\chi_{4\ell-2-\alpha} \chi_\alpha + \bar\chi_\alpha 
\chi_{4\ell-2-\alpha} )$
\\
$E_6$ & $k=10$ & $|\chi_0 + \chi_6|^2 + |\chi_3 + \chi_7|^2+ |\chi_4 +
\chi_{10}|^2 $
\\
$E_7$ & $k=16$ & $|\chi_0 + \chi_{16}|^2 + |\chi_4 + \chi_{12}|^2 +
|\chi_6 + \chi_{10}|^2 + |\chi_8|^2 $
\\
 & & $+ (\bar\chi_2 +
\bar\chi_{14})\chi_8 + \bar\chi_8 (\chi_2 + \chi_{14})$
\\
$E_8$ & $k=28$ & $|\chi_0 + \chi_{10} + \chi_{18} + \chi_{28} |^2 +
|\chi_6 + \chi_{12} + \chi_{16} + \chi_{22} |^2 $
\\
\br
\end{tabular}
\end{indented}
\end{table}
The modular invariants of the $A$ series are diagonal, 
the $D_{\rm even}$, $E_6$ and $E_8$ ones are Cardy-like, {\it i.e.} 
charge-conjugate in terms of the characters of suitably
extended algebras, and finally the $D_{\rm odd}$ and $E_7$ 
modular invariants are off-diagonal. Notice that
the $D_{\rm even}$ models present
a fixed-point ambiguity, to be resolved as in \cite{simplec} 
in order to apply the previous formalism.

For the $A$-series, one can show that there are {\it two} independent choices
for ${\cal K}$ consistent with the positivity of $\tilde{\cal K}$. 
The first corresponds to the Cardy Ansatz, and in this case
\be
{\cal K} = {\textstyle {1 \over 2}} 
\sum_{\alpha=1}^{k+1} (-1)^{\alpha-1} \chi_\alpha
\label{bcft204}
\ee
is the Frobenius-Schur indicator \cite{bert}, so that all sectors of 
integer isospin
are symmetrized, while those of half-odd-integer isospin are antisymmetrized.
The corresponding direct-channel annulus amplitudes involve $k+1$ real
charge multiplicities, and read
\be
{\cal A} =  {\textstyle {1 \over 2}} \sum_{\alpha,\beta,\gamma} \, 
{\cal N}_{\alpha\beta}{}^{\gamma} \, n^{\alpha} \, n^{\beta} \, 
\chi_{\gamma}  \, ,
\label{bcft205}
\ee
and
\be
{\cal M}  =  \pm {\textstyle {1 \over 2}}  \sum_{\alpha,\beta}\,  
(-1)^{\beta-1} \, (-1)^{{\alpha-1} \over
2} \ {\cal N}_{\beta\beta}{}^{\alpha} \, n^{\beta} \, 
\hat{\chi}_{\alpha} \,, \label{bcft206}
\ee
where for $k$ odd the overall
sign of ${\cal M}$ can actually be reversed redefining the charge 
multiplicities according to $n^\alpha \leftrightarrow  n^{k+2 - \alpha}$.
The alternative choice for the Klein-bottle projection,
\be
{\cal K}  =  {\textstyle {1 \over 2}} \sum_{\alpha=1}^{k+1} 
\, \chi_{\alpha} \, ,
\label{bcft207}
\ee
also allowed, results in the appearance of complex charges.  If $k$
is even, the model contains an odd number of characters in
$\tilde{\cal A}$, and thus an odd number of charges.  The charge
corresponding to the
middle character $\chi_{(k+2)/2}$ stays real, while those corresponding
to $\chi_\alpha$ and 
$\chi_{k+2-\alpha}$ form complex pairs.  On the other hand, if $k$
is odd all charges fall into complex pairs.  In both cases,
all signs disappear from the M\"obius
projection, and the resulting open spectra are described by
\be
{\cal A}  =  {\textstyle {1 \over 2}} \sum_{\alpha,\beta,\gamma} 
\, {\cal N}_{\alpha\beta}{}^{\gamma} \, n^{\alpha} \, n^{\beta} \,
\chi_{k+2-\gamma} 
\label{bcft208}
\ee
and
\be
{\cal M}  =  \pm {\textstyle {1 \over 2}} \sum_{\alpha,\beta} \, 
{\cal N}_{\beta\beta}{}^{\alpha} \, n^{\beta} \,
\hat{\chi}_{k+2-\alpha} \,.
\label{bcft209}
\ee
As is usually the case when complex charges are present \cite{bs},
pair-wise identifications are implicit in eqs. (\ref{bcft208}) and 
(\ref{bcft209}), and in these models $n^{k+2-\alpha} = {\bar{n}}^\alpha$. 

\begin{figure}
\begin{center}
\epsfbox{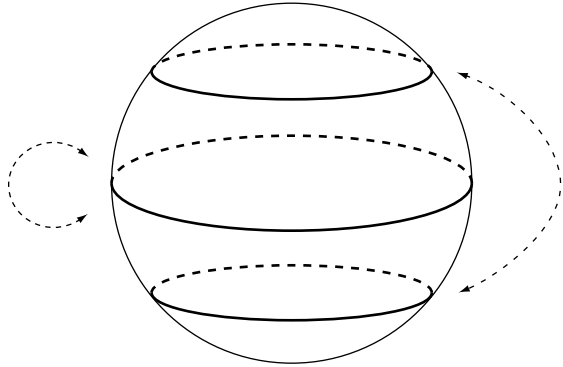}
\end{center}
\caption{D-branes in the SU(2) and SO(3) WZW models.}
\label{meridiani}
\end{figure}

As in previous examples, the
brane spectrum of the ``parent'' model can be simply recovered from
the Cardy assignment, discarding ${\cal K}$ and ${\cal M}$ and 
complexifying the charge multiplicities in ${\cal A}$. Thus, in our
case one would obtain
\be
{\cal A} =  \sum_{\alpha,\beta,\gamma} \, 
{\cal N}_{\alpha}{}^{\beta\gamma} \, n^{\alpha} \, {\bar n}_{\beta} \, 
\chi_{\gamma}  \, ,
\label{bcft210}
\ee
in terms of the multiplicities $n^\alpha$ and their 
conjugates ${\bar n}_{\alpha}$. The states of half-odd-integer isospin
occur in $\tilde{\cal A}$ with couplings of both signs, that 
reflect the $\bb{Z}_2$
symmetry of the models and mimic the R-R charges of  D-branes. The
models of the $A$-series describe strings propagating on the 
SU(2) group manifold \cite{gepwit}, the three-sphere, and
these open sectors also characterize the corresponding 
brane configurations, if all charges are taken to be complex.
These are generally D2 branes stabilized  by NS-NS fluxes on
conjugacy classes \cite{ascho}, that are depicted symbolically  in
figure \ref{meridiani} as
meridians of a two-sphere. In a similar fashion, the $D$-series
also admits a geometrical interpretation in terms of 
the propagation on the ${\rm SO}(3)={\rm SU}(2)/\bb{Z}_2$ 
group manifold, that in the figure would become symbolically 
the two-sphere with opposite points identified.  
The orientifold loci present in this and other
WZW models can be given a similar geometric interpretation \cite{orienwzw},
that also recovers rather neatly the relative signs for the various
charges present in the M\"obius amplitude in (\ref{bcft206}).
The recent literature contains several extensive analyses along these lines
of other rational conformal models and of their orientifolds \cite{mms}.  

The $D_{\rm even}$, $E_6$ and $E_8$ models are Cardy-like in terms of 
extended algebras, and as long as boundaries and crosscaps preserve them, 
the corresponding descendants can be obtained in a similar fashion, as in
\cite{pss}.
 One novelty with respect to the previous case, introduced by the extended
symmetry, is the occasional appearance in ${\cal A}$ and ${\cal M}$ of multiple
sectors with identical types of charges. Following \cite{pss,pss2}, we can
simply illustrate this phenomenon in the
$k=16$ $D_{\rm even}$ model, while retaining for brevity only the two charges
corresponding to the generalized characters $\chi_c = \chi_5 + \chi_{13}$
and $\chi_d = \chi_7 + \chi_{11}$. 
Letting $\chi_a = \chi_1 + \chi_{17}$, $\chi_b = \chi_3 + \chi_{15}$,
denoting by $\chi_e$ and $\chi_{\tilde{e}}$ the two ``resolved''
characters and choosing for definiteness an overall positive sign for the
M\"obius amplitude, one gets
\ba
{\cal A}  & = &  {{n_c^2 + n_d^2} \over 2}  \chi_a   +  
\left( {{n_c^2 + n_d^2} \over 2}  + \ n_c n_d \right) 
( \chi_b   +  \chi_e +  \chi_{\tilde{e}} )  + \nonumber \\
& & \left( {{n_c^2 + 2 n_d^2} \over 2}  +  n_c n_d \right)  \chi_c + 
\left( {{n_c^2 + 2 n_d^2} \over 2}  +  2 n_c n_d \right)  \chi_d
\label{bcft211}
\ea
and
\be
{\cal M} =   {{n_c + n_d} \over 2} ( \hat{\chi}_a - \hat{\chi}_b +
\hat{\chi}_e + \hat{\chi}_{\tilde{e}} )
 +  {{n_c + 2 n_d} \over 2}  \hat{\chi}_c  -  {{n_c} \over 2} 
\hat{\chi}_d
\,.
\label{bcft212}
\ee

This example exhibits rather neatly three types of unconventional
Chan-Paton multiplicities.  
The first presents itself in the open states
described by $\chi_c$, where factors of two occur both in the annulus and
in the M\"obius amplitude for the charges of type $d$.
There are thus {\it two families} of such states.
The others present themselves in
the open states corresponding to $\chi_d$, where the annulus amplitude
contains factors of two for both $n_d^2$ and $n_c n_d$.
Since in the M\"obius amplitude $\chi_d$ does not appear with
multiplicity $n_d$, there are {\it two
sectors} of states with a pair of charges of type $d$, described by
symmetric and antisymmetric matrices respectively. In addition, 
there are {\it two sectors} of states with
a pair of distinct charges, of types $c$ and $d$.  These multiple sets
of states reflect the occurrence in these models of multiple three point
functions, a consequence of the extended symmetry. We have already 
discussed two manifestations of the same phenomenon in subsections 
5.11 and 5.12, when we described magnetized branes 
in the presence of a quantized $B_{ab}$ and non-BPS type I branes.
If one allows the presence of boundaries that break the extended symmetry,
the analogue in this context of what we saw for magnetized orbifolds or,
more simply, for Wilson lines, the algorithm is more complicated, and
for this more general case we refer the reader to \cite{fsbreak},
where the formalism was originally developed. A more refined mathematical
framework for the whole construction, based on category theory, 
has also been recently proposed in \cite{category}, while
the link between the Abelian Chern-Simons model 
and the conformal theory of free bosons in the presence of 
boundaries and crosscaps is discussed in \cite{kogan}.

\begin{figure}
\begin{center}
\epsfbox{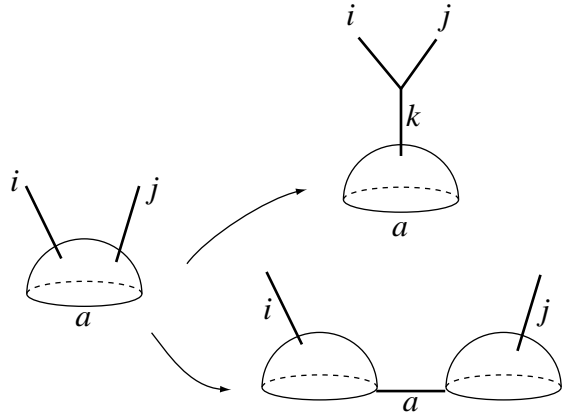}
\end{center}
\caption{Limiting behaviours of two-point functions on a disk.}
\label{twopoints}
\end{figure}

We would like to conclude with a brief discussion of the 
$D_{\rm odd}$ models, particularly interesting since their 
partition functions are
genuinely off-diagonal and contain simple currents of half-integer
spin, while their open sectors display peculiar extensions similar to
those in eq. (\ref{bcft211}). To this end, it suffices to consider the
simplest of them, the $D_5$ model with $k=6$. In this case
there are seven characters, $\chi_1, \ldots, \chi_7$, with isospins
from 0 to 3, and the partition function is
\be
{\cal T}_{D_5} =  |\chi_1|^2 +|\chi_3|^2 +|\chi_5|^2 +|\chi_7|^2 +
\chi_2 \bar{\chi}_6 + \chi_6 \bar{\chi}_2 + |\chi_4|^2 \,,
\label{tord5}
\ee
while the corresponding diagonal $A$-series model is
\be
{\cal T}_{A_6} =|\chi_1|^2 +|\chi_2|^2 +| \chi_3|^2+|\chi_4|^2 +
|\chi_5|^2 +|\chi_6|^2+|\chi_7|^2 \,. \label{tora6}
\ee
This pattern repeats for all the $D_{\rm odd}$ series,
where the
half-odd-integer isospin sectors form off-diagonal pairs
of the type $\chi_\alpha \bar{\chi}_{k+2-\alpha} +{\rm h.c.}$,
aside from the middle sector, that stays diagonal. In our
case, starting from the seven charges of the $A_6$ model, we must
end up with five charges in the $D_5$ one, as many as the sectors allowed
in $\tilde{\cal A}$ in this case. There are again two descendants,
one of which has all real charges. They were constructed with the
algorithm discussed at the beginning of this section, turning on
one charge at a time, and one type of charge was originally missed in
\cite{pss},
since this model actually presents
multiplicities in ${\cal A}$ and ${\cal M}$ similar to those in
the preceding $k=16$ example, that here can not be ascribed to bulk
extensions, and were thus excluded. 
The proper result was then obtained in  \cite{pss2}, where it was
justified in terms of
a set of polynomial equations for the rescaled boundary
one-point functions 
\be
\tilde{B}_i^a = B_i^a \, \frac{\sqrt{S_{1i}}}{S_{1a}}\, ,
\ee
a special case of the polynomial constraints that we mentioned at the
beginning of this section.
These 
reflect two limiting behaviours of the amplitude for two bulk
fields in front of a boundary of type $a$ 
in figure \ref{twopoints}, and read
\be
\tilde{B}_i^a \, \tilde{B}_j^a = \sum_k \, \epsilon_{ij}{}^k \, {\cal N}_{ij}{}^k \, \tilde{B}_k^a \, , \label{classifalg}
\ee
where $\epsilon_{ij}{}^k$ is 1 for all $(i,j,k)$ in the diagonal case. 
On the other hand, in
the off-diagonal case $\epsilon_{ij}{}^k$ is 1 when all three isospins are
integer, while if two, say $i$ and $j$, are half-integer, 
it is $(-1)^k$.
These quadratic constraints admit a number of distinct solutions, 
labelled by the index $a$, that
correspond to the allowed boundary conditions, and illustrate
the structure identified in \cite{fscl}
and termed there ``classifying algebra of boundary conditions''. They can
be simply solved for these two models yielding the 
results in table \ref{boundcoeff}, whose last line corrects a misprint
in \cite{pss2}. Notice that pairs of
$A_6$ one-point functions combine to give the $D_5$ ones, aside from 
that related to the middle field, that actually splits into two. 

One can verify that, if all the
corresponding charges are introduced, and only in this case, the
annulus coefficients satisfy \cite{pss2}
\ba
\sum_{b} {\cal A}^i{}_{a}{}^{b} \, {\cal A}^j{}_{b c}  &=&
\sum_k {\cal N}^{ij}{}_k \,  {\cal A}^k{}_{a c} \, , \label{bcft300} \\ 
\sum_{i}  {\cal A}_{i a b} \, {\cal A}^i{}_{c d}  &=& 
\sum_{i}  {\cal A}_{i a c} \, {\cal A}^i{}_{b d} \,.  \label{bcft301}
\ea
The first equation is particularly interesting for, just like the 
${\cal N}_i$'s and the ${\cal Y}_i$'s, the general annulus coefficients 
for off-diagonal amplitudes with reduced numbers of boundaries,
determined by the states allowed in $\tilde{\cal A}$, satisfy the
fusion algebra when they form a {\it complete} set. Actually,
the {\it completeness
conditions} (\ref{bcft300}) maintain the same form in more 
general settings, while the
second equation only holds in the case at hand, where each character is
associated to a different reflection coefficient. From a world-sheet
perspective, as already stressed in the discussion of the Cardy case,
eq. (\ref{bcft300}) embodies the key features of the three-point
interaction, where a pair of open strings merge into a third one
upon the disappearance of the common boundary and in fact, the
original argument in \cite{pss2}, tailored for the case of boundaries
of maximal symmetry, is a direct consequence of this picture.
The recent literature also contains
their extensions to the other amplitudes ${\cal K}$ and ${\cal M}$ 
\cite{yastesi}
\ba
\sum_b \, {\cal A}_{ia}{}^b \, {\cal M}_{jb} &=& 
\sum_k \, {\cal Y}_{ij}{}^k \, {\cal M}_{ka} 
\, , \nonumber \\
\sum_b \, {\cal M}_{ib} \, {\cal M}^{jb} &=& 
\sum_k \, {\cal Y}_{ki}{}^j \, {\cal K}^k \label{mobiuskleincompl}
\ea 
with corresponding derivations \cite{yastesi,bppz}, 
especially in view of cases where boundaries break partly
the bulk symmetry \cite{fsbreak,bppz}. One can actually show \cite{yastesi}
that the additional M\"obius-strip and Klein-bottle 
conditions (\ref{mobiuskleincompl}),
as well as similar ones for the most symmetric case,
can be deduced, under plausible assumptions, from the completeness
conditions (\ref{bcft300}). From the space-time viewpoint, these
identify complete sets of branes, and are thus the analogue, in this
context, of the modular invariance condition for bulk operators. Within
our present understanding of String Theory, however, they clearly play a
less prominent r\^ole, since branes are treated as classical
objects whose fluctuations are quantized, as is usually the case for
solitons in Field Theory. It is tempting to speculate, however, that a better
understanding of String Theory will uncover the natural similarity
between these completeness conditions for boundary operators and the 
modular invariance conditions for bulk spectra.

It is instructive to recover these assignments starting from the 
Cardy annulus for the $A_6$ model
\be
\tilde{\cal A}_6 \sim \sum_\alpha \, 
\frac{\chi_\alpha}{\sin\left( \frac{\pi \alpha}{8}\right)} \,
\left|\, \sum_\beta \, 
\sin\left ( \frac{\pi \alpha \beta}{8} \right)\, n_\beta  \, \right|^2 \ ,
\label{c5}
\ee
determined by
\be
{\cal A}_{A_6} =  \sum_{\alpha,\beta,\gamma} \, 
{{\cal N}_{\alpha}}{}^{\beta\gamma} \ n^\alpha\, \bar{n}_\beta \, 
\chi_\gamma \, , \label{c6}
\ee
and proceeding as in subsection 6.1, {\it i.e.}~ 
identifying $n_\alpha$ and $n_{8-\alpha}$ and rescaling the overall tension. 
This procedure eliminates all boundary coefficients associated to
sectors with half-odd-integer isospin, and reflects the
derivation of the $D_5$ partition function as a $\bb{Z}_2$ 
orbifold of (\ref{c6}) by $(-1)^{2 I_{\rm L}}$, where $I_{\rm L}$ denotes
the left isospin quantum number. One thus obtains
\ba
{\cal A} &=& \left[ n_1 \bar{n}_1+ n_2 \bar{n}_2 + n_3 \bar{n}_3 + 
\frac{n_4 \bar{n}_4}{2}
\right] (\chi_1 + \chi_7) \nonumber \\ 
&& + 
[ n_1 \bar{n}_2 + n_2 \bar{n}_3 + n_3 \bar{n}_4 + {\rm h.c.} ] 
(\chi_2 + \chi_6) \nonumber \\
&&+
\left[ n_2 \bar{n}_2 + 2 n_3 \bar{n}_3 + \frac{n_4 \bar{n}_4}{2} + 
(n_1 \bar{n}_3 + n_2 \bar{n}_4  + {\rm h.c.})
\right] (\chi_3 + \chi_5) \nonumber \\
&&+
[n_1 \bar{n}_4 + 2 n_2 \bar{n}_3 + n_3 \bar{n}_4 + {\rm h.c.}] \chi_4 
 \, , \label{c7}
\ea
that has apparently two problems. Firstly, there are only four charges,
while their total number is expected to be five, since five sectors are
allowed in $\tilde{\cal A}$, and moreover the terms involving
$n_4$ do not afford a proper particle interpretation, since they are
not properly normalized. This novelty originates from
the fact that the identifications have a fixed point, represented in 
this case by the $\chi_4$ sector, that indeed corresponds to a 
{\it twisted} sector of the $D_5$ model, as can be seen from 
table \ref{boundcoeff} or from the derivation of the $D_5$ torus amplitude
as an $A_6$ orbifold. The problems are indeed eliminated
if in (\ref{c7}) $n_4$ is split into a pair of charges,
\be
n_4 \to n_4 + n_5 \, ,
\ee
and a new contribution
involving $\chi_4$ and proportional to $|n_4-n_5|^2$ is added to 
$\tilde{\cal A}$.
This is allowed, since $\chi_4$ occurs diagonally in the $D_5$ model,
and plays the r\^ole of the $R_{N,D}$ breaking terms of 
section 5. The end result,
\ba
{\cal A}_{D_5} &=& \left[ n_1 \bar{n}_1+ n_2 \bar{n}_2 +  n_3 \bar{n}_3
+ n_4 \bar{n}_4+ n_5 \bar{n}_5\right] \chi_1  \nonumber \\
&&+
\left[ n_1 \bar{n}_2 + n_2 \bar{n}_3 + n_3 \bar{n}_4 + 
n_3 \bar{n}_5 + h.c. \right] (\chi_2 + \chi_6) \nonumber \\
&&+
\left[ n_2 \bar{n}_2 + 2 n_3 \bar{n}_3 +
(n_1 \bar{n}_3 + n_2 \bar{n}_4 + n_2 \bar{n}_5 + n_4 \bar{n}_5  + h.c.)
\right] \chi_3 \nonumber \\
&&+
[n_1 \bar{n}_4 + n_1 \bar{n}_5 + 2 n_2 \bar{n}_3 + n_3 \bar{n}_4 + 
n_3 \bar{n}_5 + h.c] \chi_4 \nonumber \\
&&+
\left[ n_2 \bar{n}_2 + 2 n_3 \bar{n}_3  + n_4 \bar{n}_4 + n_5 \bar{n}_5
+(n_1 \bar{n}_3 + n_2 \bar{n}_4 + n_2 \bar{n}_5  + h.c.)
\right] \chi_5 \nonumber \\
&&+ \left[ n_1 \bar{n}_1+ n_2 \bar{n}_2 + n_3 \bar{n}_3
+( n_4 \bar{n}_5 + h.c.)\right] \chi_7 
 \, , \label{c9}
\ea
coincides with the orientifold
annulus amplitude in \cite{pss2}, after the redefinitions $n_\alpha \to 
l_\alpha$, $l_1 \leftrightarrow l_2$ and $l_3 \leftrightarrow l_5$ and the
restriction to real charges.
From a geometric viewpoint, the branes in the $D_{\rm odd}$ case are
those allowed in the SO(3) manifold, {\it i.e.} pairs of SU(2) branes
with opposite latitude, aside from the equatorial one, that is fixed
and actually splits into a pair of fractional branes.
This geometric derivation of the $D_{\rm odd}$ models first appeared
in \cite{fffs}, and was later recovered in these terms in \cite{dms,matz}.
 
\begin{table}
\caption{Rescaled boundary coefficients for the $A_6$ and $D_5$ models.}
\label{boundcoeff}
{\small
\begin{tabular}{@{}crrrrrrrr}
\br
$a$ & $\tilde{B}_1^a$ & $\tilde{B}_3^a$ & $\tilde{B}_5^a$& $\tilde{B}_7^a$ &
$\tilde{B}_4^{a(A_6)}$ & $\tilde{B}_2^a$ & $\tilde{B}_6^a$ & $\tilde{B}_4^{a(D_5)}$
\\ 
\mr
$({\textstyle{1 \over 2}},{\textstyle{5 \over 2}})$ 
& $1$ & $1$ & $-1$ & $-1$ & $0$ 
& $\pm \sqrt{2}$ & $\mp \sqrt{2}$ & $0$
\\
$({\textstyle{3 \over 2}})$ & $1$ & $-1$ & $1$ & 
$-1$ & $0$ & $0$ & $0$ & $\pm 2$
\\ 
$(0,3)$ & $1$ & $1+\sqrt{2}$ & $1+\sqrt{2}$ & $1$ & 
$\pm \sqrt{2( 2+\sqrt{2})}$ & 
$\pm \sqrt{2+\sqrt{2}}$ & $\pm \sqrt{2+\sqrt{2}}$ & $0$
\\ 
$(1,2)$ & $1$ & $1-\sqrt{2}$ & $1-\sqrt{2}$ & $1$ & 
$\mp \sqrt{2(2-\sqrt{2})}$ & $\pm \sqrt{2-\sqrt{2}}$ & 
$\pm \sqrt{2 -\sqrt{2}}$ & $0$
\\
\br
\end{tabular}
}
\end{table}

\vfill\eject

\ack
The early work on open-string models described here originated during
the Ph.D. years of A.S. at Caltech, was deeply stimulated by
John H. Schwarz, and received essential contributions by Neil Marcus.
The following phase of the work, carried out
at the University of Rome ``Tor Vergata'' between the end of the 
eighties and the early nineties, rests on crucial contributions 
by Massimo Bianchi and Gianfranco Pradisi and forms the core of their
Ph.D. Dissertations. The subsequent work on Boundary
Conformal Field Theory also involved Davide Fioravanti and, later
and to a larger extent, Yassen S. Stanev.  A wide activity
on the subject followed Polchinski's paper
on D-branes, and led to many new results on lower-dimensional
models, by our group and by others. The first example of a 
four-dimensional type I vacuum with
three generations of chiral matter formed the core of the Ph.D. Thesis 
of C.A., while Fabio Riccioni, first as an undergraduate
and then as a Ph.D. student at the University of Rome ``Tor
Vergata'', contributed largely to elucidating the
nature of the generalized Green-Schwarz couplings present in these
models, as did collaborations with Sergio Ferrara and Ruben Minasian.
Our more recent work on supersymmetry breaking, done over the last
few years, emerged from stimulating interactions
with Ralph Blumenhagen, Kristin F\"orger, Matthias R. Gaberdiel,
Jihad Mourad, and especially with Ignatios Antoniadis and Emilian Dudas,
as well as with Giuseppe D'Appollonio, during his undergraduate Thesis 
at the University of Rome ``Tor Vergata''. We are most
grateful to all of them.
 
This review grew, rather slowly, out of a number of lectures given by
A.S. at CERN and at the \'Ecole Normale Sup\'erieure (1996), at the 
University of Torino and at the DESY Workshop on Conformal Field Theory
(1998), at LPT-Orsay (1999), at the University of Genova (2000), at the 
University of Firenze, at the Les Houches Summer School, at the Anttila 
Workshop on Conformal Field Theory and at the X National School on
Theoretical Physics in Parma (2001), and by C.A. at the University of 
Annecy, at
CEA-Saclay, at the \'Ecole Normale Sup\'erieure (1999) and
at the Centre Emil Borel in Paris (2000), and at the Corfu School (2001). 
In addition, it formed the core of the Andrejewski Lectures delivered 
by A.S. at the Humboldt
University in Berlin in the Fall of 1999. We are very grateful to all
the colleagues that kindly invited us to present these lectures, and to the
others that attended them and offered inspiring critical comments that
often helped us to streamline the presentation. Aside from our main
collaborators, and at the risk of
not doing justice to all, we can not but mention also C. Bachas, C. Becchi, 
A. Cappelli, E. Cremmer, P. Fr\'e, D. Friedan, 
J. Fuchs, B. Gato-Rivera, B. Julia, E. Kiritsis, C. Kounnas, 
M. Larosa, D. L\"ust, G. Parisi, P.M. Petropoulos, 
A.N. Schellekens, C. Schweigert, D. Seminara, T. Tomaras
and P. Windey, that inspired and supported our effort. We are
especially grateful to P. Bain, M. Bianchi, M. Berg, D. Clements, M. Haack,
M. Larosa, T. Maillard, M. Nicolosi, Ya.S. Stanev, and 
in particular to B. Gato-Rivera and G. Pradisi,  for a 
careful reading of the manuscript.

We would also like to acknowledge the warm hospitality of 
CERN, I.H.E.S., the \'Ecole Polytechnique, the \'Ecole Normale Sup\'erieure, 
the LPT-Orsay and the Humboldt University while this review was being 
written. C.A. is particularly grateful
to the University of Rome ``Tor Vergata'', to the \'Ecole Polytechnique and 
to the \'Ecole Normale Sup\'erieure for their warm and stimulating environments
during the last few years, and to the European Science 
Foundation, for supporting his research with a ``Marie Curie'' Fellowship.
Finally, we are grateful to the Editors and Publishers 
of ``Classical and Quantum Gravity'' for the initial invitation to write a
short review on open strings and to Prof. R. Petronzio for his kind invitation
to publish it in its present extended form in 
``Physics Reports''. This work was supported in part by
I.N.F.N., by the EC contract HPRN-CT-2000-00122, by the EC contract
HPRN-CT-2000-00148, by the INTAS contract 99-0-590 and by the MURST-COFIN
contract 2001-025492.

\secRef

\end{document}